\documentclass[notitlepage,a4paper,aps,prd,onecolumn,superscriptaddress,nofootinbib,groupedaddress]{revtex4}

\usepackage{amsmath,comment}
\usepackage{amsfonts,color}
\usepackage{amsmath}
\usepackage{amsthm}
\usepackage{pdfpages}
\usepackage{amsmath}
\usepackage{empheq}
\usepackage{amsfonts,color}
\usepackage{amssymb,float}
\usepackage{physics}
\usepackage{booktabs,multirow}
\usepackage{titlesec,multirow}

\titleformat{\paragraph}
{\normalfont\normalsize\bfseries}{\theparagraph}{1em}{}
\titlespacing*{\paragraph}
{0pt}{3.25ex plus 1ex minus .2ex}{1.5ex plus .2ex}

\usepackage{amsmath}
\usepackage[utf8]{inputenc}
\allowdisplaybreaks
\usepackage{color}
\usepackage{hyperref}
\hypersetup{
    colorlinks=true,
    linkcolor=blue,
    filecolor=magenta,      
   citecolor=blue
}

\usepackage{enumitem}

\usepackage{tikz}
\usetikzlibrary{shapes.geometric}
\usetikzlibrary{arrows.meta,arrows}

\begin{document}
\title{Stability in Cubic Metric-Affine Gravity}

\author{Sebastian Bahamonde}
\email{sbahamondebeltran@gmail.com, sebastian.bahamonde@ipmu.jp}
\affiliation{Kavli Institute for the Physics and Mathematics of the Universe (WPI), The University of Tokyo Institutes
for Advanced Study (UTIAS), The University of Tokyo, Kashiwa, Chiba 277-8583, Japan.}
\affiliation{Cosmology, Gravity, and Astroparticle Physics Group, Center for Theoretical Physics of the Universe,
Institute for Basic Science (IBS), Daejeon, 34126, Korea.}

\author{Jorge Gigante Valcarcel}
\email{jorgevalcarcel@ibs.re.kr}
\affiliation{Center for Geometry and Physics, Institute for Basic Science (IBS), Pohang 37673, Korea.}

\begin{abstract}

We analyse the stability issue of the vector and axial modes of the torsion and nonmetricity tensors around general backgrounds in the framework of cubic Metric-Affine Gravity. We show that the presence of cubic order invariants defined from the curvature, torsion and nonmetricity tensors allow the cancellation of the well-known instabilities arising in the vector and axial sectors of quadratic Metric-Affine Gravity. For the resulting theory, we also obtain Reissner-Nordström-like black hole solutions with dynamical torsion and nonmetricity, which in general include massive tensor modes for these quantities, thus avoiding further no-go theorems that potentially prevent a consistent interaction of massless higher spin fields in the quantum regime.

\end{abstract}

\maketitle

\section{Introduction}

One of the most relevant aspects to guarantee the viability of any extended theory of gravity concerns stability. \;\;\;A suitable scenario demands the absence of pathologies that may spoil the stability of the new propagating degrees of freedom of the gravitational field, which may include different types of ghostly, gradient and tachyonic instabilities, among others~\cite{Delhom:2022vae}. A noteworthy situation takes place when extending the Einstein-Hilbert action of General Relativity (GR) with quadratic curvature invariants, which renders gravity renormalisable~\cite{Stelle:1976gc}, but introduces Ostrogradsky ghosts that make the classical Hamiltonian unbounded from below~\cite{Ostrogradsky:1850fid,Buchbinder:1992rb,Woodard:2015zca}. A healthy exception is found in Lanczos-Lovelock gravity~\cite{Lanczos:1938sf,Lovelock:1971yv,Padmanabhan:2013xyr}, which in any case boils down to GR in four dimensions, up to a boundary term.

Likewise, the extension towards a post-Riemannian geometry with curvature, torsion and nonmetricity tensors leads to the formulation of Metric-Affine Gravity (MAG)~\cite{Hehl:1994ue,Blagojevic:2013xpa,Cabral:2020fax}, which in general is not free of these issues~\cite{Neville:1978bk,Sezgin:1979zf,Sezgin:1981xs,Miyamoto:1983bf,Fukui:1984gn,Fukuma:1984cz,Battiti:1985mu,Kuhfuss:1986rb,Blagojevic:1986dm,Baikov:1992uh,Yo:1999ex,Yo:2001sy,Lin:2018awc,BeltranJimenez:2019acz,Jimenez:2019qjc,Percacci:2020ddy,BeltranJimenez:2020sqf,Lin:2020phk,Marzo:2021iok,Baldazzi:2021kaf,Barker:2024dhb,Marzo:2024pyn}. In particular, the cancellation of all the possible instabilities of the spin-1 modes of the torsion and nonmetricity tensors from the quadratic action of MAG strongly restricts the parameter space of the quadratic curvature invariants from sixteen to five parameters~\cite{Jimenez-Cano:2022sds}, with the subcase of Weyl-Cartan geometry especially constrained to an Einstein-Proca theory for the Weyl vector of the nonmetricity tensor.

Following these lines, it has recently been shown that the introduction in the gravitational action of cubic order invariants defined from the curvature and torsion tensors allows the stabilisation of such modes in the Poincaré gauge framework of MAG~\cite{Bahamonde:2024sqo}. Thereby, in this work we extend these results in the presence of the nonmetricity tensor, in order to show that the vector and axial sectors of MAG can be stabilised by cubic order invariants of the curvature, torsion and nonmetricity tensors.

For this task, we organise this work as follows. In Sec.~\ref{sec:defs}, we set up our definitions and conventions for the curvature, torsion and nonmetricity tensors described in metric-affine geometry. In Sec.~\ref{sec:quadraticMAG}, we first revisit the stability issue of the vector and axial sectors of quadratic MAG. Then, in Sec.~\ref{sec:total_cubic_action}, we present the most general parity preserving cubic Lagrangian constructed from the curvature, torsion and nonmetricity tensors. A thorough stability analysis performed in Sec.~\ref{sec:stability_cubicMAG} allows us to show that the pathological terms arising in the vector and axial sectors of the quadratic action of MAG can be in fact cancelled out by an appropriate choice of the Lagrangian coefficients of the cubic action. From the resulting theory, in Sec.~\ref{sec:RN} we also find Reissner-Nordström-like black hole solutions with dynamical torsion and nonmetricity, which in contrast with the case of quadratic MAG generally include mass terms for the irreducible tensor modes of these quantities, thus avoiding further strong coupling problems and no-go theorems that would spoil their interaction in the quantum regime. Finally, we present our conclusions in Sec.~\ref{sec:conclusions}. For the sake of simplicity in the presentation, we relegate a list of cumbersome expressions and technical details to the Appendixes.

We work in natural units $c=G=1$ and we consider the metric signature $(+,-,-,-)$. In addition, we use a tilde accent to denote quantities defined from the general affine connection that includes torsion and nonmetricity, in order to distinguish them from their unaccented counterparts defined from the Levi-Civita connection. On the other hand, Latin and Greek indices run from $0$ to $3$, referring to anholonomic and coordinate bases, respectively.

\section{Definitions and conventions}\label{sec:defs}

The formulation of gravity within an affinely connected metric space-time introduces the antisymmetric part of the affine connection and the covariant derivative of the metric tensor as additional properties of the gravitational field:
\begin{equation}
    T^{\lambda}\,_{\mu \nu}=2\tilde{\Gamma}^{\lambda}\,_{[\mu \nu]}\,, \quad Q_{\lambda \mu \nu}=\tilde{\nabla}_{\lambda}g_{\mu \nu}\,,
\end{equation}
which correct the covariant derivative of an arbitrary vector $v^{\lambda}$ as
\begin{equation}
\tilde{\nabla}_{\mu}v^{\lambda}=\nabla_{\mu}v^{\lambda}+N^{\lambda}\,_{\rho\mu}v^{\rho}\,,
\end{equation}
with
\begin{equation}
    N^{\lambda}\,_{\rho\mu}=\frac{1}{2}\left(T^{\lambda}\,_{\rho \mu}-T_{\rho}\,^{\lambda}\,_{\mu}-T_{\mu}\,^{\lambda}\,_{\rho}\right)+\frac{1}{2}\left(Q^{\lambda}\,_{\rho \mu}-Q_{\rho}\,^{\lambda}\,_{\mu}-Q_{\mu}\,^{\lambda}\,_{\rho}\right)\,.
\end{equation}

Accordingly, the corresponding curvature tensor can be expressed as the sum of the Riemann tensor and further post-Riemannian corrections
\begin{equation}\label{totalcurvature}
\tilde{R}^{\lambda}\,_{\rho\mu\nu}=R^{\lambda}\,_{\rho\mu\nu}+\nabla_{\mu}N^{\lambda}\,_{\rho \nu}-\nabla_{\nu}N^{\lambda}\,_{\rho \mu}+N^{\lambda}\,_{\sigma \mu}N^{\sigma}\,_{\rho \nu}-N^{\lambda}\,_{\sigma \nu}N^{\sigma}\,_{\rho \mu}\,,
\end{equation}
whose algebraic symmetries allow the definition of three independent traces; namely, the Ricci and co-Ricci tensors
\begin{eqnarray}\label{Riccitensor}
\tilde{R}_{\mu\nu}&=&\tilde{R}^{\lambda}\,_{\mu \lambda \nu}\,,\\
\label{co-Riccitensor}
\hat{R}_{\mu\nu}&=&\tilde{R}_{\mu}\,^{\lambda}\,_{\nu\lambda}\,,
\end{eqnarray}
as well as the homothetic curvature tensor
\begin{equation}
    \tilde{R}^{\lambda}\,_{\lambda\mu\nu}=\nabla_{[\nu}Q_{\mu]\lambda}{}^{\lambda}\,.
\end{equation}
Furthermore, the trace of the Ricci and co-Ricci tensors provides a unique independent scalar curvature
\begin{equation}
    \tilde{R}=g^{\mu\nu}\tilde{R}_{\mu\nu}\,,
\end{equation}
whereas the pseudotrace of the curvature tensor gives rise to the so-called Holst pseudoscalar
\begin{equation}
\ast\tilde{R}=\varepsilon^{\lambda\rho\mu\nu}\tilde{R}_{\lambda\rho\mu\nu}\,.
\end{equation}

For stability purposes and other phenomenological aspects, it is important to consider the irreducible decomposition of torsion and nonmetricity as tensors under the four-dimensional pseudo-orthogonal group~\cite{McCrea:1992wa}:
\begin{align}
    T^{\lambda}\,_{\mu \nu}&=\frac{1}{3}\left(\delta^{\lambda}\,_{\nu}T_{\mu}-\delta^{\lambda}\,_{\mu}T_{\nu}\right)+\frac{1}{6}\,\varepsilon^{\lambda}\,_{\rho\mu\nu}S^{\rho}+t^{\lambda}\,_{\mu \nu}\,,\\
    Q_{\lambda\mu\nu}&=g_{\mu\nu}W_{\lambda}+g_{\lambda(\mu}\Lambda_{\nu)}-\frac{1}{4}g_{\mu\nu}\Lambda_{\lambda}+\frac{1}{3}\varepsilon_{\lambda\rho\sigma(\mu}\Omega_{\nu)}\,^{\rho\sigma}+q_{\lambda\mu\nu}\,,
\end{align}
which includes vector, axial and tensor modes as
\begin{align}\label{Tdec_1}
    T_{\mu}&=T^{\nu}\,_{\mu\nu}\,,\\
    W_{\mu}&=\frac{1}{4}\,Q_{\mu\nu}\,^{\nu}\,,\\
    \Lambda_{\mu}&=\frac{4}{9}\left(Q^{\nu}\,_{\mu\nu}-W_{\mu}\right)\,,\\
    S_{\mu}&=\varepsilon_{\mu\lambda\rho\nu}T^{\lambda\rho\nu}\,,\\
    t_{\lambda\mu\nu}&=T_{\lambda\mu\nu}-\frac{2}{3}g_{\lambda[\nu}T_{\mu]}-\frac{1}{6}\,\varepsilon_{\lambda\rho\mu\nu}S^{\rho}\,,\label{Tdec3}\\
    \Omega_{\lambda}\,^{\mu\nu}&=-\,\left[\varepsilon^{\mu\nu\rho\sigma}Q_{\rho\sigma\lambda}+\varepsilon^{\mu\nu\rho}\,_{\lambda}\left(\frac{3}{4}\Lambda_{\rho}-W_{\rho}\right)\right]\,,\\
    q_{\lambda\mu\nu}&=Q_{(\lambda\mu\nu)}-g_{(\mu\nu}W_{\lambda)}-\frac{3}{4}g_{(\mu\nu}\Lambda_{\lambda)}\,.\label{qtensor}
\end{align}

Thereby, in the following section we shall revisit the stability issue of the vector and axial modes of torsion and nonmetricity in quadratic MAG.

\section{Stability in quadratic MAG}\label{sec:quadraticMAG}

In order to introduce the dynamics of the torsion and nonmetricity fields, the most general parity conserving action of MAG includes 11+3+4 irreducible modes from curvature, torsion and nonmetricity, respectively. Hence, the gravitational action of quadratic MAG turns out to display a vast number of kinetics and interactions for the vector, axial and tensor modes of torsion and nonmetricity, which in general gives rise to ghostly instabilities.

Indeed, an examination on the inherent nonlinearity of the interactions involving the vector and axial modes is enough to strongly restrict the parameter space of the theory~\cite{Jimenez-Cano:2022sds}. Specifically, these pathological sectors arise from the gravitational action of MAG that is reduced to GR in the absence of torsion and nonmetricity:
\begin{align}
    S_{\rm Quad}=&\,\frac{1}{16\pi}\int
    \Bigl[
    -R-\frac{1}{2}\left(2c_{1}+c_{2}\right)\tilde{R}_{\lambda\rho\mu\nu}\tilde{R}^{\mu\nu\lambda\rho}+\left(a_{2}-c_{1}\right)\tilde{R}_{\lambda\rho\mu\nu}\tilde{R}^{\rho\lambda\mu\nu}+a_{2}\tilde{R}_{\lambda\rho\mu\nu}\tilde{R}^{\lambda\rho\mu\nu}+a_{5}\tilde{R}_{\lambda\rho\mu\nu}\tilde{R}^{\lambda\mu\rho\nu}
    \Bigr.
    \nonumber\\
    \Bigl.
    &+a_{6}\tilde{R}_{\lambda\rho\mu\nu}\tilde{R}^{\rho\mu\lambda\nu}+\left(c_{2}-a_{5}+a_{6}\right)\tilde{R}_{\lambda\rho\mu\nu}\tilde{R}^{\mu\rho\lambda\nu}+\left(d_{1}-a_{10}-a_{12}\right)\tilde{R}_{\mu\nu}\tilde{R}^{\mu\nu}+a_{9}\tilde{R}_{\mu\nu}\tilde{R}^{\nu\mu}+a_{10}\hat{R}_{\mu\nu}\hat{R}^{\mu\nu}
    \Bigr.
    \nonumber\\
    \Bigl.
    &+a_{11}\hat{R}_{\mu\nu}\hat{R}^{\nu\mu}-\left(d_{1}+a_{9}+a_{11}\right)\tilde{R}_{\mu\nu}\hat{R}^{\nu\mu}+a_{12}\tilde{R}_{\mu\nu}\hat{R}^{\mu\nu}+a_{14}\tilde{R}^{\lambda}{}_{\lambda\mu\nu}\tilde{R}^{\rho}{}_{\rho}{}^{\mu\nu}+a_{15}\tilde{R}_{\mu\nu}\tilde{R}^{\lambda}{}_{\lambda}{}^{\mu\nu}+a_{16}\hat{R}_{\mu\nu}\tilde{R}^{\lambda}{}_{\lambda}{}^{\mu\nu}
    \Bigr.
    \nonumber\\
    \Bigl.
    &+\frac{1}{2}m_{T}^{2}T_{\mu}T^{\mu}+\frac{1}{2}m_{S}^{2}S_{\mu}S^{\mu}+\frac{1}{2}m_{t}^{2}t_{\lambda\mu\nu}t^{\lambda\mu\nu}+\frac{1}{2}m_{W}^{2}W_{\mu}W^{\mu}+ \frac{1}{2}m_{\Lambda}^{2}\Lambda_{\mu}\Lambda^{\mu}+\frac{1}{2}m_{\Omega}^{2}\Omega_{\lambda\mu\nu}\Omega^{\lambda\mu\nu}
    \Bigr.
    \nonumber\\
    \Bigl.
    &+\frac{1}{2}m_{q}^{2}q_{\lambda\mu\nu}q^{\lambda\mu\nu}+\frac{1}{2}\alpha_{TW}T_{\mu}W^{\mu}+\frac{1}{2}\alpha_{T\Lambda}T_{\mu}\Lambda^{\mu}+\frac{1}{2}\alpha_{W\Lambda}W_{\mu}\Lambda^{\mu}+\frac{1}{2}\alpha_{t\Omega}\varepsilon_{\rho\mu\nu\sigma}\Omega^{\lambda\rho\mu}t^{\nu}{}_{\lambda}{}^{\sigma}\Bigr]\sqrt{-g}\,d^4x\,.\label{quadratic_action}
\end{align}
As can be seen, the action includes a total of $13$ coefficients $c_1$, $c_2$, $d_1$, $a_2$, $a_5$, $a_6$, $a_9$, $a_{10}$, $a_{11}$, $a_{12}$, $a_{14}$, $a_{15}$ and $a_{16}$ associated with the quadratic curvature invariants, as well as $11$ coefficients $m_{T}$, $m_{S}$, $m_{t}$, $m_{W}$, $m_{\Lambda}$, $m_{\Omega}$, $m_{q}$, $\alpha_{TW}$, $\alpha_{T\Lambda}$, $\alpha_{W\Lambda}$ and $\alpha_{t\Omega}$ associated with the quadratic torsion and nonmetricity invariants\footnote{Note that, in our parametrisation, only the coefficients $c_1$, $c_2$, $d_1$, $m_{T}$, $m_{S}$ and $m_{t}$ contribute to the gravitational action in the absence of nonmetricity.}.

Focusing then on the vector and axial sectors of the theory, it is a cumbersome but straightforward calculation to obtain the expression of the curvature tensor, neglecting the contribution of the tensor modes:
\begin{align}
    \tilde{R}_{\lambda\rho\mu\nu} =&\;R_{\lambda\rho\mu\nu}+\frac{1}{12}g_{\lambda[\nu}\nabla_{\mu]}\left(8T_{\rho}-12W_\rho+3\Lambda_\rho\right)-\frac{1}{12}g_{\rho[\nu}\nabla_{\mu]}\left(8T_{\lambda}-12W_\lambda+15\Lambda_\lambda\right)-g_{\lambda\rho}\nabla_{[\mu}W_{\nu]}+\frac{1}{4}g_{\lambda\rho}\nabla_{[\mu}\Lambda_{\nu]}\nonumber\\
    &+\frac{1}{6}\varepsilon_{\lambda\rho\sigma[\mu}\nabla_{\nu]}S^{\sigma}+\frac{2}{9}\left(2T_{[\lambda}g_{\rho][\nu}T_{\mu]}+g_{\lambda[\nu}g_{\mu]\rho}T_{\sigma}T^{\sigma}\right)+\frac{1}{72}\left(2S_{[\lambda}g_{\rho][\mu}S_{\nu]}+g_{\lambda[\mu}g_{\nu]\rho}S_{\sigma}S^{\sigma}\right)\nonumber\\
    &+\frac{1}{2}\left(2W_{[\lambda}g_{\rho][\nu}W_{\mu]}+g_{\lambda[\nu}g_{\mu]\rho}W_{\sigma}W^{\sigma}\right)+\frac{1}{32}\left(2\Lambda_{[\lambda}g_{\rho][\nu}\Lambda_{\mu]}+5g_{\lambda[\nu}g_{\mu]\rho}\Lambda_{\sigma}\Lambda^{\sigma}+24g_{\rho[\nu}\Lambda_{\mu]} \Lambda_\lambda\right)\nonumber\\
    &+\frac{1}{36}\left[4\varepsilon_{\mu\nu\sigma[\lambda}T_{\rho]}S^{\sigma}-6\varepsilon_{\mu\nu\sigma[\lambda}W_{\rho]}S^{\sigma}+\left(g_{\lambda[\nu}\varepsilon_{\mu]\rho\sigma\omega}-g_{\rho[\nu}\varepsilon_{\mu]\lambda\sigma\omega}\right)\left(2T^{\sigma}-3W^\sigma\right)S^{\omega}\right]+\frac{1}{48} g_{\lambda[\mu}\varepsilon_{\nu]\rho\sigma\omega}S^{\sigma}\Lambda^{\omega} \nonumber\\
    &-\frac{5}{48} g_{\rho[\mu}\varepsilon_{\nu]\lambda\sigma\omega}S^{\sigma}\Lambda^{\omega}+\frac{5}{48} \varepsilon_{\mu\nu\rho\sigma}S^{\sigma}\Lambda_{\lambda}-\frac{1}{48}\varepsilon_{\lambda\mu\nu\sigma}S^{\sigma}\Lambda_{\rho}-\frac{1}{12}g_{\rho[\mu}g_{\nu]\lambda}\left(8T_{\sigma}W^{\sigma}-6T_{\sigma}\Lambda^{\sigma}+9W_{\sigma}\Lambda^{\sigma}\right)\nonumber\\
    &-\frac{2}{3} T_{[\mu }g_{\nu] [\rho } W_{\lambda ]}  + \frac{1}{6} T_{[\mu } g_{\nu] [\rho } \Lambda_{\lambda] } -  \frac{1}{4} \Lambda_{[\mu }g_{\nu] [\rho } W_{\lambda ]}+\frac{2}{3}  W_{[\mu }g_{\nu] [\lambda } T_{\rho] } -  \frac{1}{6}\Lambda_{[\mu } g_{\nu][ \lambda } T_{\rho ]}-\frac{1}{3}g_{\rho[\mu}T_{\nu]}\Lambda_{\lambda}\nonumber\\
    &+\frac{5}{8}g_{\rho[\mu}W_{\nu]}\Lambda_{\lambda}-\frac{1}{8}g_{\lambda[\mu}W_{\nu]}\Lambda_{\rho}-\frac{1}{6}g_{\rho[\mu}\Lambda_{\nu]}\left(2T_{\lambda}-3W_{\lambda}\right)\,,\label{exp1}
\end{align}
whereas the corresponding Ricci, co-Ricci and homothetic tensors read:
\begin{align}
    \tilde{R}_{\mu\nu}=&\;R_{\mu \nu } - \frac{1}{3} \left(g_{\mu \nu } \nabla_{\lambda }T^{\lambda } + 2 \nabla_{\nu }T_{\mu }\right)+\frac{1}{12} \varepsilon_{\lambda \rho \mu \nu } \nabla^{\lambda }S^{\rho }+\frac{1}{2}\left(g_{\mu \nu } \nabla_{\lambda }W^{\lambda } -  \nabla_{\mu }W_{\nu } + 3 \nabla_{\nu }W_{\mu }\right) \nonumber\\
    &+\frac{1}{8}\left(\nabla_{\mu }\Lambda_{\nu } + \nabla_{\nu }\Lambda_{\mu }-5 g_{\mu \nu } \nabla_{\lambda }\Lambda^{\lambda }\right) + \frac{2}{9} \left(T_{\mu } T_{\nu }- g_{\mu \nu } T_{\lambda } T^{\lambda } \right)+ \frac{1}{72}\left(g_{\mu \nu } S_{\lambda } S^{\lambda } -  S_{\mu } S_{\nu }\right) \nonumber\\
    &+ \frac{1}{2}\left( W_{\mu } W_{\nu }- g_{\mu \nu } W_{\lambda } W^{\lambda }\right)+ \frac{1}{32}\left(5 g_{\mu \nu } \Lambda_{\lambda } \Lambda^{\lambda } - 11 \Lambda_{\mu } \Lambda_{\nu }\right)- \frac{2}{3} T_{(\mu } W_{\nu) } -  \frac{1}{6} T_{(\mu } \Lambda_{\nu) } \nonumber\\
    &+ \frac{1}{4}W_{(\mu}\Lambda_{\nu)}+\frac{1}{8}\varepsilon_{\mu\nu\lambda\sigma}S^{\lambda}\Lambda^{\sigma}+\frac{1}{6}g_{\mu\nu}\left(4T_{\lambda}W^{\lambda}-2T_{\lambda}\Lambda^{\lambda}+3W_{\lambda}\Lambda^{\lambda}\right)\,,\\
    \hat{R}_{\mu\nu}=&\;R_{\mu\nu}-\frac{1}{3}\left(g_{\mu\nu}\nabla_{\lambda}T^{\lambda}+2\nabla_{\nu}T_{\mu}\right)+\frac{1}{12}\varepsilon_{\lambda\rho\mu\nu}\nabla^{\lambda }S^{\rho}+\frac{1}{2}\left(g_{\mu \nu}\nabla_{\lambda}W^{\lambda}+\nabla_{\mu}W_{\nu}+\nabla_{\nu}W_{\mu}\right)\nonumber\\
    &-\frac{1}{8}\left(g_{\mu\nu}\nabla_{\lambda}\Lambda^{\lambda}+\nabla_{\mu}\Lambda_{\nu}+13\nabla_{\nu}\Lambda_{\mu}\right)+\frac{2}{9}\left(T_{\mu}T_{\nu}- g_{\mu\nu}T_{\lambda}T^{\lambda}\right)+\frac{1}{72}\left(g_{\mu\nu}S_{\lambda}S^{\lambda}-S_{\mu}S_{\nu}\right)\nonumber\\
    &+\frac{1}{2}\left(W_{\mu}W_{\nu}-g_{\mu\nu}W_{\lambda}W^{\lambda}\right)+\frac{1}{32}\left(37\Lambda_{\mu}\Lambda_{\nu}-7g_{\mu\nu}\Lambda_{\lambda}\Lambda^{\lambda}\right)-\frac{2}{3}T_{(\mu}W_{\nu)}+\frac{7}{6}T_{(\mu}\Lambda_{\nu)}\nonumber\\
    &-\frac{7}{4}W_{(\mu}\Lambda_{\nu)}-\frac{1}{8}\varepsilon_{\mu\nu\lambda\rho}S^{\lambda}\Lambda^{\rho}+\frac{1}{3}g_{\mu\nu}\left(2T_{\lambda}W^{\lambda}-2T_{\lambda}\Lambda^{\lambda}+3W_{\lambda}\Lambda^{\lambda}\right)\,,\\
    \tilde{R}^\lambda{}_{\lambda\mu\nu}=&-\,4\nabla_{[\mu}W_{\nu]}\,.\label{exp4}
\end{align}

By replacing Expressions~\eqref{exp1}-\eqref{exp4} in terms of the vector and axial modes of torsion and nonmetricity, and integrating by parts in the action, the Lagrangian density can be expressed as (up to boundary terms):
\begin{equation}
 16\pi\mathcal{L}_{\rm Quad}=-\,R+ \mathcal{L}_{1}+\mathcal{L}_{2}+\mathcal{L}_{3}+\mathcal{L}_{4}+\mathcal{L}_{5}+\mathcal{L}_{6}+\mathcal{L}_{7}+\mathcal{L}_{8}+\mathcal{L}_{9}+\mathcal{L}_{10}+\mathcal{L}_{11}+\mathcal{L}_{12}+\mathcal{L}_{13}+\mathcal{L}_{14}\,,\label{EqVectorDesQuadratic}
\end{equation}
where
\begin{align}
  \mathcal{L}_{1}=&\;p_{1}F^{(T)}_{\mu\nu}F^{(T)}{}^{\mu \nu}+\frac{1}{2}m_{T}^{2}T_{\mu}T^{\mu}\,,\\  
    \mathcal{L}_{2}=&\;p_{2}F^{(S)}_{\mu \nu } F^{(S)}{}^{\mu \nu } + \frac{1}{2} m_{S}^2 S_{\mu } S^{\mu } + p_{3}^{} R S_{\mu } S^{\mu } + p_{4}G_{\mu\nu}S^{\mu}S^{\nu}+p_{5}\nabla_{\mu}S^{\mu}\nabla_{\nu}S^{\nu}\,,\\  
      \mathcal{L}_{3}=&\;p_{6}F^{(W)}_{\mu\nu}F^{(W)}{}^{\mu\nu}+\frac{1}{2}m_{W}^{2}W_{\mu}W^{\mu}\,,\\  
        \mathcal{L}_{4}=&\;p_{7}F^{(\Lambda)}_{\mu\nu}F^{(\Lambda)}{}^{\mu\nu}+\frac{1}{2}m_{\Lambda}^2\Lambda_{\mu}\Lambda^{\mu}+p_{8}R\Lambda_{\mu}\Lambda^{\mu}+p_{9}G_{\mu\nu}\Lambda^{\mu}\Lambda^{\nu}+p_{10}\Lambda_{\mu}\Lambda^{\mu}\Lambda_{\nu}\Lambda^{\nu}+p_{11}R\nabla_{\mu}\Lambda^{\mu}+p_{12}G_{\mu}{}^{\nu}\nabla_{\nu}\Lambda^{\mu}\nonumber\\
        &+p_{13}\Lambda_{\mu}\Lambda^{\mu}\nabla_{\nu}\Lambda^{\nu}+p_{14}\nabla_{\mu}\Lambda^{\mu}\nabla_{\nu}\Lambda^{\nu}\,,\\  
          \mathcal{L}_{5}=&\;p_{15}T_{\mu}S^{\mu}T_{\nu}S^{\nu}+p_{16}T_{\mu}T^{\mu}S_{\nu}S^{\nu}+p_{17}T^{\mu}S^{\nu}\nabla_{\nu}S_{\mu}+p_{18}^{} S_{\mu } S^{\mu}\nabla_{\nu}T^{\nu}+p_{19}S^{\mu}S^{\nu}\nabla_{\mu}T_{\nu}+\frac{1}{2} p_{1}^{} \ast F^{(S)}_{\mu \nu } F^{(T)}{}^{\mu \nu }\,,\\  
            \mathcal{L}_{6}=&\;p_{20}F^{(T)}_{\mu\nu} F^{(W)}{}^{\mu\nu}+\frac{1}{2}\alpha_{TW}T_{\mu}W^{\mu}\,,\\  
              \mathcal{L}_{7}=&\;p_{21}F^{(\Lambda)}_{\mu\nu}F^{(T)}{}^{\mu\nu}+\frac{1}{2}\alpha_{T\Lambda}T_{\mu}\Lambda^{\mu}+p_{22}RT_{\mu}\Lambda^{\mu}+p_{23}G_{\mu\nu}T^{\mu}\Lambda^{\nu}+p_{24}T_{\mu}T^{\mu}T_{\nu}\Lambda^{\nu}+p_{25}T_{\mu}\Lambda^{\mu}T_{\nu}\Lambda^{\nu}\nonumber\\
        &+p_{26}T_{\mu}T^{\mu}\Lambda_{\nu}\Lambda^{\nu}+p_{27}T_{\mu}\Lambda^{\mu}\Lambda_{\nu}\Lambda^{\nu}+p_{28}T^{\mu}\Lambda^{\nu}\nabla_{\mu}T_{\nu}+p_{29}\Lambda^{\mu}\Lambda^{\nu}\nabla_{\mu}T_{\nu}+p_{30}\Lambda_{\mu}\Lambda^{\mu}\nabla_{\nu}T^{\nu}+p_{31}T^{\mu}\Lambda^{\nu}\nabla_{\nu}\Lambda_{\mu}\nonumber\\
        &+p_{32}T_{\mu}T^{\mu}\nabla_{\nu}\Lambda^{\nu}+p_{33}T^{\mu}T^{\nu}\nabla_{\mu}\Lambda_{\nu}+p_{34}\nabla_{\mu}T_{\nu}\nabla^{\mu}\Lambda^{\nu}\,,\\  
                \mathcal{L}_{8}=&\;p_{35}S_{\mu}W^{\mu}S_{\nu}W^{\nu}+p_{36}S_{\mu}S^{\mu}W_{\nu}W^{\nu}+p_{37}S^{\mu}W^{\nu}\nabla_{\mu}S_{\nu}+p_{38}S_{\mu}S^{\mu}\nabla_{\nu}W^{\nu}+p_{39}S^{\mu}S^{\nu}\nabla_{\mu}W_{\nu}\nonumber\\
                &\;+\frac{1}{4} p_{20}^{} \ast F^{(S)}_{\mu \nu }  F^{(W)}{}^{\mu \nu }\,,\\  
                  \mathcal{L}_{9}=&\;p_{40}S_{\mu}\Lambda^{\mu}S_{\nu}\Lambda^{\nu}+p_{41}S_{\mu}S^{\mu}\Lambda_{\nu}\Lambda^{\nu} + p_{42}S^{\mu}\Lambda^{\nu}\nabla_{\mu}S_{\nu}+p_{43}S_{\mu}S^{\mu}\nabla_{\nu}\Lambda^{\nu}+p_{44}S^{\mu}S^{\nu}\nabla_{\mu}\Lambda_{\nu}+p_{45}S^{\mu}\Lambda^{\nu}\ast F^{(\Lambda)}_{\mu\nu}\nonumber\\
                  &\;+\frac{1}{4} (p_{21}^{} -  \frac{1}{6} p_{74}^{})   \ast F^{(S)}_{\mu \nu }F^{(\Lambda)}{}^{\mu \nu }\,,\\  
                    \mathcal{L}_{10}=&\;p_{46}F^{(\Lambda)}_{\mu\nu}F^{(W)}{}^{\mu\nu}+\frac{1}{2} \alpha_{W\Lambda}  W_{\mu}\Lambda^{\mu}+p_{47}R W_{\mu}\Lambda^{\mu} + p_{48}W_{\mu}W^{\mu}W_{\nu}\Lambda^{\nu}+p_{49}W_{\mu}\Lambda^{\mu}W_{\nu}\Lambda^{\nu}+p_{50}G_{\mu\nu}W^{\mu}\Lambda^{\nu}\nonumber\\
        &+p_{51}W_{\mu}W^{\mu}\Lambda_{\nu}\Lambda^{\nu}+p_{52}W_{\mu}\Lambda^{\mu}\Lambda_{\nu}\Lambda^{\nu}+p_{53}W^{\mu}\Lambda^{\nu}\nabla_{\mu}W_{\nu}+p_{54}\Lambda^{\mu}\Lambda^{\nu}\nabla_{\mu}W_{\nu}+p_{55}\Lambda_{\mu}\Lambda^{\mu}\nabla_{\nu}W^{\nu}\nonumber\\
        &+p_{56}W^{\mu}W^{\nu}\nabla_{\mu}\Lambda_{\nu}+p_{57}^{} W^{\mu } \Lambda^{\nu } \nabla_{\nu }\Lambda_{\mu } + p_{58}^{} W_{\mu } W^{\mu } \nabla_{\nu }\Lambda^{\nu } + p_{59}^{} \nabla_{\nu }W_{\mu } \nabla^{\nu }\Lambda^{\mu }\,,\\  
                      \mathcal{L}_{11}=&\;p_{60}T_{\mu}S^{\mu}S_{\nu}W^{\nu}+p_{61}S_{\mu}S^{\mu}T_{\nu}W^{\nu}\,,\\  
                        \mathcal{L}_{12}=&\;p_{62}T_{\mu}S^{\mu}S_{\nu}\Lambda^{\nu}+p_{63}S_{\mu}S^{\mu}T_{\nu}\Lambda^{\nu}+p_{64}S^{\mu}\Lambda^{\nu}\ast F^{(T)}_{\mu\nu}\,,\\  
                          \mathcal{L}_{13}=&\;p_{65}T_{\mu}\Lambda^{\mu}W_{\nu}W^{\nu}+p_{66}T_{\mu}T^{\mu}W_{\nu}\Lambda^{\nu}+p_{67}T_{\mu}W^{\mu}W_{\nu}\Lambda^{\nu}+p_{68}T_{\mu}\Lambda^{\mu}W_{\nu}\Lambda^{\nu}+p_{69}T_{\mu}W^{\mu}T_{\nu}\Lambda^{\nu}\nonumber\\
        &+p_{70}T_{\mu}W^{\mu}\Lambda_{\nu}\Lambda^{\nu}+p_{71}W^{\mu}\Lambda^{\nu}\nabla_{\nu}T_{\mu}+p_{72}W_{\mu}\Lambda^{\mu}\nabla_{\nu}T^{\nu}+p_{73}T^{\mu}\Lambda^{\nu}\nabla_{\nu}W_{\mu}+p_{74}T_{\mu}\Lambda^{\mu}\nabla_{\nu}W^{\nu}\,,\\  
                            \mathcal{L}_{14}=&\;p_{75}S_{\mu }S^{\mu}W_{\nu}\Lambda^{\nu}+ p_{76}S_{\mu}W^{\mu}S_{\nu}\Lambda^{\nu}+p_{77}S^{\mu}\Lambda^{\nu}\ast F^{(W)}_{\mu\nu}\,.
\end{align}
For the sake of simplicity in the presentation, we have introduced the coefficients $\{p_{i}\}_{i=1}^{77}$ (see a list with their expressions in~\ref{appendix:p}) and defined
\begin{equation}
    F^{(X)}_{\mu\nu}=2\partial_{[\mu}X_{\nu]}\,, \quad \ast F^{(X)}_{\mu\nu}=\frac{1}{2}\varepsilon_{\mu\nu}{}^{\lambda\rho}F^{(X)}_{\lambda\rho}\,,\label{FS_compact_notation}
\end{equation}
with $X=S, T, W, \Lambda$.

Let us now analyse the contribution of the different terms to the stability issue. First, we notice that, except for the appropriate choice of signs and relations between the respective coefficients, the Lagrangian densities $ \mathcal{L}_{1},\mathcal{L}_{3},\mathcal{L}_{6},\mathcal{L}_{11},\mathcal{L}_{12}$ and $\mathcal{L}_{14}$ do not contain any dangerous term per se in the propagation of the vector and axial modes. On the other hand, the contributions of the form $\ast F^{(S)}_{\mu \nu }  F^{(T)}{}^{\mu \nu }$, $\ast F^{(S)}_{\mu \nu } F^{(W)}{}^{\mu \nu }$, $\ast F^{(S)}_{\mu \nu }  F^{(\Lambda)}{}^{\mu \nu }$ and $G_{\mu}{}^{\nu}\nabla_{\nu}\Lambda^{\mu}$ simply act as boundary terms in the action and therefore they can be neglected from the stability analysis. Conversely, the term $R\nabla_{\mu}\Lambda^{\mu}$ contributes to the field equations with higher order time derivatives of the longitudinal component of $\Lambda^{\mu}$, thus representing a ghostly interaction around general backgrounds. Further Ostrogradsky instabilities arise from terms of the form $\left(\nabla X\right)^{2}$, $\nabla X \nabla Y$, $X^2 \nabla Y$, $XY\nabla X$, $XY\nabla Z$, $RX^{2}$ and $RXY$, with $X \neq Y$. Thereby, all of these pathological terms can be eliminated from the gravitational action by setting~\cite{Jimenez-Cano:2022sds}:
\begin{align}
    p_3&=p_5=p_8=p_{11}=p_{14}=p_{17}=p_{18}=p_{19}=p_{22}=p_{28}=p_{29}=p_{30}=p_{31}=p_{32}=p_{33}=p_{34}=p_{38}=0\,,\\
p_{39}&=p_{42}=p_{43}=p_{44}=p_{47}=p_{53}=p_{54}=p_{55}=p_{56}=p_{57}=p_{58}=p_{59}=p_{71}=p_{72}=p_{73}=p_{74}=0\,,
\end{align}
or, equivalently:
\begin{equation}
    c_1=c_2=0\,,\quad a_{6}=2a_5\,,\quad a_{9}=a_{11}=-\,\frac{1}{2}\left(a_{2}+a_{5}+2a_{10}\right),\quad a_{12}=d_{1}-2 a_{10}^{}\,.\label{cond1Quad}
\end{equation}

Concerning the kinetic terms of the vector and axial modes in the Lagrangian density, they can be described as
\begin{equation}
    \mathcal{L}_{FF}=-\,\frac{1}{4}\kappa_{XY}F^{(X)}_{\mu\nu}F^{(Y)}{}^{\mu\nu}\,,\quad F^{(X)}=(F^{(S)},F^{(T)},F^{(W)},F^{(\Lambda)})\,,\label{kinetic}
\end{equation}
where the kinetic matrix becomes
\begin{eqnarray}\label{kappaQ}
    \kappa_{XY}=-\,2\left(
\begin{array}{cccc}
  2 p_2 & 0  & 0 &  0 \\
0&2 p_1  &p_{20} & p_{21} \\
  0& p_{20} & 2 p_6 & p_{46} \\
0 &  p_{21} & p_{46} & 2 p_7 \\
\end{array}
\right)\,.
\end{eqnarray}
Then, if we replace the conditions~\eqref{cond1Quad} in the kinetic matrix, it acquires the following form:
\begin{align}
    \kappa_{XY}=\left(
\begin{array}{cccc}
\frac{d_1}{18} & 0 & 0& 0 \\
 0 &  -\,\frac{8}{9}  d_1 & \frac{4}{3}(d_1+a_{15}+a_{16})  & -\,d_1 \\
0 &  \frac{4}{3}(d_1+a_{15}+a_{16}) & -\,2 (5 a_{2}+3 a_{5}+4 a_{10}+8 a_{14}+4 a_{15}) & \frac{3}{2}  (2 d_1+2 a_{16}-a_{2}-3 a_{5}-4 a_{10}) \\
 0 & -\,d_1 & \frac{3}{2}  (2 d_1+2 a_{16}-a_{2}-3 a_{5}-4 a_{10}) & -\,\frac{3}{8}  (7 a_{2}+a_{5}+12 a_{10}) \\
\end{array}
\right)\,.\label{kineticQuad}
\end{align}

In order to avoid ghostly modes with negative kinetic energies, first of all, a constraint $d_1\geq 0$ must be imposed to ensure a healthy kinetic term in the axial sector. As for the kinetics of the vector sector, the decoupling between the axial mode and the rest of vectors simplifies the analysis to the following $3 \times 3$ kinetic matrix:
\begin{align}
    \kappa_{IJ}^{\rm (vec)}=\left(
\begin{array}{ccc}
  -\,\frac{8}{9}  d_1 & \frac{4}{3}(d_1+a_{15}+a_{16})  & -\,d_1 \\
 \frac{4}{3}(d_1+a_{15}+a_{16}) & -\,2 (5 a_{2}+3 a_{5}+4 a_{10}+8 a_{14}+4 a_{15}) & \frac{3}{2}  (2 d_1+2 a_{16}-a_{2}-3 a_{5}-4 a_{10}) \\
  -\,d_1 & \frac{3}{2}  (2 d_1+2 a_{16}-a_{2}-3 a_{5}-4 a_{10}) & -\,\frac{3}{8}  (7 a_{2}+a_{5}+12 a_{10}) \\
\end{array}
\right)\,,\label{kineticQuad3D}
\end{align}
where $I,J=T, W, \Lambda$. The polynomial characteristic of this matrix then reads
\begin{eqnarray}
   p(\lambda)=-\lambda^3+s_2\lambda^2+s_1\lambda+s_0\,, \label{lambda}
\end{eqnarray}
with
\begin{align}
    s_0=&\;\frac{1}{3} \big\{16 \left(2 a_2+3 a_{14}-a_5\right) d_1^2-8 \big[8 a_2^2+2 \left(a_5+8 a_{10}+7 a_{14}+a_{15}\right) a_2-6 a_5^2+24 a_{10} a_{14}+2 a_5 \left(a_{14}-4 a_{10}-2 a_{15}\right)\nonumber\\
    &+\left(4 a_5+3 a_{15}-2 a_2\right) a_{16}\big] d_1+2 \left(7 a_2+a_5+12 a_{10}\right) \left(a_{15}+a_{16}\right)^{2}\big\}\,,\\
    s_1=&\;\frac{1}{9} \big\{106 d_1^2-a_2 \left[27 \left(2 a_5+16 a_{10}+14 a_{14}+7 a_{15}+3 a_{16}\right)+182 d_1\right]-2 \left(212 a_{10}+64 a_{14}+16 a_{15}-97 a_{16}\right) d_1\nonumber\\
    &+3 a_5 \left[72 a_{10}-9 \left(2 a_{14}+a_{15}+9 a_{16}\right)-98 d_1\right]+162 a_5^2+16 a_{15}^2+97 a_{16}^2-216 a_2^2-648 a_{10} a_{14}-324 a_{10} a_{15}\nonumber\\
    &-324 a_{10} a_{16}+32 a_{15} a_{16}\big\}\,,\\
    s_2=&-\frac{8}{9} \left(18 a_{14}+9 a_{15}+d_1\right)-\frac{101 a_2}{8}-\frac{25 a_{10}}{2}-\frac{51 a_5}{8}\,.
    \end{align}
Thereby, the requirement of nonnegative real roots for the characteristic polynomial~\eqref{lambda} is guaranteed if
  \begin{eqnarray}
    s_2\geq 0\,,\quad s_1\leq 0\,,\quad s_0\geq 0\,,
\end{eqnarray}
which in fact, due to the previous constraint $d_1 \geq 0$ arising from the axial sector, directly imposes the following restrictions on the Lagrangian coefficients:
\begin{eqnarray}
    d_1=0\,,\quad a_{16}=-\,a_{15}\,,
\end{eqnarray}
thus strongly suppressing the propagation of torsion in the theory.
On the other hand, the remaining inequalities expressed in terms of the Lagrangian coefficients read:
\begin{eqnarray}
    &&101 a_2+51 a_5+100 a_{10}+128 a_{14}+64 a_{15}\leq 0\,,\label{condMagaa}\\
     &&2\left[4 a_2^2-3a_5^2+a_{2}\left( a_5+2a_{15}\right)+a_{14}\left(7 a_2+a_5\right)+4 a_{10} \left(2 a_2+3 a_{14}-a_5\right)-4 a_5 a_{15}\right]\geq 3 a_{15}^2\,.\label{condMagbb}
\end{eqnarray}

In conclusion, it is clear that the vector and axial sectors of MAG described by the quadratic action~\eqref{quadratic_action} are generally affected by different types of pathologies. Specifically, from the $13$ Lagrangian coefficients associated with the quadratic curvature invariants, we first need to impose the six conditions~\eqref{cond1Quad} to eliminate Ostrogradsky instabilities, while the requirement of ghost-free kinetic terms involves two extra conditions $d_1=a_{15}+a_{16}=0$, as well as the inequalities~\eqref{condMagaa} and~\eqref{condMagbb}. Explicitly, the final form of the Lagrangian density with stable vector and axial sectors reads
\begin{align}
    \mathcal{L}_{\rm Quad}=&\;-R+2\tilde{R}_{(\lambda\rho)\mu\nu}\bigl(a_{2}\tilde{R}^{(\lambda\rho)\mu\nu}+2a_{5}\tilde{R}^{(\lambda\mu)\rho\nu}\bigr)+a_{14}\tilde{R}^{\lambda}{}_{\lambda\mu\nu}\tilde{R}^{\rho}{}_{\rho}{}^{\mu\nu}+a_{15}\bigl(\tilde{R}_{\mu\nu}-\hat{R}_{\mu\nu}\bigr)\tilde{R}^{\lambda}{}_{\lambda}{}^{\mu\nu}\nonumber\\
    &+a_{10}\bigl(\tilde{R}_{\mu\nu}-\hat{R}_{\mu\nu}\bigr)\bigl(\tilde{R}^{\mu\nu}-\hat{R}^{\mu\nu}\bigr)-\frac{1}{2}\left(2a_{10}+a_{2}+a_{5}\right)\bigl(\tilde{R}_{\mu\nu}-\hat{R}_{\mu\nu}\bigr)\bigl(\tilde{R}^{\nu\mu}-\hat{R}^{\nu\mu}\bigr)\nonumber\\
    &+\frac{1}{2}m_{T}^{2}T_{\mu}T^{\mu}+\frac{1}{2}m_{S}^{2}S_{\mu}S^{\mu}+\frac{1}{2}m_{t}^{2}t_{\lambda\mu\nu}t^{\lambda\mu\nu}+\frac{1}{2}m_{W}^{2}W_{\mu}W^{\mu}+ \frac{1}{2}m_{\Lambda}^{2}\Lambda_{\mu}\Lambda^{\mu}+\frac{1}{2}m_{\Omega}^{2}\Omega_{\lambda\mu\nu}\Omega^{\lambda\mu\nu}
    \Bigr.
    \nonumber\\
    \Bigl.
    &+\frac{1}{2}m_{q}^{2}q_{\lambda\mu\nu}q^{\lambda\mu\nu}+\frac{1}{2}\alpha_{TW}T_{\mu}W^{\mu}+\frac{1}{2}\alpha_{T\Lambda}T_{\mu}\Lambda^{\mu}+\frac{1}{2}\alpha_{W\Lambda}W_{\mu}\Lambda^{\mu}+\frac{1}{2}\alpha_{t\Omega}\varepsilon_{\rho\mu\nu\sigma}\Omega^{\lambda\rho\mu}t^{\nu}{}_{\lambda}{}^{\sigma}\,.\label{stablevectors_quadraticMAG}
\end{align}
Thereby, the number of coefficients associated with the quadratic curvature invariants is reduced to $5$, which especially constrains the dynamics of torsion in the theory, as reported in~\cite{Jimenez-Cano:2022sds}. In this sense, a healthy behaviour for the vector modes of nonmetricity was also remarked in~\cite{Bahamonde:2022kwg}, with a particular quadratic MAG model where the Lagrangian coefficients of Expression~\eqref{stablevectors_quadraticMAG} are restricted as follows\footnote{Note we denote the parameter $a_2$ appearing in~\cite{Bahamonde:2022kwg} for the mass term of the axial mode of torsion as $\tilde{a}_2$.}:
\begin{align}
    d_1&=a_5=a_{15}=0\,,\quad a_2=f_1\,,\quad a_{10}=-\,\frac{f_1}{4}\,,\quad a_{14}=e_1-\frac{1}{2}f_1\,,\\
    m_T&=m_t=m_W=m_\Lambda=m_\Omega=m_q=\alpha_{TW}=\alpha_{T\Lambda}=\alpha_{W\Lambda}=\alpha_{t\Omega}=0\,,\quad  m_S^2=\frac{1}{4}(2\tilde{a}_2-1)\,,
\end{align}
being $e_1\leq 0$ and  $f_1\leq 0$, in agreement with the stability constraints~\eqref{condMagaa} and~\eqref{condMagbb}.

\section{Cubic Lagrangian}\label{sec:total_cubic_action}

In this section, we shall present the general form of the cubic order of the Lagrangian of MAG. In general, the Lagrangian defined from cubic invariants of the curvature, torsion and nonmetricity tensors can include four different contributions: three different types mixing curvature and torsion, curvature and nonmetricity, as well as all of these tensors; whereas the last type is solely given by invariants of the curvature tensor and its contractions.

\subsection{Cubic invariants from curvature and torsion}

The first contribution comes from mixing terms which depend linearly on the curvature tensor and quadratically on the torsion tensor. By considering the algebraic symmetries of the curvature tensor, it is possible to define up to $46$ independent invariants, which can be split into six different types, according to the possible combinations of the irreducible parts of torsion at quadratic order\footnote{A correspondence with the cubic Lagrangian expressed in terms of the unsplit torsion tensor can be found in~\ref{appendix1}.}:
\begin{equation}
    \mathcal{L}_{\rm curv-tors}^{(3)}=\mathcal{L}^{(3)}_{\tilde{R}TT}+\mathcal{L}^{(3)}_{\tilde{R}SS}+\mathcal{L}^{(3)}_{\tilde{R}tt}+\mathcal{L}^{(3)}_{\tilde{R}TS}+\mathcal{L}^{(3)}_{\tilde{R}Tt}+\mathcal{L}^{(3)}_{\tilde{R}St}\,,\label{cubicLagIrr}
\end{equation}
where
\begin{eqnarray}
    \mathcal{L}^{(3)}_{\tilde{R}TT}&=&(h_{1}-h_{27})\tilde{R}_{\mu\nu}T^{\mu}T^{\nu}+h_{2}\tilde{R}T_{\mu}T^{\mu}+h_{27}^{} \hat{R}^{\alpha \rho } T_{\alpha } T_{\rho }\,,\\
    \mathcal{L}^{(3)}_{\tilde{R}SS}&=&(h_{3}-h_{28})\tilde{R}_{\mu\nu}S^{\mu}S^{\nu}+h_{4}\tilde{R}S_{\mu}S^{\mu}+h_{28}^{} \hat{R}^{\alpha \rho } S_{\alpha } S_{\rho }\,,\\
    \mathcal{L}^{(3)}_{\tilde{R}tt}&=&\frac{1}{2}\left(2h_5+h_{30}\right)\tilde{R}_{\lambda\rho\mu\nu}t_{\sigma}{}^{\lambda\rho}t^{\sigma\mu\nu}+h_6\tilde{R}_{\lambda\rho\mu\nu}t_{\sigma}{}^{\lambda\mu}t^{\sigma\rho\nu}+h_7\tilde{R}_{\lambda\rho\mu\nu}t^{\lambda\rho}{}_{\sigma}t^{\sigma\mu\nu}+(h_{8}{}+h_{29} )\tilde{R}_{\lambda\rho\mu\nu}t^{\lambda\mu}{}_{\sigma}t^{\sigma\rho\nu}\nonumber\\
&&+\,(h_{9}{}+h_{29})\tilde{R}_{\lambda\rho\mu\nu}t^{\lambda\mu}{}_{\sigma}t^{\rho\nu\sigma}+(h_{10}{}+h_{31}-h_{32})\tilde{R}_{\lambda\rho}t_{\mu\nu}{}^{\lambda}t^{\rho\mu\nu}+(h_{11}{}-h_{32})\tilde{R}_{\lambda\rho}t_{\mu\nu}{}^{\lambda}t^{\mu\nu\rho}+h_{12}{}\tilde{R}t_{\lambda\rho\mu}t^{\lambda\rho\mu}\nonumber\\
&&+\,h_{29}^{} \tilde{R}^{\alpha }{}_{\rho \tau \gamma } t^{\rho \tau }{}_{\mu } t^{\gamma }{}_{\alpha }{}^{\mu } + h_{30}^{} \tilde{R}^{\alpha }{}_{\rho \tau \gamma } t^{\tau \gamma }{}_{\mu } t^{\mu }{}_{\alpha }{}^{\rho } + h_{31}^{} \hat{R}^{\alpha }{}_{\tau } t^{\tau }{}_{\gamma \mu } t^{\gamma }{}_{\alpha }{}^{\mu } + h_{32}^{} \hat{R}^{\alpha }{}_{\tau } t^{\gamma }{}_{\alpha }{}^{\mu } t_{\mu }{}^{\tau }{}_{\gamma }\,,\\
    \mathcal{L}^{(3)}_{\tilde{R}TS}&=&\frac{1}{2}\left(2h_{13}{}-h_{33}\right)\varepsilon^{\lambda\rho\mu\nu}\tilde{R}_{\lambda\rho\mu\nu}T_{\sigma}S^{\sigma}+(h_{14}{}+h_{33}) \varepsilon_{\nu}{}^{\lambda\rho\sigma}\tilde{R}_{\lambda\rho\mu\sigma}T^{\mu}S^{\nu}+(h_{15}+h_{34}){}\varepsilon^{\lambda\rho\mu\nu}\tilde{R}_{\lambda\rho}T_{\mu}S_{\nu}\nonumber\\
    &&+\,h_{33}^{} \varepsilon^{\alpha \tau \gamma \mu } \tilde{R}^{\rho }{}_{\tau \gamma \mu }  T_{\rho }S_{\alpha }+ h_{34}^{} \varepsilon^{\alpha \rho }{}_{\tau }{}^{\mu } \hat{R}^{\tau }{}_{\mu }  T_{\rho }S_{\alpha }+ h_{35}^{} \varepsilon^{\alpha \rho \gamma \mu } \tilde{R}^{\omega}{}_{\omega\gamma \mu } T_{\rho } S_{\alpha }\,,\\
    \mathcal{L}^{(3)}_{\tilde{R}Tt}&=&(h_{16}{}-h_{36})\tilde{R}_{\lambda\rho\mu\nu}T^{\nu}t^{\lambda\rho\mu}+\frac{1}{2}\left(2h_{17}{}+h_{37} \right)\tilde{R}_{\lambda\rho\mu\nu}T^{\rho}t^{\lambda\mu\nu}+(h_{18}{}-h_{38}+h_{39})\tilde{R}_{\lambda\rho}T_{\mu}t^{\mu\lambda\rho}+(h_{19}{} +h_{39})\tilde{R}_{\lambda\rho}T_{\mu}t^{\lambda\rho\mu} \nonumber\\
    &&+\,h_{36}^{} \tilde{R}^{\rho }{}_{\tau }{}^{\alpha }{}_{\gamma } t^{\tau }{}_{\rho }{}^{\gamma } T_{\alpha }+ h_{37}^{} \tilde{R}^{\alpha }{}_{\rho \tau \gamma } t^{\tau \rho \gamma } T_{\alpha }+h_{38}^{} \hat{R}^{\rho }{}_{\gamma } t^{\alpha }{}_{\rho }{}^{\gamma } T_{\alpha }  + h_{39}^{} \hat{R}^{\rho }{}_{\gamma } t^{\gamma \alpha }{}_{\rho } T_{\alpha }+ h_{40}^{} \tilde{R}^{\omega}{}_{\omega\tau \gamma } t^{\tau \alpha \gamma } T_{\alpha } \,,\\
    \mathcal{L}^{(3)}_{\tilde{R}St}&=&\frac{1}{2}\left(2h_{20}{}+h_{41}\right)\varepsilon_{\alpha\rho\mu\nu}\tilde{R}_{\tau}{}^{\rho\mu\nu}S^{\gamma}t^{\alpha\tau}{}_{\gamma}+h_{21}{}\varepsilon_{\alpha\rho\mu\nu}\tilde{R}_{\tau}{}^{\rho\mu\nu}S^{\gamma}t_{\gamma}{}^{\alpha\tau}+(h_{22}{}+h_{43})\varepsilon_{\alpha\rho}{}^{\mu\nu}\tilde{R}^{\rho}{}_{\mu\tau\nu}S^{\gamma}t_{\gamma}{}^{\alpha\tau}\nonumber\\
    &&+\,(h_{23}{}-h_{42}+2h_{43})\varepsilon_{\alpha\rho}{}^{\mu\nu}\tilde{R}_{\gamma\mu\tau\nu}S^{\alpha}t^{\gamma\rho\tau}+(h_{24}{}+h_{41}-h_{42})\varepsilon_{\alpha\rho}{}^{\mu\nu}\tilde{R}_{\gamma\mu\tau\nu} S^{\alpha}t^{\rho\tau\gamma}+h_{41}^{} \varepsilon^{\alpha \gamma \mu \nu } \tilde{R}^{\rho }{}_{\tau \gamma \mu } S_{\alpha } t^{\tau }{}_{\rho \nu } \nonumber\\
    &&+\,(h_{25}{}+h_{44}+h_{45})\varepsilon_{\alpha\rho\tau\mu}\tilde{R}^{\mu}{}_{\gamma}S^{\alpha}t^{\rho\tau\gamma}+\frac{1}{2}\left(2h_{26}{}-h_{44}\right)\varepsilon_{\lambda\rho\mu\nu} \tilde{R}^{\lambda\rho}S_{\sigma}t^{\sigma\mu\nu}+h_{42}^{} \varepsilon^{\alpha }{}_{\rho }{}^{\mu \nu } \tilde{R}^{\rho }{}_{\tau \gamma \mu } S_{\alpha } t^{\gamma \tau }{}_{\nu }\nonumber\\
    &&+\, h_{43}^{} \varepsilon^{\alpha }{}_{\rho }{}^{\tau \nu } \tilde{R}^{\rho }{}_{\tau \gamma \mu } S_{\alpha } t^{\gamma \mu }{}_{\nu } + h_{44}^{} \varepsilon^{\alpha \gamma \mu \nu } \hat{R}^{\rho }{}_{\gamma } S_{\alpha } t_{\mu \rho \nu } + h_{45}^{} \varepsilon^{\alpha }{}_{\rho }{}^{\mu \nu } \hat{R}^{\rho }{}_{\gamma } S_{\alpha } t_{\mu }{}^{\gamma }{}_{\nu }+h_{46}^{} \varepsilon^{\alpha \gamma \mu \nu } \tilde{R}^{\omega}{}_{\omega\tau \gamma } S_{\alpha } t_{\mu }{}^{\tau }{}_{\nu}\,.
\end{eqnarray}
Note that, for convenience, we have introduced the Lagrangian coefficients of this branch to match the notation used in~\cite{Bahamonde:2024sqo}, in the absence of nonmetricity.

Thereby, there is a large number of cubic invariants in this branch, while for different geometrical restrictions this number is reduced. In particular, in the case of Weyl-Cartan geometry, there exist $29$ independent invariants, whereas in a Riemann-Cartan space-time the number is reduced to $26$~\cite{Bahamonde:2024sqo}.

\subsection{Cubic invariants from curvature and nonmetricity}

Similarly, the second contribution accounts for the mixing terms that are linear in the curvature tensor and quadratic in the nonmetricity tensor. In this branch, it is then possible to define $61$ independent invariants, which can be split into ten different types\footnote{See~\ref{appendix2} for a correspondence with the cubic Lagrangian expressed in terms of the unsplit nonmetricity tensor.}:
\begin{equation}
    \mathcal{L}_{\rm curv-nonm}^{(3)}=\mathcal{L}^{(3)}_{\tilde{R}WW}+\mathcal{L}^{(3)}_{\tilde{R}\Lambda\Lambda}+\mathcal{L}^{(3)}_{\tilde{R}\Omega\Omega}+\mathcal{L}^{(3)}_{\tilde{R}qq}+\mathcal{L}^{(3)}_{\tilde{R}W\Lambda}+\mathcal{L}^{(3)}_{\tilde{R}W\Omega}+\mathcal{L}^{(3)}_{\tilde{R}Wq}+\mathcal{L}^{(3)}_{\tilde{R}\Lambda \Omega}+\mathcal{L}^{(3)}_{\tilde{R}\Lambda q}+\mathcal{L}^{(3)}_{\tilde{R}\Omega q}\,,\label{cubicLagIrrQ}
\end{equation}
where
\begin{eqnarray}
    \mathcal{L}^{(3)}_{\tilde{R}WW}&=& h_{47}^{} \tilde{R}_{\alpha \rho } W^{\alpha } W^{\rho } + h_{48}^{} \hat{R}_{\alpha \rho } W^{\alpha } W^{\rho }+h_{49}^{} \tilde{R} W_{\alpha } W^{\alpha }\,,\\
     \mathcal{L}^{(3)}_{\tilde{R}\Lambda\Lambda}&=& h_{50}^{} \tilde{R}_{\alpha \rho } \Lambda^{\alpha } \Lambda^{\rho } + h_{51}^{} \hat{R}_{\alpha \rho } \Lambda^{\alpha } \Lambda^{\rho }+h_{52}^{} \tilde{R} \Lambda_{\alpha } \Lambda^{\alpha } \,,\\
      \mathcal{L}^{(3)}_{\tilde{R}\Omega\Omega}&=&h_{53}^{} \Omega_{\alpha }{}^{\rho }{}_{\tau } \Omega_{\gamma \rho \mu } \tilde{R}^{\alpha \tau \gamma \mu } + h_{54}^{} \Omega_{\alpha }{}^{\rho }{}_{\tau } \Omega_{\gamma \rho \mu } \tilde{R}^{\alpha \gamma \tau \mu } + h_{55}^{} \Omega_{\alpha }{}^{\rho }{}_{\tau } \Omega_{\gamma \rho \mu } \tilde{R}^{\alpha \mu \tau \gamma } + h_{56}^{} \Omega_{\alpha }{}^{\rho }{}_{\tau } \Omega_{\gamma \rho \mu } \tilde{R}^{\tau \alpha \gamma \mu } \nonumber\\
      &&+ \,h_{57}^{} \Omega_{\alpha }{}^{\rho }{}_{\tau } \Omega_{\gamma \rho \mu } \tilde{R}^{\tau \gamma \alpha \mu } + h_{58}^{} \Omega_{\alpha }{}^{\rho }{}_{\tau } \Omega_{\gamma \rho \mu } \tilde{R}^{\tau \mu \alpha \gamma }  + h_{59}^{} \Omega_{\tau }{}^{\alpha }{}_{\gamma } \Omega^{\tau \rho \gamma } \tilde{R}_{\alpha \rho }+ h_{60}^{} \Omega^{\alpha }{}_{\tau \gamma } \Omega^{\rho \tau \gamma } \tilde{R}_{\alpha \rho }\nonumber\\
      &&+\,h_{61}^{} \Omega_{\tau }{}^{\alpha }{}_{\gamma } \Omega^{\tau \rho \gamma } \hat{R}_{\alpha \rho }+ h_{62}^{} \Omega^{\alpha }{}_{\tau \gamma } \Omega^{\rho \tau \gamma } \hat{R}_{\alpha \rho }  + h_{63}^{} \Omega_{\alpha \rho \tau } \Omega^{\alpha \rho \tau } \tilde{R} \,,\\
       \mathcal{L}^{(3)}_{\tilde{R}qq}&=&h_{64}^{} q_{\alpha }{}^{\tau \mu } q^{\rho \gamma }{}_{\mu } \tilde{R}^{\alpha }{}_{\rho \tau \gamma } + h_{65}^{} q^{\alpha }{}_{\tau \gamma } q^{\rho \tau \gamma } \tilde{R}_{\alpha \rho } + h_{66}^{} q^{\alpha }{}_{\tau \gamma } q^{\rho \tau \gamma } \hat{R}_{\alpha \rho } + h_{67}^{} q_{\alpha \rho \tau } q^{\alpha \rho \tau } \tilde{R}\,,\\
       \mathcal{L}^{(3)}_{\tilde{R}W\Lambda}&=& h_{68}^{} \tilde{R}_{\alpha \rho } W^{\rho } \Lambda^{\alpha } + h_{69}^{} \tilde{R}_{\alpha \rho } W^{\alpha } \Lambda^{\rho } + h_{70}^{} \hat{R}_{\alpha \rho } W^{\alpha } \Lambda^{\rho }  + h_{71}^{} \hat{R}_{\alpha \rho } W^{\rho } \Lambda^{\alpha }+ h_{72}^{} \tilde{R}^{\omega}{}_{\omega\alpha \rho } W^{\alpha } \Lambda^{\rho }+h_{73}^{} \tilde{R} W^{\alpha } \Lambda_{\alpha } \,,\\
       \mathcal{L}^{(3)}_{\tilde{R}W\Omega}&=& h_{74}^{} \varepsilon^{\tau \gamma \mu \nu } \Omega_{\rho \tau \gamma } \tilde{R}^{\rho }{}_{\mu \alpha \nu } W^{\alpha } +h_{75}^{} \varepsilon^{\tau \gamma \mu \nu } \Omega_{\rho \tau \gamma } \tilde{R}_{\alpha \mu }{}^{\rho }{}_{\nu } W^{\alpha } + h_{76}^{} \varepsilon^{\tau \gamma }{}_{\mu }{}^{\nu } \Omega_{\rho \tau \gamma } \tilde{R}^{\mu }{}_{\alpha }{}^{\rho }{}_{\nu } W^{\alpha } + h_{77}^{} \varepsilon^{\tau \gamma }{}_{\mu }{}^{\nu } \Omega_{\rho \tau \gamma } \tilde{R}^{\mu \rho }{}_{\alpha \nu } W^{\alpha } \nonumber\\
       &&+\, h_{78}^{} \varepsilon^{\rho }{}_{\tau }{}^{\gamma \mu } \Omega^{\alpha }{}_{\gamma \mu } \tilde{R}_{\alpha \rho } W^{\tau } + h_{79}^{} \varepsilon^{\alpha }{}_{\tau }{}^{\gamma \mu } \Omega^{\rho }{}_{\gamma \mu } \tilde{R}_{\alpha \rho } W^{\tau } + h_{80}^{} \varepsilon^{\rho }{}_{\tau }{}^{\gamma \mu } \Omega^{\alpha }{}_{\gamma \mu } \hat{R}_{\alpha \rho } W^{\tau } + h_{81}^{} \varepsilon^{\alpha }{}_{\tau }{}^{\gamma \mu } \Omega^{\rho }{}_{\gamma \mu } \hat{R}_{\alpha \rho } W^{\tau }\nonumber\\
       &&+\, h_{82}^{} \varepsilon_{\alpha }{}^{\tau \gamma \mu } \Omega_{\rho \tau \gamma }\tilde{R}^{\omega}{}_{\omega}{}^{\rho }{}_{\mu } W^{\alpha } \,,\\
       \mathcal{L}^{(3)}_{\tilde{R}Wq}&=&h_{83}^{} q_{\rho }{}^{\tau \gamma } \tilde{R}^{\rho }{}_{\tau \alpha \gamma } W^{\alpha } + h_{84}^{} q^{\alpha \rho }{}_{\tau } \tilde{R}_{\alpha \rho } W^{\tau } + h_{85}^{} q^{\alpha \rho }{}_{\tau } \hat{R}_{\alpha \rho } W^{\tau }\,,\\
       \mathcal{L}^{(3)}_{\tilde{R}\Lambda \Omega}&=&h_{86}^{} \varepsilon^{\tau \gamma \mu \nu } \Omega_{\rho \tau \gamma } \tilde{R}_{\alpha \mu }{}^{\rho }{}_{\nu } \Lambda^{\alpha } + h_{87}^{} \varepsilon^{\tau \gamma \mu \nu } \Omega_{\rho \tau \gamma } \tilde{R}^{\rho }{}_{\mu \alpha \nu } \Lambda^{\alpha } + h_{88}^{} \varepsilon^{\tau \gamma }{}_{\mu }{}^{\nu } \Omega_{\rho \tau \gamma } \tilde{R}^{\mu }{}_{\alpha }{}^{\rho }{}_{\nu } \Lambda^{\alpha }+ h_{89}^{} \varepsilon^{\tau \gamma }{}_{\mu }{}^{\nu } \Omega_{\rho \tau \gamma } \tilde{R}^{\mu \rho }{}_{\alpha \nu } \Lambda^{\alpha }\nonumber\\
       && + \,h_{90}^{} \varepsilon^{\rho }{}_{\tau }{}^{\gamma \mu } \Omega^{\alpha }{}_{\gamma \mu } \tilde{R}_{\alpha \rho } \Lambda^{\tau } + h_{91}^{} \varepsilon^{\alpha }{}_{\tau }{}^{\gamma \mu } \Omega^{\rho }{}_{\gamma \mu } \tilde{R}_{\alpha \rho } \Lambda^{\tau } + h_{92}^{} \varepsilon^{\rho }{}_{\tau }{}^{\gamma \mu } \Omega^{\alpha }{}_{\gamma \mu } \hat{R}_{\alpha \rho } \Lambda^{\tau }+ h_{93}^{} \varepsilon^{\alpha }{}_{\tau }{}^{\gamma \mu } \Omega^{\rho }{}_{\gamma \mu } \hat{R}_{\alpha \rho } \Lambda^{\tau } \nonumber\\
       &&+\, h_{94}^{} \varepsilon_{\alpha }{}^{\tau \gamma \mu } \Omega_{\rho \tau \gamma }\tilde{R}^{\omega}{}_{\omega}{}^{\rho }{}_{\mu } \Lambda^{\alpha } \,,\\
       \mathcal{L}^{(3)}_{\tilde{R}\Lambda q}&=&h_{95}^{} q_{\rho }{}^{\tau \gamma } \tilde{R}^{\rho }{}_{\tau \alpha \gamma } \Lambda^{\alpha } + h_{96}^{} q^{\alpha \rho }{}_{\tau } \tilde{R}_{\alpha \rho } \Lambda^{\tau } + h_{97}^{} q^{\alpha \rho }{}_{\tau } \hat{R}_{\alpha \rho } \Lambda^{\tau }\,,\\
       \mathcal{L}^{(3)}_{\tilde{R}\Omega q}&=&h_{98}^{} \varepsilon^{\rho \tau }{}_{\mu \sigma } \Omega_{\alpha \rho \tau } q_{\gamma }{}^{\nu \sigma } \tilde{R}^{\gamma \alpha \mu }{}_{\nu } + h_{99}^{} \varepsilon^{\rho \tau }{}_{\gamma \sigma } \Omega_{\alpha \rho \tau } q^{\mu \nu \sigma } \tilde{R}^{\gamma }{}_{\mu }{}^{\alpha }{}_{\nu } + h_{100}^{} \varepsilon^{\rho \tau }{}_{\nu \sigma } \Omega_{\alpha \rho \tau } q_{\gamma }{}^{\mu \sigma } \tilde{R}^{\gamma }{}_{\mu }{}^{\alpha \nu } \nonumber\\
       && +\, h_{101}^{} \varepsilon^{\rho \tau }{}_{\gamma \nu } \Omega_{\alpha \rho \tau } q^{\alpha \mu \sigma } \tilde{R}^{\gamma }{}_{\mu }{}^{\nu }{}_{\sigma } + h_{102}^{} \varepsilon^{\rho \tau }{}_{\mu \nu } \Omega_{\alpha \rho \tau } q^{\alpha }{}_{\gamma }{}^{\sigma } \tilde{R}^{\gamma \mu \nu }{}_{\sigma }+ h_{103}^{} \varepsilon^{\rho \tau \mu }{}_{\nu } \Omega_{\alpha \rho \tau } q^{\alpha \gamma \nu } \tilde{R}_{\mu \gamma } \nonumber\\
       &&+\, h_{104}^{} \varepsilon^{\rho \tau \gamma }{}_{\nu } \Omega_{\alpha \rho \tau } q^{\alpha \mu \nu } \tilde{R}_{\mu \gamma } + h_{105}^{} \varepsilon^{\rho \tau \gamma }{}_{\nu } \Omega_{\alpha \rho \tau } q^{\alpha \mu \nu } \hat{R}_{\gamma \mu } +h_{106}^{} \varepsilon^{\rho \tau \mu }{}_{\nu } \Omega_{\alpha \rho \tau } q^{\alpha \gamma \nu } \hat{R}_{\gamma \mu }  \nonumber\\
       &&+\, h_{107}^{} \varepsilon^{\rho \tau }{}_{\mu \nu } \Omega_{\alpha \rho \tau } q^{\alpha \gamma \nu }\tilde{R}^{\omega}{}_{\omega}{}^{\mu }{}_{\gamma }\,.
\end{eqnarray}

Therefore, in comparison with the previous branch, the richer algebraic structure of the nonmetricity tensor gives rise to a higher number of cubic invariants. In any case, it is worthwhile to note that in the realm of Weyl-Cartan geometry this number is significantly reduced from $61$ to $2$, whereas it identically vanishes in Riemann-Cartan geometry.

\subsection{Cubic invariants from curvature, torsion and nonmetricity}

The third branch mixes all of the curvature, torsion and nonmetricity tensors, which in general provides a very large number of cubic invariants, in comparison with the two previous branches. Specifically, in this branch it is possible to define $102$ independent invariants, which are included into $12$ different types as follows\footnote{See~\ref{appendix3} for a correspondence with the cubic Lagrangian expressed in terms of the unsplit torsion and nonmetricity tensors.}:
\begin{align}
    \mathcal{L}_{\rm curv-tor-nonm}^{(3)}=&\;\mathcal{L}^{(3)}_{\tilde{R}TW}+\mathcal{L}^{(3)}_{\tilde{R}T\Lambda}+\mathcal{L}^{(3)}_{\tilde{R}T\Omega}+\mathcal{L}^{(3)}_{\tilde{R}Tq}+\mathcal{L}^{(3)}_{\tilde{R}SW}+\mathcal{L}^{(3)}_{\tilde{R}S\Lambda}+\mathcal{L}^{(3)}_{\tilde{R}S\Omega}+\mathcal{L}^{(3)}_{\tilde{R}Sq}+\mathcal{L}^{(3)}_{\tilde{R}tW}+\mathcal{L}^{(3)}_{\tilde{R}t\Lambda}\nonumber\\
    &+\mathcal{L}^{(3)}_{\tilde{R}t\Omega}+\mathcal{L}^{(3)}_{\tilde{R}tq}\,,\label{cubicLagIrrQT}
\end{align}
where
\begin{eqnarray}
\mathcal{L}^{(3)}_{\tilde{R}TW}&=&h_{108}\tilde{R}_{\alpha\rho}T^{\rho}W^{\alpha}+h_{109}\tilde{R}_{\alpha\rho}T^{\alpha}W^{\rho}+h_{110} \hat{R}_{\alpha\rho}T^{\rho}W^{\alpha}+h_{111}\hat{R}_{\alpha\rho}T^{\alpha}W^{\rho}+h_{112}\tilde{R}^{\omega}{}_{\omega}{}^{\alpha}{}_{\rho}T_{\alpha}W^{\rho}+h_{113}\tilde{R}T_{\alpha}W^{\alpha},\\
\mathcal{L}^{(3)}_{\tilde{R}T\Lambda}&=& h_{114}^{} \tilde{R}_{\alpha \rho } T^{\rho } \Lambda^{\alpha }+ h_{115}^{} \tilde{R}_{\alpha \rho } T^{\alpha } \Lambda^{\rho }  + h_{116}^{} \hat{R}_{\alpha \rho } T^{\rho } \Lambda^{\alpha }  + h_{117}^{} \hat{R}_{\alpha \rho } T^{\alpha } \Lambda^{\rho }+ h_{118}^{}\tilde{R}^{\omega}{}_{\omega}{}^{\alpha }{}_{\rho } T_{\alpha } \Lambda^{\rho }+h_{119}^{} \tilde{R} T_{\alpha } \Lambda^{\alpha }\,,\\
\mathcal{L}^{(3)}_{\tilde{R}T\Omega}&=&h_{120}^{} \varepsilon^{\tau \gamma \mu \nu } \Omega_{\rho \tau \gamma } \tilde{R}^{\alpha }{}_{\mu }{}^{\rho }{}_{\nu } T_{\alpha } + h_{121}^{} \varepsilon^{\tau \gamma \mu \nu } \Omega_{\rho \tau \gamma } \tilde{R}^{\rho }{}_{\mu }{}^{\alpha }{}_{\nu } T_{\alpha } + h_{122}^{} \varepsilon^{\tau \gamma }{}_{\mu }{}^{\nu } \Omega_{\rho \tau \gamma } \tilde{R}^{\mu \alpha \rho }{}_{\nu } T_{\alpha } + h_{123}^{} \varepsilon^{\tau \gamma }{}_{\mu }{}^{\nu } \Omega_{\rho \tau \gamma } \tilde{R}^{\mu \rho \alpha }{}_{\nu } T_{\alpha }\nonumber\\
&&+ \,h_{124}^{} \varepsilon^{\rho \tau \gamma \mu } \Omega^{\alpha }{}_{\gamma \mu } \tilde{R}_{\alpha \rho } T_{\tau } + h_{125}^{} \varepsilon^{\alpha \tau \gamma \mu } \Omega^{\rho }{}_{\gamma \mu } \tilde{R}_{\alpha \rho } T_{\tau } + h_{126}^{} \varepsilon^{\alpha \tau \gamma \mu } \Omega^{\rho }{}_{\gamma \mu } \hat{R}_{\alpha \rho } T_{\tau } + h_{127}^{} \varepsilon^{\alpha \rho \gamma \mu } \Omega^{\tau }{}_{\gamma \mu } \hat{R}_{\alpha \rho } T_{\tau }\nonumber\\
&&+\,h_{128}^{} \varepsilon^{\alpha \tau \gamma \mu } \Omega_{\rho \tau \gamma }\tilde{R}^{\omega}{}_{\omega}{}^{\rho }{}_{\mu } T_{\alpha }\,,\\
\mathcal{L}^{(3)}_{\tilde{R}Tq}&=&h_{129}^{} q_{\rho }{}^{\tau \gamma } \tilde{R}^{\rho }{}_{\tau }{}^{\alpha }{}_{\gamma } T_{\alpha } + h_{130}^{} q^{\alpha \rho \tau } \tilde{R}_{\alpha \rho } T_{\tau } + h_{131}^{} q_{\alpha }{}^{\rho \tau } \hat{R}^{\alpha }{}_{\rho } T_{\tau }\,,\\
\mathcal{L}^{(3)}_{\tilde{R}SW}&=&h_{132}^{} \varepsilon^{\alpha \tau \gamma \mu } \tilde{R}_{\rho \tau \gamma \mu } S_{\alpha } W^{\rho } + h_{133}^{} \varepsilon^{\alpha }{}_{\tau }{}^{\gamma \mu } \tilde{R}^{\tau }{}_{\rho \gamma \mu } S_{\alpha } W^{\rho } + h_{134}^{} \varepsilon^{\alpha }{}_{\tau }{}^{\gamma \mu } \tilde{R}^{\tau }{}_{\gamma \rho \mu } S_{\alpha } W^{\rho } + h_{135}^{} \varepsilon^{\alpha \rho \tau }{}_{\gamma } \tilde{R}_{\alpha \rho } S_{\tau } W^{\gamma } \nonumber\\
&&+\, h_{136}^{} \varepsilon^{\alpha \rho \tau }{}_{\gamma } \hat{R}_{\alpha \rho } S_{\tau } W^{\gamma } + h_{137}^{} \varepsilon^{\alpha }{}_{\rho }{}^{\tau \gamma } \tilde{R}^{\omega}{}_{\omega\tau \gamma } S_{\alpha } W^{\rho }\,,\\
\mathcal{L}^{(3)}_{\tilde{R}S\Lambda}&=&h_{138}^{} \varepsilon^{\alpha \tau \gamma \mu } \tilde{R}_{\rho \tau \gamma \mu } S_{\alpha } \Lambda^{\rho } + h_{139}^{} \varepsilon^{\alpha }{}_{\tau }{}^{\gamma \mu } \tilde{R}^{\tau }{}_{\rho \gamma \mu } S_{\alpha } \Lambda^{\rho } + h_{140}^{} \varepsilon^{\alpha }{}_{\tau }{}^{\gamma \mu } \tilde{R}^{\tau }{}_{\gamma \rho \mu } S_{\alpha } \Lambda^{\rho }  + h_{141}^{} \varepsilon^{\alpha \rho \tau }{}_{\gamma } \tilde{R}_{\alpha \rho } S_{\tau } \Lambda^{\gamma }\nonumber\\
&&+\,h_{142}^{} \varepsilon^{\alpha \rho \tau }{}_{\gamma } \hat{R}_{\alpha \rho } S_{\tau } \Lambda^{\gamma }+ h_{143}^{} \varepsilon^{\alpha }{}_{\rho }{}^{\tau \gamma } \tilde{R}^{\omega}{}_{\omega\tau \gamma } S_{\alpha } \Lambda^{\rho }\,,\\
\mathcal{L}^{(3)}_{\tilde{R}S\Omega}&=&h_{144}^{} \Omega_{\rho \tau \gamma } \tilde{R}^{\alpha \rho \tau \gamma } S_{\alpha } + h_{145}^{} \Omega_{\rho \tau \gamma } \tilde{R}^{\alpha \tau \rho \gamma } S_{\alpha } + h_{146}^{} \Omega_{\rho \tau \gamma } \tilde{R}^{\rho \alpha \tau \gamma } S_{\alpha } + h_{147}^{} \Omega_{\rho \tau \gamma } \tilde{R}^{\rho \tau \alpha \gamma } S_{\alpha } + h_{148}^{} \Omega_{\rho \tau \gamma } \tilde{R}^{\tau \alpha \rho \gamma } S_{\alpha } \nonumber\\
&&+\, h_{149}^{} \Omega_{\rho \tau \gamma } \tilde{R}^{\tau \rho \alpha \gamma } S_{\alpha } + h_{150}^{} \Omega_{\rho \tau \gamma } \tilde{R}^{\tau \gamma \alpha \rho } S_{\alpha }+ h_{151}^{} \Omega^{\alpha \rho \tau } \tilde{R}_{\alpha \rho } S_{\tau } + h_{152}^{} \Omega^{\rho \alpha \tau } \tilde{R}_{\alpha \rho } S_{\tau } + h_{153}^{} \Omega^{\alpha \rho \tau } \hat{R}_{\alpha \rho } S_{\tau } \nonumber\\
&&+ \,h_{154}^{} \Omega^{\rho \alpha \tau } \hat{R}_{\alpha \rho } S_{\tau } + h_{155}^{} \Omega_{\rho }{}^{\alpha }{}_{\tau }\tilde{R}^{\omega}{}_{\omega}{}^{\rho \tau } S_{\alpha } \,,\\
\mathcal{L}^{(3)}_{\tilde{R}Sq}&=&h_{156}^{} \varepsilon^{\alpha \gamma \mu \nu } q_{\rho }{}^{\tau }{}_{\nu } \tilde{R}^{\rho }{}_{\tau \gamma \mu } S_{\alpha } + h_{157}^{} \varepsilon^{\alpha \tau \mu \nu } q_{\rho }{}^{\gamma }{}_{\nu } \tilde{R}^{\rho }{}_{\tau \gamma \mu } S_{\alpha } + h_{158}^{} \varepsilon^{\alpha }{}_{\rho }{}^{\mu \nu } q^{\tau \gamma }{}_{\nu } \tilde{R}^{\rho }{}_{\tau \gamma \mu } S_{\alpha }\,,\\
\mathcal{L}^{(3)}_{\tilde{R}tW}&=&h_{159}^{} \tilde{R}^{\rho }{}_{\tau \alpha \gamma } t_{\rho }{}^{\tau \gamma } W^{\alpha }  + h_{160}^{} \tilde{R}^{\rho }{}_{\alpha \tau \gamma } t^{\tau }{}_{\rho }{}^{\gamma } W^{\alpha } + h_{161}^{} \tilde{R}^{\rho }{}_{\tau \alpha \gamma } t^{\tau }{}_{\rho }{}^{\gamma } W^{\alpha } + h_{162}^{} \tilde{R}_{\alpha \rho \tau \gamma } t^{\tau \rho \gamma } W^{\alpha } + h_{163}^{} \tilde{R}_{\alpha \rho } t^{\alpha \rho }{}_{\tau } W^{\tau } \nonumber\\
&& +\, h_{164}^{} \tilde{R}_{\alpha \rho } t_{\tau }{}^{\alpha \rho } W^{\tau }+ h_{165}^{} \hat{R}_{\alpha \rho } t^{\alpha \rho }{}_{\tau } W^{\tau } + h_{166}^{} \hat{R}_{\alpha \rho } t_{\tau }{}^{\alpha \rho } W^{\tau }+ h_{167}^{} \tilde{R}^{\omega}{}_{\omega\rho \tau } t^{\rho }{}_{\alpha }{}^{\tau } W^{\alpha }\,,\\
\mathcal{L}^{(3)}_{\tilde{R}t\Lambda}&=&h_{168}^{} \tilde{R}^{\rho }{}_{\tau \alpha \gamma } t_{\rho }{}^{\tau \gamma } \Lambda^{\alpha }+ h_{169}^{} \tilde{R}^{\rho }{}_{\alpha \tau \gamma } t^{\tau }{}_{\rho }{}^{\gamma } \Lambda^{\alpha } + h_{170}^{} \tilde{R}^{\rho }{}_{\tau \alpha \gamma } t^{\tau }{}_{\rho }{}^{\gamma } \Lambda^{\alpha } + h_{171}^{} \tilde{R}_{\alpha \rho \tau \gamma } t^{\tau \rho \gamma } \Lambda^{\alpha } \nonumber\\
&&+ \,h_{172}^{} \tilde{R}_{\alpha \rho } t^{\alpha \rho }{}_{\tau } \Lambda^{\tau } + h_{173}^{} \tilde{R}_{\alpha \rho } t_{\tau }{}^{\alpha \rho } \Lambda^{\tau }+ h_{174}^{} \hat{R}_{\alpha \rho } t^{\alpha \rho }{}_{\tau } \Lambda^{\tau }  + h_{175}^{} \hat{R}_{\alpha \rho } t_{\tau }{}^{\alpha \rho } \Lambda^{\tau } + h_{176}^{} \tilde{R}^{\omega}{}_{\omega\rho \tau } t^{\rho }{}_{\alpha }{}^{\tau } \Lambda^{\alpha } \,,\\
\mathcal{L}^{(3)}_{\tilde{R}t\Omega}&=&  h_{177}^{} \varepsilon^{\rho \tau }{}_{\nu \sigma } \Omega_{\alpha \rho \tau } \tilde{R}^{\gamma }{}_{\mu }{}^{\alpha \nu } t_{\gamma }{}^{\mu \sigma } + h_{178}^{} \varepsilon^{\rho \tau }{}_{\mu \sigma } \Omega_{\alpha \rho \tau } \tilde{R}^{\gamma \alpha \mu }{}_{\nu } t_{\gamma }{}^{\nu \sigma } + h_{179}^{} \varepsilon^{\rho \tau }{}_{\mu \sigma } \Omega_{\alpha \rho \tau } \tilde{R}^{\gamma \mu \alpha }{}_{\nu } t_{\gamma }{}^{\nu \sigma } \nonumber\\
&& +\, h_{180}^{} \varepsilon^{\rho \tau }{}_{\mu \sigma } \Omega_{\alpha \rho \tau } \tilde{R}^{\alpha }{}_{\gamma }{}^{\mu }{}_{\nu } t^{\gamma \nu \sigma }+ h_{181}^{} \varepsilon^{\rho \tau }{}_{\nu \sigma } \Omega_{\alpha \rho \tau } \tilde{R}^{\gamma }{}_{\mu }{}^{\alpha \nu } t^{\mu }{}_{\gamma }{}^{\sigma } + h_{182}^{} \varepsilon^{\rho \tau }{}_{\gamma \sigma } \Omega_{\alpha \rho \tau } \tilde{R}^{\alpha \gamma }{}_{\mu \nu } t^{\mu \nu \sigma }\nonumber\\
&& +\, h_{183}^{} \varepsilon^{\rho \tau }{}_{\gamma \sigma } \Omega_{\alpha \rho \tau } \tilde{R}^{\gamma \alpha }{}_{\mu \nu } t^{\mu \nu \sigma } + h_{184}^{} \varepsilon^{\rho \tau }{}_{\gamma \sigma } \Omega_{\alpha \rho \tau } \tilde{R}^{\gamma }{}_{\mu }{}^{\alpha }{}_{\nu } t^{\mu \nu \sigma }+ h_{185}^{} \varepsilon^{\rho \tau }{}_{\mu \sigma } \Omega_{\alpha \rho \tau } \tilde{R}^{\gamma \alpha \mu }{}_{\nu } t^{\nu }{}_{\gamma }{}^{\sigma }  \nonumber\\
&&+\, h_{186}^{} \varepsilon^{\rho \tau }{}_{\mu \sigma } \Omega_{\alpha \rho \tau } \tilde{R}^{\gamma \mu \alpha }{}_{\nu } t^{\nu }{}_{\gamma }{}^{\sigma }+ h_{187}^{} \varepsilon^{\rho \tau }{}_{\mu \sigma } \Omega_{\alpha \rho \tau } \tilde{R}^{\alpha }{}_{\gamma }{}^{\mu }{}_{\nu } t^{\nu \gamma \sigma } + h_{188}^{} \varepsilon^{\rho \tau }{}_{\gamma \sigma } \Omega_{\alpha \rho \tau } \tilde{R}^{\gamma }{}_{\mu }{}^{\alpha }{}_{\nu } t^{\nu \mu \sigma }  \nonumber\\
&&+\,h_{189}^{} \varepsilon^{\rho \gamma \mu }{}_{\nu } \Omega_{\tau \gamma \mu } \tilde{R}_{\alpha \rho } t^{\alpha \tau \nu }  + h_{190}^{} \varepsilon^{\alpha \gamma \mu }{}_{\nu } \Omega_{\tau \gamma \mu } \tilde{R}_{\alpha \rho } t^{\rho \tau \nu }+ h_{191}^{} \varepsilon^{\rho \gamma \mu }{}_{\nu } \Omega_{\tau \gamma \mu } \tilde{R}_{\alpha \rho } t^{\nu \alpha \tau } + h_{192}^{} \varepsilon^{\alpha \gamma \mu }{}_{\nu } \Omega_{\tau \gamma \mu } \tilde{R}_{\alpha \rho } t^{\nu \rho \tau } \nonumber\\
&&+ \,h_{193}^{} \varepsilon^{\rho \gamma \mu }{}_{\nu } \Omega_{\tau \gamma \mu } \hat{R}_{\alpha \rho } t^{\alpha \tau \nu }+ h_{194}^{} \varepsilon^{\alpha \gamma \mu }{}_{\nu } \Omega_{\tau \gamma \mu } \hat{R}_{\alpha \rho } t^{\rho \tau \nu }+ h_{195}^{} \varepsilon^{\rho \gamma \mu }{}_{\nu } \Omega_{\tau \gamma \mu } \hat{R}_{\alpha \rho } t^{\nu \alpha \tau }  + h_{196}^{} \varepsilon^{\alpha \gamma \mu }{}_{\nu } \Omega_{\tau \gamma \mu } \hat{R}_{\alpha \rho } t^{\nu \rho \tau }  \nonumber\\
&&+ \,h_{197}^{} \varepsilon^{\rho \tau }{}_{\gamma \nu } \Omega_{\alpha \rho \tau }\tilde{R}^{\omega}{}_{\omega}{}^{\gamma }{}_{\mu } t^{\nu \alpha \mu }+ h_{198}^{} \varepsilon^{\rho \tau }{}_{\gamma \nu } \Omega_{\alpha \rho \tau }\tilde{R}^{\omega}{}_{\omega}{}^{\gamma }{}_{\mu } t^{\mu \alpha \nu }+ h_{199}^{} \varepsilon^{\rho \tau }{}_{\gamma \mu } \Omega_{\alpha \rho \tau } \tilde{R} t^{\gamma \alpha \mu }\,,\\
\mathcal{L}^{(3)}_{\tilde{R}tq}&=&h_{200}^{} q_{\alpha }{}^{\gamma \mu } \tilde{R}^{\alpha }{}_{\rho \tau \gamma } t^{\rho \tau }{}_{\mu }  + h_{201}^{} q^{\rho \gamma \mu } \tilde{R}^{\alpha }{}_{\rho \tau \gamma } t^{\tau }{}_{\alpha \mu }+ h_{202}^{} q_{\alpha }{}^{\rho \mu } \tilde{R}^{\alpha }{}_{\rho \tau \gamma } t^{\tau \gamma }{}_{\mu } + h_{203}^{} q^{\rho \gamma \mu } \tilde{R}^{\alpha }{}_{\rho \tau \gamma } t_{\mu \alpha }{}^{\tau }+ h_{204}^{} q_{\alpha }{}^{\gamma \mu } \tilde{R}^{\alpha }{}_{\rho \tau \gamma } t^{\tau \rho }{}_{\mu }\nonumber\\
&&+\,h_{205}^{} q^{\rho \tau \gamma } \tilde{R}_{\alpha \rho } t_{\tau }{}^{\alpha }{}_{\gamma } + h_{206}^{} q^{\alpha \tau \gamma } \tilde{R}_{\alpha \rho } t_{\tau }{}^{\rho }{}_{\gamma } + h_{207}^{} q^{\rho \tau \gamma } \hat{R}_{\alpha \rho } t_{\tau }{}^{\alpha }{}_{\gamma } + h_{208}^{} q^{\alpha \tau \gamma } \hat{R}_{\alpha \rho } t_{\tau }{}^{\rho }{}_{\gamma } + h_{209}^{} q^{\rho \tau \gamma } \tilde{R}^{\omega}{}_{\omega\alpha \rho } t_{\tau }{}^{\alpha }{}_{\gamma } \,.
\end{eqnarray}

In the particular case of Weyl-Cartan geometry, this number is reduced to $13$, while in the absence of nonmetricity all of the cubic invariants of this branch are also identically zero.

\subsection{Cubic invariants from curvature}\label{sec:cubicR}

The last branch only contains cubic invariants constructed from the curvature tensor and its contractions. In any case, concerning the stability problem, it is worthwhile to stress that this branch does not contain counter terms to cancel out the pathological interactions arising in the vector and axial
sectors of quadratic MAG, while it also breaks the quasilinearity of the field equations. Thereby, the present branch is not relevant in the stability analysis and it will not be included in our calculations, but we only provide all of its cubic invariants for completeness. 

In general, there exist $178$ independent terms constructed from the curvature tensor and its contractions, which can be split into $27$ different types:
\begin{align}
    \mathcal{L}_{\rm curv}^{(3)}&=\mathcal{L}^{(3)}_{\tilde{R}_{\lambda\rho\mu\nu}^3}+\mathcal{L}^{(3)}_{\tilde{R}_{\lambda\rho}^3}+\mathcal{L}^{(3)}_{\hat{R}_{\lambda\rho}^3}+\mathcal{L}^{(3)}_{\tilde{R}^3}+\mathcal{L}^{(3)}_{\tilde{R}_{\lambda\rho\mu\nu}^2\tilde{R}_{\alpha\beta}}+\mathcal{L}^{(3)}_{\tilde{R}_{\lambda\rho\mu\nu}^2\hat{R}_{\alpha\beta}}+\mathcal{L}^{(3)}_{\tilde{R}_{\lambda\rho\mu\nu}^2\tilde{R}^{\omega}{}_{\omega\alpha\beta}}+\mathcal{L}^{(3)}_{\tilde{R}_{\lambda\rho\mu\nu}^2\tilde{R}}\nonumber\\
    &+\mathcal{L}^{(3)}_{\tilde{R}_{\lambda\rho\mu\nu}\tilde{R}_{\alpha\beta}^2}+\mathcal{L}^{(3)}_{\tilde{R}_{\lambda\rho\mu\nu}\hat{R}_{\alpha\beta}^2}+\mathcal{L}^{(3)}_{\tilde{R}_{\lambda\rho\mu\nu}(\tilde{R}^{\omega}{}_{\omega\alpha\beta})^2}+\mathcal{L}^{(3)}_{\tilde{R}_{\lambda\rho\mu\nu}\tilde{R}_{\alpha\beta}\hat{R}_{\alpha\beta}}+\mathcal{L}^{(3)}_{\tilde{R}_{\lambda\rho\mu\nu}\tilde{R}_{\alpha\beta}\tilde{R}^{\omega}{}_{\omega\alpha\beta}}+\mathcal{L}^{(3)}_{\tilde{R}_{\lambda\rho\mu\nu}\hat{R}_{\alpha\beta}\tilde{R}^{\omega}{}_{\omega\alpha\beta}}\nonumber\\
    &+\mathcal{L}^{(3)}_{\tilde{R}_{\lambda\rho}^2\hat{R}_{\lambda\rho}}+\mathcal{L}^{(3)}_{\tilde{R}_{\lambda\rho}^2\tilde{R}^{\omega}{}_{\omega\lambda\rho}}+\mathcal{L}^{(3)}_{\tilde{R}_{\lambda\rho}^2\tilde{R}}+\mathcal{L}^{(3)}_{\tilde{R}_{\lambda\rho}\hat{R}_{\lambda\rho}^2}+\mathcal{L}^{(3)}_{\tilde{R}_{\lambda\rho}(\tilde{R}^{\omega}{}_{\omega\lambda\rho})^2}+\mathcal{L}^{(3)}_{\tilde{R}_{\lambda\rho}\hat{R}_{\lambda\rho}\tilde{R}^{\omega}{}_{\omega\lambda\rho}}+\mathcal{L}^{(3)}_{\tilde{R}_{\lambda\rho}\hat{R}_{\lambda\rho}\tilde{R}}+\mathcal{L}^{(3)}_{\tilde{R}_{\lambda\rho}\tilde{R}^{\omega}{}_{\omega\lambda\rho}\tilde{R}}\nonumber\\
    &+\mathcal{L}^{(3)}_{\hat{R}_{\lambda\rho}^2\tilde{R}^{\omega}{}_{\omega\lambda\rho}}+\mathcal{L}^{(3)}_{\hat{R}_{\lambda\rho}^2\tilde{R}}+\mathcal{L}^{(3)}_{\hat{R}_{\lambda\rho}(\tilde{R}^{\omega}{}_{\omega\lambda\rho})^2}+\mathcal{L}^{(3)}_{\hat{R}_{\lambda\rho}\tilde{R}^{\omega}{}_{\omega\lambda\rho}\tilde{R}}+\mathcal{L}^{(3)}_{\hat{R}^2\tilde{R}}\,,\label{cubicLagPureCurv}
\end{align}
with
\small{
\begin{eqnarray}
\mathcal{L}^{(3)}_{\tilde{R}_{\lambda\rho\mu\nu}^3}&=&(k_{1}^{}- k_{32}^{} + k_{48}^{} -  k_{49}^{}) \tilde{R}_{\alpha \tau }{}^{\mu \nu } \tilde{R}^{\alpha \rho 
\tau \gamma } \tilde{R}_{\rho \gamma \mu \nu } + (k_{2}^{}-  k_{34}^{} -  k_{36}^{} -  k_{39}^{} -  k_{42}^{} -  k_{43}^{} + k_{44}^{} -  k_{51}^{}) \tilde{R}_{
\alpha \tau }{}^{\mu \nu } \tilde{R}^{\alpha \rho \tau \gamma } \tilde{R}_{\rho \mu \gamma \nu }  \nonumber\\
&&+ \,(k_{3}^{}-k_{35}) \tilde{R}_{\alpha }{}^{\mu }{}_{\tau }{}^{\nu } \tilde{R}^{\alpha \rho \tau \gamma } \tilde{R}_{\rho \mu \gamma \nu }+ ( k_{4}^{}- k_{37}^{} + k_{38}^{}  -  k_{50}^{})\tilde{R}_{\alpha }{}^{\mu }{}_{\tau }{}^{\nu } \tilde{R}^{\alpha \rho \tau \gamma } \tilde{R}_{\rho \nu \gamma \mu }+(k_{5}^{} -k_{59})\tilde{R}_{\alpha \rho }{}^{\mu \nu } \tilde{R}^{\alpha \rho \tau \gamma } \tilde{R}_{\tau \gamma \mu \nu } \nonumber\\
&&+ \,(k_{6}^{}+k_{55}) \tilde{R}_{\alpha \rho }{}^{\mu \nu } \tilde{R}^{\alpha \rho \tau \gamma } \tilde{R}_{\tau \mu \gamma \nu } + (k_{7}^{}+k_{33}^{} -  k_{41}^{} -  k_{45}^{} + k_{47}^{} -  k_{54}^{} -  k_{58}^{} + k_{61}^{} ) \tilde{R}_{\alpha \tau }{}^{\mu \nu } \tilde{R}^{\alpha \rho \tau \gamma } \tilde{R}_{\gamma \mu \rho \nu }\nonumber\\
&&+\, (k_{8}^{}- k_{40}^{} -  k_{52}^{} -  k_{56}^{}) \tilde{R}_{\alpha }{}^{\mu }{}_{\tau }{}^{\nu } \tilde{R}^{\alpha \rho \tau \gamma } \tilde{R}_{\gamma \nu \rho \mu } +(k_{9}^{} +k_{53})\tilde{R}^{\alpha \rho \tau \gamma } \tilde{R}_{\tau }{}^{\mu }{}_{\alpha }{}^{\nu } \tilde{R}_{\gamma \nu \rho \mu }+ k_{10}^{} \tilde{R}^{\alpha \rho \tau \gamma } \tilde{R}_{\tau \gamma }{}^{\mu \nu } \tilde{R}_{\mu \nu \alpha \rho }\nonumber\\
&&+\, (k_{11}^{}-  k_{46}^{} + k_{57}^{} -  k_{60}^{}) \tilde{R}_{\alpha \tau }{}^{\mu \nu } \tilde{R}^{\alpha \rho \tau \gamma } \tilde{R}_{\mu \nu \rho \gamma }  +k_{32}^{} \tilde{R}^{\alpha \rho }{}_{\tau \gamma } \tilde{R}^{\tau }{}_{\alpha }{}^{\mu \nu } \tilde{R}^{\gamma }{}_{\rho \mu \nu }+ k_{33}^{} \tilde{R}^{\alpha \rho }{}_{\tau \gamma } \tilde{R}^{\tau }{}_{\alpha }{}^{\mu \nu } \tilde{R}^{\gamma }{}_{\mu \rho \nu }\nonumber\\
&& +\, k_{34}^{} \tilde{R}^{\alpha }{}_{\rho }{}^{\tau \gamma } \tilde{R}^{\rho }{}_{\tau \mu \nu } \tilde{R}^{\mu }{}_{\alpha \gamma }{}^{\nu } + k_{35}^{} \tilde{R}^{\alpha }{}_{\rho }{}^{\tau \gamma } \tilde{R}^{\rho }{}_{\mu \tau \nu } \tilde{R}^{\mu }{}_{\alpha \gamma }{}^{\nu } + k_{36}^{} \tilde{R}^{\alpha \rho }{}_{\tau \gamma } \tilde{R}^{\tau }{}_{\mu }{}^{\gamma \nu } \tilde{R}^{\mu }{}_{\rho \alpha \nu } + k_{37}^{} \tilde{R}^{\alpha \rho \tau }{}_{\gamma } \tilde{R}^{\gamma }{}_{\mu \alpha }{}^{\nu } \tilde{R}^{\mu }{}_{\rho \tau \nu }\nonumber\\
&& + \,k_{38}^{} \tilde{R}^{\alpha \rho \tau \gamma } \tilde{R}_{\mu \gamma \alpha }{}^{\nu } \tilde{R}^{\mu }{}_{\rho \tau \nu } + k_{39}^{} \tilde{R}^{\alpha \rho \tau \gamma } \tilde{R}_{\mu }{}^{\nu }{}_{\alpha \gamma } \tilde{R}^{\mu }{}_{\rho \tau \nu } + k_{40}^{} \tilde{R}^{\alpha \rho \tau }{}_{\gamma } \tilde{R}^{\gamma }{}_{\mu \alpha }{}^{\nu } \tilde{R}^{\mu }{}_{\tau \rho \nu } + k_{41}^{} \tilde{R}^{\alpha \rho \tau \gamma } \tilde{R}_{\mu }{}^{\nu }{}_{\alpha \gamma } \tilde{R}^{\mu }{}_{\tau \rho \nu }\nonumber\\
&& + \,k_{42}^{} \tilde{R}^{\alpha }{}_{\rho }{}^{\tau \gamma } \tilde{R}^{\rho }{}_{\mu \alpha }{}^{\nu } \tilde{R}^{\mu }{}_{\tau \gamma \nu } + k_{43}^{} \tilde{R}^{\alpha \rho \tau \gamma } \tilde{R}_{\mu \alpha \rho }{}^{\nu } \tilde{R}^{\mu }{}_{\tau \gamma \nu } + k_{44}^{} \tilde{R}^{\alpha \rho \tau \gamma } \tilde{R}_{\mu \rho \alpha }{}^{\nu } \tilde{R}^{\mu }{}_{\tau \gamma \nu }+ k_{45}^{} \tilde{R}^{\alpha }{}_{\rho }{}^{\tau \gamma } \tilde{R}^{\rho }{}_{\tau \mu \nu } \tilde{R}^{\mu }{}_{\gamma \alpha }{}^{\nu }\nonumber\\
&&  + \,k_{46}^{} \tilde{R}^{\alpha \rho }{}_{\tau \gamma } \tilde{R}^{\tau }{}_{\mu }{}^{\gamma \nu } \tilde{R}^{\mu }{}_{\nu \alpha \rho } + k_{47}^{} \tilde{R}^{\alpha \rho \tau }{}_{\gamma } \tilde{R}^{\gamma }{}_{\mu \alpha }{}^{\nu } \tilde{R}^{\mu }{}_{\nu \rho \tau } + k_{48}^{} \tilde{R}^{\alpha }{}_{\rho }{}^{\tau \gamma } \tilde{R}^{\rho }{}_{\mu \alpha }{}^{\nu } \tilde{R}^{\mu }{}_{\nu \tau \gamma }+ k_{49}^{} \tilde{R}^{\alpha \rho \tau \gamma } \tilde{R}_{\mu \rho \alpha }{}^{\nu } \tilde{R}^{\mu }{}_{\nu \tau \gamma } \nonumber\\
&& +\, k_{50}^{} \tilde{R}^{\alpha }{}_{\rho }{}^{\tau \gamma } \tilde{R}^{\rho }{}_{\mu \tau \nu } \tilde{R}^{\nu }{}_{\alpha \gamma }{}^{\mu } + k_{51}^{} \tilde{R}^{\alpha \rho \tau }{}_{\gamma } \tilde{R}^{\gamma }{}_{\alpha }{}^{\mu }{}_{\nu } \tilde{R}^{\nu }{}_{\rho \tau \mu } + k_{52}^{} \tilde{R}^{\alpha \rho \tau }{}_{\gamma } \tilde{R}^{\gamma \mu }{}_{\alpha \nu } \tilde{R}^{\nu }{}_{\rho \tau \mu }+ k_{53}^{} \tilde{R}^{\alpha \rho \tau }{}_{\gamma } \tilde{R}^{\gamma }{}_{\mu \rho \nu } \tilde{R}^{\nu }{}_{\tau \alpha }{}^{\mu } \nonumber\\
&& + \,k_{54}^{} \tilde{R}^{\alpha \rho \tau }{}_{\gamma } \tilde{R}^{\gamma }{}_{\alpha }{}^{\mu }{}_{\nu } \tilde{R}^{\nu }{}_{\tau \rho \mu } + k_{55}^{} \tilde{R}^{\alpha }{}_{\rho }{}^{\tau \gamma } \tilde{R}^{\rho }{}_{\alpha }{}^{\mu }{}_{\nu } \tilde{R}^{\nu }{}_{\tau \gamma \mu } + k_{56}^{} \tilde{R}^{\alpha }{}_{\rho }{}^{\tau \gamma } \tilde{R}^{\rho }{}_{\mu \tau \nu } \tilde{R}^{\nu }{}_{\gamma \alpha }{}^{\mu } + k_{57}^{} \tilde{R}^{\alpha \rho \tau }{}_{\gamma } \tilde{R}^{\gamma }{}_{\alpha }{}^{\mu }{}_{\nu } \tilde{R}^{\nu }{}_{\mu \rho \tau }\nonumber\\
&& + \,k_{58}^{} \tilde{R}^{\alpha \rho \tau }{}_{\gamma } \tilde{R}^{\gamma \mu }{}_{\alpha \nu } \tilde{R}^{\nu }{}_{\mu \rho \tau } + k_{59}^{} \tilde{R}^{\alpha }{}_{\rho }{}^{\tau \gamma } \tilde{R}^{\rho }{}_{\alpha }{}^{\mu }{}_{\nu } \tilde{R}^{\nu }{}_{\mu \tau \gamma } + k_{60}^{} \tilde{R}^{\alpha \rho \tau }{}_{\gamma } \tilde{R}^{\gamma }{}_{\mu \tau \nu } \tilde{R}^{\nu \mu }{}_{\alpha \rho }+ k_{61}^{} \tilde{R}^{\alpha \rho \tau }{}_{\gamma } \tilde{R}^{\gamma }{}_{\mu \rho \nu } \tilde{R}^{\nu \mu }{}_{\alpha \tau }\,,
\end{eqnarray}}
\normalsize
\small{\begin{align}
\mathcal{L}^{(3)}_{\tilde{R}_{\lambda\rho}^3}&=(k_{12}^{}-  k_{156}^{} -  k_{158}^{} -  k_{159}^{} -  k_{161}^{} -  k_{163}^{} -  k_{164}^{} -  k_{63}^{}) \tilde{R}_{\rho }{}^{\mu } \tilde{R}^{\rho \tau } \tilde{R}_{\tau \mu } \nonumber\\
&\,\,\,\,\,+ (k_{13}^{}-  k_{157}^{} -  k_{162}^{} -  k_{62}^{}) \tilde{R}^{\rho \tau } \tilde{R}_{\tau }{}^{\mu } \tilde{R}_{\mu \rho }\,,\\
\mathcal{L}^{(3)}_{\hat{R}_{\lambda\rho}^3}&=k_{62}^{} \hat{R}^{\alpha \rho } \hat{R}_{\rho }{}^{\tau } \hat{R}_{\tau \alpha } + k_{63}^{} \hat{R}^{\alpha \rho } \hat{R}_{\tau \alpha } \hat{R}^{\tau }{}_{\rho }\,,\\
\mathcal{L}^{(3)}_{\tilde{R}^3}&=k_{14}^{} \tilde{R}^3\,,\\
\mathcal{L}^{(3)}_{\tilde{R}_{\lambda\rho\mu\nu}^2\tilde{R}_{\alpha\beta}}&=(k_{15}^{}-  k_{64}^{} + k_{66}^{} + k_{68}^{} -  k_{82}^{} -  k_{84}^{} + 
k_{86}^{} + k_{88}^{}) \tilde{R}^{\rho \tau } \tilde{R}_{\rho }{}^{\gamma \mu \nu 
} \tilde{R}_{\tau \gamma \mu \nu }  \nonumber\\
&\,\,\,\,\,\,+( k_{16}^{}+ k_{69}^{} -  k_{70}^{} + k_{71}^{} -  k_{83}^{} + k_{93}^{} -  
k_{94}^{} + k_{95}^{}) \tilde{R}^{\rho \tau } 
\tilde{R}_{\rho }{}^{\gamma \mu \nu } \tilde{R}_{\tau \mu \gamma \nu 
}\nonumber\\
&\,\,\,\,\,\,+ (k_{17}^{} +k_{67}^{} -  k_{72}^{} + k_{75}^{} + k_{87}^{} -  k_{91}^{} -  
k_{96}^{} + k_{99}^{})\tilde{R}^{\rho \tau } \tilde{R}_{\tau }{}^{\gamma \mu 
\nu } \tilde{R}_{\gamma \mu \rho \nu } \nonumber\\
&\,\,\,\,\,\,+ (k_{18}^{}+k_{100}^{} + k_{65}^{} -  k_{73}^{} + k_{76}^{} + k_{85}^{} -  k_{92}^{} -  k_{97}^{}) \tilde{R}^{\rho \tau } \tilde{R}_{\rho }{}^{\gamma \mu \nu } \tilde{R}_{\gamma \mu \tau 
\nu } \nonumber\\
&\,\,\,\,\,\,+(k_{19}^{}+ k_{74}^{} -  k_{89}^{} + k_{98}^{}) \tilde{R}^{\rho \tau } \tilde{R}_{\gamma \mu \tau \nu } \tilde{R}^{\gamma \mu }{}_{\rho }{}^{\nu } \nonumber\\
&\,\,\,\,\,\,+ (k_{20}^{}+k_{101}^{} + k_{106}^{} -  k_{107}^{}+ k_{77}^{} + 
k_{80}^{} -  k_{81}^{} -  k_{90}^{}) \tilde{R}^{\rho \tau } \tilde{R}_{\gamma \nu \tau \mu } \tilde{R}^{\gamma \mu }{}_{\rho }{}^{\nu } \nonumber\\
&\,\,\,\,\,\,+ (k_{21}^{}+k_{102}^{} -  k_{103}^{} + k_{78}^{}) \tilde{R}^{\rho \tau } \tilde{R}_{\tau }{}^{\gamma \mu \nu } \tilde{R}_{\mu \nu \rho \gamma } + ( k_{22}^{}+k_{104}^{} -  k_{105}^{}  + k_{79}^{}) \tilde{R}^{\rho \tau } \tilde{R}_{\rho }{}^{\gamma \mu \nu } \tilde{R}_{\mu \nu \tau \gamma }\nonumber \\
&\,\,\,\,\,\,+ k_{64}^{} \tilde{R}_{\tau }{}^{\rho \gamma \mu } \tilde{R}^{\tau \alpha }{}_{\gamma \mu } \tilde{R}_{\alpha \rho } + k_{65}^{} \tilde{R}_{\tau }{}^{\gamma \rho \mu } \tilde{R}^{\tau \alpha }{}_{\gamma \mu } \tilde{R}_{\alpha \rho } + k_{66}^{} \tilde{R}^{\rho }{}_{\tau \gamma \mu } \tilde{R}^{\tau \alpha \gamma \mu } \tilde{R}_{\alpha \rho } + k_{67}^{} \tilde{R}_{\tau }{}^{\gamma \alpha \mu } \tilde{R}^{\tau \rho }{}_{\gamma \mu } \tilde{R}_{\alpha \rho }\nonumber\\
&\,\,\,\,\,\,+ k_{68}^{} \tilde{R}^{\alpha }{}_{\tau \gamma \mu } \tilde{R}^{\tau \rho \gamma \mu } \tilde{R}_{\alpha \rho } + k_{69}^{} \tilde{R}^{\rho }{}_{\tau \gamma \mu } \tilde{R}^{\gamma \alpha \tau \mu } \tilde{R}_{\alpha \rho } + k_{70}^{} \tilde{R}^{\tau \alpha }{}_{\gamma \mu } \tilde{R}^{\gamma \rho }{}_{\tau }{}^{\mu } \tilde{R}_{\alpha \rho } + k_{71}^{} \tilde{R}^{\alpha }{}_{\tau \gamma \mu } \tilde{R}^{\gamma \rho \tau \mu } \tilde{R}_{\alpha \rho }\nonumber \\
&\,\,\,\,\,\,+k_{72}^{} \tilde{R}^{\tau \rho }{}_{\gamma \mu } \tilde{R}^{\gamma }{}_{\tau }{}^{\alpha \mu } \tilde{R}_{\alpha \rho } + k_{73}^{} \tilde{R}^{\tau \alpha }{}_{\gamma \mu } \tilde{R}^{\gamma }{}_{\tau }{}^{\rho \mu } \tilde{R}_{\alpha \rho } + k_{74}^{} \tilde{R}^{\tau }{}_{\gamma }{}^{\alpha }{}_{\mu } \tilde{R}^{\gamma }{}_{\tau }{}^{\rho \mu } \tilde{R}_{\alpha \rho } + k_{75}^{} \tilde{R}^{\rho }{}_{\tau \gamma \mu } \tilde{R}^{\gamma \tau \alpha \mu } \tilde{R}_{\alpha \rho } \nonumber \\
&\,\,\,\,\,\,+ k_{76}^{} \tilde{R}^{\alpha }{}_{\tau \gamma \mu } \tilde{R}^{\gamma \tau \rho \mu } \tilde{R}_{\alpha \rho } + k_{77}^{} \tilde{R}^{\tau }{}_{\gamma }{}^{\alpha \mu } \tilde{R}^{\gamma }{}_{\mu }{}^{\rho }{}_{\tau } \tilde{R}_{\alpha \rho } + k_{78}^{} \tilde{R}^{\tau \rho }{}_{\gamma \mu } \tilde{R}^{\gamma \mu \alpha }{}_{\tau } \tilde{R}_{\alpha \rho } + k_{79}^{} \tilde{R}^{\tau \alpha }{}_{\gamma \mu } \tilde{R}^{\gamma \mu \rho }{}_{\tau } \tilde{R}_{\alpha \rho } \nonumber \\
&\,\,\,\,\,\,+ k_{80}^{} \tilde{R}^{\tau }{}_{\gamma }{}^{\alpha }{}_{\mu } \tilde{R}^{\mu }{}_{\tau }{}^{\rho \gamma } \tilde{R}_{\alpha \rho } + k_{81}^{} \tilde{R}^{\tau \gamma \rho }{}_{\mu } \tilde{R}^{\mu }{}_{\gamma }{}^{\alpha }{}_{\tau } \tilde{R}_{\alpha \rho }\,,\\
\mathcal{L}^{(3)}_{\tilde{R}_{\lambda\rho\mu\nu}^2\hat{R}_{\alpha\beta}}&=k_{82}^{} \tilde{R}^{\alpha }{}_{\tau \gamma \mu } \tilde{R}^{\rho \tau \gamma \mu } \hat{R}_{\alpha \rho } + k_{83}^{} \tilde{R}^{\alpha }{}_{\tau \gamma \mu } \tilde{R}^{\rho \gamma \tau \mu } \hat{R}_{\alpha \rho } + k_{84}^{} \tilde{R}_{\tau }{}^{\rho \gamma \mu } \tilde{R}^{\tau \alpha }{}_{\gamma \mu } \hat{R}_{\alpha \rho } + k_{85}^{} \tilde{R}_{\tau }{}^{\gamma \rho \mu } \tilde{R}^{\tau \alpha }{}_{\gamma \mu } \hat{R}_{\alpha \rho } \nonumber \\
&\,\,\,\,\,\,+ k_{86}^{} \tilde{R}^{\rho }{}_{\tau \gamma \mu } \tilde{R}^{\tau \alpha \gamma \mu } \hat{R}_{\alpha \rho } + k_{87}^{} \tilde{R}_{\tau }{}^{\gamma \alpha \mu } \tilde{R}^{\tau \rho }{}_{\gamma \mu } \hat{R}_{\alpha \rho } + k_{88}^{} \tilde{R}^{\alpha }{}_{\tau \gamma \mu } \tilde{R}^{\tau \rho \gamma \mu } \hat{R}_{\alpha \rho } + k_{89}^{} \tilde{R}_{\tau }{}^{\gamma \rho \mu } \tilde{R}^{\tau }{}_{\gamma }{}^{\alpha }{}_{\mu } \hat{R}_{\alpha \rho } \nonumber \\
&\,\,\,\,\,\,+ k_{90}^{} \tilde{R}_{\tau }{}^{\mu \rho \gamma } \tilde{R}^{\tau }{}_{\gamma }{}^{\alpha }{}_{\mu } \hat{R}_{\alpha \rho } + k_{91}^{} \tilde{R}^{\rho }{}_{\tau \gamma \mu } \tilde{R}^{\tau \gamma \alpha \mu } \hat{R}_{\alpha \rho } + k_{92}^{} \tilde{R}^{\alpha }{}_{\tau \gamma \mu } \tilde{R}^{\tau \gamma \rho \mu } \hat{R}_{\alpha \rho } + k_{93}^{} \tilde{R}^{\rho }{}_{\tau \gamma \mu } \tilde{R}^{\gamma \alpha \tau \mu } \hat{R}_{\alpha \rho }\nonumber \\
&\,\,\,\,\,\, + k_{94}^{} \tilde{R}^{\tau \alpha }{}_{\gamma \mu } \tilde{R}^{\gamma \rho }{}_{\tau }{}^{\mu } \hat{R}_{\alpha \rho } + k_{95}^{} \tilde{R}^{\alpha }{}_{\tau \gamma \mu } \tilde{R}^{\gamma \rho \tau \mu } \hat{R}_{\alpha \rho } + k_{96}^{} \tilde{R}^{\tau \rho }{}_{\gamma \mu } \tilde{R}^{\gamma }{}_{\tau }{}^{\alpha \mu } \hat{R}_{\alpha \rho } + k_{97}^{} \tilde{R}^{\tau \alpha }{}_{\gamma \mu } \tilde{R}^{\gamma }{}_{\tau }{}^{\rho \mu } \hat{R}_{\alpha \rho }\nonumber \\
& \,\,\,\,\,\,+ k_{98}^{} \tilde{R}^{\tau }{}_{\gamma }{}^{\alpha }{}_{\mu } \tilde{R}^{\gamma }{}_{\tau }{}^{\rho \mu } \hat{R}_{\alpha \rho } + k_{99}^{} \tilde{R}^{\rho }{}_{\tau \gamma \mu } \tilde{R}^{\gamma \tau \alpha \mu } \hat{R}_{\alpha \rho } + k_{100}^{} \tilde{R}^{\alpha }{}_{\tau \gamma \mu } \tilde{R}^{\gamma \tau \rho \mu } \hat{R}_{\alpha \rho } + k_{101}^{} \tilde{R}^{\tau }{}_{\gamma }{}^{\alpha \mu } \tilde{R}^{\gamma }{}_{\mu }{}^{\rho }{}_{\tau } \hat{R}_{\alpha \rho } \nonumber \\
&\,\,\,\,\,\,+ k_{102}^{} \tilde{R}^{\tau \rho }{}_{\gamma \mu } \tilde{R}^{\gamma \mu \alpha }{}_{\tau } \hat{R}_{\alpha \rho } + k_{103}^{} \tilde{R}^{\rho }{}_{\tau \gamma \mu } \tilde{R}^{\gamma \mu \alpha \tau } \hat{R}_{\alpha \rho } + k_{104}^{} \tilde{R}^{\tau \alpha }{}_{\gamma \mu } \tilde{R}^{\gamma \mu \rho }{}_{\tau } \hat{R}_{\alpha \rho } + k_{105}^{} \tilde{R}^{\alpha }{}_{\tau \gamma \mu } \tilde{R}^{\gamma \mu \rho \tau } \hat{R}_{\alpha \rho }\nonumber \\&
\,\,\,\,\,\,+ k_{106}^{} \tilde{R}^{\tau }{}_{\gamma }{}^{\alpha }{}_{\mu } \tilde{R}^{\mu }{}_{\tau }{}^{\rho \gamma } \hat{R}_{\alpha \rho } + k_{107}^{} \tilde{R}^{\tau \gamma \rho }{}_{\mu } \tilde{R}^{\mu }{}_{\gamma }{}^{\alpha }{}_{\tau } \hat{R}_{\alpha \rho }\,,\\
\mathcal{L}^{(3)}_{\tilde{R}_{\lambda\rho\mu\nu}^2\tilde{R}^{\omega}{}_{\omega\alpha\beta}}&=k_{108}^{} \tilde{R}^{\alpha }{}_{\rho }{}^{\tau \gamma } \tilde{R}^{\rho }{}_{\tau \gamma \mu } \tilde{R}^{\omega}{}_{\omega\alpha }{}^{\mu } + k_{109}^{} \tilde{R}^{\alpha }{}_{\rho }{}^{\tau \gamma } \tilde{R}^{\rho }{}_{\mu \tau \gamma } \tilde{R}^{\omega}{}_{\omega\alpha }{}^{\mu } + k_{110}^{} \tilde{R}^{\alpha \rho \tau }{}_{\gamma } \tilde{R}^{\gamma }{}_{\rho \tau \mu } \tilde{R}^{\omega}{}_{\omega\alpha }{}^{\mu } + k_{111}^{} \tilde{R}^{\alpha \rho \tau }{}_{\gamma } \tilde{R}^{\gamma }{}_{\tau \rho \mu } \tilde{R}^{\omega}{}_{\omega\alpha }{}^{\mu }\nonumber \\
&\,\,\,\,\,\,+ k_{112}^{} \tilde{R}^{\alpha \rho \tau }{}_{\gamma } \tilde{R}^{\gamma }{}_{\mu \rho \tau } \tilde{R}^{\omega}{}_{\omega\alpha }{}^{\mu } + k_{113}^{} \tilde{R}_{\alpha }{}^{\tau \gamma }{}_{\mu } \tilde{R}^{\alpha }{}_{\rho \tau \gamma }\tilde{R}^{\omega}{}_{\omega}{}^{\rho \mu } + k_{114}^{} \tilde{R}^{\alpha }{}_{\rho }{}^{\tau }{}_{\gamma } \tilde{R}^{\rho }{}_{\tau \alpha \mu }\tilde{R}^{\omega}{}_{\omega}{}^{\gamma \mu } + k_{115}^{} \tilde{R}^{\alpha }{}_{\rho }{}^{\tau }{}_{\gamma } \tilde{R}^{\rho }{}_{\mu \alpha \tau }\tilde{R}^{\omega}{}_{\omega}{}^{\gamma \mu } \nonumber \\
&\,\,\,\,\,\,+ k_{116}^{} \tilde{R}^{\alpha \rho }{}_{\tau \gamma } \tilde{R}^{\tau }{}_{\mu \alpha \rho }\tilde{R}^{\omega}{}_{\omega}{}^{\gamma \mu }\,,\\
\mathcal{L}^{(3)}_{\tilde{R}_{\lambda\rho\mu\nu}\tilde{R}_{\alpha\beta}^2}&=(k_{23}^{}+k_{120}^{} -  k_{122}^{} + k_{125}^{} -  k_{130}^{} + k_{133}^{} -  
k_{138}^{} + k_{141}^{}) \tilde{R}^{\rho \tau } \tilde{R}_{\gamma \mu } \tilde{R}_{\rho \tau }{}^{\gamma \mu } \nonumber\\
&\,\,\,\,\,\,+ (k_{24}^{}- k_{123}^{} -  k_{131}^{} + k_{136}^{}) \tilde{R}^{\rho \tau } \tilde{R}_{\gamma \mu } \tilde{R}_{\rho }{}^{\gamma }{}_{\tau }{}^{\mu } \nonumber\\
&\,\,\,\,\,\,+ (k_{25}^{}- k_{121}^{} -  k_{124}^{} -  k_{126}^{} -  k_{132}^{} -  k_{134}^{} 
+ k_{137}^{} + k_{139}^{}) \tilde{R}^{\rho \tau } \tilde{R}_{\mu \gamma } \tilde{R}_{\rho }{}^{\gamma }{}_{\tau }{}^{\mu } \nonumber\\
&\,\,\,\,\,\,+( k_{26}^{}- k_{127}^{} -  k_{135}^{} + k_{140}^{}) \tilde{R}^{\rho \tau } \tilde{R}_{\mu \gamma } \tilde{R}_{\tau }{}^{\gamma }{}_{\rho }{}^{\mu }+k_{120}^{} \tilde{R}^{\rho \alpha \tau \gamma } \tilde{R}_{\alpha \rho } \tilde{R}_{\tau \gamma } + k_{121}^{} \tilde{R}^{\rho \tau \alpha \gamma } \tilde{R}_{\alpha \rho } \tilde{R}_{\tau \gamma }\,,\\
\mathcal{L}^{(3)}_{\tilde{R}_{\lambda\rho\mu\nu}^2\tilde{R}}&=(k_{27}^{} +k_{117})\tilde{R}_{\tau \gamma \mu \nu } \tilde{R}^{\tau \gamma \mu \nu } \tilde{R} + (k_{28}^{} +k_{118}^{} + k_{119}^{})\tilde{R}_{\tau \mu \gamma \nu } \tilde{R}^{\tau \gamma \mu \nu } \tilde{R} + k_{29}^{} \tilde{R}^{\tau \gamma \mu \nu } \tilde{R}_{\mu \nu \tau \gamma } \tilde{R}+k_{117}^{} \tilde{R}^{\alpha }{}_{\rho }{}^{\tau \gamma } \tilde{R}^{\rho }{}_{\alpha \tau \gamma } \tilde{R}\nonumber\\
&\,\,\,\,\,\,+ k_{118}^{} \tilde{R}^{\alpha }{}_{\rho }{}^{\tau \gamma } \tilde{R}^{\rho }{}_{\tau \alpha \gamma } \tilde{R} + k_{119}^{} \tilde{R}^{\alpha \rho \tau }{}_{\gamma } \tilde{R}^{\gamma }{}_{\rho \alpha \tau } \tilde{R}\,,\\
\mathcal{L}^{(3)}_{\tilde{R}_{\lambda\rho\mu\nu}\hat{R}_{\alpha\beta}^2}&=k_{122}^{} \tilde{R}^{\alpha \rho \tau \gamma } \hat{R}_{\alpha \rho } \hat{R}_{\tau \gamma } + k_{123}^{} \tilde{R}^{\alpha \tau \rho \gamma } \hat{R}_{\alpha \rho } \hat{R}_{\tau \gamma } + k_{124}^{} \tilde{R}^{\alpha \gamma \rho \tau } \hat{R}_{\alpha \rho } \hat{R}_{\tau \gamma } + k_{125}^{} \tilde{R}^{\rho \alpha \tau \gamma } \hat{R}_{\alpha \rho } \hat{R}_{\tau \gamma }\nonumber\\
& \,\,\,\,\,\,+ k_{126}^{} \tilde{R}^{\rho \tau \alpha \gamma } \hat{R}_{\alpha \rho } \hat{R}_{\tau \gamma }+ k_{127}^{} \tilde{R}^{\rho \gamma \alpha \tau } \hat{R}_{\alpha \rho } \hat{R}_{\tau \gamma }\,,\\
\mathcal{L}^{(3)}_{\tilde{R}_{\lambda\rho\mu\nu}(\tilde{R}^{\omega}{}_{\omega\alpha\beta})^2}&=k_{128}^{} \tilde{R}^{\alpha }{}_{\rho \tau \gamma } \tilde{R}^{\lambda}{}_{\lambda\alpha }{}^{\tau }\tilde{R}^{\omega}{}_{\omega}{}^{\rho \gamma } + k_{129}^{} \tilde{R}^{\alpha }{}_{\rho \tau \gamma } \tilde{R}^{\lambda}{}_{\lambda\alpha }{}^{\rho }\tilde{R}^{\omega}{}_{\omega}{}^{\tau \gamma }\,,\\
\mathcal{L}^{(3)}_{\tilde{R}_{\lambda\rho\mu\nu}\tilde{R}_{\alpha\beta}\hat{R}_{\alpha\beta}}&=k_{130}^{} \tilde{R}^{\alpha \rho \tau \gamma } \tilde{R}_{\alpha \rho } \hat{R}_{\tau \gamma } + k_{131}^{} \tilde{R}^{\alpha \tau \rho \gamma } \tilde{R}_{\alpha \rho } \hat{R}_{\tau \gamma } + k_{132}^{} \tilde{R}^{\alpha \gamma \rho \tau } \tilde{R}_{\alpha \rho } \hat{R}_{\tau \gamma } + k_{133}^{} \tilde{R}^{\rho \alpha \tau \gamma } \tilde{R}_{\alpha \rho } \hat{R}_{\tau \gamma } \nonumber\\
&\,\,\,\,\,\,+ k_{134}^{} \tilde{R}^{\rho \tau \alpha \gamma } \tilde{R}_{\alpha \rho } \hat{R}_{\tau \gamma } + k_{135}^{} \tilde{R}^{\rho \gamma \alpha \tau } \tilde{R}_{\alpha \rho } \hat{R}_{\tau \gamma } + k_{136}^{} \tilde{R}^{\tau \alpha \rho \gamma } \tilde{R}_{\alpha \rho } \hat{R}_{\tau \gamma } + k_{137}^{} \tilde{R}^{\tau \rho \alpha \gamma } \tilde{R}_{\alpha \rho } \hat{R}_{\tau \gamma } \nonumber\\
&\,\,\,\,\,\,+ k_{138}^{} \tilde{R}^{\tau \gamma \alpha \rho } \tilde{R}_{\alpha \rho } \hat{R}_{\tau \gamma } + k_{139}^{} \tilde{R}^{\gamma \alpha \rho \tau } \tilde{R}_{\alpha \rho } \hat{R}_{\tau \gamma } + k_{140}^{} \tilde{R}^{\gamma \rho \alpha \tau } \tilde{R}_{\alpha \rho } \hat{R}_{\tau \gamma } + k_{141}^{} \tilde{R}^{\gamma \tau \alpha \rho } \tilde{R}_{\alpha \rho } \hat{R}_{\tau \gamma }\,,\\
\mathcal{L}^{(3)}_{\tilde{R}_{\lambda\rho\mu\nu}\tilde{R}_{\alpha\beta}\tilde{R}^{\omega}{}_{\omega\alpha\beta}}&=k_{142}^{} \tilde{R}^{\tau \alpha \rho }{}_{\gamma } \tilde{R}_{\alpha \rho } \tilde{R}^{\omega}{}_{\omega\tau }{}^{\gamma } + k_{143}^{} \tilde{R}^{\tau \rho \alpha }{}_{\gamma } \tilde{R}_{\alpha \rho } \tilde{R}^{\omega}{}_{\omega\tau }{}^{\gamma } + k_{144}^{} \tilde{R}^{\tau }{}_{\gamma }{}^{\alpha \rho } \tilde{R}_{\alpha \rho } \tilde{R}^{\omega}{}_{\omega\tau }{}^{\gamma } + k_{145}^{} \tilde{R}^{\alpha \rho }{}_{\tau \gamma } \tilde{R}_{\alpha \rho }\tilde{R}^{\omega}{}_{\omega}{}^{\tau \gamma } \nonumber\\
&\,\,\,\,\,\,+ k_{146}^{} \tilde{R}^{\alpha }{}_{\tau }{}^{\rho }{}_{\gamma } \tilde{R}_{\alpha \rho }\tilde{R}^{\omega}{}_{\omega}{}^{\tau \gamma } + k_{147}^{} \tilde{R}^{\rho \alpha }{}_{\tau \gamma } \tilde{R}_{\alpha \rho }\tilde{R}^{\omega}{}_{\omega}{}^{\tau \gamma } + k_{148}^{} \tilde{R}^{\rho }{}_{\tau }{}^{\alpha }{}_{\gamma } \tilde{R}_{\alpha \rho }\tilde{R}^{\omega}{}_{\omega}{}^{\tau \gamma }\,,\\
\mathcal{L}^{(3)}_{\tilde{R}_{\lambda\rho\mu\nu}\hat{R}_{\alpha\beta}\tilde{R}^{\omega}{}_{\omega\alpha\beta}}&=k_{149}^{} \tilde{R}^{\tau \alpha \rho }{}_{\gamma } \hat{R}_{\alpha \rho } \tilde{R}^{\omega}{}_{\omega\tau }{}^{\gamma } + k_{150}^{} \tilde{R}^{\tau \rho \alpha }{}_{\gamma } \hat{R}_{\alpha \rho } \tilde{R}^{\omega}{}_{\omega\tau }{}^{\gamma } + k_{151}^{} \tilde{R}^{\tau }{}_{\gamma }{}^{\alpha \rho } \hat{R}_{\alpha \rho } \tilde{R}^{\omega}{}_{\omega\tau }{}^{\gamma } + k_{152}^{} \tilde{R}^{\alpha \rho }{}_{\tau \gamma } \hat{R}_{\alpha \rho }\tilde{R}^{\omega}{}_{\omega}{}^{\tau \gamma } \nonumber\\
&\,\,\,\,\,\,+ k_{153}^{} \tilde{R}^{\alpha }{}_{\tau }{}^{\rho }{}_{\gamma } \hat{R}_{\alpha \rho }\tilde{R}^{\omega}{}_{\omega}{}^{\tau \gamma } + k_{154}^{} \tilde{R}^{\rho \alpha }{}_{\tau \gamma } \hat{R}_{\alpha \rho }\tilde{R}^{\omega}{}_{\omega}{}^{\tau \gamma } + k_{155}^{} \tilde{R}^{\rho }{}_{\tau }{}^{\alpha }{}_{\gamma } \hat{R}_{\alpha \rho }\tilde{R}^{\omega}{}_{\omega}{}^{\tau \gamma }\,,\\
\mathcal{L}^{(3)}_{\tilde{R}_{\lambda\rho}^2\hat{R}_{\lambda\rho}}&=k_{156}^{} \tilde{R}_{\alpha }{}^{\rho } \tilde{R}_{\rho \tau } \hat{R}^{\alpha \tau } + k_{157}^{} \tilde{R}_{\alpha }{}^{\rho } \tilde{R}_{\rho \tau } \hat{R}^{\tau \alpha } + k_{158}^{} \tilde{R}_{\alpha }{}^{\rho } \tilde{R}_{\tau \rho } \hat{R}^{\tau \alpha } + k_{159}^{} \tilde{R}_{\alpha \tau } \tilde{R}^{\alpha }{}_{\rho } \hat{R}^{\tau \rho }\,,\\
\mathcal{L}^{(3)}_{\tilde{R}_{\lambda\rho}^2\tilde{R}^{\omega}{}_{\omega\lambda\rho}}&=k_{160}^{} \tilde{R}_{\alpha }{}^{\rho } \tilde{R}_{\rho \tau }\tilde{R}^{\omega}{}_{\omega}{}^{\alpha \tau }\,,\\
\mathcal{L}^{(3)}_{\tilde{R}_{\lambda\rho}^2\tilde{R}}&=(k_{30}^{} - k_{170}^{} -  k_{174}^{})\tilde{R}_{\gamma \mu } \tilde{R}^{\gamma\mu}\tilde{R}+ 
(k_{31}^{}- k_{171}^{} -  k_{175}^{}) \tilde{R}^{\gamma \mu } \tilde{R}_{\mu \gamma } \tilde{R}\,, \\
\mathcal{L}^{(3)}_{\tilde{R}_{\lambda\rho}\hat{R}_{\lambda\rho}^2}&=k_{161}^{} \tilde{R}_{\alpha \rho } \hat{R}^{\alpha }{}_{\tau } \hat{R}^{\rho \tau } + k_{162}^{} \tilde{R}_{\alpha \rho } \hat{R}^{\rho \tau } \hat{R}_{\tau }{}^{\alpha } + k_{163}^{} \tilde{R}_{\alpha \rho } \hat{R}^{\alpha }{}_{\tau } \hat{R}^{\tau \rho } + k_{164}^{} \tilde{R}_{\alpha \rho } \hat{R}_{\tau }{}^{\alpha } \hat{R}^{\tau \rho } \,,\\
\mathcal{L}^{(3)}_{\tilde{R}_{\lambda\rho}(\tilde{R}^{\omega}{}_{\omega\lambda\rho})^2}&= k_{165}^{} \tilde{R}_{\alpha \rho }\tilde{R}^{\lambda}{}_{\lambda}{}^{\alpha }{}_{\tau }\tilde{R}^{\omega}{}_{\omega}{}^{\rho \tau }\,,\\
\mathcal{L}^{(3)}_{\tilde{R}_{\lambda\rho}\hat{R}_{\lambda\rho}\tilde{R}^{\omega}{}_{\omega\lambda\rho}}&= k_{166}^{} \tilde{R}_{\alpha \rho } \hat{R}^{\rho }{}_{\tau }\tilde{R}^{\omega}{}_{\omega}{}^{\alpha \tau } + k_{167}^{} \tilde{R}_{\alpha \rho } \hat{R}_{\tau }{}^{\rho }\tilde{R}^{\omega}{}_{\omega}{}^{\alpha \tau } + k_{168}^{} \tilde{R}_{\alpha \rho } \hat{R}^{\alpha }{}_{\tau }\tilde{R}^{\omega}{}_{\omega}{}^{\rho \tau } + k_{169}^{} \tilde{R}_{\alpha \rho } \hat{R}_{\tau }{}^{\alpha }\tilde{R}^{\omega}{}_{\omega}{}^{\rho \tau }\,,\\
\mathcal{L}^{(3)}_{\tilde{R}_{\lambda\rho}\hat{R}_{\lambda\rho}\tilde{R}}&=k_{170}^{} \tilde{R}_{\alpha \rho } \hat{R}^{\alpha \rho } \tilde{R} + k_{171}^{} \tilde{R}_{\alpha \rho } \hat{R}^{\rho \alpha } \tilde{R}\,,\\
\mathcal{L}^{(3)}_{\tilde{R}_{\lambda\rho}\tilde{R}^{\omega}{}_{\omega\lambda\rho}\tilde{R}}&=k_{172}^{} \tilde{R}_{\alpha \rho }\tilde{R}^{\omega}{}_{\omega}{}^{\alpha \rho } \tilde{R}\,,\\
\mathcal{L}^{(3)}_{\hat{R}_{\lambda\rho}^2\tilde{R}^{\omega}{}_{\omega\lambda\rho}}&=k_{173}^{} \hat{R}_{\alpha }{}^{\rho } \hat{R}_{\rho \tau }\tilde{R}^{\omega}{}_{\omega}{}^{\alpha \tau }\,,\\
\mathcal{L}^{(3)}_{\hat{R}_{\lambda\rho}^2\tilde{R}}&=k_{174}^{} \hat{R}_{\alpha \rho } \hat{R}^{\alpha \rho } \tilde{R} + k_{175}^{} \hat{R}^{\alpha \rho } \hat{R}_{\rho \alpha } \tilde{R}\,,\\ 
\mathcal{L}^{(3)}_{\hat{R}_{\lambda\rho}(\tilde{R}^{\omega}{}_{\omega\lambda\rho})^2}&=k_{176}^{} \hat{R}_{\alpha \rho }\tilde{R}^{\lambda}{}_{\lambda}{}^{\alpha }{}_{\tau }\tilde{R}^{\omega}{}_{\omega}{}^{\rho \tau }\,,\\
\mathcal{L}^{(3)}_{\hat{R}_{\lambda\rho}\tilde{R}^{\omega}{}_{\omega\lambda\rho}\tilde{R}}&=k_{177}^{} \hat{R}_{\alpha \rho }\tilde{R}^{\omega}{}_{\omega}{}^{\alpha \rho } \tilde{R}\,,\\
\mathcal{L}^{(3)}_{\hat{R}^2\tilde{R}}&=k_{178}^{} \hat{R}_{\alpha \rho } \hat{R}^{\alpha \rho } \tilde{R}\,.
\end{align}}{\normalsize It should be noted that we have parametrised the 178 terms in such a way that, in the absence of nonmetricity, we obtain exactly the same Lagrangian densities as the ones presented in~\cite{Bahamonde:2024sqo}}. \normalsize As can be seen, this branch includes the highest number of independent cubic invariants, although in the particular case of Weyl-Cartan geometry the total number of independent invariants is $43$, and $31$ in Riemann-Cartan geometry~\cite{Bahamonde:2024sqo}.

\section{Stability analysis for the vector and axial sectors at cubic order}\label{sec:stability_cubicMAG}

By including all the possible mixing terms of cubic order between the curvature, torsion and nonmetricity tensors in the gravitational action of MAG that is reduced to GR in Riemannian geometry, the full Lagrangian density reads:
\begin{align}
    16\pi\mathcal{L}_{\rm Cubic}=&-R-\frac{1}{2}\left(2c_{1}+c_{2}\right)\tilde{R}_{\lambda\rho\mu\nu}\tilde{R}^{\mu\nu\lambda\rho}+\left(a_{2}-c_{1}\right)\tilde{R}_{\lambda\rho\mu\nu}\tilde{R}^{\rho\lambda\mu\nu}+a_{2}\tilde{R}_{\lambda\rho\mu\nu}\tilde{R}^{\lambda\rho\mu\nu}+a_{5}\tilde{R}_{\lambda\rho\mu\nu}\tilde{R}^{\lambda\mu\rho\nu}\nonumber\\
    &+a_{6}\tilde{R}_{\lambda\rho\mu\nu}\tilde{R}^{\rho\mu\lambda\nu}+\left(c_{2}-a_{5}+a_{6}\right)\tilde{R}_{\lambda\rho\mu\nu}\tilde{R}^{\mu\rho\lambda\nu}+\left(d_{1}-a_{10}-a_{12}\right)\tilde{R}_{\mu\nu}\tilde{R}^{\mu\nu}+a_{9}\tilde{R}_{\mu\nu}\tilde{R}^{\nu\mu}\nonumber\\
    &+a_{10}\hat{R}_{\mu\nu}\hat{R}^{\mu\nu}+a_{11}\hat{R}_{\mu\nu}\hat{R}^{\nu\mu}-\left(d_{1}+a_{9}+a_{11}\right)\tilde{R}_{\mu\nu}\hat{R}^{\nu\mu}+a_{12}\tilde{R}_{\mu\nu}\hat{R}^{\mu\nu}+a_{14}\tilde{R}^{\lambda}{}_{\lambda\mu\nu}\tilde{R}^{\rho}{}_{\rho}{}^{\mu\nu}
    \nonumber\\
    &+a_{15}\tilde{R}_{\mu\nu}\tilde{R}^{\lambda}{}_{\lambda}{}^{\mu\nu}+a_{16}\hat{R}_{\mu\nu}\tilde{R}^{\lambda}{}_{\lambda}{}^{\mu\nu}+\frac{1}{2}m_{T}^{2}T_{\mu}T^{\mu}+\frac{1}{2}m_{S}^{2}S_{\mu}S^{\mu}+\frac{1}{2}m_{t}^{2}t_{\lambda\mu\nu}t^{\lambda\mu\nu}+\frac{1}{2}m_{W}^{2}W_{\mu}W^{\mu}
    \nonumber\\
    &+ \frac{1}{2}m_{\Lambda}^{2}\Lambda_{\mu}\Lambda^{\mu}+\frac{1}{2}m_{\Omega}^{2}\Omega_{\lambda\mu\nu}\Omega^{\lambda\mu\nu}+\frac{1}{2}m_{q}^{2}q_{\lambda\mu\nu}q^{\lambda\mu\nu}+\frac{1}{2}\alpha_{TW}T_{\mu}W^{\mu}+\frac{1}{2}\alpha_{T\Lambda}T_{\mu}\Lambda^{\mu}+\frac{1}{2}\alpha_{W\Lambda}W_{\mu}\Lambda^{\mu}\nonumber\\
    &+\frac{1}{2}\alpha_{t\Omega}\varepsilon_{\rho\mu\nu\sigma}\Omega^{\lambda\rho\mu}t^{\nu}{}_{\lambda}{}^{\sigma}+\mathcal{L}_{\rm curv-tors}^{(3)}+ \mathcal{L}_{\rm curv-nonm}^{(3)}+ \mathcal{L}_{\rm curv-tor-nonm}^{(3)}\,.\label{Theory}
\end{align}
Then, focusing on the vector and axial modes of torsion and nonmetricity for the gravitational corrections to GR, the Lagrangian density includes the following contributions:
\begin{equation}
  16\pi\mathcal{L}_{\rm Cubic}=-\,R+\mathcal{L}_{1}+\mathcal{L}_{2}+\mathcal{L}_{3}+\mathcal{L}_{4}+\mathcal{L}_{5}+\mathcal{L}_{6}+\mathcal{L}_{7}+\mathcal{L}_{8}+\mathcal{L}_{9}+\mathcal{L}_{10}+\mathcal{L}_{11}+\mathcal{L}_{12}+\mathcal{L}_{13}+\mathcal{L}_{14}+\mathcal{L}_{15}\,,\label{EqVectorDes}
\end{equation}
where
\begin{align}
    \mathcal{L}_{1}=&\;l_{1}{} F^{(T)}_{\alpha \beta } F^{(T)}{}^{\alpha \beta } + \frac{1}{2} m_{T}{}^2 T_{\alpha } T^{\alpha } + l_{2}{} R T_{\alpha } T^{\alpha } + l_{3}{} G_{\alpha \beta } T^{\alpha } T^{\beta } + l_{4}{} T_{\alpha } T^{\alpha } T_{\beta } T^{\beta } + l_{5}{} T_{\alpha } T^{\alpha } \nabla_{\beta }T^{\beta }\,,\label{L1}\\
    \mathcal{L}_{2}=&\;l_{6}{} F^{(S)}_{\alpha \beta } F^{(S)}{}^{\alpha \beta } + \frac{1}{2} m_{S}{}^2 S_{\alpha } S^{\alpha } + l_{7}{} R S_{\alpha } S^{\alpha } + l_{8}^{} G_{\alpha \beta } S^{\alpha } S^{\beta } + l_{9}^{} S_{\alpha } S^{\alpha } S_{\beta } S^{\beta } + l_{10}^{} \nabla_{\alpha }S^{\alpha } \nabla_{\beta }S^{\beta }\,,\label{L2}\\
    \mathcal{L}_{3}=&\;l_{11}^{} F^{(W)}_{\alpha \beta } F^{(W)}{}^{\alpha \beta } + \frac{1}{2} m_{W}{}^2 W_{\alpha } W^{\alpha } + l_{12}^{} R W_{\alpha } W^{\alpha } + l_{13}^{} G_{\alpha \beta } W^{\alpha } W^{\beta }+l_{14}^{} W_{\alpha } W^{\alpha } W_{\beta } W^{\beta }\nonumber\\
    &+l_{15}^{} W_{\alpha } W^{\alpha } \nabla_{\beta }W^{\beta }\,,\label{L3}\\
    \mathcal{L}_{4}=&\;l_{16}^{} F^{(\Lambda)}_{\alpha \beta } F^{(\Lambda)}{}^{\alpha \beta } + \frac{1}{2} m_{\Lambda}{}^2 \Lambda_{\alpha } \Lambda^{\alpha } + l_{17}^{} R \Lambda_{\alpha } \Lambda^{\alpha } + l_{18}^{} G_{\alpha \beta } \Lambda^{\alpha } \Lambda^{\beta } + l_{19}^{} \Lambda_{\alpha } \Lambda^{\alpha } \Lambda_{\beta } \Lambda^{\beta } + l_{20}^{} R \nabla_{\alpha }\Lambda^{\alpha }\nonumber\\
    &+l_{21}^{} G_{\alpha}{}^{\beta}\nabla_{\beta}\Lambda^{\alpha}+l_{22}^{} \Lambda_{\alpha } \Lambda^{\alpha } \nabla_{\beta }\Lambda^{\beta }+ l_{23}^{} \nabla_{\alpha }\Lambda^{\alpha } \nabla_{\beta }\Lambda^{\beta } \,,\label{L4}\\
    \mathcal{L}_{5}=&\;l_{24}^{} S^{\alpha } S^{\beta } T_{\alpha } T_{\beta } + l_{25}^{} S_{\alpha } S^{\alpha } T_{\beta } T^{\beta } + l_{26}^{} S^{\alpha } T^{\beta } \nabla_{\alpha }S_{\beta } + l_{27}^{} S_{\alpha } S^{\alpha } \nabla_{\beta }T^{\beta }+ l_{28}^{} S^{\alpha } S^{\beta } \nabla_{\beta }T_{\alpha }  + l_{29}\ast{} F^{(T)}_{\alpha \rho } S^{\alpha } T^{\rho }\nonumber\\
    &\;+\frac{1}{2} l_{1}{} \ast F^{(S)}_{\mu \nu }F^{(T)}{}^{\mu \nu } \,,\label{L5}\\
    \mathcal{L}_{6}=&\;l_{30}^{} F^{(T)}_{\alpha \beta }F^{(W)}{}^{\alpha \beta } + \frac{1}{2} \alpha_{TW} T^{\alpha } W_{\alpha } + l_{31}^{} R T^{\alpha } W_{\alpha }+ l_{32}^{} G_{\alpha \beta } T^{\alpha } W^{\beta }  + l_{33}^{} T_{\alpha } T^{\alpha } T^{\beta } W_{\beta } + l_{34}^{} T^{\alpha } T^{\beta } W_{\alpha } W_{\beta } \nonumber\\
    &+l_{35}^{} T_{\alpha } T^{\alpha } W_{\beta } W^{\beta } + l_{36}^{} T^{\alpha } W_{\alpha } W_{\beta } W^{\beta } + l_{37}^{} T^{\alpha } W^{\beta } \nabla_{\alpha }T_{\beta } + l_{38}^{} W^{\alpha } W^{\beta } \nabla_{\beta }T_{\alpha } + l_{39}^{} W_{\alpha } W^{\alpha } \nabla_{\beta }T^{\beta }\nonumber\\
    &+ l_{40}^{} T^{\alpha } W^{\beta } \nabla_{\beta }W_{\alpha }+l_{41}^{} T_{\alpha } T^{\alpha } \nabla_{\beta }W^{\beta }+ l_{42}^{} T^{\alpha } T^{\beta } \nabla_{\beta }W_{\alpha } \,,\label{L6}\\
    \mathcal{L}_{7}=&\;l_{43}^{}F^{(T)}_{\alpha \rho } F^{(\Lambda)}{}^{\alpha \rho }+ \frac{1}{2}   \alpha_{T\Lambda}  \Lambda^{\alpha } + l_{44}^{} R T_{\alpha } \Lambda^{\alpha } + l_{45}^{} G^{\alpha }{}_{\rho } T_{\alpha } \Lambda^{\rho } + l_{46}^{} T_{\alpha } T^{\alpha } T_{\rho } \Lambda^{\rho } + l_{47}^{} T_{\alpha } T_{\rho } \Lambda^{\alpha } \Lambda^{\rho } + l_{48}^{} T_{\alpha } T^{\alpha } \Lambda_{\rho } \Lambda^{\rho } \nonumber\\
    &+l_{49}^{} T_{\alpha } \Lambda^{\alpha } \Lambda_{\rho } \Lambda^{\rho } + l_{50}^{} T_{\alpha } \Lambda^{\rho } \nabla^{\alpha }T_{\rho } + l_{51}^{} \Lambda^{\alpha } \Lambda^{\rho } \nabla_{\rho }T_{\alpha } + l_{52}^{} \Lambda_{\alpha } \Lambda^{\alpha } \nabla_{\rho }T^{\rho } + l_{53}^{} T_{\alpha } \Lambda^{\rho } \nabla_{\rho }\Lambda^{\alpha } + l_{54}^{} T_{\alpha } T^{\alpha } \nabla_{\rho }\Lambda^{\rho } \nonumber\\
    &+l_{55}^{} \nabla_{\alpha }T^{\alpha } \nabla_{\rho }\Lambda^{\rho } + l_{56}^{} T_{\alpha } T_{\rho } \nabla^{\rho }\Lambda^{\alpha } + l_{57}^{} \nabla_{\rho }T_{\alpha } \nabla^{\rho }\Lambda^{\alpha }\,,\label{L7}\\
    \mathcal{L}_{8}=&\;l_{58}^{} S_{\alpha } S_{\rho } W^{\alpha } W^{\rho } + l_{59}^{} S_{\alpha } S^{\alpha } W_{\rho } W^{\rho } + l_{60}^{} S_{\alpha } W^{\rho } \nabla^{\alpha }S_{\rho } + l_{61}^{} S_{\alpha } S^{\alpha } \nabla_{\rho }W^{\rho } + l_{62}^{} S_{\alpha } S_{\rho } \nabla^{\rho }W^{\alpha } \nonumber\\
    &+l_{63}\ast F^{(W)}_{\alpha\rho } S^{\alpha } W^{\rho }+\frac{1}{4} l_{30} \ast F^{(S)}_{\mu \nu } F^{(W)}{}^{\mu \nu } \,,\label{L8}\\
    \mathcal{L}_{9}=&\;l_{64}^{} S_{\alpha } S_{\rho } \Lambda^{\alpha } \Lambda^{\rho } + l_{65}^{} S_{\alpha } S^{\alpha } \Lambda_{\rho } \Lambda^{\rho } + l_{66}^{} S_{\alpha } \Lambda^{\rho } \nabla^{\alpha }S_{\rho } + l_{67}^{} S_{\alpha } S^{\alpha } \nabla_{\rho }\Lambda^{\rho }+l_{68}^{} S_{\alpha } S_{\rho } \nabla^{\rho }\Lambda^{\alpha }+l_{69}\ast F^{(\Lambda)}_{\alpha\rho } S^{\alpha } \Lambda^{\rho }\nonumber\\
    &\;+\frac{1}{4} (l_{43} -  \frac{1}{6} l_{84}) \ast F^{(S)}_{\mu \nu } F^{(\Lambda)}{}^{\mu \nu } \,,\label{L9}\\
    \mathcal{L}_{10}=&\;l_{70}^{}F^{(W)}_{\alpha \rho }F^{(\Lambda)}{}^{\alpha \rho } + \frac{1}{2} \alpha_{W\Lambda}  W^{\alpha } \Lambda_{\alpha } + l_{71}^{} R W^{\alpha } \Lambda_{\alpha } + l_{72}^{} W_{\alpha } W^{\alpha } W^{\rho } \Lambda_{\rho } + l_{73}^{} W^{\alpha } W^{\rho } \Lambda_{\alpha } \Lambda_{\rho } + l_{74}^{} G_{\alpha \rho } W^{\alpha } \Lambda^{\rho } \nonumber\\
    &+l_{75}^{} W_{\alpha } W^{\alpha } \Lambda_{\rho } \Lambda^{\rho } + l_{76}^{} W^{\alpha } \Lambda_{\alpha } \Lambda_{\rho } \Lambda^{\rho } + l_{77}^{} W^{\alpha } \Lambda^{\rho } \nabla_{\alpha }W_{\rho } + l_{78}^{} \Lambda^{\alpha } \Lambda^{\rho } \nabla_{\rho }W_{\alpha } + l_{79}^{} \Lambda_{\alpha } \Lambda^{\alpha } \nabla_{\rho }W^{\rho }\nonumber\\
    &+l_{80}^{} W^{\alpha } W^{\rho } \nabla_{\rho }\Lambda_{\alpha }+l_{81}^{} W^{\alpha } \Lambda^{\rho } \nabla_{\rho }\Lambda_{\alpha } + l_{82}^{} W_{\alpha } W^{\alpha } \nabla_{\rho }\Lambda^{\rho } + l_{83}^{} \nabla_{\alpha }W^{\alpha } \nabla_{\rho }\Lambda^{\rho } + l_{84}^{} \nabla_{\rho }W_{\alpha } \nabla^{\rho }\Lambda^{\alpha }\,,\label{L10}\\
    \mathcal{L}_{11}=&\;l_{85}^{} S^{\alpha } S^{\beta } T_{\alpha } W_{\beta } + l_{86}^{} S_{\alpha } S^{\alpha } T^{\beta } W_{\beta }  + l_{87}\ast F^{(S)}_{\alpha\rho } T^{\alpha } W^{\rho }+l_{88}\ast F^{(T)}_{\alpha\rho } S^{\alpha } W^{\rho }\,,\label{L11}\\
    \mathcal{L}_{12}=&\;l_{89}^{} S^{\alpha } S^{\beta } T_{\alpha } \Lambda_{\beta } + l_{90}^{} S_{\alpha } S^{\alpha } T^{\beta } \Lambda_{\beta }  + l_{91}\ast F^{(S)}_{\alpha \rho } T^{\alpha } \Lambda^{\rho }+l_{92}\ast F^{(T)}_{\alpha\rho } S^{\alpha } \Lambda^{\rho }\,,\label{L12}\\
    \mathcal{L}_{13}=&\;l_{93}^{} T_{\alpha } W_{\rho } W^{\rho } \Lambda^{\alpha } + l_{94}^{} T_{\alpha } T^{\alpha } W^{\rho } \Lambda_{\rho } + l_{95}^{} T_{\alpha } W^{\alpha } W^{\rho } \Lambda_{\rho } + l_{96}^{} T_{\alpha } W^{\rho } \Lambda^{\alpha } \Lambda_{\rho } + l_{97}^{} T_{\alpha } T_{\rho } W^{\alpha } \Lambda^{\rho } + l_{98}^{} T_{\alpha } W^{\alpha } \Lambda_{\rho } \Lambda^{\rho }  \nonumber\\
    &+l_{99}^{} W^{\alpha } \Lambda^{\rho } \nabla_{\alpha }T_{\rho }+ l_{100}^{} T_{\alpha } \Lambda^{\rho } \nabla^{\alpha }W_{\rho } + l_{101}^{} W^{\alpha } \Lambda^{\rho } \nabla_{\rho }T_{\alpha } + l_{102}^{} W^{\alpha } \Lambda_{\alpha } \nabla_{\rho }T^{\rho } + l_{103}^{} T_{\alpha } \Lambda^{\rho } \nabla_{\rho }W^{\alpha } \nonumber\\
    &+l_{104}^{} T_{\alpha } \Lambda^{\alpha } \nabla_{\rho }W^{\rho }\,,\label{L13}\\
    \mathcal{L}_{14}=&\;l_{105}^{} S^{\alpha } S^{\beta } W_{\alpha } \Lambda_{\beta } + l_{106}^{} S_{\alpha } S^{\alpha } W^{\beta } \Lambda_{\beta }  + l_{107} \ast F^{(S)}_{\alpha \rho } W^{\alpha } \Lambda^{\rho }+l_{108}{} \ast F^{(W)}_{\alpha\rho}S^{\alpha}\Lambda^{\rho}\,,\label{L14}\\
    \mathcal{L}_{15}=&\;l_{109}^{} \varepsilon_{\alpha \beta \tau \gamma } S^{\alpha } T^{\beta } W^{\tau } \Lambda^{\gamma }\,.\label{L15}
\end{align}
Note that, for simplicity, we have also introduced coefficients $\{l_{i}\}_{i=1}^{109}$, whose definitions can be found in~\ref{appendix:l}.

With this parametrisation, the kinetic matrix~\eqref{kinetic} for the vector and axial modes of the theory simply reads
\begin{eqnarray}\label{kappa}
    \kappa_{XY}=-\,2\left(
\begin{array}{cccc}
 2 l_{6}  & 0 & 0 & 0\\
 0 & 2 l_1& l_{30} &  l_{43}\\
 0 &l_{30} & 2 l_{11} & l_{70} \\
 0 & l_{43} & l_{70} & 2 l_{16} \\
\end{array}
\right)\,,
\end{eqnarray}
where the coefficients $l_i$ appearing in the matrix are yet associated with the quadratic curvature invariants of MAG (i.e. they do not depend on the parameters $h_{i}$), but they will be less constrained under the stability conditions due to the presence of cubic invariants that can cancel out the different Ostrogradsky instabilities of the theory. Likewise, it is also convenient to stress that the decoupling between the axial mode of torsion and the rest of vector modes allows their kinetic terms to be analysed in a separated way. Thus, while the axial mode can propagate safely if $l_6\leq 0$, the study of the vector modes focuses on  the $3 \times 3$ matrix: \begin{eqnarray}\label{kappa3D}
    \kappa_{IJ}^{\rm (vec)}=-\,2\left(
\begin{array}{ccc}
 2 l_1& l_{30} &  l_{43}\\
 l_{30} & 2 l_{11} & l_{70} \\
 l_{43} & l_{70} & 2 l_{16} \\
\end{array}
\right)\,.
\end{eqnarray}

Since the Lagrangian density~\eqref{Theory} constitutes a rather complicated expression involving a large number of interactions between the vector and axial modes of torsion and nonmetricity, it is first worthwhile to proceed with the stability analysis in different geometrical setups, in order to obtain a more comprehensive picture of the different cases of interest included in the theory. Thereby, in the next subsections we shall revisit first the restricted case of Riemann-Cartan geometry, heading then to the analysis in Weyl-Cartan geometry and in the torsion-free case. Finally, we shall consider the general metric-affine geometry.

\subsection{Riemann-Cartan geometry ($Q_{\lambda\mu\nu}=0\Longrightarrow W_\mu=\Lambda_\mu=0$)}\label{RCsector}

In Riemann-Cartan geometry, apart from the Einstein-Hilbert Lagrangian, only the Lagrangian densities~\eqref{L1},~\eqref{L2} and~\eqref{L5} are nonvanishing. Then, by demanding that both vector and axial modes of torsion propagate with kinetic terms $F^{(T)}_{\mu\nu}F^{(T)\mu\nu}$ and $F^{(S)}_{\mu\nu}F^{(S)\mu\nu}$, the kinetic matrix is simply reduced to
\begin{eqnarray}\label{kappa_RC}
    \kappa_{XY}=-\,4\left(
\begin{array}{cc}
 l_{6}  & 0 \\
 0 & l_1 \\
\end{array}
\right)\,,
\end{eqnarray}
with $X,Y=S,T$, while the terms of the form $\left(\nabla S\right)^2$, $S^{2}\nabla T$, $TS\nabla S$, $RS^{2}$ and $RT^{2}$ generate Ostrogradsky instabilities. These can be directly cancelled out from the action if:
\begin{equation}
    l_2=l_7=l_{10}=l_{26}=l_{27}=l_{28}=0\,,\label{EC}
\end{equation}
which provides the following conditions for the Lagrangian coefficients:
\begin{equation}\label{condEC}
    c_{2}=2c_{1}\,, \quad h_{2}=-\,\frac{h_{1}}{2}\,,\quad h_{3}= -\,\frac{1}{6}\left(c_{1}+6h_{13}\right),\quad h_{4}= \frac{h_{13}}{2}\,,\quad h_{14}= -\,2h_{13}\,,\quad h_{15}=4h_{13}\,.
\end{equation}

Likewise, the respective eigenvalues of the kinetic matrix~\eqref{kappa_RC} must be nonnegative, leading to:
\begin{equation}
    -\,\frac{d_1}{6} \leq c_{1} \leq -\,\frac{d_{1}}{3}\,, \quad d_1 \leq 0\,,\label{kincond}
\end{equation}
as found in Ref.~\cite{Bahamonde:2024sqo}.

\subsection{Weyl-Cartan geometry (${\nearrow\!\!\!\!\!\!\!Q}_{\lambda\mu\nu}=0\Longrightarrow\Lambda_\mu=0$)}

In Weyl-Cartan geometry, on top of the vector and axial modes of torsion, the Weyl vector of nonmetricity is also included in the stability analysis. Then, besides the conditions~\eqref{EC} involving only the aforementioned modes of torsion, we must also remove from the Lagrangian densities~\eqref{L3},~\eqref{L6} and~\eqref{L8} additional pathological terms of the form $S^2 \nabla W$, $T^2 \nabla W$, $W^2 \nabla T$, $SW\nabla S$, $TW\nabla T$, $TW\nabla W$, $RW^{2}$ and $RTW$, which requires:
\begin{eqnarray}
    l_{12}=l_{31}=l_{37}=l_{38}=l_{39}=l_{40}=l_{41}=l_{42}=l_{60}=l_{61}=l_{62}=0\,.\label{WC}
\end{eqnarray}
Overall, the Lagrangian coefficients must then satisfy the following conditions:
\begin{align}\label{weylc1a}
   c_{2}=&\;2c_{1}\,, \quad h_{2}=-\,\frac{h_{1}}{2}\,,\quad h_{3}= -\,\frac{1}{6}\left(c_{1}+6h_{13}\right),\quad h_{4}= \frac{h_{13}}{2}\,,\quad h_{14}= -\,2h_{13}\,,\quad h_{15}=4h_{13}\,,\\ 
   h_{48}=&\;\frac{9}{4}h_1-h_{47}\,,\quad h_{49}=\frac{3}{2}h_{113}=-\,\frac{9}{8}h_1\,,\quad h_{110}=-\,\frac{3}{2} h_{1}^{} -  h_{108}^{}\,,\quad h_{111} =3h_1-h_{109}\,, \label{weylc1b}\\
 h_{112}=&\;\frac{1}{8} \left(4 h_{108}^{}-9 h_{1}^{} - 4 h_{109}^{}\right),\quad h_{133}=3 h_{13}^{} + h_{132}^{}\,,\quad h_{134}=0\,,\quad h_{136}=6 h_{13}^{} -  h_{135}^{}\,.\label{weylc1c}
\end{align}

On the other hand, the kinetic matrix in this case reads
\begin{eqnarray}
    \kappa_{XY}=-\,2\left(
\begin{array}{ccc}
2 l_{6}& 0 & 0 \\
 0 &  2 l_1  & l_{30} \\
 0 &l_{30} & 2 l_{11} \\
\end{array}
\right)\,,
\end{eqnarray}
with $X,Y=S,T,W$. After diagonalisation, it acquires the form
\begin{eqnarray}
 \kappa_{XY}^{\rm diag}=   \left(
\begin{array}{ccc}
-\,4 l_{6} & 0 & 0 \\
 0 & -\,2 \sqrt{(l_1-l_{11})^2+l_{30}^2}-2 \left(l_{1}+l_{11}\right) 
  & 0 \\
 0 & 0 & 2 \sqrt{(l_1-l_{11})^2+l_{30}^2}-2 \left(l_{1}+l_{11}\right) \\
\end{array}
\right)\,,\label{WCkinetic}
\end{eqnarray}
which straightforwardly leads to the following constraints by demanding nonnegative eigenvalues:
\begin{equation}
     l_6\leq 0\,,\quad l_1\leq 0\,,\quad l_{11}\leq 0\,,\quad -\,2\sqrt{l_{1}l_{11}}\leq l_{30}\leq 2\sqrt{l_{1}l_{11}}\,.\label{WCcond1}
\end{equation}
Hence, besides the conditions~\eqref{weylc1a}-\eqref{weylc1c}, the Lagrangian coefficients must satisfy the following inequalities:
\begin{equation}
     -\,\frac{d_1}{6} \leq c_{1} \leq -\,\frac{d_{1}}{3}\,, \quad d_1 \leq 0\,, \quad a_{15}\leq \frac{1}{2} \left(a_9+a_{10}+a_{12}-2 a_2-a_5-4 a_{14}-d_1\right),\label{ineq_WC1}
\end{equation}
and
\begin{eqnarray}
    &&-\,2\sqrt{\left(6 c_1+2 d_1\right)\left[d_1+a_5+2\left(a_2+2 a_{14}+a_{15}\right)-a_9-a_{10}-a_{12}\right]}\leq \nonumber\\
    &&a_6+a_9-a_{11}+a_{12}-3 d_1-2 \left(2 c_1+a_{15}+a_{16}+a_5-a_{10}\right)\leq\nonumber\\
    &&2\sqrt{\left(6 c_1+2 d_1\right)\left[d_1+a_5+2\left(a_2+2 a_{14}+a_{15}\right)-a_9-a_{10}-a_{12}\right]}\,.\label{ineq_WC2}
\end{eqnarray}

In summary, it is also possible to obtain a theory with healthy vector and axial modes for torsion and nonmetricity in Weyl-Cartan geometry, if the conditions~\eqref{weylc1a}-\eqref{weylc1c},~\eqref{ineq_WC1} and~\eqref{ineq_WC2} hold.

A strong but still meaningful restriction within this geometry is the case of Weyl geometry, where torsion identically vanishes and the nonmetricity tensor is completely determined by the Weyl vector. In this geometry, the Lagrangian density of the Weyl vector acquires the form~\eqref{L3}, where the possible instabilities arise only around general backgrounds due to the presence of the term $RW^{2}$, which can be directly removed from the action if:
\begin{equation}
    l_{12}=0\,,
\end{equation}
namely:
\begin{eqnarray}
    h_{49}=-\,\frac{1}{2}\left(h_{47}+h_{48}\right).\label{cond1_W}
\end{eqnarray}

Furthermore, the appropriate sign in the kinetic term of the Weyl vector requires $l_{11}\leq 0$, which means:
\begin{eqnarray}
    a_{15}\leq \frac{1}{2}\left(a_{10}+a_{12}-4 a_{14}-2 a_{2}-a_{5}+a_{9}-d_1\right).\label{cond2_W}
\end{eqnarray}
Then, under the conditions~\eqref{cond1_W} and~\eqref{cond2_W}, the Lagrangian density for the MAG theory with a Weyl vector reads
\begin{align}
  \mathcal{L}=&-R+\left(d_{1}^{}+ 2 a_{2}^{}+ a_{5}^{}+ 4 a_{14}^{} + 2 a_{15}^{} -  a_{9}^{}- a_{10}^{} -  a_{12}^{}\right) F^{(W)}_{\alpha \rho } F^{(W)}{}^{\alpha \rho } + \frac{1}{2} m_{W}{}^2 W_{\alpha } W^{\alpha }  \nonumber\\
  &+\left(h_{47}^{} + h_{48}^{}\right) G_{\alpha \rho } W^{\alpha } W^{\rho }+ \frac{3}{4} \left(h_{47}^{} + h_{48}^{}\right) W_{\alpha } W^{\alpha } W_{\rho } W^{\rho } -  \frac{3}{2} \left(h_{47}^{} + h_{48}^{}\right) W_{\alpha } W^{\alpha } \nabla_{\rho }W^{\rho }\,,\label{MAG_W}
\end{align}
which generalises the Einstein-Proca Lagrangian with a quartic potential, as well as with healthy Galileon interactions and nonminimal couplings to the Riemannian gravity sector. In fact, it is worthwhile to stress that similar generalisations were also found in different geometrical setups, such as the one characterised by a vector distortion,  where the torsion and nonmetricity tensors are assumed to be proportional to a vector field~\cite{BeltranJimenez:2015pnp,BeltranJimenez:2016rff,BeltranJimenez:2016wxw}, while in the case of Weyl geometry they naturally arise when considering cubic order invariants in the gravitational action.

\subsection{Torsion-free case ($T_{\lambda\mu\nu}=0\Longrightarrow T_\mu=S_\mu=0$)}

In the absence of torsion, only the Lagrangians densities~\eqref{L3},~\eqref{L4} and~\eqref{L10} provide the dynamics of the respective vector modes of the nonmetricity tensor. In particular, the Lagrangians densities~\eqref{L3} and~\eqref{L4} include pathological terms of the form $\left(\nabla \Lambda\right)^2$, $RW^{2}$, $R\Lambda^{2}$ and $R\nabla\Lambda$, which can be eliminated by setting:
\begin{equation}
    l_{12}=l_{17}=l_{20}=l_{23}=0\,.\label{casel1}
\end{equation}
As for the Lagrangian density~\eqref{L10}, the identity
\begin{equation}
    \int\left(\nabla_{\mu}\Lambda^{\mu}\nabla_{\nu}W^{\nu}-\nabla_{\mu}\Lambda^{\nu}\nabla_{\nu}W^{\mu}\right)\sqrt{-\,g}d^{4}x = \int (G_{\mu\nu}+\frac{1}{2} g_{\mu\nu}R)W^{\mu}\Lambda^{\nu}\sqrt{-\,g}d^{4}x\,,\label{identity1}
\end{equation}
allows us to rewrite it in the action as (up to boundary terms):
\begin{align}
     \mathcal{L}_{10}=&\;\frac{1}{2}(2l_{70}^{}-l_{83}) F^{(\Lambda)}_{\alpha \rho} F^{(W)}{}^{\alpha \rho } +  \frac{1}{2} \alpha_{W\Lambda} W^{\alpha } \Lambda_{\alpha } + \frac{1}{2}(2l_{71}^{}+l_{83}) R W^{\alpha } \Lambda_{\alpha } + l_{72}^{} W_{\alpha } W^{\alpha } W^{\rho } \Lambda_{\rho } + l_{73}^{} W^{\alpha } W^{\rho } \Lambda_{\alpha } \Lambda_{\rho } \nonumber\\
  &+ (l_{74}^{}+l_{83}) G_{\alpha \rho } W^{\alpha } \Lambda^{\rho } + l_{75}^{} W_{\alpha } W^{\alpha } \Lambda_{\rho } \Lambda^{\rho } + l_{76}^{} W^{\alpha } \Lambda_{\alpha } \Lambda_{\rho } \Lambda^{\rho } + l_{77}^{} W^{\alpha } \Lambda^{\rho } \nabla_{\alpha }W_{\rho } + l_{78}^{} \Lambda^{\alpha } \Lambda^{\rho } \nabla_{\rho }W_{\alpha }  \nonumber\\
  &+ l_{79}^{} \Lambda_{\alpha } \Lambda^{\alpha } \nabla_{\rho }W^{\rho }+ l_{80}^{} W^{\alpha } W^{\rho } \nabla_{\rho }\Lambda_{\alpha } + l_{81}^{} W^{\alpha } \Lambda^{\rho } \nabla_{\rho }\Lambda_{\alpha } + l_{82}^{} W_{\alpha } W^{\alpha } \nabla_{\rho }\Lambda^{\rho } +(l_{83}+l_{84})\nabla_\mu \Lambda^\nu \nabla^\mu W_\nu\,.\label{L10b}
 \end{align}
As can be seen, the Lagrangian displays pathological terms of the form $\nabla W \nabla\Lambda$, $W^{2}\nabla\Lambda$, $\Lambda^{2}\nabla W$, $W\Lambda\nabla W$, $W\Lambda\nabla\Lambda$ and $RW\Lambda$, which can be directly removed from the action if:
 \begin{equation}
    l_{77}=l_{78}=l_{79}=l_{80}=l_{81}=l_{82}=2l_{71}+l_{83}=l_{83}+l_{84}=0\,.\label{casel2}
 \end{equation}
Thereby, by combining the conditions~\eqref{casel1} and~\eqref{casel2}, the Lagrangian coefficients must satisfy:
\begin{align}\label{noT1}
    c_2=&\;2 c_{1}{} -2 \left(4 a_{10}^{} + 4 a_{11}^{} + 2 a_{2}^{} + a_{6}^{}\right),\quad a_{12}=  -\,2 a_{10}^{} -  a_{11}^{} + a_{9}^{} + d_{1}^{}\,,\\
    h_{48}=&\,-h_{49}=-\,\frac{16}{9}h_{52}=-\,\frac{4}{3}h_{70}=\frac{2}{3}h_{71}=-\,\frac{4}{3}h_{73}=h_{47}\,,\\
    h_{51}=&\;\frac{9}{8} h_{47}^{} -  h_{50}^{}\,,\quad h_{69}=\frac{3}{4} h_{47}^{} -  h_{68}^{}\,,\quad h_{72}=\frac{3}{4} h_{47}+h_{68}\,.\label{noT3}
\end{align}

Furthermore, the kinetic matrix for this geometry can also be diagonalised as
\begin{eqnarray}
 \kappa_{XY}^{\rm diag}=\left(
\begin{array}{cc}
 -\,2 \sqrt{(l_{11}-l_{16})^2+l_{70}^2}-2\left(l_{11}+l_{16}\right) & 0 \\
 0 & 2 \sqrt{(l_{11}-l_{16})^2+l_{70}^2}-2\left(l_{11}+l_{16}\right) \\
\end{array}
\right),\label{kineticTfree}
 \end{eqnarray}
with $X, Y = W,\Lambda$. Hence, it is clear that the eigenvalues of the kinetic matrix are nonnegative if:
\begin{equation}
   l_{11}\leq 0\,,\quad l_{16}\leq0 \,,\quad -\,2\sqrt{l_{11}l_{16}}\leq l_{70}\leq 2\sqrt{l_{11}l_{16}}\,,\label{condTfree}
\end{equation}
which, using the conditions~\eqref{noT1}-\eqref{noT3}, provides the following inequalities for the Lagrangian coefficients:
\begin{eqnarray}
&&a_{10}\leq 2 a_9-2 a_2-a_5-a_{11}-4 a_{14}-2 a_{15}\,,\quad a_{11}\leq -\,\frac{1}{5} \left(6 a_2+3 a_5+11 a_{10}\right), \quad \textrm{and}\label{torsionfree_ineq1}\\
&&-\,\sqrt{3\left(6 a_2+3 a_5+11 a_{10}+5 a_{11}\right) \left(2 a_2+a_5-2 a_9+a_{10}+a_{11}+4 a_{14}+2 a_{15}\right)}\leq \nonumber\\
&&-\,3 \left(2 a_2+a_5+a_9+3 a_{10}+4 a_{11}+a_{16}+d_1\right)\leq \nonumber\\
&&   \sqrt{3\left(6 a_2+3 a_5+11 a_{10}+5 a_{11}\right) \left(2 a_2+a_5-2 a_9+a_{10}+a_{11}+4 a_{14}+2 a_{15}\right)}\,.\label{torsionfree_ineq2}
\end{eqnarray}
\normalsize 
Thus, in general it is possible to avoid gravitational instabilities in the vector sector of the torsion-free case if the Lagrangian coefficients satisfy the relations~\eqref{noT1}-\eqref{noT3}, as well as the inequalities~\eqref{torsionfree_ineq1} and \eqref{torsionfree_ineq2}.

\subsection{General metric-affine geometry}

For the case with general torsion and nonmetricity tensors, all of the vector and axial modes of these tensors are present in the gravitational action. Hence, on top of the conditions~\eqref{EC},~\eqref{WC},~\eqref{casel1} and~\eqref{casel2}, further constraints must be imposed on the Lagrangian coefficients to eliminate Ostrogradsky instabilities associated with terms of the form $\nabla T \nabla\Lambda$, $T^{2}\nabla\Lambda$, $S^{2}\nabla\Lambda$, $\Lambda^{2}\nabla T$, $T\Lambda\nabla T$, $T\Lambda\nabla\Lambda$, $S\Lambda\nabla S$, $T\Lambda\nabla W$, $W\Lambda\nabla T$ and $RT\Lambda$, which appear in the Lagrangian densities~\eqref{L7},~\eqref{L9} and~\eqref{L13}. In this sense, by using the identity
\begin{equation}
    \int\left(\nabla_{\mu}\Lambda^{\mu}\nabla_{\nu}T^{\nu}-\nabla_{\mu}\Lambda^{\nu}\nabla_{\nu}T^{\mu}\right)\sqrt{-\,g}\,d^{4}x = \int (G_{\mu\nu}+\frac{1}{2} g_{\mu\nu}R)T^{\mu}\Lambda^{\nu}\sqrt{-\,g}\,d^{4}x\,,\label{identity2}
\end{equation}
we can first rewrite the Lagrangian density~\eqref{L7} in the action as (up to boundary terms):
\begin{align}
  \mathcal{L}_{7}=&\;\frac{1}{2} (2l_{43}^{} -  l_{55}^{}) F^{(T)}_{\alpha \rho }F^{(\Lambda)}{}^{\alpha \rho } +\frac{1}{2}   \alpha_{T\Lambda} T_{\alpha } \Lambda^{\alpha } +\frac{1}{2} (2l_{44}^{} + l_{55}^{}) R T_{\alpha } \Lambda^{\alpha } + (l_{45}^{} + l_{55}^{}) G^{\alpha }{}_{\rho } T_{\alpha } \Lambda^{\rho } + l_{46}^{} T_{\alpha } T^{\alpha } T_{\rho } \Lambda^{\rho }\nonumber\\
  &+ l_{47}^{} T_{\alpha } T_{\rho } \Lambda^{\alpha } \Lambda^{\rho } + l_{48}^{} T_{\alpha } T^{\alpha } \Lambda_{\rho } \Lambda^{\rho } + l_{49}^{} T_{\alpha } \Lambda^{\alpha } \Lambda_{\rho } \Lambda^{\rho } + l_{50}^{} T_{\alpha } \Lambda^{\rho } \nabla^{\alpha }T_{\rho } + l_{51}^{} \Lambda^{\alpha } \Lambda^{\rho } \nabla_{\rho }T_{\alpha } + l_{52}^{} \Lambda_{\alpha } \Lambda^{\alpha } \nabla_{\rho }T^{\rho } \nonumber\\
  &+ l_{53}^{} T_{\alpha } \Lambda^{\rho } \nabla_{\rho }\Lambda^{\alpha } + l_{54}^{} T_{\alpha } T^{\alpha } \nabla_{\rho }\Lambda^{\rho } + l_{56}^{} T_{\alpha } T_{\rho } \nabla^{\rho }\Lambda^{\alpha } + (l_{55}^{} + l_{57}^{}) \nabla_{\rho }T_{\alpha } \nabla^{\rho }\Lambda^{\alpha }\,. \label{L7B}
\end{align}
Then, in order to eliminate the aforementioned pathological terms appearing in Expression~\eqref{L7B},~\eqref{L9} and~\eqref{L13}, we require:
\begin{align}
    l_{50}=&\;l_{51}=l_{52}=l_{53}=l_{54}=l_{56}=2l_{44}+l_{55}=l_{55}+l_{57}=0\,,\label{cond1generalMAG}\\
  l_{66}=&\;l_{67}=l_{68}=l_{99}=l_{100}=l_{101}=l_{102}=l_{103}=l_{104}=0\,.\label{cond2generalMAG}
\end{align}
Overall, in terms of the Lagrangian coefficients, the conditions~\eqref{EC},~\eqref{WC},~\eqref{casel1},~\eqref{casel2},~\eqref{cond1generalMAG} and~\eqref{cond2generalMAG} can be expressed as follows:
\begin{align}
    c_{2}&=2c_{1}\,, \quad a_{11}= -\,\frac{1}{4} \left(2 a_{2}^{} +  a_{6}^{}+4 a_{10}^{}\right),\quad a_{12}=\frac{1}{4} \left(4 d_{1}^{}+2 a_{2}^{} + a_{6}^{} + 4 a_{9}^{}-4 a_{10}^{}\right),\label{MAG1}\\
    h_{3}&= -\,\frac{1}{6}\left(c_{1}+6h_{13}\right),\quad h_{4}= \frac{h_{13}}{2}\,,\quad h_{14}= -\,2h_{13}\,,\quad h_{15}=4h_{13}\,,\quad h_{51}=-\,h_{50}\,,\\
    h_{133}&=3h_{13}+h_{132}\,,\quad h_{136}=6 h_{13}^{} -  h_{135}^{}\,,\quad h_{139}=\frac{1}{12} \left(2 a_{5}^{} -  a_{6}^{} + c_{1}{} - 9 h_{13}^{} + 12 h_{138}^{} + 36 h_{28}^{}\right),\\
    h_{140}&=\frac{1}{12} \left(2 a_{5}^{} -  a_{6}^{} - 8 c_{1}{} - 36 h_{13}^{} - 72 h_{28}^{}\right), \\
    h_{142}&=\frac{1}{24} \left(12 a_{10}^{} + 6 a_{2}^{} + 4 a_{5}^{} + a_{6}^{} + 12 a_{9}^{} + 20 c_{1}{} + 36 h_{13}^{} - 24 h_{141}^{} + 288 h_{28}^{}\right),\\
    h_1&=h_2=h_{27}=h_{47}=h_{48}=h_{49}=h_{52}=h_{68}=h_{69}=h_{70}=h_{71}=h_{72}=h_{73}=h_{108}=0\,,\\
    h_{109}&=h_{110}=h_{111}=h_{112}=h_{113}=h_{114}=h_{115}=h_{116}=h_{117}=h_{118}=h_{119}=h_{134}=0\,.\label{MAG6}
\end{align}
Therefore, these constraints reduce the parameter space of the present cubic MAG theory from $233$ to $194$ parameters.

On the other hand, as for the requirement of ghost-free kinetic terms for the vector and axial modes, we will assume that in general all of them can propagate. In that case, the kinetic term of the axial mode is healthy if
\begin{align}
    l_6\leq 0\quad \Longleftrightarrow \quad d_1 \geq -\,6 c_1\,.\label{condKinetic1}
\end{align}
For the rest of vector modes, we need to impose the $3 \times 3$ kinetic matrix~\eqref{kappa3D} to be positive semidefinite. Explicitly, under the previous stability conditions~\eqref{MAG1}-\eqref{MAG6}, the coefficients $l_i$ appearing in the matrix read:
\begin{align}
    l_1&=\frac{2}{9} \left(3 c_1+d_1\right),\quad l_{11}=\frac{3 a_2}{2}+a_5-2 a_9+4 a_{14}+2 a_{15}-\frac{a_6}{4}\,,\quad l_{16}=\frac{3}{64} \left(14 a_2+12 a_5-5 a_6+24 a_{10}\right),\\
    l_{30}&=\frac{1}{6} \left(2 a_2-4 a_5+3 a_6+4 a_9+4 a_{10}-4 a_{15}-4 a_{16}-8 c_1-4 d_1\right),\\
    l_{43}&=\frac{1}{8} \left(2 a_2+4 a_5-a_6+4 a_9+4 a_{10}+8 c_1+4 d_1\right),\quad
    l_{70}=-\frac{3}{2}  \left(a_5-a_6+a_9-a_{10}+a_{16}+d_1\right).
\end{align}
Hence, the corresponding characteristic polynomial acquires the form
\begin{equation}
    p(\lambda)=-\,\lambda^{3}+s_{2}\lambda^{2}+s_{1}\lambda+s_{0}\,,\label{poly}
\end{equation}where\begin{align}
    s_{0}&=16 \left[l_{16} l_{30}^2+l_{43} \left(l_{11} l_{43}-l_{30} l_{70}\right)+l_1 \left(l_{70}^2-4 l_{11} l_{16}\right)\right],\label{s0}\\
    s_{1}&=4 \left[l_{30}^2+l_{43}^2+l_{70}^2-4 l_1 l_{11}-4 \left(l_1+l_{11}\right) l_{16}\right],\label{s1}\\
    s_{2}&=-\,4 \left(l_1+l_{11}+l_{16}\right).\label{s2}
\end{align}
Accordingly, in comparison with the case of quadratic MAG, the coefficients~\eqref{s0}-\eqref{s2} are less constrained by the removal of the Ostrogradsky instabilities of the theory. Thereby, in order then to have three positive real roots in~\eqref{poly}, we must simply demand the three inequalities
\begin{eqnarray}
    s_2\geq 0\,,\quad s_1\leq 0\,,\quad s_0\geq 0\,,
\end{eqnarray}
namely 
\begin{eqnarray}
 &&   l_1+l_{11}+l_{16}\leq 0\,,\quad 4 l_{11} l_{16}+4 l_1 \left(l_{11}+l_{16}\right)-l_{30}^2-l_{43}^2-l_{70}^2\geq 0\,,\label{condKinetic2}\\
    &&l_{30}l_{43}l_{70}+4 l_1 l_{11} l_{16}-l_{16} l_{30}^2-l_{11} l_{43}^2-l_1 l_{70}^2\leq 0\,.\label{condKinetic3}
\end{eqnarray}

In conclusion, to avoid Ostrogradsky instabilities in the vector and axial sectors of the general metric-affine geometry, we require first the conditions~\eqref{MAG1}-\eqref{MAG6}, while to ensure ghost-free kinetic terms in these sectors the conditions~\eqref{condKinetic1}, \eqref{condKinetic2} and \eqref{condKinetic3} must also hold. Thus, in contrast with quadratic MAG, in cubic MAG the vector and axial modes of the torsion and nonmetricity tensors find a safer context to propagate in a healthy way.

\section{Reissner-Nordström-like black hole solutions}\label{sec:RN}

In analogy to quadratic MAG, it turns out that the action given by the Lagrangian density~\eqref{Theory} with cubic order invariants constructed from the curvature, torsion and nonmetricity tensors admits Reissner-Nordström-like black hole solutions with dynamical torsion and nonmetricity. Specifically, by considering a static and spherically symmetric setup~\cite{Hohmann:2019fvf}:
\begin{equation}\label{sph_metric}
    g_{tt}=\Psi_{1}(r)\,, \quad g_{rr}=-\,\frac{1}{\Psi_{2}(r)}\,, \quad g_{\vartheta\vartheta}=g_{\varphi\varphi}\csc^2\vartheta=-\,r^2\,,
\end{equation}
\begin{align}\label{sph_torsion}
    T^t\,_{t r} &= t_{1}(r)\,, \quad T^r\,_{t r} = t_{2}(r)\,, \quad T^\vartheta\,_{t \vartheta} = T^\varphi\,_{t \varphi} = t_{3}(r)\,, \quad T^\vartheta\,_{r \vartheta} = T^\varphi\,_{r \varphi} =  t_{4}(r)\,, \\
    T^\vartheta\,_{t \varphi} &= T^\varphi\,_{\vartheta t} \sin^{2}\vartheta = t_{5}(r) \sin{\vartheta}\,, \quad T^\vartheta\,_{r \varphi} = T^\varphi\,_{\vartheta r} \sin^{2}\vartheta = t_{6}(r) \sin{\vartheta}\,, \\
    T^t\,_{\vartheta \varphi} &= t_{7}(r) \sin\vartheta\,, \quad T^r\,_{\vartheta \varphi} = t_{8}(r) \sin \vartheta\,,
\end{align}
\begin{align}\label{sph_nonmetricity}
    Q_{t t t} &= q_1(r)\,, \quad Q_{t r r}=q_2(r)\,, \quad Q_{t t r}= q_3(r)\,, \quad Q_{t \vartheta \vartheta}=Q_{t \varphi \varphi}\csc^2\vartheta=q_4(r)\,, \\
    Q_{r t t}&=q_5(r)\,, \quad Q_{r r r}=q_6(r)\,, \quad Q_{r t r}= q_7(r)\,, \quad Q_{r \vartheta \vartheta}=Q_{r \varphi \varphi}\csc^2\vartheta=q_8(r)\,, \\
    Q_{\vartheta t \vartheta}&= Q_{\varphi t \varphi}\csc^2\vartheta =  q_9(r)\,, \quad
    Q_{\vartheta r \vartheta}= Q_{\varphi r \varphi}\csc^2\vartheta = q_{10}(r)\,, \\
    Q_{\vartheta t \varphi}&= -\,Q_{\varphi t \vartheta} = q_{11}(r) \sin \vartheta\,, \quad Q_{\vartheta r \varphi}= -\,Q_{\varphi r \vartheta} = q_{12}(r) \sin \vartheta\,,
\end{align}
as well as the stability conditions~\eqref{MAG1}-\eqref{MAG6} for the vector and axial sectors of the theory, a general solution of the Euler-Lagrange equations takes the form:
\begin{align}
    t_1(r)=&\;\frac{\Psi'(r)}{2\Psi(r)}+\frac{wr}{\Psi(r)}+\frac{\kappa_{\rm d}}{2r\Psi(r)}+\frac{\kappa_{\rm sh}}{2r\Psi(r)}\,,\quad t_4(r)=-\,\frac{1}{2r}-\frac{wr}{2\Psi(r)}-\frac{\kappa_{\rm d}}{2r\Psi(r)}+\frac{\kappa_{\rm sh}}{2r\Psi(r)}\,, \quad t_{6}(r)=\frac{\kappa_{\rm s}}{r\Psi(r)}\,,\\
    t_{2}(r)=&\;t_{1}(r)\Psi(r)\,,\quad t_{3}(r)=-\,t_{4}(r)\Psi(r)\,, \quad t_{5}(r)=-\,t_{6}(r)\Psi(r)\,, \quad t_{7}(r)=t_{8}(r)=0\,,\label{t7}\\
    q_{1}(r)=&\;q_7(r) \Psi^{2} (r)+\frac{(\kappa_{\rm d}-3 \kappa_{\rm sh}) \Psi (r)}{r}\,,\quad    q_2(r)=q_7(r)-\frac{\kappa_{\rm d}+\kappa_{\rm sh}}{r \Psi (r)}\,, \quad q_3(r)=\frac{2 \kappa_{\rm sh}}{r}-q_7(r) \Psi (r)\,,\\
    q_4(r)=&-r\left(\kappa_{\rm d}+\kappa_{\rm sh}\right)\,, \quad q_{5}(r)=\frac{\kappa_{\rm sh}-\kappa_{\rm d}}{r}-q_7(r) \Psi (r)\,,\quad  q_{6}(r)=\frac{\kappa_{\rm d}-\kappa_{\rm sh}}{r\Psi^{2}(r)}-\frac{q_7(r)}{\Psi (r)}\,, \quad q_9(r)=-\,2\kappa_{\rm sh} r\,,\\
    q_{7}(r)=&\;\frac{r^2}{\Psi^{2}(r)}\int \tilde{q}_{7}(r)dr\,,\quad q_{8}(r)=\frac{r \left(\kappa_{\rm d}+\kappa_{\rm sh}\right)}{\Psi (r)}\,, \quad q_{10}(r)=\frac{2 \kappa_{\rm sh} r}{\Psi (r)}\,,\quad q_{11}(r)=q_{12}(r)=0\,,\\
    \tilde{q}_7(r)=&\;\frac{2 \kappa_{\rm sh}}{15r^3}\Big\{45 N_3 \Psi'(r)+\frac{wr}{2 a_2+a_6}\Big[ 18 a_2 \left(46 N_3-31\right)+9 a_6 \left(46 N_3-31\right)+4 \big(36 h_{53}-36 h_{55}+36 h_{56}+36 h_{57}\nonumber\\
    &+108 h_{59}-108 h_{61}+1179 h_{65}-605 h_{66}+24 h_{91}+24 h_{92}+24 h_{94}+16 h_{95}-23 h_{96}+51 h_{97}+156 h_{103}\nonumber\\
    &+246 h_{104}+90 h_{105}-45 (h_{99}+h_{101}+h_{102})\big)\Big]\Big\}\,,\\
    \Psi(r)=&\;1-\frac{2 m}{r}+\frac{1}{r^2}\bigg\{\frac{1}{3} \kappa_{\rm s}^2 \left(d_1+4 h_{25}\right)+\frac{1}{2}\kappa_{\rm d}^2 \left(2a_{6}-4a_{2}-8 a_{14}-d_{1}\right)+\frac{1}{15} \kappa_{\rm sh}^2\big[6 a_2 \left(216 N_3-221\right)+a_6 \left(648 N_3-663\right)\nonumber\\
    &+8\left(36 h_{53}-36 h_{55}+36 h_{56}+36 h_{57}+108 h_{59}-108 h_{61}+1179 h_{65}-605 h_{66}+24 h_{91}+24 h_{92}+24 h_{94}+16 h_{95}
    \right.
    \nonumber\\
    &
    \left.-\,23 h_{96}+51 h_{97}-45 h_{99}-45 h_{101}-45 h_{102}+156 h_{103}+246 h_{104}+90 h_{105}\right)\big]\bigg\}\,.
\end{align}
where $m$, $\kappa_{\rm s},\kappa_{\rm d}$ and $\kappa_{\rm sh}$ represent the parameters related to the mass, spin, dilation and shear charges of the solution, $w$ provides mass terms for the tensor modes of torsion and nonmetricity, and $N_1$ and $N_3$ are proportionality constants between the Lagrangian coefficients\footnote{For simplicity in the presentation, we omit another proportionality constant $N_2$ providing a nonvanishing torsion function $t_7(r)=N_2\kappa_{\rm s} r/\Psi (r)$ and further corrections in the metric function $\Psi(r)$. In addition, there are several special cases arising for certain relations between $N_1$ and $N_2$, which can also introduce a mass for the axial mode of torsion, in line with previous results found in Ref.~\cite{Bahamonde:2024sqo}.}. In this regard, on top of the stability conditions~\eqref{MAG1}-\eqref{MAG6}, the remaining Lagrangian coefficients satisfy additional relations, which for those associated with the quadratic curvature, torsion and nonmetricity invariants read:
\begin{align}
c_2=&\;2c_1=-\,\frac{d_1}{2}\,, \quad a_{2} \neq -\,\frac{a_{6}}{2}\,,\quad a_5=\frac{1}{4} \left(2 a_6-d_1\right), \quad a_9=a_{11}=-\,\frac{1}{16}\left[3\left(d_1+2 a_6\right)+4 a_2\right],\\
a_{10}=&\;\frac{1}{16}\left(3 d_1-4 a_2+2 a_6\right), \quad a_{12}=\frac{1}{8}\left(4 a_2-2 a_6+5 d_1\right),\quad a_{15}=\frac{1}{4}\left(d_1-4 a_6\right),\quad a_{16}=a_6-\frac{d_1}{4}\,,\\
m_T^2=&\;m_S^2=m_W^2=m_{\Lambda }^2=\alpha _{TW}=\alpha _{T\Lambda}=\alpha_{W\Lambda}= 0\,,\quad w=\frac{m_t^2}{3\left(2 h_6-2 h_{25}-d_1\right)}\,,\\
m_{\Omega}^2=&\;\frac{w}{180}\big\{24786 a_2 \left(N_3-1\right)+12393 a_6 \left(N_3-1\right)+2\big[2484 h_{53}-270 h_{54}-2754 h_{55}+3024 h_{56}+2754 h_{57}-270 h_{58}\nonumber\\
&+8082 h_{59}+360 h_{60}-8802 h_{61}+87216 h_{65}-49290 h_{66}+240 h_{90}+2136 h_{91}+1896 h_{92}+1896 h_{94}\nonumber\\
&+7\bigl(172 h_{95}-241 h_{96}+557 h_{97}-375 h_{99}-33\left(15 h_{101}+15 h_{102}-44 h_{103}-74 h_{104}-30 h_{105}\right)\bigr)\big]\big\}\,,\\
m_q^2=&\;\frac{2w}{35}\big[486 a_2 \left(N_3-1\right)+243 a_6 \left(N_3-1\right)+108 h_{53}-108 h_{55}+108 h_{56}+108 h_{57}+324 h_{59}-324 h_{61}\nonumber\\
&+3527 h_{65}-1665 h_{66}+72 h_{91}+72 h_{92}+72 h_{94}+48 h_{95}-59 h_{96}+3 (51 h_{97}-35 h_{99}-45 h_{101}\nonumber\\
&-45 h_{102}+136 h_{103}+226 h_{104}+90 h_{105})\big]\,,\\
\alpha _{t\Omega}=&\;\frac{w}{675}\big\{5346 a_2 \left(N_3-1\right)+2673 a_6 \left(N_3-1\right)-600 d_1+800 h_5+1600 h_6+200 h_9+3\big[40 h_{41}-20 h_{23}\nonumber\\
&-500 h_{25}+40 h_{42}+360 h_{43}+396 h_{53}-396 h_{55}+396 h_{56}+396 h_{57}+1188 h_{59}-1188 h_{61}+12969 h_{65}\nonumber\\
&-6655 h_{66}+264 h_{91}+264 h_{92}+264 h_{94}+176 h_{95}-253 h_{96}+561 h_{97}-495 h_{99}-495 h_{101}-495 h_{102}\nonumber\\
&+1716 h_{103}+2706 h_{104}+10 (99 h_{105}+180 h_{151}+180 h_{152}+180 h_{153}+180 h_{154}+137 h_{178}+125 h_{179}\nonumber\\
&-145 h_{180}+26 h_{181}-38 h_{182}-115 h_{184}+133 h_{185}+155 h_{186}-111 h_{187}-123 h_{188}+62 h_{189}+82 h_{190}\nonumber\\
&+38 h_{191}+22 h_{192}+24 h_{193}+12 h_{194})\big]\big\}\,,
\end{align}
while the expressions for the Lagrangian coefficients associated with the cubic invariants acquire a more cumbersome form and are displayed in~\ref{appendix:RN}.

As can be seen, the mass terms of the tensor modes of torsion and nonmetricity included in the action are nonvanishing, whereas the vector and axial modes remain massless. This result contrasts with the case of quadratic MAG, where the solution displays a mass term for the axial mode of torsion alone~\cite{Bahamonde:2022kwg}. In this sense, it is worthwhile to stress that the presence of massive tensor modes in the model of the solution additionally guarantees the avoidance of further strong coupling problems and no-go theorems constraining the interaction of massless higher-spin modes in the quantum regime~\cite{Loebbert:2008zz,Aoki:2023sum}.

As for the vector and axial sectors, it turns out that the Lagrangian density~\eqref{EqVectorDes} includes the following kinetic and interaction terms when fixing the Lagrangian coefficients of the solution:
\begin{align}
    16\pi \mathcal{L}=&-R+ \frac{1}{18}d_{1}{} F^{(T)}_{\mu  \nu  } F^{(T)}{}^{\mu  \nu  }+ \frac{1}{144}d_{1}{} F^{(S)}_{\mu  \nu  } F^{(S)}{}^{\mu  \nu  }+\frac{1}{8}\left(16a_{2}-8a_{6}+32a_{14}{}+5d_{1}\right)F^{(W)}_{\mu\nu} F^{(W)}{}^{\mu\nu}\nonumber\\
    &  + \frac{3}{128}\left(16 a_{2}{} + 8 a_{6}{} + 3 d_{1}{}\right)F^{(\Lambda)}_{\mu  \nu}F^{(\Lambda)}{}^{\mu\nu}+ \frac{1}{8}d_{1}F^{(T)}_{\mu  \nu}F^{(\Lambda)}{}^{\mu  \nu} - \frac{3}{16}  d_{1}F^{(W)}_{\mu\nu}F^{(\Lambda)}{}^{\mu\nu} - \frac{1}{6} d_{1}{} F^{(T)}_{\mu  \nu  } F^{(W)}{}^{\mu  \nu  }\nonumber\\
    &+\left(2 h_{34}{}- h_{33}{}  -4 h_{35}{}\right)*F^{(T)}_{\mu\nu}S^{\mu}W^{\nu} -2\left(h_{132}{} + h_{135}{} + 2 h_{137}{}\right) *F^{(W)}_{\mu\nu}S^{\mu} W^{\nu}\nonumber\\
    &+\left(h_{33}{} -2 h_{34}{} + 4 h_{35}\right)*F^{(S)}_{\mu\nu}T^{\mu}W^{\nu} + \frac{1}{6} (5 h_{132}{}  -3 h_{135}{}) S_{\mu  } S^{\mu  } W^{\nu  } \Lambda _{\nu  } - \frac{3}{2}\left(h_{132}{} - h_{135}\right)*F^{(S)}_{\mu\nu}W^{\mu}\Lambda^{\nu}\nonumber\\
    &+ \frac{1}{8}\big[6\left(8 h_{28}{} + h_{33}{} + 2 h_{34}{}\right)- d_{1}{}\bigr] *F^{(T)}_{\mu\nu}S^{\mu}\Lambda^{\nu} + \frac{1}{6} (3 h_{135}{}-2 h_{132}{}) S_{\mu  } S_{\nu  } W^{\mu  } \Lambda ^{\nu  }  \nonumber\\
    &+ \frac{1}{128}\left(288 h_{28}{} + 96 h_{141}{}-24 a_{2}{} -4 a_{6}{} -5 d_{1}-576P\right)*F^{(\Lambda )}_{\mu\nu}S^{\mu}\Lambda^{\nu}+ P S_{\mu  } S_{\nu  } \Lambda ^{\mu  } \Lambda ^{\nu  }\nonumber\\
    &+\frac{1}{96}\left(576P-40a_{2}-28a_{6}-43d_{1}+2016h_{28}+144h_{132}-144h_{135}-480h_{141}-384h_{143}\right)*F^{(W)}_{\mu\nu}S^{\mu}\Lambda^{\nu}\nonumber\\
    &-\frac{3}{4}\left(h_{33}{} + 2 h_{34}{}\right)*F^{(S)}_{\mu\nu}T^{\mu}\Lambda^{\nu} - \frac{1}{6}\left(h_{33}{}+3h_{34}{}\right)S_{\mu}S_{\nu}T^{\mu}\Lambda^{\nu} + \frac{1}{12}\left(5 h_{33}{} + 6 h_{34}{}\right)S_{\mu}S^{\mu}T_{\nu}\Lambda^{\nu}\nonumber\\
    &+\frac{1}{128}\left(8 a_{2}{} + 12 a_{6}{} + 13 d_{1}{}+96h_{141}{}-576 h_{28}{}-320 P\right) S_{\mu  } S^{\mu  } \Lambda _{\nu  } \Lambda ^{\nu  } \nonumber\\
    &+\frac{1}{64}\bigl(48 h_{90}{} + 60 h_{91}{} + 12 h_{92}{} + 12 h_{94}{} + 21 h_{95}{} + 63 h_{96}{} + 7 h_{97}{}\bigr)\big(12 W^{\mu}\Lambda_{\mu}\Lambda_{\nu}\Lambda^{\nu}- 8T_{\mu}\Lambda^{\mu}\Lambda_{\nu}\Lambda ^{\nu}\nonumber\\
    &-9\Lambda_{\mu}\Lambda^{\mu}\Lambda_{\nu}\Lambda^{\nu}-12\Lambda_{\mu}\Lambda^{\mu}\nabla_{\nu}\Lambda^{\nu}\big)\,,
\end{align}
with
\begin{align}
    P=&\;\frac{1}{129600} \big(1100448 a_{2}{} + 600624 a_{6}{} -24375 d_{1}{} + 2800 h_{5}{} + 5600 h_{6}{} + 700 h_{9}{} -34770 h_{23}{} + 69450 h_{25}{} + 388800 h_{28}{} \nonumber\\
    &+ 69540 h_{41}{} + 69540 h_{42}{}  -65340 h_{43}{} + 140400 h_{44}{} + 140400 h_{45}{}  -244944 h_{53}{} + 27000 h_{54}{} + 261144 h_{55}{} \nonumber\\
    &-244944 h_{56}{}  -228744 h_{57}{} + 21600 h_{58}{} -761832 h_{59}{} -43200 h_{60}{} + 751032 h_{61}{} -8231616 h_{65}{} + 4044720 h_{66}{} 
    \nonumber\\
    &-21600 h_{90}{}  -199296 h_{91}{}  -163296 h_{92}{}  -170496 h_{94}{}  -106464 h_{95}{} + 143592 h_{96}{}  -359304 h_{97}{} + 283680 h_{99}{} \nonumber\\
    &+ 359280 h_{101}{} + 258480 h_{102}{}  -975024 h_{103}{} -1517184 h_{104}{} -693360 h_{105}{} + 64800 h_{141}{} -48600 h_{144}{}  -24300 h_{145}{} \nonumber\\
    &+ 81000 h_{146}{} + 48600 h_{147}{} + 40500 h_{148}{} + 48600 h_{149}{}  -78300 h_{151}{}  -94500 h_{152}{} + 67500 h_{153}{} + 18900 h_{154}{} \nonumber\\
    &-226800 h_{158}{} + 3600 h_{171}{} + 10800 h_{174}{} + 19920 h_{178}{} + 30000 h_{179}{}  -13200 h_{180}{}  -21840 h_{181}{} -480 h_{182}{} \nonumber\\
    & -16800 h_{184}{} + 1680 h_{185}{}  -6000 h_{186}{} -9360 h_{187}{} + 720 h_{188}{} + 23520 h_{189}{} + 6720 h_{190}{}  -10320 h_{191}{} + 3120 h_{192}{}\nonumber\\
    &+ 1440 h_{193}{} + 11520 h_{194}{}  -1134648 a_{2}{} N_{3}{} -567324 a_{6}{} N_{3}{}\big)\,.
\end{align}
Hence, it is shown that the vector and axial modes of the model are free of Ostrogradsky instabilities, while the fulfillment of the constraints~\eqref{condKinetic1}, \eqref{condKinetic2} and \eqref{condKinetic3} also involves the fact that their kinetic terms are ghost-free if
\begin{align}
    d_1\leq 0\,,\quad a_6\leq -\,2 a_2\,,\quad a_{14}\leq \frac{1}{8} \left(2 a_6-4 a_2-d_1\right)\,.
\end{align}

\section{Conclusions}\label{sec:conclusions}

It is well-known that the vector and axial sectors of quadratic MAG generally contain pathologies that spoil the stability of the theory~\cite{Neville:1978bk,Sezgin:1979zf,Sezgin:1981xs,Miyamoto:1983bf,Fukui:1984gn,Fukuma:1984cz,Battiti:1985mu,Kuhfuss:1986rb,Blagojevic:1986dm,Baikov:1992uh,Yo:1999ex,Yo:2001sy,Lin:2018awc,BeltranJimenez:2019acz,Jimenez:2019qjc,Percacci:2020ddy,BeltranJimenez:2020sqf,Lin:2020phk,Marzo:2021iok,Baldazzi:2021kaf,Jimenez-Cano:2022sds,Barker:2024dhb,Marzo:2024pyn}. In this work, we have explicitly shown that the extension of the gravitational action of MAG in the presence of cubic order invariants defined from the curvature, torsion and nonmetricity tensors can provide a safe framework for the propagation of the vector and axial modes of the torsion and nonmetricity tensors around any general curved background.

For this task, we have first introduced the most general parity preserving cubic Lagrangian defined from the invariants of the curvature, torsion and nonmetricity tensors, which initially includes $209$ mixing terms between the curvature, torsion and nonmetricity tensors. In particular, under an appropriate choice of the Lagrangian coefficients, such terms allow the cancellation of the gravitational instabilities provided by the quadratic curvature invariants that affect the vector and axial sectors of the theory. The conditions arising from the different types of geometries described in MAG are collected in Table \ref{tab:condition}.

\begin{table}[H]
    \centering
    \renewcommand{\arraystretch}{2.7}
    \resizebox{18cm}{!}{\begin{tabular}{|c|c|c|}
        \hline
      \multirow{2}{*}{\textbf{Geometry}} &   \multirow{2}{*}{\textbf{Avoidance of Ostrogradsky instabilities}} & \multirow{2}{*}{\textbf{Well-posed kinetic terms} }\\ 
      &  & \\
        \hline
         \multirow{1}{*}{Riemann-Cartan} & $c_{2}=\;2c_{1}\,,  h_{2}=-\,\displaystyle\frac{h_{1}}{2}\,, h_{3}= -\,\displaystyle\frac{1}{6}\left(c_{1}+6h_{13}\right),$ &  \multirow{2}{*}{$  -\,\displaystyle\frac{d_1}{6} \leq c_{1} \leq -\,\displaystyle\frac{d_{1}}{3}\,, \quad d_1 \leq 0$}\\
   \multirow{1}{*}{($Q_{\lambda\mu\nu}=0$)}  & $h_{4}= \displaystyle \frac{h_{13}}{2}\,, h_{14}= -\,2h_{13}\,, h_{15}=4h_{13}$& \\
        \hline
      \multirow{2}{*}{Weyl-Cartan} & $c_{2}=\;2c_{1}\,,  h_{2}=-\,\displaystyle\frac{h_{1}}{2}\,, h_{3}= -\,\displaystyle\frac{1}{6}\left(c_{1}+6h_{13}\right), h_{4}= \displaystyle\frac{h_{13}}{2}\,, h_{14}= -\,2h_{13}\,, h_{15}=4h_{13}\,,
$ & $ -\,\displaystyle\frac{d_1}{6} \leq c_{1} \leq -\,\displaystyle\frac{d_{1}}{3}\,, d_1 \leq 0\,, a_{15}\leq \displaystyle\frac{1}{2} \left(a_9+a_{10}+a_{12}-2 a_2-a_5-4 a_{14}-d_1\right),$ \\
     \multirow{2}{*}{(${\nearrow\!\!\!\!\!\!\!Q}_{\lambda\mu\nu}=0\Longrightarrow\Lambda_\mu=0$)} & $h_{48}=\;\displaystyle\frac{9}{4}h_1-h_{47}\,, h_{49}=\displaystyle\frac{3}{2}h_{113}=-\,\displaystyle\frac{9}{8}h_1\,, h_{110}=-\,\displaystyle\frac{3}{2} h_{1}^{} -  h_{108}^{}\,, h_{111} =3h_1-h_{109}\,,$ & $-\,2\sqrt{\left(6 c_1+2 d_1\right)\left[d_1+a_5+2\left(a_2+2 a_{14}+a_{15}\right)-a_9-a_{10}-a_{12}\right]}\leq$\\
      & $ h_{112}=\;\displaystyle\frac{1}{8} \left(4 h_{108}^{}-9 h_{1}^{} - 4 h_{109}^{}\right), h_{133}=3 h_{13}^{} + h_{132}^{}\,, h_{134}=0\,, h_{136}=6 h_{13}^{} -  h_{135}^{}$ & $a_6+a_9-a_{11}+a_{12}-3 d_1-2 \left(2 c_1+a_{15}+a_{16}+a_5-a_{10}\right)\leq$\\ 
      & & $2\sqrt{\left(6 c_1+2 d_1\right)\left[d_1+a_5+2\left(a_2+2 a_{14}+a_{15}\right)-a_9-a_{10}-a_{12}\right]}$\\
        \hline
      \multirow{3}{*}{Torsion-free} & $c_2=\;2 c_{1}{} -2 \left(4 a_{10}^{} + 4 a_{11}^{} + 2 a_{2}^{} + a_{6}^{}\right), a_{12}=  -\,2 a_{10}^{} -  a_{11}^{} + a_{9}^{} + d_{1}^{}\,,$ & $a_{10}\leq 2 a_9-2 a_2-a_5-a_{11}-4 a_{14}-2 a_{15}\,, a_{11}\leq -\,\frac{1}{5} \left(6 a_2+3 a_5+11 a_{10}\right)\,,$ \\  
     &  $h_{48}=\,-h_{49}=-\,\displaystyle\frac{16}{9}h_{52}=-\,\displaystyle\frac{4}{3}h_{70}=\displaystyle\frac{2}{3}h_{71}=-\,\displaystyle\frac{4}{3}h_{73}=h_{47}\,,$ & $-\,\sqrt{3\left(6 a_2+3 a_5+11 a_{10}+5 a_{11}\right) \left(2 a_2+a_5-2 a_9+a_{10}+a_{11}+4 a_{14}+2 a_{15}\right)}\leq$  \\
  ($T^\lambda{}_{\mu\nu}=0$)    &$h_{51}=\;\displaystyle\frac{9}{8} h_{47}^{} -  h_{50}^{}\,, h_{69}=\displaystyle\frac{3}{4} h_{47}^{} -  h_{68}^{}\,, h_{72}=\displaystyle\frac{3}{4} h_{47}+h_{68}$ & $-\,3 \left(2 a_2+a_5+a_9+3 a_{10}+4 a_{11}+a_{16}+d_1\right)\leq $\\ 
    & & $\sqrt{3\left(6 a_2+3 a_5+11 a_{10}+5 a_{11}\right) \left(2 a_2+a_5-2 a_9+a_{10}+a_{11}+4 a_{14}+2 a_{15}\right)}$
    \\
        \hline
       \multirow{7}{*}{General} & $c_{2}=2c_{1}\,,  a_{11}= -\,\displaystyle \frac{1}{4} \left(2 a_{2}^{} +  a_{6}^{}+4 a_{10}^{}\right), a_{12}=\displaystyle \frac{1}{4} \left(4 d_{1}^{}+2 a_{2}^{} + a_{6}^{} + 4 a_{9}^{}-4 a_{10}^{}\right),$ & \multirow{3}{*}{$l_6\leq 0, l_1+l_{11}+l_{16}\leq 0\,, 4 l_{11} l_{16}+4 l_1 \left(l_{11}+l_{16}\right)-l_{30}^2-l_{43}^2-l_{70}^2\geq 0\,,$}\\
       & $ h_{3}= -\,\displaystyle \frac{1}{6}\left(c_{1}+6h_{13}\right), h_{4}= \displaystyle \frac{h_{13}}{2}\,, h_{14}= -\,2h_{13}\,, h_{15}=4h_{13}\,, h_{51}=-\,h_{50}\,,$ & \multirow{3}{*}{$l_{30}l_{43}l_{70}+4 l_1 l_{11} l_{16}-l_{16} l_{30}^2-l_{11} l_{43}^2-l_1 l_{70}^2\leq 0$} \\
        & $h_{133}=3h_{13}+h_{132}\,, h_{136}=6 h_{13}^{} -  h_{135}^{}\,, h_{139}=\displaystyle \frac{1}{12} \left(2 a_{5}^{} -  a_{6}^{} + c_{1}{} - 9 h_{13}^{} + 12 h_{138}^{} + 36 h_{28}^{}\right),$ & \\
         & $h_{140}=\displaystyle \frac{1}{12} \left(2 a_{5}^{} -  a_{6}^{} - 8 c_{1}{} - 36 h_{13}^{} - 72 h_{28}^{}\right),$ & (See~\ref{appendix:l} for the definitions of $l_i$) \\
          & $h_{142}=\displaystyle \frac{1}{24} \left(12 a_{10}^{} + 6 a_{2}^{} + 4 a_{5}^{} + a_{6}^{} + 12 a_{9}^{} + 20 c_{1}{} + 36 h_{13}^{} - 24 h_{141}^{} + 288 h_{28}^{}\right),$ & \\
           & $h_1=h_2=h_{27}=h_{47}=h_{48}=h_{49}=h_{52}=h_{68}=h_{69}=h_{70}=h_{71}=h_{72}=h_{73}=h_{108}=0\,,$ & \\
            & $h_{109}=h_{110}=h_{111}=h_{112}=h_{113}=h_{114}=h_{115}=h_{116}=h_{117}=h_{118}=h_{119}=h_{134}=0$ & \\
        \hline
    \end{tabular}}
    \caption{Stability conditions of the vector and axial sectors for the different types of geometries in MAG.}
    \label{tab:condition}
\end{table}

On the other hand, a further restriction on the Lagrangian coefficients is also shown to contain Reissner-Nordström-like black hole solutions with spin, dilation and shear charges, which in contrast with the case of quadratic MAG include massive tensor modes, and thus avoid additional strong coupling problems and no-go theorems for a consistent interaction of these modes in the quantum regime. Overall, the resulting action depends on $5$ Lagrangian coefficients associated with the quadratic curvature, torsion and nonmetricity invariants, and $86$ associated with the cubic ones. Therefore, the stability analysis of the vector and axial sectors reduces the parameter space of the theory, while the remaining coupling constants are expected to be fixed by a further examination on the stability of the tensor sector (including the kinetics and interactions with the vector and axial modes of torsion and nonmetricity).  Further research in this direction will be addressed in future works.

\noindent
\section*{Acknowledgements}

We would like to thank Jose Beltrán Jiménez for helpful discussions. S.B. is supported by “Agencia Nacional de Investigación y Desarrollo” (ANID), Grant “Becas Chile postdoctorado al extranjero” No. 74220006. The work of J.G.V. is supported by the Institute for Basic Science (IBS-R003-D1).

\newpage
\appendix

\renewcommand{\thesection}{Appendix \Alph{section}}
\renewcommand{\theequation}{\Alph{section}\arabic{equation}}
\setcounter{equation}{0} 

\section{Coefficients of the vector and axial sectors in the quadratic MAG Lagrangian}\label{appendix:p}

The coefficients $\{p_i\}_{i=1}^{77}$ displayed in the Lagrangian~\eqref{EqVectorDesQuadratic} of the vector and axial sectors of quadratic MAG are defined as 
\begin{align}
p_{1}^{}={}&\frac{1}{9} (4 c_{1}{}
 + c_{2}{}
 + 2 d_{1}^{}),\\
p_{2}^{}={}&\frac{1}{72} (-4 c_{1}{}
 -  c_{2}{}
 -  d_{1}^{}),\\
p_{3}^{}={}&\frac{1}{72} (4 c_{1}{}
 + c_{2}{}),\\
p_{4}^{}={}&\frac{1}{36} (4 c_{1}{}
 + c_{2}{}),\\
p_{5}^{}={}&\frac{1}{24} (-2 c_{1}{}
 + c_{2}{}),\\
p_{6}^{}={}&- a_{10}^{}
 -  a_{12}^{}
 + 4 a_{14}^{}
 + 2 a_{15}^{}
 + 2 a_{2}^{}
 + a_{5}^{}
 -  a_{9}^{}
 + d_{1}^{},\\
p_{7}^{}={}&\frac{1}{16} (21 a_{10}^{}
 + 5 a_{11}^{}
 - 2 a_{12}^{}
 + 14 a_{2}^{}
 + 9 a_{5}^{}
 - 2 a_{6}^{}
 + 2 a_{9}^{}
 + 2 c_{1}{}
 -  c_{2}{}
 + 2 d_{1}^{}),\\
p_{8}^{}={}&\frac{1}{8} (-6 a_{10}^{}
 - 11 a_{11}^{}
 + 5 a_{12}^{}
 - 8 a_{2}^{}
 - 4 a_{6}^{}
 - 5 a_{9}^{}
 + 4 c_{1}{}
 - 2 c_{2}{}
 - 5 d_{1}^{}),\\
p_{9}^{}={}&-2 a_{11}^{}
 + 2 a_{12}^{}
 - 2 a_{2}^{}
 -  a_{6}^{}
 - 2 a_{9}^{}
 + c_{1}{}
 -  \frac{1}{2} c_{2}{}
 - 2 d_{1}^{},\\
p_{10}^{}={}&\frac{9}{256} (26 a_{10}^{}
 + 37 a_{11}^{}
  - 11 a_{12}^{}
 + 24 a_{2}^{}
 + 12 a_{6}^{}
 + 11 a_{9}^{}
 - 12 c_{1}{}
 + 6 c_{2}{}
 + 11 d_{1}^{}),\\
p_{11}^{}={}&\frac{1}{2} (-2 a_{10}^{}
 -  a_{11}^{}
 -  a_{12}^{}
 + a_{9}^{}
 + d_{1}^{}),\\
p_{12}^{}={}&-2 (2 a_{10}^{}
 + a_{11}^{}
 + a_{12}^{}
 -  a_{9}^{}
 -  d_{1}^{}),\\
p_{13}^{}={}&\frac{3}{64} (62 a_{10}^{}
 + 79 a_{11}^{}
 - 17 a_{12}^{}
 + 48 a_{2}^{}
 + 24 a_{6}^{}
 + 17 a_{9}^{}
 - 24 c_{1}{}
 + 12 c_{2}{}
 + 17 d_{1}^{}),\\
p_{14}^{}={}&\frac{3}{8} (6 a_{10}^{}
 + 7 a_{11}^{}
 -  a_{12}^{}
 + 4 a_{2}^{}
 + 2 a_{6}^{}
 + a_{9}^{}
 - 2 c_{1}{}
 + c_{2}{}
 + d_{1}^{}),\\
p_{15}^{}={}&\frac{1}{81} (- c_{1}{}
 + 2 c_{2}{}),\\
p_{16}^{}={}&\frac{1}{162} (-4 c_{1}{}
 -  c_{2}{}),\\
p_{17}^{}={}&\frac{1}{18} (2 c_{1}{}
 -  c_{2}{}),\\
p_{18}^{}={}&\frac{1}{108} (-4 c_{1}{}
 -  c_{2}{}),\\
p_{19}^{}={}&\frac{1}{27} (c_{1}{}
 - 2 c_{2}{}),\\
p_{20}^{}={}&\frac{1}{3} (2 a_{10}^{}
 -  a_{11}^{}
 + a_{12}^{}
 - 2 a_{15}^{}
 - 2 a_{16}^{}
 - 2 a_{5}^{}
 + a_{6}^{}
 + a_{9}^{}
 - 4 c_{1}{}
 - 3 d_{1}^{}),\\
p_{21}^{}={}&\frac{1}{12} (2 a_{10}^{}
 - 5 a_{11}^{}
 + a_{12}^{}
 + 6 a_{5}^{}
 - 3 a_{6}^{}
 + 5 a_{9}^{}
 + 12 c_{1}{}
 + 5 d_{1}^{}),\\
p_{22}^{}={}&\frac{1}{6} (2 a_{10}^{}
 + a_{11}^{}
 + a_{12}^{}
 -  a_{9}^{}
 -  d_{1}^{}),\\
p_{23}^{}={}&2 a_{10}^{}
 + a_{11}^{}
 + a_{12}^{}
 -  a_{9}^{}
 -  d_{1}^{},\\
p_{24}^{}={}&\frac{2}{9} (2 a_{10}^{}
 + a_{11}^{}
 + a_{12}^{}
 -  a_{9}^{}
 -  d_{1}^{}),\\
p_{25}^{}={}&\frac{1}{18} (18 a_{10}^{}
 + 13 a_{11}^{}
 + 5 a_{12}^{}
 + 4 a_{2}^{}
 + 2 a_{6}^{}
 - 5 a_{9}^{}
 - 2 c_{1}{}
 + c_{2}{}
 - 5 d_{1}^{}),\\
p_{26}^{}={}&\frac{1}{36} (18 a_{10}^{}
 + 25 a_{11}^{}
 - 7 a_{12}^{}
 + 16 a_{2}^{}
 + 8 a_{6}^{}
 + 7 a_{9}^{}
 - 8 c_{1}{}
 + 4 c_{2}{}
 + 7 d_{1}^{}),\\
p_{27}^{}={}&\frac{1}{32} (62 a_{10}^{}
 + 79 a_{11}^{}
 - 17 a_{12}^{}
 + 48 a_{2}^{}
 + 24 a_{6}^{}
 + 17 a_{9}^{}
 - 24 c_{1}{}
 + 12 c_{2}{}
 + 17 d_{1}^{}),\\
p_{28}^{}={}&- \frac{2}{3} (2 a_{10}^{}
 + a_{11}^{}
 + a_{12}^{}
 -  a_{9}^{}
 -  d_{1}^{}),\\
p_{29}^{}={}&\frac{1}{6} (-18 a_{10}^{}
 - 13 a_{11}^{}
 - 5 a_{12}^{}
 - 4 a_{2}^{}
 - 2 a_{6}^{}
 + 5 a_{9}^{}
 + 2 c_{1}{}
 -  c_{2}{}
 + 5 d_{1}^{}),\\
p_{30}^{}={}&\frac{1}{12} (18 a_{10}^{}
 + 17 a_{11}^{}
 + a_{12}^{}
 + 8 a_{2}^{}
 + 4 a_{6}^{}
 -  a_{9}^{}
 - 4 c_{1}{}
 + 2 c_{2}{}
 -  d_{1}^{}),\\
p_{31}^{}={}&\frac{1}{2} (-6 a_{10}^{}
 - 7 a_{11}^{}
 + a_{12}^{}
 - 4 a_{2}^{}
 - 2 a_{6}^{}
 -  a_{9}^{}
 + 2 c_{1}{}
 -  c_{2}{}
 -  d_{1}^{}),\\
p_{32}^{}={}&\frac{1}{3} (2 a_{10}^{}
 + a_{11}^{}
 + a_{12}^{}
 -  a_{9}^{}
 -  d_{1}^{}),\\
p_{33}^{}={}&p_{28},\\
p_{34}^{}={}&p_{23},\\
p_{35}^{}={}&\frac{1}{36} (- c_{1}{}
 + 2 c_{2}{}),\\
p_{36}^{}={}&\frac{1}{72} (-4 c_{1}{}
 -  c_{2}{}),\\
p_{37}^{}={}&\frac{1}{12} (-2 c_{1}{}
 + c_{2}{}),\\
p_{38}^{}={}&p_{3},\\
p_{39}^{}={}&\frac{1}{18} (- c_{1}{}
 + 2 c_{2}{}),\\
p_{40}^{}={}&\frac{1}{576} (-24 a_{10}^{}
 - 48 a_{11}^{}
 - 48 a_{12}^{}
 + 40 a_{2}^{}
 + 12 a_{5}^{}
 - 14 a_{6}^{}
 - 24 a_{9}^{}
 - 29 c_{1}{}
 + 8 c_{2}{}+12d_{1}^{}),\\
p_{41}^{}={}&\frac{1}{1152} (12 a_{10}^{}
 + 78 a_{11}^{}
 + 78 a_{12}^{}
 - 128 a_{2}^{}
 + 48 a_{5}^{}
 + 16 a_{6}^{}
 + 66 a_{9}^{}
 + 28 c_{1}{}
 - 13 c_{2}{}- 6 d_{1}^{}),\\
p_{42}^{}={}&\frac{1}{48} (4 a_{10}^{}
 - 2 a_{11}^{}
 + 2 a_{12}^{}
 + 2 a_{9}^{}
 + 6 c_{1}{}
 - 3 c_{2}{}
 - 2 d_{1}^{}),\\
p_{43}^{}={}&\frac{1}{288} (8 a_{10}^{}
 - 8 a_{11}^{}
 + 4 a_{12}^{}
 + 20 a_{5}^{}
 - 10 a_{6}^{}
 + 8 a_{9}^{}
 - 12 c_{1}{}
 - 13 c_{2}{}
 - 4 d_{1}^{}),\\
p_{44}^{}={}&\frac{1}{72} (4 a_{10}^{}
 + 2 a_{11}^{}
 + 2 a_{12}^{}
 - 2 a_{5}^{}
 + a_{6}^{}
 - 2 a_{9}^{}
 + 3 c_{1}{}
 - 5 c_{2}{}
 - 2 d_{1}^{}),\\
p_{45}^{}={}&-\frac{1}{16} (6 a_{10}^{}
 - 9 a_{11}^{}
 - 3 a_{12}^{}
 + 8 a_{2}^{}
 + 6 a_{5}^{}
 - 5 a_{6}^{}
 - 3 a_{9}^{}
 - 4 c_{1}{}
 - 4 c_{2}{}
 - 3 d_{1}^{}),\\
p_{46}^{}={}&\frac{1}{4} (2 a_{10}^{}
 - 2 a_{11}^{}
 - 2 a_{12}^{}
 - 6 a_{16}^{}
 - 6 a_{5}^{}
 + 6 a_{6}^{}
 - 4 a_{9}^{}
 - 6 c_{1}{}
 + 3 c_{2}{}
 - 4 d_{1}^{}),\\
p_{47}^{}={}&\frac{1}{4} (-2 a_{10}^{}
 -  a_{11}^{}
 -  a_{12}^{}
 + a_{9}^{}
 + d_{1}^{}),\\
p_{48}^{}={}&- \frac{3}{4} (2 a_{10}^{}
 + a_{11}^{}
 + a_{12}^{}
 -  a_{9}^{}
 -  d_{1}^{}),\\
p_{49}^{}={}&\frac{1}{8} (18 a_{10}^{}
 + 13 a_{11}^{}
 + 5 a_{12}^{}
 + 4 a_{2}^{}
 + 2 a_{6}^{}
 - 5 a_{9}^{}
 - 2 c_{1}{}
 + c_{2}{}
 - 5 d_{1}^{}),\\
p_{50}^{}={}&- \frac{3}{2} (2 a_{10}^{}
 + a_{11}^{}
 + a_{12}^{}
 -  a_{9}^{}
 -  d_{1}^{}),\\
p_{51}^{}={}&\frac{1}{16} (18 a_{10}^{}
 + 25 a_{11}^{}
 - 7 a_{12}^{}
 + 16 a_{2}^{}
 + 8 a_{6}^{}
 + 7 a_{9}^{}
 - 8 c_{1}{}
 + 4 c_{2}{}
 + 7 d_{1}^{}),\\
p_{52}^{}={}&- \frac{3}{64} (62 a_{10}^{}
 + 79 a_{11}^{}
 - 17 a_{12}^{}
 + 48 a_{2}^{}
 + 24 a_{6}^{}
 + 17 a_{9}^{}
 - 24 c_{1}{}
 + 12 c_{2}{}
 + 17 d_{1}^{}),\\
p_{53}^{}={}&p_{50},\\
p_{54}^{}={}&\frac{1}{4} (18 a_{10}^{}
 + 13 a_{11}^{}
 + 5 a_{12}^{}
 + 4 a_{2}^{}
 + 2 a_{6}^{}
 - 5 a_{9}^{}
 - 2 c_{1}{}
 + c_{2}{}
 - 5 d_{1}^{}),\\
p_{55}^{}={}&\frac{1}{8} (-18 a_{10}^{}
 - 17 a_{11}^{}
 -  a_{12}^{}
 - 8 a_{2}^{}
 - 4 a_{6}^{}
 + a_{9}^{}
 + 4 c_{1}{}
 - 2 c_{2}{}
 + d_{1}^{}),\\
p_{56}^{}={}&p_{50},\\
p_{57}^{}={}&\frac{3}{4} (6 a_{10}^{}
 + 7 a_{11}^{}
 -  a_{12}^{}
 + 4 a_{2}^{}
 + 2 a_{6}^{}
 + a_{9}^{}
 - 2 c_{1}{}
 + c_{2}{}
 + d_{1}^{}),\\
p_{58}^{}={}&\frac{3}{4} (2 a_{10}^{}
 + a_{11}^{}
 + a_{12}^{}
 -  a_{9}^{}
 -  d_{1}^{}),\\
p_{59}^{}={}&p_{50},\\
p_{60}^{}={}&p_{19},\\
p_{61}^{}={}&\frac{1}{54} (4 c_{1}{}
 + c_{2}{}),\\
p_{62}^{}={}&\frac{1}{108} (-4 a_{10}^{}
 - 2 a_{11}^{}
 - 2 a_{12}^{}
 + 2 a_{5}^{}
 -  a_{6}^{}
 + 2 a_{9}^{}
 - 3 c_{1}{}
 + 5 c_{2}{}
 + 2 d_{1}^{}),\\
p_{63}^{}={}&\frac{1}{216} (2 a_{10}^{}
 + a_{11}^{}
 + a_{12}^{}
 + 8 a_{5}^{}
 - 4 a_{6}^{}
 -  a_{9}^{}
 - 12 c_{1}{}
 - 7 c_{2}{}
 -  d_{1}^{}),\\
p_{64}^{}={}&-\frac{1}{6} (2 a_{10}^{}
 -  a_{11}^{}
 + a_{12}^{}
 + 2 a_{5}^{}
 -  a_{6}^{}
 + a_{9}^{}
 -  c_{2}{}
 -  d_{1}^{}),\\
p_{65}^{}={}&\frac{1}{2} (2 a_{10}^{}
 + a_{11}^{}
 + a_{12}^{}
 -  a_{9}^{}
 -  d_{1}^{}),\\
p_{66}^{}={}&\frac{1}{3} (-2 a_{10}^{}
 -  a_{11}^{}
 -  a_{12}^{}
 + a_{9}^{}
 + d_{1}^{}),\\
p_{67}^{}={}&p_{23},\\
p_{68}^{}={}&p_{29}^{},\\
p_{69}^{}={}&p_{28},\\
p_{70}^{}={}&\frac{1}{12} (-18 a_{10}^{}
 - 25 a_{11}^{}
 + 7 a_{12}^{}
 - 16 a_{2}^{}
 - 8 a_{6}^{}
 - 7 a_{9}^{}
 + 8 c_{1}{}
 - 4 c_{2}{}
 - 7 d_{1}^{}),\\
p_{71}^{}={}&p_{23}^{},\\
p_{72}^{}={}&-2 a_{10}^{}
 -  a_{11}^{}
 -  a_{12}^{}
 + a_{9}^{}
 + d_{1}^{},\\
p_{73}^{}={}&p_{23}^{},\\
p_{74}^{}={}&p_{72},\\
p_{75}^{}={}&\frac{1}{144} (-2 a_{10}^{}
 -  a_{11}^{}
 -  a_{12}^{}
 - 8 a_{5}^{}
 + 4 a_{6}^{}
 + a_{9}^{}
 + 12 c_{1}{}
 + 7 c_{2}{}
 + d_{1}^{}),\\
p_{76}^{}={}&p_{44},\\
p_{77}^{}={}&-\frac{1}{4} (-2 a_{10}^{}
 -  a_{11}^{}
 - 3 a_{12}^{}
 + 2 a_{15}^{}
 - 2 a_{16}^{}
 - 2 a_{5}^{}
 + 3 a_{6}^{}
 - 3 a_{9}^{}
 + 2 c_{2}{}
 + d_{1}^{}).
\end{align}

\section{First branch of the cubic Lagrangian in terms of the torsion tensor}\label{appendix1}

The Lagrangian~\eqref{cubicLagIrr} containing cubic invariants from curvature and torsion can be written in terms of the torsion tensor as:
\begin{eqnarray}
    \mathcal{\bar{L}}_{\rm curv-tor}^{(3)}=\mathcal{\bar{L}}^{(3)}_{1,\rm tor}+\mathcal{\bar{L}}^{(3)}_{2,\rm tor}+\mathcal{\bar{L}}^{(3)}_{3,\rm tor}+\mathcal{\bar{L}}^{(3)}_{4,\rm tor}+\mathcal{\bar{L}}^{(3)}_{5,\rm tor}\,,
\end{eqnarray}
where we have defined:
\begin{eqnarray}
\mathcal{\bar{L}}^{(3)}_{1,\rm tor}&=&\bar{h}_{1}^{} \tilde{R}_{\rho \tau \gamma \mu } T_{\alpha }{}^{\gamma \mu \
} T^{\alpha \rho \tau } + \bar{h}_{2}^{} \tilde{R}_{\rho \gamma \tau \mu } 
T_{\alpha }{}^{\gamma \mu } T^{\alpha \rho \tau } + \bar{h}_{3}^{} \tilde{R}_{\alpha \tau \gamma \mu } T^{\alpha \rho \tau } T_{\rho }{}^{
\gamma \mu } + \bar{h}_{4}^{} \tilde{R}_{\alpha \gamma \tau \mu } T^{\alpha 
\rho \tau } T_{\rho }{}^{\gamma \mu } + \bar{h}_{5}^{} \tilde{R}_{\tau \gamma \alpha \mu } T^{\alpha \rho \tau } T_{\rho }{}^{\gamma \mu }\nonumber\\
&&+ \bar{h}_{6}^{} \tilde{R}_{\gamma \mu \alpha \tau } T^{\alpha \rho \tau } T_{\rho }{}^{\gamma \mu }+ \bar{h}_{7}^{} \tilde{R}_{\rho \tau \gamma \mu } T^{\tau \gamma \mu } T^{\rho } + \bar{h}_{8}^{} \tilde{R}_{\rho \gamma \tau \mu } T^{\tau \gamma \mu } T^{\rho } + \bar{h}_{9}^{} \tilde{R}_{\tau \gamma \rho \mu } T^{\tau \gamma \mu } T^{\rho } + \bar{h}_{10}^{} \tilde{R}_{\gamma \mu \rho \tau } T^{\tau \gamma \mu } T^{\rho }\nonumber\\
&&+ \bar{h}_{11}^{} \tilde{R}_{\alpha \tau \gamma \mu } T^{\alpha \rho \tau } T^{\gamma }{}_{\rho }{}^{\mu } + \bar{h}_{12}^{} \tilde{R}_{\alpha \gamma \tau \mu } T^{\alpha \rho \tau } T^{\gamma }{}_{\rho }{}^{\mu } + \bar{h}_{13}^{} \tilde{R}_{\alpha \mu \tau \gamma } T^{\alpha \rho \tau } T^{\gamma }{}_{\rho }{}^{\mu } + \bar{h}_{14}^{} \tilde{R}_{\tau \mu \alpha \gamma } T^{\alpha \rho \tau } T^{\gamma }{}_{\rho }{}^{\mu }\nonumber\\
&&+\bar{h}_{27}{} \tilde{R}^{\alpha }{}_{\rho \tau \gamma } T^{\rho }{}_{\alpha \mu } T^{\tau \gamma \mu } + \bar{h}_{28}{} \tilde{R}^{\alpha }{}_{\rho \tau \gamma } T^{\rho \tau }{}_{\mu } T^{\gamma }{}_{\alpha }{}^{\mu } + \bar{h}_{29}{} \tilde{R}^{\alpha }{}_{\rho \tau \gamma } T^{\rho \tau }{}_{\mu } T^{\mu }{}_{\alpha }{}^{\gamma } + \bar{h}_{30}{} \tilde{R}^{\alpha }{}_{\rho \tau \gamma } T^{\tau \rho }{}_{\mu } T^{\mu }{}_{\alpha }{}^{\gamma } \nonumber\\
&&+ \bar{h}_{31}{} \tilde{R}^{\alpha }{}_{\rho \tau \gamma } T^{\rho }{}_{\alpha \mu } T^{\mu \tau \gamma } + \bar{h}_{32}{} \tilde{R}^{\rho \alpha }{}_{\tau \gamma } T_{\rho }{}^{\tau \gamma } T_{\alpha } + \bar{h}_{33}{} \tilde{R}^{\rho \alpha }{}_{\tau \gamma } T^{\tau }{}_{\rho }{}^{\gamma } T_{\alpha } + \bar{h}_{34}{} \tilde{R}^{\rho }{}_{\tau }{}^{\alpha }{}_{\gamma } T^{\tau }{}_{\rho }{}^{\gamma } T_{\alpha } \,,\\
\mathcal{\bar{L}}^{(3)}_{2,\rm tor}&=&\bar{h}_{15}^{} \tilde{R}^{\alpha \rho } T_{\alpha }{}^{\tau \gamma } T_{\rho \tau \gamma } + \bar{h}_{16}^{} \tilde{R}^{\alpha \rho } T_{\rho }{}^{\tau \gamma } T_{\tau \alpha \gamma } + \bar{h}_{17}^{} \tilde{R}^{\alpha \rho } T_{\alpha }{}^{\tau \gamma } T_{\tau \rho \gamma } + \bar{h}_{18}^{} \tilde{R}^{\alpha \rho } T_{\tau \rho \gamma } T^{\tau }{}_{\alpha }{}^{\gamma } + \bar{h}_{19}^{} \tilde{R}^{\alpha \rho } T^{\tau }{}_{\alpha }{}^{\gamma } T_{\gamma \rho \tau } \nonumber\\
&&+ \bar{h}_{20}^{} \tilde{R}^{\alpha \rho } T_{\alpha } T_{\rho } + \bar{h}_{21}^{} \tilde{R}^{\alpha \rho } T_{\alpha \rho }{}^{\tau } T_{\tau } + \bar{h}_{22}^{} \tilde{R}^{\alpha \rho } T_{\rho \alpha }{}^{\tau } T_{\tau } + \bar{h}_{23}^{} \tilde{R}^{\alpha \rho } T^{\tau }{}_{\alpha \rho } T_{\tau }\,,\\
\mathcal{\bar{L}}^{(3)}_{3,\rm tor}&=&\bar{h}_{35}{} \hat{R}_{\alpha \rho } T^{\alpha }{}_{\tau \gamma } T^{\rho \tau \gamma } + \bar{h}_{36}{} \hat{R}_{\alpha \rho } T_{\tau }{}^{\rho \gamma } T^{\tau \alpha }{}_{\gamma } + \bar{h}_{37}{} \hat{R}_{\alpha \rho } T^{\rho }{}_{\tau \gamma } T^{\tau \alpha \gamma } + \bar{h}_{38}{} \hat{R}_{\alpha \rho } T^{\alpha }{}_{\tau \gamma } T^{\tau \rho \gamma } + \bar{h}_{39}{} \hat{R}_{\alpha \rho } T^{\tau \alpha }{}_{\gamma } T^{\gamma \rho }{}_{\tau } \nonumber\\
&&+ \bar{h}_{40}{} \hat{R}_{\alpha \rho } T^{\alpha } T^{\rho } + \bar{h}_{41}{} \hat{R}_{\alpha \rho } T^{\alpha \rho \tau } T_{\tau } + \bar{h}_{42}{} \hat{R}_{\alpha \rho } T^{\rho \alpha \tau } T_{\tau } + \bar{h}_{43}{} \hat{R}_{\alpha \rho } T^{\tau \alpha \rho } T_{\tau }\\
\mathcal{\bar{L}}^{(3)}_{4,\rm tor}&=&\bar{h}_{44}{} \tilde{R}^{\omega}{}_{\omega\alpha \rho } T^{\alpha }{}_{\tau \gamma } T^{\tau \rho \gamma } + \bar{h}_{45}{} \tilde{R}^{\omega}{}_{\omega\rho \tau } T^{\alpha \rho \tau } T_{\alpha } + \bar{h}_{46}{} \tilde{R}^{\omega}{}_{\omega\rho \tau } T^{\rho \alpha \tau } T_{\alpha }\\
\mathcal{\bar{L}}^{(3)}_{5,\rm tor}&=&\bar{h}_{24}^{} \tilde{R} T_{\alpha \rho \tau } T^{\alpha \rho \tau } + 
\bar{h}_{25}^{} \tilde{R} T^{\alpha \rho \tau } T_{\rho \alpha \tau } + 
\bar{h}_{26}^{} \tilde{R} T_{\rho } T^{\rho }\,,
\end{eqnarray}
and where the respective Lagrangian coefficients are related as\footnote{Note that we have chosen our parametrisation to obtain the same Lagrangian coefficients as in~\cite{Bahamonde:2024sqo}, in the absence of nonmetricity.}:
\begin{align}
\bar{h}_{1}{}={}&2 h_{20}^{}
 -  \frac{1}{6} h_{30}^{}
 + h_{41}^{}
 + \frac{1}{3} h_{5}^{}
 -  \frac{1}{3} h_{7}^{},\\
\bar{h}_{2}{}={}&\frac{1}{3} (h_{6}^{}-12 h_{20}^{}
 + 6 h_{24}^{}
 -  h_{29}^{}
 - 6 h_{42}^{}
 -  h_{8}^{}),\\
\bar{h}_{3}{}={}&\frac{1}{6} (h_{30}^{}
 + 2 h_{5}^{}
 - 2 h_{7}^{}),\\
\bar{h}_{4}{}={}&\frac{1}{3} (6 h_{22}^{}
 - 6 h_{23}^{}
 + 6 h_{42}^{}
 - 6 h_{43}^{}
 + h_{6}^{}
 -  h_{8}^{}
 + h_{9}^{}),\\
\bar{h}_{5}{}=&-\frac{1}{3} (h_{6}^{}
 +  h_{8}^{}+12 h_{21}^{}
 + 2 h_{29}^{}),\\
\bar{h}_{6}{}={}&\frac{1}{3} (2 h_{5}^{}+ h_{30}^{}-12 h_{20}^{}
 - 6 h_{22}^{}
 - 6 h_{41}^{}),\\
\bar{h}_{7}{}={}&\frac{1}{9} (18 h_{21}^{}-18 h_{13}^{}
 - 18 h_{20}^{}
 + 6 h_{24}^{}
 - 2 h_{29}^{}
 - 9 h_{33}^{}
 + 3 h_{37}^{}
 + 3 h_{41}^{}
 - 6 h_{42}^{}- 2 h_{5}^{}
 -  h_{6}^{}),\\
\bar{h}_{8}{}={}&\frac{1}{9} (36 h_{13}^{}
 + 36 h_{20}^{}
 - 36 h_{21}^{}
 - 12 h_{24}^{}
 - 2 h_{29}^{}
 + 18 h_{33}^{}
 + 3 h_{37}^{}
 - 6 h_{41}^{}
 + 12 h_{42}^{}- 2 h_{5}^{}
 -  h_{6}^{}),\\
\bar{h}_{9}{}={}&\frac{1}{9} (18 h_{22}^{}-36 h_{13}^{}
 - 18 h_{14}^{}
 - 6 h_{16}^{}
 - 12 h_{20}^{}
 - 6 h_{23}^{}
 + 6 h_{24}^{}
 + 9 h_{36}^{}
 + 2 h_{5}^{}+ h_{6}^{}
 - 4 h_{7}^{}
 -  h_{9}^{}),\\
\bar{h}_{10}{}={}&\frac{1}{9} (2 h_{7}^{}
 + 2 h_{9}^{}+3 h_{16}^{}-36 h_{13}^{}
 - 18 h_{14}^{}
 - 12 h_{20}^{}
 + 18 h_{22}^{}
 - 6 h_{23}^{}
 + 6 h_{24}^{}
 - 4 h_{5}^{}
 - 2 h_{6}^{}),\\
\bar{h}_{11}{}={}&\frac{1}{3} (6 h_{22}^{}
 -  h_{30}^{}
 - 2 h_{7}^{}),\\
\bar{h}_{12}{}={}&\frac{1}{3} (h_{9}^{}+ 6 h_{23}^{}-  h_{8}^{}-12 h_{22}^{}
 -  h_{29}^{}
 - 6 h_{42}^{}),\\
\bar{h}_{13}{}={}&\frac{1}{3} (6 h_{22}^{}
 - 6 h_{23}^{}
 + 2 h_{29}^{}
 + 12 h_{42}^{}
 - 6 h_{43}^{}
 + h_{8}^{}
 + h_{9}^{}),\\
\bar{h}_{14}{}=&-4 h_{21}^{}
 -  \frac{1}{3} h_{29}^{}
 - 2 h_{42}^{},\\
\bar{h}_{15}{}={}&\frac{1}{18} (2 h_{11}^{}-4 h_{10}^{}
 - 6 h_{23}^{}
 - 6 h_{24}^{}
 + 12 h_{25}^{}
 - 36 h_{28}^{}
 + 3 h_{29}^{}
 + 36 h_{3}^{}
 + 3 h_{30}^{}- 4 h_{31}^{}\nonumber\\
& + 2 h_{32}^{}
 - 18 h_{41}^{}
 + 18 h_{42}^{}
 + 12 h_{44}^{}
 + 12 h_{45}^{}
 + 2 h_{5}^{}
 + h_{6}^{}
 + 2 h_{7}^{}
 + h_{8}^{}+ h_{9}^{}),\\
\bar{h}_{16}{}={}&\frac{1}{9} (h_{11}^{}-2 h_{10}^{}
 + 6 h_{23}^{}
 + 6 h_{24}^{}
 - 12 h_{25}^{}
 - 36 h_{26}^{}
 + 36 h_{28}^{}
 - 3 h_{29}^{}
 - 36 h_{3}^{}- 3 h_{30}^{}\nonumber\\
&- 2 h_{31}^{}
 + h_{32}^{}
 + 18 h_{41}^{}
 - 18 h_{42}^{}
 + 6 h_{44}^{}
 - 12 h_{45}^{}
 - 2 h_{5}^{}
 -  h_{6}^{}
 - 2 h_{7}^{}-  h_{8}^{}
 -  h_{9}^{}),\\
\bar{h}_{17}{}={}&\frac{1}{9} (h_{11}^{}-2 h_{10}^{}
 + 6 h_{23}^{}
 + 6 h_{24}^{}
 + 6 h_{25}^{}
 + 36 h_{26}^{}
 + 36 h_{28}^{}
 - 3 h_{29}^{}
 - 36 h_{3}^{}- 3 h_{30}^{}\nonumber\\
& - 2 h_{31}^{}
 + h_{32}^{}
 + 18 h_{41}^{}
 - 18 h_{42}^{}
 - 12 h_{44}^{}
 + 6 h_{45}^{}
 - 2 h_{5}^{}
 -  h_{6}^{}
 - 2 h_{7}^{}-h_{8}^{}
 -  h_{9}^{}),\\
\bar{h}_{18}{}={}&\frac{1}{9} (5 h_{11}^{}- h_{10}^{}
 - 6 h_{23}^{}
 - 6 h_{24}^{}
 - 6 h_{25}^{}
 - 36 h_{28}^{}
 + 3 h_{29}^{}
 + 36 h_{3}^{}
 + 3 h_{30}^{}-  h_{31}^{}\nonumber\\
& - 4 h_{32}^{}
 - 18 h_{41}^{}
 + 18 h_{42}^{}
 - 6 h_{44}^{}
 - 6 h_{45}^{}
 + 2 h_{5}^{}
 + h_{6}^{}
 + 2 h_{7}^{}
 + h_{8}^{}+ h_{9}^{}),\\
\bar{h}_{19}{}={}&\frac{1}{9} (h_{10}^{}
 + 4 h_{11}^{}
 + 6 h_{23}^{}
 + 6 h_{24}^{}
 + 6 h_{25}^{}
 + 36 h_{28}^{}
 - 3 h_{29}^{}
 - 36 h_{3}^{}
 - 3 h_{30}^{}
 + h_{31}^{}\nonumber\\
& - 5 h_{32}^{}
 + 18 h_{41}^{}
 - 18 h_{42}^{}
 + 6 h_{44}^{}
 + 6 h_{45}^{}
 - 2 h_{5}^{}
 -  h_{6}^{}
 - 2 h_{7}^{}
 -  h_{8}^{}
 -  h_{9}^{}),\\
\bar{h}_{20}{}={}&\frac{1}{9} (9 h_{1}^{}
 + h_{10}^{}
 - 4 h_{11}^{}
 - 3 h_{16}^{}
 + 6 h_{17}^{}
 - 3 h_{19}^{}
 - 9 h_{27}^{}
 -  h_{29}^{}
 + h_{31}^{}
 + 3 h_{32}^{}\nonumber\\
& + 3 h_{36}^{}
 + 3 h_{37}^{}
 - 3 h_{39}^{}
 + 2 h_{5}^{}
 + h_{6}^{}
 - 2 h_{7}^{}
 -  h_{8}^{}
 -  h_{9}^{}),\\
\bar{h}_{21}{}={}&\frac{1}{9} (3 h_{11}^{}-3 h_{10}^{}
 + 18 h_{15}^{}
 - 3 h_{18}^{}
 + 6 h_{19}^{}
 + 12 h_{21}^{}
 - 6 h_{22}^{}
 + 6 h_{23}^{}
 + 6 h_{25}^{}+ 24 h_{26}^{}
 \nonumber\\
& -  h_{29}^{}- 3 h_{31}^{}
 + 18 h_{34}^{}
 + 3 h_{38}^{}
 + 3 h_{39}^{}
 - 6 h_{42}^{}
 + 6 h_{43}^{}
 - 6 h_{44}^{}+ 6 h_{45}^{}
 -  h_{8}^{}
 + 2 h_{9}^{}),\\
\bar{h}_{22}{}={}&\frac{1}{9} (3 h_{11}^{}-3 h_{10}^{}
 - 18 h_{15}^{}
 + 3 h_{18}^{}
 + 3 h_{19}^{}
 - 12 h_{21}^{}
 + 6 h_{22}^{}
 - 6 h_{23}^{}
 - 6 h_{25}^{}- 24 h_{26}^{}\nonumber\\
& + h_{29}^{} - 3 h_{31}^{}
 - 18 h_{34}^{}
 - 3 h_{38}^{}
 + 6 h_{39}^{}
 + 6 h_{42}^{}
 - 6 h_{43}^{}
 + 6 h_{44}^{}- 6 h_{45}^{}
 + h_{8}^{}
 + h_{9}^{}),\\
\bar{h}_{23}{}={}&\frac{1}{9} (18 h_{15}^{}
 + 6 h_{18}^{}
 - 3 h_{19}^{}
 + 12 h_{21}^{}
 - 6 h_{22}^{}
 + 6 h_{23}^{}
 + 6 h_{25}^{}
 + 24 h_{26}^{}
 + 2 h_{29}^{}\nonumber\\
& + 18 h_{34}^{}
 - 6 h_{38}^{}
 + 3 h_{39}^{}
 - 6 h_{42}^{}
 + 6 h_{43}^{}
 - 6 h_{44}^{}
 + 6 h_{45}^{}
 + 2 h_{8}^{}
 -  h_{9}^{}),\\
\bar{h}_{24}{}={}&\frac{1}{18} (12 h_{12}^{}
 + 6 h_{23}^{}
 + 6 h_{24}^{}
 - 3 h_{29}^{}
 - 36 h_{3}^{}
 - 3 h_{30}^{}
 - 36 h_{4}^{}
 + 18 h_{41}^{}
 - 18 h_{42}^{}- 2 h_{5}^{}
 -  h_{6}^{}
 - 2 h_{7}^{}
 -  h_{8}^{}
 -  h_{9}^{}),\\
\bar{h}_{25}{}={}&\frac{1}{9} (6 h_{12}^{}
 - 6 h_{23}^{}
 - 6 h_{24}^{}
 + 3 h_{29}^{}
 + 36 h_{3}^{}
 + 3 h_{30}^{}
 + 36 h_{4}^{}
 - 18 h_{41}^{}
 + 18 h_{42}^{}+ 2 h_{5}^{}
 + h_{6}^{}
 + 2 h_{7}^{}
 + h_{8}^{}
 + h_{9}^{}),\\
\bar{h}_{26}{}={}&\frac{1}{9} (h_{11}^{}- h_{10}^{}
 - 6 h_{12}^{}
 + 3 h_{19}^{}
 + 9 h_{2}^{}
 + h_{9}^{}),\\
\bar{h}_{27}{}={}&\frac{1}{3} h_{30}^{}-4 h_{21}^{}
 - 2 h_{22}^{}
 - 4 h_{41}^{},\\
\bar{h}_{28}{}=&-\frac{1}{3}h_{9}^{}-4h_{21}^{}
 - 2 h_{22}^{}
 - 2 h_{43}^{},\\
\bar{h}_{29}{}={}&\frac{1}{3} (12 h_{20}^{}
 + 6 h_{22}^{}
 - 6 h_{24}^{}
 + 2 h_{29}^{}
 + 6 h_{42}^{}
 + 6 h_{43}^{}
 + h_{6}^{}
 + h_{9}^{}),\\
\bar{h}_{30}{}={}&2 h_{24}^{}-4 h_{20}^{}
 - 4 h_{42}^{}
 -  \frac{1}{3} h_{6}^{},\\
\bar{h}_{31}{}={}&\frac{1}{6} \bigl[h_{30}^{}-12 h_{21}^{}
 + 2 (h_{5}^{}
 + h_{7}^{}-6 h_{41}^{})\bigr],\\
\bar{h}_{32}{}={}&\frac{1}{9} (18 h_{13}^{}
 + 6 h_{17}^{}
 + 6 h_{20}^{}
 - 2 h_{29}^{}
 - 9 h_{33}^{}
 + 3 h_{37}^{}
 + 3 h_{41}^{}
 - 6 h_{42}^{}
 + 2 h_{5}^{}
 + h_{6}^{}- 2 h_{7}^{}
 - 2 h_{8}^{}),\\
\bar{h}_{33}{}={}&\frac{1}{9} (6 h_{17}^{}-36 h_{13}^{}
 - 12 h_{20}^{}
 - 2 h_{29}^{}
 + 18 h_{33}^{}
 + 3 h_{37}^{}
 - 6 h_{41}^{}
 + 12 h_{42}^{}
 + 2 h_{5}^{}+ h_{6}^{}
 - 2 h_{7}^{}
 - 2 h_{8}^{}),\\
\bar{h}_{34}{}={}&\frac{1}{9} (36 h_{13}^{}
 + 18 h_{14}^{}
 - 3 h_{16}^{}
 + 12 h_{20}^{}
 - 18 h_{22}^{}
 + 6 h_{23}^{}
 - 6 h_{24}^{}
 + 9 h_{36}^{}
 - 2 h_{5}^{}-  h_{6}^{}
 - 2 h_{7}^{}
 + h_{9}^{}),\\
\bar{h}_{35}{}={}&\frac{1}{18} (2 h_{5}^{}
 + h_{6}^{}
 + 2 h_{7}^{}
 + h_{8}^{}
 + h_{9}^{}-6 h_{23}^{}
 - 6 h_{24}^{}
 + 36 h_{28}^{}
 + 3 h_{29}^{}+ 3 h_{30}^{}
 + 4 h_{31}^{}
 - 2 h_{32}^{}\nonumber\\
& 
 - 18 h_{41}^{}
 + 18 h_{42}^{}- 12 h_{44}^{}
 - 12 h_{45}^{}),\\
\bar{h}_{36}{}={}&\frac{1}{9} (2 h_{5}^{}
 + h_{6}^{}
 + 2 h_{7}^{}
 + h_{8}^{}
 + h_{9}^{}-6 h_{23}^{}
 - 6 h_{24}^{}
 + 36 h_{28}^{}
 + 3 h_{29}^{}+ 3 h_{30}^{}
 + h_{31}^{}
 + 4 h_{32}^{}\nonumber\\
& 
 - 18 h_{41}^{}
 + 18 h_{42}^{}+ 6 h_{44}^{}
 + 6 h_{45}^{}),\\
\bar{h}_{37}{}={}&\frac{1}{9} (6 h_{23}^{}
 + 6 h_{24}^{}
 - 36 h_{28}^{}
 - 3 h_{29}^{}
 - 3 h_{30}^{}
 + 2 h_{31}^{}
 -  h_{32}^{}
 + 18 h_{41}^{}
 - 18 h_{42}^{} - 6 h_{44}^{}
 + 12 h_{45}^{}
 - 2 h_{5}^{}\nonumber\\
&
 -  h_{6}^{}
 - 2 h_{7}^{}
 -  h_{8}^{}
 -  h_{9}^{}),\\
\bar{h}_{38}{}={}&\frac{1}{9} (6 h_{23}^{}
 + 6 h_{24}^{}
 - 36 h_{28}^{}
 - 3 h_{29}^{}
 - 3 h_{30}^{}
 + 2 h_{31}^{}
 -  h_{32}^{}
 + 18 h_{41}^{}
 - 18 h_{42}^{} + 12 h_{44}^{}
 - 6 h_{45}^{}
 - 2 h_{5}^{}\nonumber\\
&
 -  h_{6}^{}
 - 2 h_{7}^{}
 -  h_{8}^{}
 -  h_{9}^{}),\\
\bar{h}_{39}{}={}&\frac{1}{9} (6 h_{23}^{}
 + 6 h_{24}^{}
 - 36 h_{28}^{}
 - 3 h_{29}^{}
 - 3 h_{30}^{}
 -  h_{31}^{}
 + 5 h_{32}^{}
 + 18 h_{41}^{}
 - 18 h_{42}^{}- 6 h_{44}^{}
 - 6 h_{45}^{}
 - 2 h_{5}^{}\nonumber\\
& 
 -  h_{6}^{}
 - 2 h_{7}^{}
 -  h_{8}^{}
 -  h_{9}^{}),\\
\bar{h}_{40}{}={}&\frac{1}{9} (9 h_{27}^{}
 + h_{29}^{}
 -  h_{31}^{}
 - 3 h_{32}^{}
 - 3 h_{36}^{}
 - 3 h_{37}^{}
 + 3 h_{39}^{}
 + 2 h_{5}^{}
 + h_{6}^{}
 -  h_{9}^{}),\\
\bar{h}_{41}{}={}&\frac{1}{9} (2 h_{9}^{}-6 h_{22}^{}
 + h_{29}^{}
 + 3 h_{31}^{}
 - 18 h_{34}^{}
 - 3 h_{38}^{}
 - 3 h_{39}^{}
 + 6 h_{42}^{}
 - 6 h_{43}^{}
 + 6 h_{44}^{}- 6 h_{45}^{}),\\
\bar{h}_{42}{}={}&\frac{1}{9} (6 h_{22}^{}
 -  h_{29}^{}
 + 3 h_{31}^{}
 + 18 h_{34}^{}
 + 3 h_{38}^{}
 - 6 h_{39}^{}
 - 6 h_{42}^{}
 + 6 h_{43}^{}
 - 6 h_{44}^{}
 + 6 h_{45}^{}+ h_{9}^{}),\\
\bar{h}_{43}{}={}&\frac{1}{9} (6 h_{38}^{}-6 h_{22}^{}
 - 2 h_{29}^{}
 - 18 h_{34}^{}
 - 3 h_{39}^{}
 + 6 h_{42}^{}
 - 6 h_{43}^{}
 + 6 h_{44}^{}
 - 6 h_{45}^{}-  h_{9}^{}),\\
\bar{h}_{44}{}={}&2 h_{46}^{},\\
\bar{h}_{45}{}={}&\frac{1}{9} (2 h_{7}^{}-6 h_{21}^{}
 - 18 h_{35}^{}
 + 3 h_{40}^{}
 - 6 h_{41}^{}
 + 6 h_{46}^{}),\\
\bar{h}_{46}{}={}&\frac{1}{9} (2 h_{7}^{}+12 h_{21}^{}
 + 36 h_{35}^{}
 + 3 h_{40}^{}
 + 12 h_{41}^{}
 - 12 h_{46}^{}).
\end{align}

\section{Second branch of the cubic Lagrangian in terms of the nonmetricity tensor}\label{appendix2}

The Lagrangian~\eqref{cubicLagIrrQ} containing cubic invariants from curvature and nonmetricity can be written in terms of the nonmetricity tensor as:
\begin{equation}
    \mathcal{\bar{L}}_{\rm curv-nonm}^{(3)}=\mathcal{\bar{L}}^{(3)}_{1,\rm non}+\mathcal{\bar{L}}^{(3)}_{2,\rm non}+\mathcal{\bar{L}}^{(3)}_{3,\rm non}+\mathcal{\bar{L}}^{(3)}_{4,\rm non}+\mathcal{\bar{L}}^{(3)}_{5,\rm non}\,,
\end{equation}
where we have defined:
\begin{eqnarray}
    \mathcal{\bar{L}}^{(3)}_{1,\rm non}&=&\bar{h}_{47}{} Q^{\alpha \rho \tau } Q^{\gamma }{}_{\rho }{}^{\mu } \tilde{R}_{\alpha \tau \gamma \mu } + \bar{h}_{48}{} Q^{\alpha \rho \tau } Q_{\rho }{}^{\gamma \mu } \tilde{R}_{\alpha \gamma \tau \mu }  + \bar{h}_{49}{} Q^{\alpha \rho \tau } Q^{\gamma }{}_{\rho }{}^{\mu } \tilde{R}_{\alpha \gamma \tau \mu } + \bar{h}_{50}{} Q^{\alpha \rho \tau } Q^{\gamma }{}_{\rho }{}^{\mu } \tilde{R}_{\alpha \mu \tau \gamma } \nonumber\\
    && + \bar{h}_{51}{} Q_{\alpha }{}^{\gamma \mu } Q^{\alpha \rho \tau } \tilde{R}_{\rho \gamma \tau \mu } + \bar{h}_{52}{} Q^{\alpha }{}_{\alpha }{}^{\rho } Q^{\tau \gamma \mu } \tilde{R}_{\rho \gamma \tau \mu } + \bar{h}_{53}{} Q^{\alpha \rho \tau } Q^{\gamma }{}_{\rho }{}^{\mu } \tilde{R}_{\tau \alpha \gamma \mu }+ \bar{h}_{54}{} Q^{\alpha \rho \tau } Q_{\rho }{}^{\gamma \mu } \tilde{R}_{\tau \gamma \alpha \mu }   \nonumber\\
    && + \bar{h}_{55}{} Q^{\alpha \rho \tau } Q^{\gamma }{}_{\rho }{}^{\mu } \tilde{R}_{\tau \gamma \alpha \mu } + \bar{h}_{56}{} Q^{\alpha }{}_{\alpha }{}^{\rho } Q^{\tau \gamma \mu } \tilde{R}_{\tau \gamma \rho \mu }+ \bar{h}_{57}{} Q^{\alpha \rho \tau } Q^{\gamma }{}_{\rho }{}^{\mu } \tilde{R}_{\tau \mu \alpha \gamma } + \bar{h}_{58}{} Q^{\alpha \rho \tau } Q_{\rho }{}^{\gamma \mu } \tilde{R}_{\gamma \alpha \tau \mu }\nonumber\\
    &&+ \bar{h}_{59}{} Q^{\alpha }{}_{\alpha }{}^{\rho } Q^{\tau \gamma \mu } \tilde{R}_{\gamma \rho \tau \mu }+ \bar{h}_{60}{} Q^{\alpha \rho \tau } Q_{\rho }{}^{\gamma \mu } \tilde{R}_{\gamma \tau \alpha \mu }  + \bar{h}_{61}{} Q^{\alpha }{}_{\alpha }{}^{\rho } Q^{\tau \gamma \mu } \tilde{R}_{\gamma \tau \rho \mu } + \bar{h}_{62}{} Q^{\alpha \rho \tau } Q_{\rho }{}^{\gamma \mu } \tilde{R}_{\gamma \mu \alpha \tau } \nonumber\\
    &&+  \bar{h}_{63}{} Q^{\alpha }{}_{\alpha }{}^{\rho } Q^{\tau \gamma \mu } \tilde{R}_{\gamma \mu \rho \tau }+\bar{h}_{64}{} Q^{\alpha }{}_{\rho }{}^{\rho } Q^{\tau \gamma \mu } \tilde{R}_{\alpha \gamma \tau \mu }+ \bar{h}_{65}{} Q^{\alpha }{}_{\rho }{}^{\rho } Q^{\tau \gamma \mu } \tilde{R}_{\tau \gamma \alpha \mu } + \bar{h}_{66}{} Q^{\alpha }{}_{\rho }{}^{\rho } Q^{\tau \gamma \mu } \tilde{R}_{\gamma \alpha \tau \mu }\nonumber\\
    &&+ \bar{h}_{67}{} Q^{\alpha }{}_{\rho }{}^{\rho } Q^{\tau \gamma \mu } \tilde{R}_{\gamma \tau \alpha \mu }+ \bar{h}_{68}{} Q^{\alpha }{}_{\rho }{}^{\rho } Q^{\tau \gamma \mu } \tilde{R}_{\gamma \mu \alpha \tau }\,,\\
     \mathcal{\bar{L}}^{(3)}_{2,\rm non}&=&\bar{h}_{69}{} Q_{\alpha }{}^{\tau \gamma } Q_{\rho \tau \gamma } \tilde{R}^{\alpha \rho } + \bar{h}_{70}{} Q_{\rho }{}^{\tau \gamma } Q_{\tau \alpha \gamma } \tilde{R}^{\alpha \rho } + \bar{h}_{71}{} Q_{\alpha }{}^{\tau \gamma } Q_{\tau \rho \gamma } \tilde{R}^{\alpha \rho }+ \bar{h}_{72}{} Q_{\tau \rho \gamma } Q^{\tau }{}_{\alpha }{}^{\gamma } \tilde{R}^{\alpha \rho }+ \bar{h}_{73}{} Q^{\tau }{}_{\alpha }{}^{\gamma } Q_{\gamma \rho \tau } \tilde{R}^{\alpha \rho }\nonumber\\
     &&  + \bar{h}_{74}{} Q^{\tau }{}_{\alpha \tau } Q^{\gamma }{}_{\rho \gamma } \tilde{R}^{\alpha \rho } + \bar{h}_{75}{} Q_{\alpha \rho }{}^{\tau } Q^{\gamma }{}_{\tau \gamma } \tilde{R}^{\alpha \rho } + \bar{h}_{76}{} Q_{\rho \alpha }{}^{\tau } Q^{\gamma }{}_{\tau \gamma } \tilde{R}^{\alpha \rho } + \bar{h}_{77}{} Q^{\tau }{}_{\alpha \rho } Q^{\gamma }{}_{\tau \gamma } \tilde{R}^{\alpha \rho }+ \bar{h}_{78}{} Q_{\alpha \tau }{}^{\tau } Q^{\gamma }{}_{\rho \gamma } \tilde{R}^{\alpha \rho }\nonumber\\
     &&+ \bar{h}_{79}{} Q_{\rho \tau }{}^{\tau } Q^{\gamma }{}_{\alpha \gamma } \tilde{R}^{\alpha \rho } + \bar{h}_{80}{} Q_{\alpha \tau }{}^{\tau } Q_{\rho \gamma }{}^{\gamma } \tilde{R}^{\alpha \rho } + \bar{h}_{81}{} Q_{\alpha \rho }{}^{\tau } Q_{\tau \gamma }{}^{\gamma } \tilde{R}^{\alpha \rho }+ \bar{h}_{82}{} Q_{\rho \alpha }{}^{\tau } Q_{\tau \gamma }{}^{\gamma } \tilde{R}^{\alpha \rho } + \bar{h}_{83}{} Q_{\tau \gamma }{}^{\gamma } Q^{\tau }{}_{\alpha \rho } \tilde{R}^{\alpha \rho }\,,\nonumber \\ \\
      \mathcal{\bar{L}}^{(3)}_{3,\rm non}&=&\bar{h}_{84}{} Q_{\alpha }{}^{\tau \gamma } Q_{\rho \tau \gamma } \hat{R}^{\alpha \rho }  + \bar{h}_{85}{} Q_{\rho }{}^{\tau \gamma } Q_{\tau \alpha \gamma } \hat{R}^{\alpha \rho } + \bar{h}_{86}{} Q_{\alpha }{}^{\tau \gamma } Q_{\tau \rho \gamma } \hat{R}^{\alpha \rho }  + \bar{h}_{87}{} Q_{\tau \rho \gamma } Q^{\tau }{}_{\alpha }{}^{\gamma } \hat{R}^{\alpha \rho } + \bar{h}_{88}{} Q^{\tau }{}_{\alpha }{}^{\gamma } Q_{\gamma \rho \tau } \hat{R}^{\alpha \rho }  \nonumber\\
      &&+ \bar{h}_{89}{} Q^{\tau }{}_{\alpha \tau } Q^{\gamma }{}_{\rho \gamma } \hat{R}^{\alpha \rho }+ \bar{h}_{90}{} Q_{\alpha \rho }{}^{\tau } Q^{\gamma }{}_{\tau \gamma } \hat{R}^{\alpha \rho } + \bar{h}_{91}{} Q_{\rho \alpha }{}^{\tau } Q^{\gamma }{}_{\tau \gamma } \hat{R}^{\alpha \rho } + \bar{h}_{92}{} Q^{\tau }{}_{\alpha \rho } Q^{\gamma }{}_{\tau \gamma } \hat{R}^{\alpha \rho }+ \bar{h}_{93}{} Q_{\alpha \tau }{}^{\tau } Q^{\gamma }{}_{\rho \gamma } \hat{R}^{\alpha \rho }\nonumber\\
      &&+ \bar{h}_{94}{} Q_{\rho \tau }{}^{\tau } Q^{\gamma }{}_{\alpha \gamma } \hat{R}^{\alpha \rho }  + \bar{h}_{95}{} Q_{\alpha \tau }{}^{\tau } Q_{\rho \gamma }{}^{\gamma } \hat{R}^{\alpha \rho }+ \bar{h}_{96}{} Q_{\alpha \rho }{}^{\tau } Q_{\tau \gamma }{}^{\gamma } \hat{R}^{\alpha \rho } + \bar{h}_{97}{} Q_{\rho \alpha }{}^{\tau } Q_{\tau \gamma }{}^{\gamma } \hat{R}^{\alpha \rho } + \bar{h}_{98}{} Q_{\tau \gamma }{}^{\gamma } Q^{\tau }{}_{\alpha \rho } \hat{R}^{\alpha \rho }\,,\nonumber \\ \\
       \mathcal{\bar{L}}^{(3)}_{4,\rm non}&=&\bar{h}_{99}{} Q^{\alpha \rho \tau } Q_{\rho \tau }{}^{\gamma } \tilde{R}^{\omega}{}_{\omega\alpha \gamma } + \bar{h}_{100}{} Q^{\alpha }{}_{\alpha }{}^{\rho } Q^{\tau }{}_{\rho }{}^{\gamma } \tilde{R}^{\omega}{}_{\omega\tau \gamma }+ \bar{h}_{101}{} Q^{\alpha }{}_{\rho }{}^{\rho } Q^{\tau }{}_{\alpha }{}^{\gamma } \tilde{R}^{\omega}{}_{\omega\tau \gamma }+ \bar{h}_{102}{} Q^{\alpha }{}_{\alpha }{}^{\rho } Q^{\tau }{}_{\gamma }{}^{\gamma } \tilde{R}^{\omega}{}_{\omega\rho \tau }  \,,\\
        \mathcal{\bar{L}}^{(3)}_{5,\rm non}&=&\bar{h}_{103}{} Q_{\alpha \rho \tau } Q^{\alpha \rho \tau } \tilde{R}  + \bar{h}_{104}{} Q^{\alpha \rho \tau } Q_{\rho \alpha \tau } \tilde{R}+ \bar{h}_{105}{} Q^{\alpha }{}_{\alpha }{}^{\rho } Q^{\tau }{}_{\rho \tau } \tilde{R} + \bar{h}_{106}{} Q_{\alpha \rho }{}^{\rho } Q^{\alpha }{}_{\tau }{}^{\tau } \tilde{R}+ \bar{h}_{107}{} Q^{\alpha }{}_{\alpha }{}^{\rho } Q_{\rho \tau }{}^{\tau } \tilde{R} \,,
\end{eqnarray}
and where the respective Lagrangian coefficients are related as:
\begin{align}
\bar{h}_{47}{}=&-\frac{2}{3} h_{100}^{}
 -  h_{57}^{},\\
\bar{h}_{48}{}={}&\frac{1}{9} (6 h_{101}^{}
 - 9 h_{56}^{}
 + 9 h_{58}^{}
 + h_{64}^{}
 + 6 h_{98}^{}
 + 6 h_{99}^{}),\\
\bar{h}_{49}{}={}&\frac{1}{9} (6 h_{101}^{}
 - 6 h_{102}^{}
 - 9 h_{58}^{}
 + h_{64}^{}
 + 6 h_{99}^{}),\\
\bar{h}_{50}{}={}&\frac{1}{9} (6 h_{101}^{}
 + 6 h_{102}^{}
 + 9 h_{56}^{}
 + h_{64}^{}
 - 6 h_{98}^{}
 + 6 h_{99}^{}),\\
\bar{h}_{51}{}={}&\frac{1}{9} (-9 h_{53}^{}
 - 9 h_{54}^{}
 + 9 h_{56}^{}
 - 9 h_{58}^{}
 + h_{64}^{}
 + 6 h_{98}^{}
 - 6 h_{99}^{}),\\
\bar{h}_{52}{}={}&\frac{1}{9} (2 h_{100}^{}
 + 2 h_{102}^{}
 - 6 h_{56}^{}
 - 3 h_{57}^{}
 + 3 h_{58}^{}
 + 8 h_{86}^{}
 - 4 h_{98}^{}),\\
\bar{h}_{53}{}={}&- \frac{2}{3} h_{100}^{}
 + h_{55}^{},\\
\bar{h}_{54}{}={}&\frac{1}{9} (-6 h_{101}^{}
 + 9 h_{53}^{}
 + 9 h_{54}^{}
 + h_{64}^{}
 + 6 h_{98}^{}),\\
\bar{h}_{55}{}={}&\frac{1}{9} (-6 h_{101}^{}
 - 6 h_{102}^{}
 - 9 h_{53}^{}
 + h_{64}^{}),\\
\bar{h}_{56}{}={}&\frac{1}{27} (-6 h_{100}^{}
 - 6 h_{102}^{}
 + 9 h_{53}^{}
 + 9 h_{54}^{}
 - 9 h_{56}^{}
 - 9 h_{57}^{}
 + 24 h_{89}^{}
 + 4 h_{95}^{}
 + 6 h_{99}^{}),\\
\bar{h}_{57}{}={}&\frac{1}{9} (-6 h_{101}^{}
 + 6 h_{102}^{}
 - 9 h_{54}^{}
 + h_{64}^{}
 - 6 h_{98}^{}),\\
\bar{h}_{58}{}={}&\frac{1}{9} (6 h_{102}^{}
 - 9 h_{53}^{}
 - 9 h_{58}^{}
 -  h_{64}^{}
 + 6 h_{99}^{}),\\
\bar{h}_{59}{}={}&\frac{1}{9} (2 h_{100}^{}
 + 2 h_{101}^{}
 - 6 h_{53}^{}
 + 3 h_{55}^{}
 - 3 h_{58}^{}
 + 8 h_{88}^{}
 - 2 h_{99}^{}),\\
\bar{h}_{60}{}={}&\frac{1}{9} (-6 h_{102}^{}
 - 9 h_{54}^{}
 + 9 h_{56}^{}
 -  h_{64}^{}
 + 6 h_{98}^{}
 + 6 h_{99}^{}),\\
\bar{h}_{61}{}={}&\frac{1}{27} (-6 h_{100}^{}
 - 6 h_{101}^{}
 - 9 h_{53}^{}
 - 9 h_{54}^{}
 + 9 h_{55}^{}
 + 9 h_{56}^{}
 + 24 h_{87}^{}
 + 4 h_{95}^{}
 + 6 h_{98}^{}- 6 h_{99}^{}),\\
\bar{h}_{62}{}={}&- \frac{2}{3} h_{100}^{}
 -  h_{55}^{}
 + h_{57}^{},\\
\bar{h}_{63}{}={}&\frac{1}{27} (-6 h_{100}^{}
 - 12 h_{101}^{}
 - 12 h_{102}^{}
 - 9 h_{55}^{}
 + 9 h_{57}^{}
 - 24 h_{87}^{}
 - 24 h_{89}^{}
 + 4 h_{95}^{} + 12 h_{98}^{}),\\
\bar{h}_{64}{}={}&\frac{1}{18} (2 h_{100}^{}
 + 2 h_{102}^{}
 + 12 h_{56}^{}
 + 6 h_{57}^{}
 - 6 h_{58}^{}
 + 9 h_{75}^{}
 - 4 h_{86}^{}
 - 4 h_{98}^{}),\\
\bar{h}_{65}{}={}&\frac{1}{108} (24 h_{100}^{}
 + 36 h_{101}^{}
 + 24 h_{102}^{}
 - 36 h_{53}^{}
 - 36 h_{54}^{}
 + 36 h_{56}^{}
 + 36 h_{57}^{}
 + 54 h_{77}^{}\nonumber\\
& + 9 h_{83}^{}
 - 24 h_{89}^{}
 - 4 h_{95}^{}
 - 36 h_{98}^{}
 + 12 h_{99}^{}),\\
\bar{h}_{66}{}={}&\frac{1}{18} (2 h_{100}^{}
 + 2 h_{101}^{}
 + 12 h_{53}^{}
 - 6 h_{55}^{}
 + 6 h_{58}^{}
 + 9 h_{76}^{}
 - 4 h_{88}^{}
 - 2 h_{99}^{}),\\
\bar{h}_{67}{}={}&\frac{1}{108} (24 h_{100}^{}
 + 24 h_{101}^{}
 + 36 h_{102}^{}
 + 36 h_{53}^{}
 + 36 h_{54}^{}
 - 36 h_{55}^{}
 - 36 h_{56}^{}
 + 54 h_{74}^{}\nonumber\\
& + 9 h_{83}^{}
 - 24 h_{87}^{}
 - 4 h_{95}^{}
 - 24 h_{98}^{}
 - 12 h_{99}^{}),\\
\bar{h}_{68}{}={}&\frac{1}{108} (24 h_{100}^{}
 + 12 h_{101}^{}
 + 12 h_{102}^{}
 + 36 h_{55}^{}
 - 36 h_{57}^{}
 - 54 h_{74}^{}
 - 54 h_{77}^{}
 + 9 h_{83}^{}\nonumber\\
& + 24 h_{87}^{}
 + 24 h_{89}^{}
 - 4 h_{95}^{}
 - 12 h_{98}^{}),\\
\bar{h}_{69}{}={}&\frac{1}{9} (6 h_{103}^{}
 + 6 h_{104}^{}
 + 9 h_{59}^{}
 + h_{65}^{}),\\
\bar{h}_{70}{}={}&\frac{1}{9} (-6 h_{103}^{}
 + 12 h_{104}^{}
 - 9 h_{59}^{}
 + 2 h_{65}^{}),\\
\bar{h}_{71}{}={}&\frac{1}{9} (12 h_{103}^{}
 - 6 h_{104}^{}
 - 9 h_{59}^{}
 + 2 h_{65}^{}),\\
\bar{h}_{72}{}={}&- \frac{2}{3} h_{103}^{}
 -  \frac{2}{3} h_{104}^{}
 + h_{56}^{}
 + h_{57}^{}
 + h_{59}^{}
 - 2 h_{60}^{}
 + \frac{2}{9} h_{65}^{},\\
\bar{h}_{73}{}={}&\frac{1}{9} (-6 h_{103}^{}
 - 6 h_{104}^{}
 - 9 h_{56}^{}
 - 9 h_{57}^{}
 + 18 h_{60}^{}
 + 2 h_{65}^{}),\\
\bar{h}_{74}{}={}&\frac{1}{81} (-6 h_{100}^{}
 - 6 h_{101}^{}
 - 6 h_{102}^{}
 + 30 h_{103}^{}
 + 30 h_{104}^{}
 + 16 h_{50}^{}
 + 18 h_{53}^{}
 + 9 h_{54}^{}\nonumber\\
& - 9 h_{55}^{}
 - 18 h_{56}^{}
 - 9 h_{57}^{}
 + 9 h_{58}^{}
 - 36 h_{59}^{}
 + 18 h_{60}^{}
 + h_{64}^{}
 - 2 h_{65}^{}
 - 24 h_{87}^{}\nonumber\\
& - 24 h_{88}^{}
 + 24 h_{90}^{}
 + 24 h_{91}^{}
 + 4 h_{95}^{}
 - 8 h_{96}^{}
 + 6 h_{98}^{}
 + 12 h_{99}^{}),\\
\bar{h}_{75}{}={}&\frac{1}{27} (6 h_{102}^{}
 - 18 h_{103}^{}
 + 9 h_{53}^{}
 - 9 h_{56}^{}
 - 9 h_{57}^{}
 + 9 h_{58}^{}
 - 9 h_{59}^{}
 + 18 h_{60}^{}
 -  h_{64}^{}\nonumber\\
& - 4 h_{65}^{}
 + 24 h_{91}^{}
 + 4 h_{96}^{}
 - 6 h_{99}^{}),\\
\bar{h}_{76}{}={}&\frac{1}{27} (-6 h_{102}^{}
 - 18 h_{104}^{}
 + 9 h_{54}^{}
 - 18 h_{56}^{}
 - 9 h_{57}^{}
 - 9 h_{59}^{}
 + 18 h_{60}^{}
 -  h_{64}^{}
 - 4 h_{65}^{}\nonumber\\
& + 24 h_{90}^{}
 + 4 h_{96}^{}
 + 6 h_{98}^{}
 - 6 h_{99}^{}),\\
\bar{h}_{77}{}={}&\frac{1}{27} (-9 h_{53}^{}
 - 9 h_{54}^{}
 + 27 h_{56}^{}
 + 18 h_{57}^{}
 - 9 h_{58}^{}
 + 18 h_{59}^{}
 - 36 h_{60}^{}
 -  h_{64}^{}
 - 4 h_{65}^{}\nonumber\\
& - 24 h_{90}^{}
 - 24 h_{91}^{}
 + 4 h_{96}^{}
 - 6 h_{98}^{}
 - 6 h_{99}^{}),\\
\bar{h}_{78}{}={}&\frac{1}{324} (-12 h_{100}^{}
 - 12 h_{101}^{}
 - 12 h_{102}^{}
 - 84 h_{103}^{}
 + 24 h_{104}^{}
 - 16 h_{50}^{}
 - 72 h_{53}^{}\nonumber\\
& - 36 h_{54}^{}
 + 36 h_{55}^{}
 + 72 h_{56}^{}
 + 36 h_{57}^{}
 - 36 h_{58}^{}
 + 144 h_{59}^{}
 - 72 h_{60}^{}
 + 2 h_{64}^{}\nonumber\\
& - 4 h_{65}^{}
 + 36 h_{69}^{}
 - 54 h_{76}^{}
 + 54 h_{78}^{}
 - 9 h_{84}^{}
 + 96 h_{87}^{}
 + 24 h_{88}^{}
 - 24 h_{90}^{}\nonumber\\
& - 96 h_{91}^{}
 + 8 h_{95}^{}
 - 4 h_{96}^{}
 + 12 h_{98}^{}
 - 12 h_{99}^{}),\\
\bar{h}_{79}{}={}&\frac{1}{324} (24 h_{100}^{}
 + 24 h_{101}^{}
 + 24 h_{102}^{}
 + 24 h_{103}^{}
 - 84 h_{104}^{}
 - 16 h_{50}^{}
 - 72 h_{53}^{}\nonumber\\
& - 36 h_{54}^{}
 + 36 h_{55}^{}
 + 72 h_{56}^{}
 + 36 h_{57}^{}
 - 36 h_{58}^{}
 + 144 h_{59}^{}
 - 72 h_{60}^{}
 + 2 h_{64}^{}\nonumber\\
& - 4 h_{65}^{}
 + 36 h_{68}^{}
 - 54 h_{74}^{}
 + 54 h_{79}^{}
 + 9 h_{83}^{}
 - 9 h_{84}^{}
 + 24 h_{87}^{}
 + 96 h_{88}^{}\nonumber\\
& - 96 h_{90}^{}
 - 24 h_{91}^{}
 - 4 h_{95}^{}
 - 4 h_{96}^{}
 - 24 h_{98}^{}
 - 12 h_{99}^{}),\\
\bar{h}_{80}{}={}&\frac{1}{1296} (48 h_{100}^{}
 + 48 h_{101}^{}
 + 48 h_{102}^{}
 - 240 h_{103}^{}
 - 240 h_{104}^{}
 + 81 h_{47}^{}
 + 16 h_{50}^{}\nonumber\\
& + 288 h_{53}^{}
 + 144 h_{54}^{}
 - 144 h_{55}^{}
 - 288 h_{56}^{}
 - 144 h_{57}^{}
 + 144 h_{58}^{}
 - 576 h_{59}^{}\nonumber\\
& + 288 h_{60}^{}
 + 4 h_{64}^{}
 - 8 h_{65}^{}
 - 36 h_{68}^{}
 - 36 h_{69}^{}
 + 216 h_{74}^{}
 + 216 h_{76}^{}
 - 216 h_{78}^{}\nonumber\\
& - 216 h_{79}^{}
 + 18 h_{83}^{}
 - 36 h_{84}^{}
 - 96 h_{87}^{}
 - 96 h_{88}^{}
 + 96 h_{90}^{}
 + 96 h_{91}^{}
 - 8 h_{95}^{}\nonumber\\
& + 16 h_{96}^{}
 - 48 h_{98}^{}
 - 96 h_{99}^{}),\\
\bar{h}_{81}{}={}&\frac{1}{108} (12 h_{102}^{}
 + 36 h_{104}^{}
 - 36 h_{53}^{}
 + 36 h_{56}^{}
 + 36 h_{57}^{}
 - 36 h_{58}^{}
 + 36 h_{59}^{}
 - 72 h_{60}^{}\nonumber\\
& - 2 h_{64}^{}
 - 8 h_{65}^{}
 + 54 h_{79}^{}
 + 9 h_{84}^{}
 - 24 h_{91}^{}
 - 4 h_{96}^{}
 + 24 h_{99}^{}),\\
\bar{h}_{82}{}={}&\frac{1}{108} (-12 h_{102}^{}
 + 36 h_{103}^{}
 - 36 h_{54}^{}
 + 72 h_{56}^{}
 + 36 h_{57}^{}
 + 36 h_{59}^{}
 - 72 h_{60}^{}
 - 2 h_{64}^{}\nonumber\\
& - 8 h_{65}^{}
 + 54 h_{78}^{}
 + 9 h_{84}^{}
 - 24 h_{90}^{}
 - 4 h_{96}^{}
 + 12 h_{98}^{}
 + 24 h_{99}^{}),\\
\bar{h}_{83}{}={}&\frac{1}{108} (36 h_{103}^{}
 + 36 h_{104}^{}
 + 36 h_{53}^{}
 + 36 h_{54}^{}
 - 108 h_{56}^{}
 - 72 h_{57}^{}
 + 36 h_{58}^{}
 - 72 h_{59}^{}\nonumber\\
& + 144 h_{60}^{}
 - 2 h_{64}^{}
 - 8 h_{65}^{}
 - 54 h_{78}^{}
 - 54 h_{79}^{}
 + 9 h_{84}^{}
 + 24 h_{90}^{}
 + 24 h_{91}^{}\nonumber\\
& - 4 h_{96}^{}
 - 12 h_{98}^{}
 + 24 h_{99}^{}),\\
\bar{h}_{84}{}={}&\frac{1}{9} (6 h_{105}^{}
 + 6 h_{106}^{}
 + 9 h_{61}^{}
 + h_{66}^{}),\\
\bar{h}_{85}{}={}&\frac{1}{9} (-6 h_{105}^{}
 + 12 h_{106}^{}
 - 9 h_{61}^{}
 + 2 h_{66}^{}),\\
\bar{h}_{86}{}={}&\frac{1}{9} (12 h_{105}^{}
 - 6 h_{106}^{}
 - 9 h_{61}^{}
 + 2 h_{66}^{}),\\
\bar{h}_{87}{}={}&- \frac{2}{3} h_{105}^{}
 -  \frac{2}{3} h_{106}^{}
 -  h_{53}^{}
 + h_{55}^{}
 + h_{61}^{}
 - 2 h_{62}^{}
 + \frac{2}{9} h_{66}^{},\\
\bar{h}_{88}{}={}&- \frac{2}{3} h_{105}^{}
 -  \frac{2}{3} h_{106}^{}
 + h_{53}^{}
 -  h_{55}^{}
 + 2 h_{62}^{}
 + \frac{2}{9} h_{66}^{},\\
\bar{h}_{89}{}={}&\frac{1}{81} (6 h_{100}^{}
 + 6 h_{101}^{}
 + 6 h_{102}^{}
 + 30 h_{105}^{}
 + 30 h_{106}^{}
 + 16 h_{51}^{}
 + 18 h_{53}^{}
 + 9 h_{54}^{}\nonumber\\
& - 9 h_{55}^{}
 - 18 h_{56}^{}
 - 9 h_{57}^{}
 + 9 h_{58}^{}
 - 36 h_{61}^{}
 + 18 h_{62}^{}
 + h_{64}^{}
 - 2 h_{66}^{}
 + 24 h_{86}^{}\nonumber\\
& + 24 h_{89}^{}
 + 24 h_{92}^{}
 + 24 h_{93}^{}
 - 4 h_{95}^{}
 - 8 h_{97}^{}
 - 18 h_{98}^{}),\\
\bar{h}_{90}{}={}&\frac{1}{27} (-6 h_{101}^{}
 - 18 h_{105}^{}
 + 9 h_{53}^{}
 - 9 h_{55}^{}
 - 9 h_{56}^{}
 + 9 h_{58}^{}
 - 9 h_{61}^{}
 + 18 h_{62}^{}
 -  h_{64}^{}\nonumber\\
& - 4 h_{66}^{}
 + 24 h_{93}^{}
 + 4 h_{97}^{}
 + 6 h_{98}^{}
 - 6 h_{99}^{}),\\
\bar{h}_{91}{}={}&\frac{1}{27} (6 h_{101}^{}
 - 18 h_{106}^{}
 + 18 h_{53}^{}
 + 9 h_{54}^{}
 - 9 h_{55}^{}
 - 9 h_{61}^{}
 + 18 h_{62}^{}
 -  h_{64}^{}
 - 4 h_{66}^{}\nonumber\\
& + 24 h_{92}^{}
 + 4 h_{97}^{}
 + 6 h_{98}^{}),\\
\bar{h}_{92}{}={}&\frac{1}{27} (-27 h_{53}^{}
 - 9 h_{54}^{}
 + 18 h_{55}^{}
 + 9 h_{56}^{}
 - 9 h_{58}^{}
 + 18 h_{61}^{}
 - 36 h_{62}^{}
 -  h_{64}^{}
 - 4 h_{66}^{}\nonumber\\
& - 24 h_{92}^{}
 - 24 h_{93}^{}
 + 4 h_{97}^{}
 + 6 h_{98}^{}
 + 6 h_{99}^{}),\\
\bar{h}_{93}{}={}&\frac{1}{324} (12 h_{100}^{}
 + 12 h_{101}^{}
 + 12 h_{102}^{}
 - 84 h_{105}^{}
 + 24 h_{106}^{}
 - 16 h_{51}^{}
 - 72 h_{53}^{}\nonumber\\
& - 36 h_{54}^{}
 + 36 h_{55}^{}
 + 72 h_{56}^{}
 + 36 h_{57}^{}
 - 36 h_{58}^{}
 + 144 h_{61}^{}
 - 72 h_{62}^{}
 + 2 h_{64}^{}\nonumber\\
& - 4 h_{66}^{}
 + 36 h_{70}^{}
 + 54 h_{75}^{}
 + 54 h_{80}^{}
 - 9 h_{85}^{}
 - 24 h_{86}^{}
 - 96 h_{89}^{}
 - 24 h_{92}^{}\nonumber\\
& - 96 h_{93}^{}
 - 8 h_{95}^{}
 - 4 h_{97}^{}),\\
\bar{h}_{94}{}={}&\frac{1}{324} (-24 h_{100}^{}
 - 24 h_{101}^{}
 - 24 h_{102}^{}
 + 24 h_{105}^{}
 - 84 h_{106}^{}
 - 16 h_{51}^{}
 - 72 h_{53}^{}\nonumber\\
& - 36 h_{54}^{}
 + 36 h_{55}^{}
 + 72 h_{56}^{}
 + 36 h_{57}^{}
 - 36 h_{58}^{}
 + 144 h_{61}^{}
 - 72 h_{62}^{}
 + 2 h_{64}^{}\nonumber\\
& - 4 h_{66}^{}
 + 36 h_{71}^{}
 + 54 h_{77}^{}
 + 54 h_{81}^{}
 - 9 h_{83}^{}
 - 9 h_{85}^{}
 - 96 h_{86}^{}
 - 24 h_{89}^{}\nonumber\\
& - 96 h_{92}^{}
 - 24 h_{93}^{}
 + 4 h_{95}^{}
 - 4 h_{97}^{}
 + 36 h_{98}^{}),\\
\bar{h}_{95}{}={}&\frac{1}{1296} (-48 h_{100}^{}
 - 48 h_{101}^{}
 - 48 h_{102}^{}
 - 240 h_{105}^{}
 - 240 h_{106}^{}
 + 81 h_{48}^{}
 + 16 h_{51}^{}\nonumber\\
& + 288 h_{53}^{}
 + 144 h_{54}^{}
 - 144 h_{55}^{}
 - 288 h_{56}^{}
 - 144 h_{57}^{}
 + 144 h_{58}^{}
 - 576 h_{61}^{}\nonumber\\
& + 288 h_{62}^{}
 + 4 h_{64}^{}
 - 8 h_{66}^{}
 - 36 h_{70}^{}
 - 36 h_{71}^{}
 - 216 h_{75}^{}
 - 216 h_{77}^{}
 - 216 h_{80}^{}\nonumber\\
& - 216 h_{81}^{}
 - 18 h_{83}^{}
 - 36 h_{85}^{}
 + 96 h_{86}^{}
 + 96 h_{89}^{}
 + 96 h_{92}^{}
 + 96 h_{93}^{}
 + 8 h_{95}^{}\nonumber\\
& + 16 h_{97}^{}
 + 144 h_{98}^{}),\\
\bar{h}_{96}{}={}&\frac{1}{108} (-12 h_{101}^{}
 + 36 h_{106}^{}
 - 36 h_{53}^{}
 + 36 h_{55}^{}
 + 36 h_{56}^{}
 - 36 h_{58}^{}
 + 36 h_{61}^{}\nonumber\\
& - 72 h_{62}^{}
 - 2 h_{64}^{}
 - 8 h_{66}^{}
 + 54 h_{81}^{}
 + 9 h_{85}^{}
 - 24 h_{93}^{}
 - 4 h_{97}^{}
 - 24 h_{98}^{}
 - 12 h_{99}^{}),\\
\bar{h}_{97}{}={}&\frac{1}{108} (12 h_{101}^{}
 + 36 h_{105}^{}
 - 72 h_{53}^{}
 - 36 h_{54}^{}
 + 36 h_{55}^{}
 + 36 h_{61}^{}
 - 72 h_{62}^{}
 - 2 h_{64}^{}\nonumber\\
& - 8 h_{66}^{}
 + 54 h_{80}^{}
 + 9 h_{85}^{}
 - 24 h_{92}^{}
 - 4 h_{97}^{}
 - 24 h_{98}^{}),\\
\bar{h}_{98}{}={}&\frac{1}{108} (36 h_{105}^{}
 + 36 h_{106}^{}
 + 108 h_{53}^{}
 + 36 h_{54}^{}
 - 72 h_{55}^{}
 - 36 h_{56}^{}
 + 36 h_{58}^{}
 - 72 h_{61}^{}\nonumber\\
& + 144 h_{62}^{}
 - 2 h_{64}^{}
 - 8 h_{66}^{}
 - 54 h_{80}^{}
 - 54 h_{81}^{}
 + 9 h_{85}^{}
 + 24 h_{92}^{}
 + 24 h_{93}^{}\nonumber\\
& - 4 h_{97}^{}
 - 24 h_{98}^{}
 + 12 h_{99}^{}),\\
\bar{h}_{99}{}={}&2 h_{107}^{},\\
\bar{h}_{100}{}={}&\frac{1}{9} (2 h_{100}^{}
 - 6 h_{107}^{}
 - 3 h_{55}^{}
 + 3 h_{57}^{}
 + 8 h_{94}^{}),\\
\bar{h}_{101}{}={}&\frac{1}{18} (2 h_{100}^{}
 - 6 h_{107}^{}
 + 6 h_{55}^{}
 - 6 h_{57}^{}
 + 9 h_{82}^{}
 - 4 h_{94}^{}),\\
\bar{h}_{102}{}={}&\frac{1}{108} (12 h_{100}^{}
 + 12 h_{101}^{}
 + 12 h_{102}^{}
 + 36 h_{107}^{}
 - 12 h_{72}^{}
 + 18 h_{74}^{}
 + 18 h_{77}^{}\nonumber\\
& + 18 h_{82}^{}
 + 3 h_{83}^{}
 + 24 h_{87}^{}
 + 24 h_{89}^{}
 + 24 h_{94}^{}
 - 4 h_{95}^{}
 - 12 h_{98}^{}),\\
\bar{h}_{103}{}={}&\frac{1}{3} (-3 h_{59}^{}
 - 3 h_{61}^{}
 - 6 h_{63}^{}
 + h_{67}^{}),\\
\bar{h}_{104}{}={}&h_{59}^{}
 + h_{61}^{}
 + 2 h_{63}^{}
 + \frac{2}{3} h_{67}^{},\\
\bar{h}_{105}{}={}&\frac{1}{81} (6 h_{103}^{}
 + 6 h_{104}^{}
 + 6 h_{105}^{}
 + 6 h_{106}^{}
 + 16 h_{52}^{}
 - 18 h_{53}^{}
 - 9 h_{54}^{}
 + 9 h_{55}^{}\nonumber\\
& + 18 h_{56}^{}
 + 9 h_{57}^{}
 - 9 h_{58}^{}
 + 36 h_{59}^{}
 - 18 h_{60}^{}
 + 36 h_{61}^{}
 - 18 h_{62}^{}
 + 54 h_{63}^{}
 + h_{64}^{}\nonumber\\
& + 2 h_{65}^{}
 + 2 h_{66}^{}
 - 18 h_{67}^{}
 - 24 h_{90}^{}
 - 24 h_{91}^{}
 - 24 h_{92}^{}
 - 24 h_{93}^{}
 - 4 h_{96}^{}
 - 4 h_{97}^{}\nonumber\\
& - 6 h_{98}^{}
 + 6 h_{99}^{}),\\
\bar{h}_{106}{}={}&\frac{1}{1296} (-48 h_{103}^{}
 - 48 h_{104}^{}
 - 48 h_{105}^{}
 - 48 h_{106}^{}
 + 81 h_{49}^{}
 + 16 h_{52}^{}
 - 288 h_{53}^{}\nonumber\\
& - 144 h_{54}^{}
 + 144 h_{55}^{}
 + 288 h_{56}^{}
 + 144 h_{57}^{}
 - 144 h_{58}^{}
 + 576 h_{59}^{}
 - 288 h_{60}^{}\nonumber\\
& + 576 h_{61}^{}
 - 288 h_{62}^{}
 + 864 h_{63}^{}
 + 4 h_{64}^{}
 + 8 h_{65}^{}
 + 8 h_{66}^{}
 - 72 h_{67}^{}
 - 36 h_{73}^{}\nonumber\\
& + 216 h_{78}^{}
 + 216 h_{79}^{}
 + 216 h_{80}^{}
 + 216 h_{81}^{}
 - 18 h_{84}^{}
 - 18 h_{85}^{}
 - 96 h_{90}^{}
 - 96 h_{91}^{}\nonumber\\
& - 96 h_{92}^{}
 - 96 h_{93}^{}
 + 8 h_{96}^{}
 + 8 h_{97}^{}
 + 48 h_{98}^{}
 - 48 h_{99}^{}),\\
\bar{h}_{107}{}={}&\frac{1}{324} (-12 h_{103}^{}
 - 12 h_{104}^{}
 - 12 h_{105}^{}
 - 12 h_{106}^{}
 - 32 h_{52}^{}
 + 144 h_{53}^{}
 + 72 h_{54}^{}\nonumber\\
& - 72 h_{55}^{}
 - 144 h_{56}^{}
 - 72 h_{57}^{}
 + 72 h_{58}^{}
 - 288 h_{59}^{}
 + 144 h_{60}^{}
 - 288 h_{61}^{}\nonumber\\
& + 144 h_{62}^{}
 - 432 h_{63}^{}
 + 4 h_{64}^{}
 + 8 h_{65}^{}
 + 8 h_{66}^{}
 - 72 h_{67}^{}
 + 36 h_{73}^{}
 - 54 h_{78}^{}\nonumber\\
& - 54 h_{79}^{}
 - 54 h_{80}^{}
 - 54 h_{81}^{}
 - 9 h_{84}^{}
 - 9 h_{85}^{}
 + 120 h_{90}^{}
 + 120 h_{91}^{}
 + 120 h_{92}^{}\nonumber\\
& + 120 h_{93}^{}
 - 4 h_{96}^{}
 - 4 h_{97}^{}
 + 12 h_{98}^{}
 - 12 h_{99}^{}).
\end{align}

\section{Third branch of the cubic Lagrangian in terms of the torsion and nonmetricity tensors}\label{appendix3}

The Lagrangian~\eqref{cubicLagIrrQT} containing cubic invariants from curvature, torsion and nonmetricity can be written in terms of the torsion and nonmetricity tensors as:
\begin{eqnarray}
    \mathcal{\bar{L}}_{\rm curv-tor-nonm}^{(3)}=\mathcal{\bar{L}}^{(3)}_{1,\rm tor-non}+\mathcal{\bar{L}}^{(3)}_{2,\rm tor-non}+\mathcal{\bar{L}}^{(3)}_{3,\rm  tor-non}+\mathcal{\bar{L}}^{(3)}_{4,\rm  tor-non}+\mathcal{\bar{L}}^{(3)}_{5,\rm tor-non}\,,
\end{eqnarray}
where we have defined:
\begin{eqnarray}
  \mathcal{\bar{L}}^{(3)}_{1,\rm tor-non}&=&\bar{h}_{108}^{} Q^{\alpha \rho \tau } \tilde{R}_{\rho \tau }{}^{\gamma \mu } T_{\alpha \gamma \mu } + \bar{h}_{109}^{} Q^{\alpha \rho \tau } \tilde{R}_{\rho }{}^{\gamma }{}_{\tau }{}^{\mu } T_{\alpha \gamma \mu } + \bar{h}_{110}^{} Q^{\alpha \rho \tau } \tilde{R}^{\gamma }{}_{\rho \tau }{}^{\mu } T_{\alpha \gamma \mu } + \bar{h}_{111}^{} Q^{\alpha \rho \tau } \tilde{R}_{\alpha \rho }{}^{\gamma \mu } T_{\tau \gamma \mu }\nonumber\\
  && + \bar{h}_{112}^{} Q^{\alpha \rho \tau } \tilde{R}_{\alpha }{}^{\gamma }{}_{\rho }{}^{\mu } T_{\tau \gamma \mu } + \bar{h}_{113}^{} Q^{\alpha \rho \tau } \tilde{R}_{\rho \alpha }{}^{\gamma \mu } T_{\tau \gamma \mu } + \bar{h}_{114}^{} Q^{\alpha }{}_{\alpha }{}^{\rho } \tilde{R}_{\rho }{}^{\tau \gamma \mu } T_{\tau \gamma \mu }+ \bar{h}_{115}^{} Q^{\alpha \rho \tau } \tilde{R}_{\rho }{}^{\gamma }{}_{\alpha }{}^{\mu } T_{\tau \gamma \mu }\nonumber\\
  &&+ \bar{h}_{116}^{} Q^{\alpha }{}_{\alpha }{}^{\rho } \tilde{R}^{\tau }{}_{\rho }{}^{\gamma \mu } T_{\tau \gamma \mu } + \bar{h}_{117}^{} Q^{\alpha }{}_{\alpha }{}^{\rho } \tilde{R}^{\tau \gamma }{}_{\rho }{}^{\mu } T_{\tau \gamma \mu }+ \bar{h}_{118}^{} Q^{\alpha \rho \tau } \tilde{R}^{\gamma }{}_{\alpha \rho }{}^{\mu } T_{\tau \gamma \mu } + \bar{h}_{119}^{} Q^{\alpha \rho \tau } \tilde{R}^{\gamma }{}_{\rho \alpha }{}^{\mu } T_{\tau \gamma \mu }\nonumber\\
  &&+ \bar{h}_{120}^{} Q^{\alpha \rho \tau } \tilde{R}^{\gamma \mu }{}_{\alpha \rho } T_{\tau \gamma \mu } + \bar{h}_{121}^{} Q^{\alpha \rho \tau } \tilde{R}_{\rho \tau }{}^{\gamma \mu } T_{\gamma \alpha \mu } + \bar{h}_{122}^{} Q^{\alpha \rho \tau } \tilde{R}_{\rho }{}^{\gamma }{}_{\tau }{}^{\mu } T_{\gamma \alpha \mu } + \bar{h}_{123}^{} Q^{\alpha \rho \tau } \tilde{R}^{\gamma }{}_{\rho \tau }{}^{\mu } T_{\gamma \alpha \mu }  \nonumber\\
  &&+ \bar{h}_{124}^{} Q^{\alpha \rho \tau } \tilde{R}_{\alpha \rho }{}^{\gamma \mu } T_{\gamma \tau \mu }+ \bar{h}_{125}^{} Q^{\alpha \rho \tau } \tilde{R}_{\alpha }{}^{\gamma }{}_{\rho }{}^{\mu } T_{\gamma \tau \mu } + \bar{h}_{126}^{} Q^{\alpha \rho \tau } \tilde{R}_{\rho \alpha }{}^{\gamma \mu } T_{\gamma \tau \mu } + \bar{h}_{127}^{} Q^{\alpha }{}_{\alpha }{}^{\rho } \tilde{R}_{\rho }{}^{\tau \gamma \mu } T_{\gamma \tau \mu } \nonumber\\
  &&+ \bar{h}_{128}^{} Q^{\alpha \rho \tau } \tilde{R}_{\rho }{}^{\gamma }{}_{\alpha }{}^{\mu } T_{\gamma \tau \mu }  + \bar{h}_{129}^{} Q^{\alpha }{}_{\alpha }{}^{\rho } \tilde{R}^{\tau }{}_{\rho }{}^{\gamma \mu } T_{\gamma \tau \mu }  + \bar{h}_{130}^{} Q^{\alpha }{}_{\alpha }{}^{\rho } \tilde{R}^{\tau \gamma }{}_{\rho }{}^{\mu } T_{\gamma \tau \mu } + \bar{h}_{131}^{} Q^{\alpha \rho \tau } \tilde{R}^{\gamma }{}_{\alpha \rho }{}^{\mu } T_{\gamma \tau \mu }\nonumber\\
  && + \bar{h}_{132}^{} Q^{\alpha \rho \tau } \tilde{R}^{\gamma }{}_{\rho \alpha }{}^{\mu } T_{\gamma \tau \mu } + \bar{h}_{133}^{} Q^{\alpha \rho \tau } \tilde{R}^{\gamma \mu }{}_{\alpha \rho } T_{\gamma \tau \mu } + \bar{h}_{134}^{} Q^{\alpha \rho \tau } \tilde{R}_{\rho }{}^{\gamma }{}_{\tau }{}^{\mu } T_{\mu \alpha \gamma } + \bar{h}_{135}^{} Q^{\alpha \rho \tau } \tilde{R}^{\gamma }{}_{\rho \tau }{}^{\mu } T_{\mu \alpha \gamma } \nonumber\\
  &&+ \bar{h}_{136}^{} Q^{\alpha \rho \tau } \tilde{R}_{\alpha }{}^{\gamma }{}_{\rho }{}^{\mu } T_{\mu \tau \gamma }+ \bar{h}_{137}^{} Q^{\alpha \rho \tau } \tilde{R}_{\rho }{}^{\gamma }{}_{\alpha }{}^{\mu } T_{\mu \tau \gamma }+ \bar{h}_{138}^{} Q^{\alpha }{}_{\alpha }{}^{\rho } \tilde{R}^{\tau \gamma }{}_{\rho }{}^{\mu } T_{\mu \tau \gamma } + \bar{h}_{139}^{} Q^{\alpha \rho \tau } \tilde{R}^{\gamma }{}_{\alpha \rho }{}^{\mu } T_{\mu \tau \gamma } \nonumber\\
   &&  + \bar{h}_{140}^{} Q^{\alpha \rho \tau } \tilde{R}^{\gamma }{}_{\rho \alpha }{}^{\mu } T_{\mu \tau \gamma }+ \bar{h}_{141}^{} Q^{\alpha \rho \tau } \tilde{R}^{\gamma \mu }{}_{\alpha \rho } T_{\mu \tau \gamma } + \bar{h}_{142}^{} Q^{\alpha \rho \tau } \tilde{R}_{\alpha \rho \tau }{}^{\gamma } T_{\gamma } + \bar{h}_{143}^{} Q^{\alpha \rho \tau } \tilde{R}_{\rho \alpha \tau }{}^{\gamma } T_{\gamma }\nonumber\\
    &&  + \bar{h}_{144}^{} Q^{\alpha \rho \tau } \tilde{R}_{\rho \tau \alpha }{}^{\gamma } T_{\gamma }+ \bar{h}_{145}^{} Q^{\alpha \rho \tau } \tilde{R}_{\rho }{}^{\gamma }{}_{\alpha \tau } T_{\gamma } + \bar{h}_{146}^{} Q^{\alpha \rho \tau } \tilde{R}^{\gamma }{}_{\rho \alpha \tau } T_{\gamma } + \bar{h}_{147}^{} Q^{\alpha }{}_{\rho }{}^{\rho } \tilde{R}_{\alpha }{}^{\tau \gamma \mu } T_{\tau \gamma \mu }\nonumber\\
  && + \bar{h}_{148}^{} Q^{\alpha }{}_{\rho }{}^{\rho } \tilde{R}^{\tau }{}_{\alpha }{}^{\gamma \mu } T_{\tau \gamma \mu }  + \bar{h}_{149}^{} Q^{\alpha }{}_{\rho }{}^{\rho } \tilde{R}^{\tau \gamma }{}_{\alpha }{}^{\mu } T_{\tau \gamma \mu }+ \bar{h}_{150}^{} Q^{\alpha }{}_{\rho }{}^{\rho } \tilde{R}_{\alpha }{}^{\tau \gamma \mu } T_{\gamma \tau \mu } + \bar{h}_{151}^{} Q^{\alpha }{}_{\rho }{}^{\rho } \tilde{R}^{\tau }{}_{\alpha }{}^{\gamma \mu } T_{\gamma \tau \mu }\nonumber \\
   &&+ \bar{h}_{152}^{} Q^{\alpha }{}_{\rho }{}^{\rho } \tilde{R}^{\tau \gamma }{}_{\alpha }{}^{\mu } T_{\gamma \tau \mu }+ \bar{h}_{153}^{} Q^{\alpha }{}_{\rho }{}^{\rho } \tilde{R}^{\tau \gamma }{}_{\alpha }{}^{\mu } T_{\mu \tau \gamma }\,,\\
     \mathcal{\bar{L}}^{(3)}_{2,\rm tor-non}&=& \bar{h}_{154}^{} Q^{\tau }{}_{\tau }{}^{\gamma } \tilde{R}^{\alpha \rho } T_{\alpha \rho \gamma } + \bar{h}_{155}^{} Q^{\tau }{}_{\rho }{}^{\gamma } \tilde{R}^{\alpha \rho } T_{\alpha \tau \gamma } + \bar{h}_{156}^{} Q^{\tau }{}_{\tau }{}^{\gamma } \tilde{R}^{\alpha \rho } T_{\rho \alpha \gamma } + \bar{h}_{157}^{} Q^{\tau }{}_{\alpha }{}^{\gamma } \tilde{R}^{\alpha \rho } T_{\rho \tau \gamma } + \bar{h}_{158}^{} Q_{\rho }{}^{\tau \gamma } \tilde{R}^{\alpha \rho } T_{\tau \alpha \gamma }\nonumber\\
     && + \bar{h}_{159}^{} Q^{\tau }{}_{\rho }{}^{\gamma } \tilde{R}^{\alpha \rho } T_{\tau \alpha \gamma } + \bar{h}_{160}^{} Q_{\alpha }{}^{\tau \gamma } \tilde{R}^{\alpha \rho } T_{\tau \rho \gamma }+ \bar{h}_{161}^{} Q^{\tau }{}_{\alpha }{}^{\gamma } \tilde{R}^{\alpha \rho } T_{\tau \rho \gamma } + \bar{h}_{162}^{} Q^{\tau }{}_{\tau }{}^{\gamma } \tilde{R}^{\alpha \rho } T_{\gamma \alpha \rho } \nonumber\\
     &&    + \bar{h}_{163}^{} Q^{\tau }{}_{\rho }{}^{\gamma } \tilde{R}^{\alpha \rho } T_{\gamma \alpha \tau }  + \bar{h}_{164}^{} Q^{\tau }{}_{\alpha }{}^{\gamma } \tilde{R}^{\alpha \rho } T_{\gamma \rho \tau } + \bar{h}_{165}^{} Q^{\tau }{}_{\rho \tau } \tilde{R}^{\alpha \rho } T_{\alpha } + \bar{h}_{166}^{} Q^{\tau }{}_{\alpha \tau } \tilde{R}^{\alpha \rho } T_{\rho } + \bar{h}_{167}^{} Q_{\alpha \rho }{}^{\tau } \tilde{R}^{\alpha \rho } T_{\tau }  \nonumber\\
   &&  + \bar{h}_{168}^{} Q_{\rho \alpha }{}^{\tau } \tilde{R}^{\alpha \rho } T_{\tau }+ \bar{h}_{169}^{} Q^{\tau }{}_{\alpha \rho } \tilde{R}^{\alpha \rho } T_{\tau }  + \bar{h}_{170}^{} Q_{\alpha \tau }{}^{\tau } \tilde{R}^{\alpha \rho } T_{\rho }+ \bar{h}_{171}^{} Q_{\rho \tau }{}^{\tau } \tilde{R}^{\alpha \rho } T_{\alpha }+\bar{h}_{172}^{} Q^{\tau }{}_{\gamma }{}^{\gamma } \tilde{R}^{\alpha \rho } T_{\alpha \rho \tau }\nonumber\\
   &&  + \bar{h}_{173}^{} Q^{\tau }{}_{\gamma }{}^{\gamma } \tilde{R}^{\alpha \rho } T_{\rho \alpha \tau }+ \bar{h}_{174}^{} Q^{\tau }{}_{\gamma }{}^{\gamma } \tilde{R}^{\alpha \rho } T_{\tau \alpha \rho } \,,\\
      \mathcal{\bar{L}}^{(3)}_{3,\rm tor-non}&=& \bar{h}_{175}^{} Q^{\tau }{}_{\tau }{}^{\gamma } \hat{R}^{\alpha \rho } T_{\alpha \rho \gamma } + \bar{h}_{176}^{} Q^{\tau }{}_{\rho }{}^{\gamma } \hat{R}^{\alpha \rho } T_{\alpha \tau \gamma }  + \bar{h}_{177}^{} Q^{\tau }{}_{\tau }{}^{\gamma } \hat{R}^{\alpha \rho } T_{\rho \alpha \gamma } + \bar{h}_{178}^{} Q^{\tau }{}_{\alpha }{}^{\gamma } \hat{R}^{\alpha \rho } T_{\rho \tau \gamma } + \bar{h}_{179}^{} Q_{\rho }{}^{\tau \gamma } \hat{R}^{\alpha \rho } T_{\tau \alpha \gamma } \nonumber\\
      && + \bar{h}_{180}^{} Q^{\tau }{}_{\rho }{}^{\gamma } \hat{R}^{\alpha \rho } T_{\tau \alpha \gamma } + \bar{h}_{181}^{} Q_{\alpha }{}^{\tau \gamma } \hat{R}^{\alpha \rho } T_{\tau \rho \gamma }+ \bar{h}_{182}^{} Q^{\tau }{}_{\alpha }{}^{\gamma } \hat{R}^{\alpha \rho } T_{\tau \rho \gamma } + \bar{h}_{183}^{} Q^{\tau }{}_{\tau }{}^{\gamma } \hat{R}^{\alpha \rho } T_{\gamma \alpha \rho } \nonumber\\
      &&+ \bar{h}_{184}^{} Q^{\tau }{}_{\rho }{}^{\gamma } \hat{R}^{\alpha \rho } T_{\gamma \alpha \tau } + \bar{h}_{185}^{} Q^{\tau }{}_{\alpha }{}^{\gamma } \hat{R}^{\alpha \rho } T_{\gamma \rho \tau }+ \bar{h}_{186}^{} Q^{\tau }{}_{\rho \tau } \hat{R}^{\alpha \rho } T_{\alpha }+ \bar{h}_{187}^{} Q^{\tau }{}_{\alpha \tau } \hat{R}^{\alpha \rho } T_{\rho } + \bar{h}_{188}^{} Q_{\alpha \rho }{}^{\tau } \hat{R}^{\alpha \rho } T_{\tau }
      \nonumber\\
      &&+ \bar{h}_{189}^{} Q_{\rho \alpha }{}^{\tau } \hat{R}^{\alpha \rho } T_{\tau }   + \bar{h}_{190}^{} Q^{\tau }{}_{\alpha \rho } \hat{R}^{\alpha \rho } T_{\tau }+ \bar{h}_{191}^{} Q_{\alpha \tau }{}^{\tau } \hat{R}^{\alpha \rho } T_{\rho } + \bar{h}_{192}^{} Q_{\rho \tau }{}^{\tau } \hat{R}^{\alpha \rho } T_{\alpha }+\bar{h}_{193}^{} Q^{\tau }{}_{\gamma }{}^{\gamma } \hat{R}^{\alpha \rho } T_{\alpha \rho \tau } \nonumber\\
      && + \bar{h}_{194}^{} Q^{\tau }{}_{\gamma }{}^{\gamma } \hat{R}^{\alpha \rho } T_{\rho \alpha \tau }+ \bar{h}_{195}^{} Q^{\tau }{}_{\gamma }{}^{\gamma } \hat{R}^{\alpha \rho } T_{\tau \alpha \rho }\,,\\
      \mathcal{\bar{L}}^{(3)}_{4,\rm tor-non}&=&\bar{h}_{196}^{} Q^{\alpha \rho \tau } \tilde{R}^{\omega}{}_{\omega\rho }{}^{\gamma } T_{\alpha \tau \gamma }  + \bar{h}_{197}^{} Q^{\alpha \rho \tau } \tilde{R}^{\omega}{}_{\omega\alpha }{}^{\gamma } T_{\rho \tau \gamma } + \bar{h}_{198}^{} Q^{\alpha }{}_{\alpha }{}^{\rho }\tilde{R}^{\omega}{}_{\omega}{}^{\tau \gamma } T_{\rho \tau \gamma } + \bar{h}_{199}^{} Q^{\alpha \rho \tau } \tilde{R}^{\omega}{}_{\omega\rho }{}^{\gamma } T_{\tau \alpha \gamma } \nonumber\\
      &&+ \bar{h}_{200}^{} Q^{\alpha }{}_{\alpha }{}^{\rho }\tilde{R}^{\omega}{}_{\omega}{}^{\tau \gamma } T_{\tau \rho \gamma }+ \bar{h}_{201}^{} Q^{\alpha \rho \tau } \tilde{R}^{\omega}{}_{\omega\rho }{}^{\gamma } T_{\gamma \alpha \tau } + \bar{h}_{202}^{} Q^{\alpha \rho \tau } \tilde{R}^{\omega}{}_{\omega\alpha \rho } T_{\tau } + \bar{h}_{203}^{} Q^{\alpha }{}_{\alpha }{}^{\rho } \tilde{R}^{\omega}{}_{\omega\rho }{}^{\tau } T_{\tau }\nonumber\\
      &&
     + \bar{h}_{204}^{} Q^{\alpha }{}_{\rho }{}^{\rho }\tilde{R}^{\omega}{}_{\omega}{}^{\tau \gamma } T_{\alpha \tau \gamma }+ \bar{h}_{205}^{} Q^{\alpha }{}_{\rho }{}^{\rho }\tilde{R}^{\omega}{}_{\omega}{}^{\tau \gamma } T_{\tau \alpha \gamma }+ \bar{h}_{206}^{} Q^{\alpha }{}_{\rho }{}^{\rho } \tilde{R}^{\omega}{}_{\omega\alpha }{}^{\tau } T_{\tau }  \,,\\
       \mathcal{\bar{L}}^{(3)}_{5,\rm tor-non}&=&\bar{h}_{207}^{} Q^{\alpha \rho \tau } \tilde{R} T_{\rho \alpha \tau } + \bar{h}_{208}^{} Q^{\alpha }{}_{\alpha }{}^{\rho } \tilde{R} T_{\rho } + \bar{h}_{209}^{} Q^{\alpha }{}_{\rho }{}^{\rho } \tilde{R} T_{\alpha }\,.
\end{eqnarray}
and where the respective Lagrangian coefficients are related as:
\begin{align}
\bar{h}_{108}^{}={}&\frac{1}{9} (-18 h_{144}^{}
 - 18 h_{146}^{}
 - 6 h_{156}^{}
 + 6 h_{182}^{}
 + 6 h_{183}^{}
 -  h_{202}^{}),\\
\bar{h}_{109}^{}={}&\frac{1}{9} (-18 h_{145}^{}
 - 18 h_{147}^{}
 - 6 h_{157}^{}
 + 6 h_{180}^{}
 + 6 h_{184}^{}
 - 6 h_{187}^{}
 - 6 h_{188}^{}
 + h_{200}^{} -  h_{204}^{}),\\
\bar{h}_{110}^{}={}&\frac{1}{9} (-18 h_{148}^{}
 - 18 h_{149}^{}
 - 6 h_{158}^{}
 + 6 h_{178}^{}
 + 6 h_{179}^{}
 - 6 h_{185}^{}
 - 6 h_{186}^{}
 -  h_{201}^{}- 2 h_{203}^{}),\\
\bar{h}_{111}^{}={}&\frac{1}{9} (18 h_{144}^{}
 - 6 h_{156}^{}
 - 6 h_{183}^{}
 -  h_{202}^{}),\\
\bar{h}_{112}^{}={}&\frac{1}{9} (18 h_{145}^{}
 - 6 h_{157}^{}
 - 6 h_{184}^{}
 + 6 h_{188}^{}
 + h_{200}^{}
 -  h_{204}^{}),\\
\bar{h}_{113}^{}={}&\frac{1}{9} (18 h_{146}^{}
 - 6 h_{156}^{}
 - 6 h_{182}^{}
 -  h_{202}^{}),\\
\bar{h}_{114}^{}={}&\frac{1}{27} (-24 h_{138}^{}
 + 18 h_{144}^{}
 - 18 h_{145}^{}
 + 6 h_{156}^{}
 - 6 h_{157}^{}
 + 4 h_{171}^{}
 - 6 h_{183}^{}- 12 h_{184}^{}
\nonumber\\
&  - 6 h_{188}^{}
 - 2 h_{200}^{}+ h_{202}^{}
 -  h_{204}^{}),\\
\bar{h}_{115}^{}={}&\frac{1}{9} (18 h_{147}^{}
 - 6 h_{157}^{}
 - 6 h_{180}^{}
 + 6 h_{187}^{}
 + h_{200}^{}
 -  h_{204}^{}),\\
\bar{h}_{116}^{}={}&\frac{1}{27} (-24 h_{139}^{}
 + 18 h_{146}^{}
 - 18 h_{148}^{}
 + 6 h_{156}^{}
 - 6 h_{158}^{}
 + 4 h_{169}^{}
 - 12 h_{179}^{} - 6 h_{182}^{}
 - 6 h_{186}^{}\nonumber\\
& -  h_{201}^{}
 + h_{202}^{}
 + h_{203}^{}),\\
\bar{h}_{117}^{}={}&\frac{1}{27} (-24 h_{140}^{}
 + 18 h_{147}^{}
 - 18 h_{149}^{}
 + 18 h_{150}^{}
 + 6 h_{157}^{}
 - 6 h_{158}^{}
 + 8 h_{168}^{}+ 4 h_{170}^{}
 - 12 h_{177}^{}
 - 12 h_{178}^{}\nonumber\\
&  - 6 h_{180}^{}
 - 6 h_{181}^{}
 - 6 h_{185}^{}
 + 6 h_{187}^{}
 -  h_{200}^{} -  h_{201}^{}
 + h_{203}^{}
 + h_{204}^{}),\\
\bar{h}_{118}^{}={}&\frac{1}{9} (18 h_{148}^{}
 - 6 h_{158}^{}
 - 6 h_{179}^{}
 + 6 h_{186}^{}
 -  h_{201}^{}
 - 2 h_{203}^{}),\\
\bar{h}_{119}^{}={}&\frac{1}{9} (18 h_{149}^{}
 - 6 h_{158}^{}
 - 6 h_{178}^{}
 + 6 h_{185}^{}
 -  h_{201}^{}
 - 2 h_{203}^{}),\\
\bar{h}_{120}^{}={}&\frac{2}{3} (3 h_{150}^{}
 + h_{177}^{}
 -  h_{181}^{}),\\
\bar{h}_{121}^{}={}&\frac{1}{9} (36 h_{144}^{}
 + 36 h_{146}^{}
 + 12 h_{156}^{}
 + 6 h_{182}^{}
 + 6 h_{183}^{}
 -  h_{202}^{}),\\
\bar{h}_{122}^{}={}&\frac{1}{9} (18 h_{145}^{}
 + 18 h_{147}^{}
 + 6 h_{157}^{}
 + 12 h_{180}^{}
 + 12 h_{184}^{}
 + 6 h_{187}^{}
 + 6 h_{188}^{}
 + 2 h_{200}^{}+ h_{204}^{}),\\
\bar{h}_{123}^{}={}&\frac{1}{9} (18 h_{148}^{}
 + 18 h_{149}^{}
 + 6 h_{158}^{}
 + 12 h_{178}^{}
 + 12 h_{179}^{}
 + 6 h_{185}^{}
 + 6 h_{186}^{}
 + h_{201}^{} -  h_{203}^{}),\\
\bar{h}_{124}^{}={}&\frac{1}{9} (-36 h_{144}^{}
 + 12 h_{156}^{}
 - 6 h_{183}^{}
 -  h_{202}^{}),\\
\bar{h}_{125}^{}={}&\frac{1}{9} (-18 h_{145}^{}
 + 6 h_{157}^{}
 - 12 h_{184}^{}
 - 6 h_{188}^{}
 + 2 h_{200}^{}
 + h_{204}^{}),\\
\bar{h}_{126}^{}={}&\frac{1}{9} (-36 h_{146}^{}
 + 12 h_{156}^{}
 - 6 h_{182}^{}
 -  h_{202}^{}),\\
\bar{h}_{127}^{}={}&\frac{1}{27} (48 h_{138}^{}
 - 36 h_{144}^{}
 + 36 h_{145}^{}
 - 12 h_{156}^{}
 + 12 h_{157}^{}
 + 4 h_{171}^{}
 - 6 h_{183}^{} - 12 h_{184}^{}
 - 6 h_{188}^{}\nonumber\\
& - 2 h_{200}^{}
 + h_{202}^{}
 -  h_{204}^{}),\\
\bar{h}_{128}^{}={}&\frac{1}{9} (-18 h_{147}^{}
 + 6 h_{157}^{}
 - 12 h_{180}^{}
 - 6 h_{187}^{}
 + 2 h_{200}^{}
 + h_{204}^{}),\\
\bar{h}_{129}^{}={}&\frac{1}{27} (48 h_{139}^{}
 - 36 h_{146}^{}
 + 36 h_{148}^{}
 - 12 h_{156}^{}
 + 12 h_{158}^{}
 + 4 h_{169}^{}
 - 12 h_{179}^{}- 6 h_{182}^{}
 - 6 h_{186}^{}\nonumber\\
&  -  h_{201}^{}
 + h_{202}^{}
 + h_{203}^{}),\\
\bar{h}_{130}^{}={}&\frac{1}{27} (24 h_{140}^{}
 - 18 h_{147}^{}
 + 18 h_{149}^{}
 - 18 h_{150}^{}
 - 6 h_{157}^{}
 + 6 h_{158}^{}
 + 4 h_{168}^{}
 + 8 h_{170}^{}- 6 h_{177}^{}\nonumber\\
& 
 - 6 h_{178}^{}
 - 12 h_{180}^{}
 - 12 h_{181}^{}
 + 6 h_{185}^{}
 - 6 h_{187}^{}
 - 2 h_{200}^{}
 + h_{201}^{}+ 2 h_{203}^{}
 -  h_{204}^{}),\\
\bar{h}_{131}^{}={}&\frac{1}{9} (-18 h_{148}^{}
 + 6 h_{158}^{}
 - 12 h_{179}^{}
 - 6 h_{186}^{}
 + h_{201}^{}
 -  h_{203}^{}),\\
\bar{h}_{132}^{}={}&\frac{1}{9} (-18 h_{149}^{}
 + 6 h_{158}^{}
 - 12 h_{178}^{}
 - 6 h_{185}^{}
 + h_{201}^{}
 -  h_{203}^{}),\\
\bar{h}_{133}^{}={}&\frac{2}{3} (-3 h_{150}^{}
 + 2 h_{177}^{}
 + h_{181}^{}),\\
\bar{h}_{134}^{}={}&\frac{1}{9} (-18 h_{145}^{}
 - 18 h_{147}^{}
 - 6 h_{157}^{}
 + 6 h_{180}^{}
 + 6 h_{184}^{}
 + 12 h_{187}^{}
 + 12 h_{188}^{}
 + h_{200}^{}+ 2 h_{204}^{}),\\
\bar{h}_{135}^{}={}&\frac{1}{9} (-18 h_{148}^{}
 - 18 h_{149}^{}
 - 6 h_{158}^{}
 + 6 h_{178}^{}
 + 6 h_{179}^{}
 + 12 h_{185}^{}
 + 12 h_{186}^{}
 + 2 h_{201}^{} + h_{203}^{}),\\
\bar{h}_{136}^{}={}&\frac{1}{9} (18 h_{145}^{}
 - 6 h_{157}^{}
 - 6 h_{184}^{}
 - 12 h_{188}^{}
 + h_{200}^{}
 + 2 h_{204}^{}),\\
\bar{h}_{137}^{}={}&\frac{1}{9} (18 h_{147}^{}
 - 6 h_{157}^{}
 - 6 h_{180}^{}
 - 12 h_{187}^{}
 + h_{200}^{}
 + 2 h_{204}^{}),\\
\bar{h}_{138}^{}={}&\frac{1}{27} (-24 h_{140}^{}
 + 18 h_{147}^{}
 - 18 h_{149}^{}
 + 18 h_{150}^{}
 + 6 h_{157}^{}
 - 6 h_{158}^{}
 - 4 h_{168}^{} + 4 h_{170}^{} + 6 h_{177}^{}\nonumber\\
& + 6 h_{178}^{}
 - 6 h_{180}^{}
 - 6 h_{181}^{}
 + 12 h_{185}^{}
 - 12 h_{187}^{}
 -  h_{200}^{}+ 2 h_{201}^{}
 + h_{203}^{}
 - 2 h_{204}^{}),\\
\bar{h}_{139}^{}={}&\frac{1}{9} (18 h_{148}^{}
 - 6 h_{158}^{}
 - 6 h_{179}^{}
 - 12 h_{186}^{}
 + 2 h_{201}^{}
 + h_{203}^{}),\\
\bar{h}_{140}^{}={}&\frac{1}{9} (18 h_{149}^{}
 - 6 h_{158}^{}
 - 6 h_{178}^{}
 - 12 h_{185}^{}
 + 2 h_{201}^{}
 + h_{203}^{}),\\
\bar{h}_{141}^{}={}&\frac{2}{3} (3 h_{150}^{}
 + h_{177}^{}
 + 2 h_{181}^{}),\\
\bar{h}_{142}^{}={}&\frac{1}{9} (-18 h_{123}^{}
 - 3 h_{129}^{}
 + 6 h_{178}^{}
 + 6 h_{179}^{}
 - 6 h_{183}^{}
 - 6 h_{184}^{}
 + h_{200}^{}
 -  h_{202}^{} -  h_{203}^{}),\\
\bar{h}_{143}^{}={}&\frac{1}{9} (-18 h_{121}^{}
 - 3 h_{129}^{}
 - 6 h_{179}^{}
 + 6 h_{180}^{}
 - 6 h_{182}^{}
 + 6 h_{184}^{}
 + h_{200}^{}
 -  h_{202}^{}-  h_{203}^{}),\\
\bar{h}_{144}^{}={}&\frac{1}{9} (18 h_{121}^{}
 + 18 h_{123}^{}
 - 3 h_{129}^{}
 - 6 h_{178}^{}
 - 6 h_{180}^{}
 + 6 h_{182}^{}
 + 6 h_{183}^{}
 + h_{200}^{} -  h_{202}^{}
 -  h_{203}^{}),\\
\bar{h}_{145}^{}={}&\frac{2}{3} (3 h_{122}^{}
 + h_{177}^{}
 - 2 h_{187}^{}
 -  h_{188}^{}),\\
\bar{h}_{146}^{}={}&\frac{2}{3} (3 h_{120}^{}
 + h_{181}^{}
 - 2 h_{185}^{}
 -  h_{186}^{}),\\
\bar{h}_{147}^{}={}&\frac{1}{108} (-54 h_{132}^{}
 + 24 h_{138}^{}
 - 72 h_{144}^{}
 + 72 h_{145}^{}
 + 12 h_{156}^{}
 - 12 h_{157}^{}
 + 9 h_{162}^{}- 4 h_{171}^{}
 + 24 h_{183}^{}
 + 48 h_{184}^{}\nonumber\\
&  + 24 h_{188}^{}
 - 4 h_{200}^{}
 + 2 h_{202}^{}
 - 2 h_{204}^{}),\\
\bar{h}_{148}^{}={}&\frac{1}{108} (-54 h_{133}^{}
 + 24 h_{139}^{}
 - 72 h_{146}^{}
 + 72 h_{148}^{}
 + 12 h_{156}^{}
 - 12 h_{158}^{}
 + 9 h_{160}^{}- 4 h_{169}^{}
 + 48 h_{179}^{}
 + 24 h_{182}^{}\nonumber\\
&  + 24 h_{186}^{}
 - 2 h_{201}^{}
 + 2 h_{202}^{}
 + 2 h_{203}^{}),\\
\bar{h}_{149}^{}={}&\frac{1}{108} (-54 h_{134}^{}
 + 24 h_{140}^{}
 - 72 h_{147}^{}
 + 72 h_{149}^{}
 - 72 h_{150}^{}
 + 12 h_{157}^{}
 - 12 h_{158}^{}+ 18 h_{159}^{}
 + 9 h_{161}^{}
 - 8 h_{168}^{}\nonumber\\
&  - 4 h_{170}^{}
 + 48 h_{177}^{}
 + 48 h_{178}^{}
 + 24 h_{180}^{}
 + 24 h_{181}^{}+ 24 h_{185}^{}
 - 24 h_{187}^{}
 - 2 h_{200}^{}
 - 2 h_{201}^{}
 + 2 h_{203}^{}
 + 2 h_{204}^{}),\\
\bar{h}_{150}^{}={}&\frac{1}{108} (108 h_{132}^{}
 - 48 h_{138}^{}
 + 144 h_{144}^{}
 - 144 h_{145}^{}
 - 24 h_{156}^{}
 + 24 h_{157}^{}
 + 9 h_{162}^{}- 4 h_{171}^{}
 + 24 h_{183}^{}
 + 48 h_{184}^{}\nonumber\\
&  + 24 h_{188}^{}
 - 4 h_{200}^{}
 + 2 h_{202}^{}
 - 2 h_{204}^{}),\\
\bar{h}_{151}^{}={}&\frac{1}{108} (108 h_{133}^{}
 - 48 h_{139}^{}
 + 144 h_{146}^{}
 - 144 h_{148}^{}
 - 24 h_{156}^{}
 + 24 h_{158}^{}
 + 9 h_{160}^{}- 4 h_{169}^{}
 + 48 h_{179}^{}
 + 24 h_{182}^{}\nonumber\\
&  + 24 h_{186}^{}
 - 2 h_{201}^{}
 + 2 h_{202}^{}
 + 2 h_{203}^{}),\\
\bar{h}_{152}^{}={}&\frac{1}{108} (54 h_{134}^{}
 - 24 h_{140}^{}
 + 72 h_{147}^{}
 - 72 h_{149}^{}
 + 72 h_{150}^{}
 - 12 h_{157}^{}
 + 12 h_{158}^{} + 9 h_{159}^{}
 + 18 h_{161}^{}
 - 4 h_{168}^{} - 8 h_{170}^{}\nonumber\\
& + 24 h_{177}^{}
 + 24 h_{178}^{}
 + 48 h_{180}^{}
 + 48 h_{181}^{}- 24 h_{185}^{}
 + 24 h_{187}^{}
 - 4 h_{200}^{} + 2 h_{201}^{}
 + 4 h_{203}^{}
 - 2 h_{204}^{}),\\
\bar{h}_{153}^{}={}&\frac{1}{108} (-54 h_{134}^{}
 + 24 h_{140}^{}
 - 72 h_{147}^{}
 + 72 h_{149}^{}
 - 72 h_{150}^{}
 + 12 h_{157}^{}
 - 12 h_{158}^{} - 9 h_{159}^{}
 + 9 h_{161}^{}
 + 4 h_{168}^{}
 - 4 h_{170}^{}\nonumber\\
& - 24 h_{177}^{}
 - 24 h_{178}^{}
 + 24 h_{180}^{}
 + 24 h_{181}^{} - 48 h_{185}^{}
 + 48 h_{187}^{}
 - 2 h_{200}^{}
 + 4 h_{201}^{}
 + 2 h_{203}^{}
 - 4 h_{204}^{}),\\
\bar{h}_{154}^{}={}&\frac{1}{27} (-24 h_{141}^{}
 + 36 h_{144}^{}
 - 36 h_{145}^{}
 - 18 h_{147}^{}
 + 18 h_{149}^{}
 - 18 h_{150}^{}
 - 18 h_{151}^{}+ 18 h_{152}^{}
 + 6 h_{157}^{}
 + 8 h_{172}^{}
 - 4 h_{173}^{}\nonumber\\
&  - 12 h_{180}^{}
 - 12 h_{184}^{}
 - 6 h_{187}^{}
 - 6 h_{188}^{} - 12 h_{189}^{}
 - 6 h_{190}^{}
 + 6 h_{191}^{}
 - 6 h_{192}^{}
 + 2 h_{200}^{}
 + h_{204}^{}
 - 3 h_{206}^{}),\\
\bar{h}_{155}^{}={}&\frac{2}{3} (3 h_{145}^{}
 + 3 h_{150}^{}
 - 3 h_{152}^{}
 + 2 h_{189}^{}
 -  h_{191}^{}),\\
\bar{h}_{156}^{}={}&\frac{1}{27} (24 h_{141}^{}
 - 36 h_{144}^{}
 + 36 h_{145}^{}
 + 18 h_{147}^{}
 - 18 h_{149}^{}
 + 18 h_{150}^{}
 + 18 h_{151}^{} - 18 h_{152}^{}
 - 6 h_{157}^{}
 + 4 h_{172}^{}
 + 4 h_{173}^{}\nonumber\\
& - 6 h_{180}^{}
 - 6 h_{184}^{}
 - 12 h_{187}^{}
 - 12 h_{188}^{} - 6 h_{189}^{}
 - 12 h_{190}^{}
 - 6 h_{191}^{}
 + 6 h_{192}^{}
 + h_{200}^{}
 + 2 h_{204}^{}
 - 3 h_{205}^{}),\\
\bar{h}_{157}^{}={}&\frac{2}{3} (6 h_{144}^{}
 + 3 h_{149}^{}
 - 3 h_{151}^{}
 + 2 h_{190}^{}
 -  h_{192}^{}),\\
\bar{h}_{158}^{}={}&\frac{1}{3} (6 h_{191}^{}
 + h_{205}^{}),\\
\bar{h}_{159}^{}={}&\frac{1}{3} (-6 h_{145}^{}
 - 6 h_{150}^{}
 + 6 h_{152}^{}
 + 2 h_{189}^{}
 - 4 h_{191}^{}
 + h_{205}^{}),\\
\bar{h}_{160}^{}={}&\frac{1}{3} (6 h_{192}^{}
 + h_{206}^{}),\\
\bar{h}_{161}^{}={}&\frac{1}{3} (-12 h_{144}^{}
 - 6 h_{149}^{}
 + 6 h_{151}^{}
 + 2 h_{190}^{}
 - 4 h_{192}^{}
 + h_{206}^{}),\\
\bar{h}_{162}^{}={}&\frac{1}{27} (-24 h_{141}^{}
 + 36 h_{144}^{}
 - 36 h_{145}^{}
 - 18 h_{147}^{}
 + 18 h_{149}^{}
 - 18 h_{150}^{}
 - 18 h_{151}^{}+ 18 h_{152}^{}
 + 6 h_{157}^{}
 - 4 h_{172}^{}
 + 8 h_{173}^{}\nonumber\\
&  + 6 h_{180}^{}
 + 6 h_{184}^{}
 - 6 h_{187}^{}
 - 6 h_{188}^{}+ 6 h_{189}^{}
 - 6 h_{190}^{}
 - 12 h_{191}^{}
 + 12 h_{192}^{}
 -  h_{200}^{}
 + h_{204}^{}
 - 3 h_{205}^{}
 + 3 h_{206}^{}),\\
\bar{h}_{163}^{}={}&\frac{1}{3} (6 h_{145}^{}
 + 6 h_{150}^{}
 - 6 h_{152}^{}
 - 2 h_{189}^{}
 - 2 h_{191}^{}
 + h_{205}^{}),\\
\bar{h}_{164}^{}={}&\frac{1}{3} (12 h_{144}^{}
 + 6 h_{149}^{}
 - 6 h_{151}^{}
 - 2 h_{190}^{}
 - 2 h_{192}^{}
 + h_{206}^{}),\\
\bar{h}_{165}^{}={}&\frac{1}{27} (12 h_{115}^{}
 - 18 h_{122}^{}
 + 18 h_{124}^{}
 - 3 h_{130}^{}
 + 4 h_{168}^{}
 - 4 h_{173}^{}
 - 6 h_{177}^{}
 - 6 h_{178}^{}- 6 h_{185}^{}
 + 12 h_{187}^{}
 + 6 h_{188}^{}\nonumber\\
&  + 6 h_{190}^{}
 + 24 h_{191}^{}
 - 6 h_{192}^{}
 -  h_{201}^{}
 - 4 h_{205}^{} -  h_{206}^{}),\\
\bar{h}_{166}^{}={}&\frac{1}{27} (12 h_{114}^{}
 - 18 h_{121}^{}
 + 18 h_{125}^{}
 + 3 h_{129}^{}
 - 3 h_{130}^{}
 + 4 h_{169}^{}
 - 4 h_{172}^{}
 + 4 h_{173}^{}- 12 h_{179}^{}
 + 6 h_{180}^{}
 - 6 h_{182}^{}\nonumber\\
&  + 6 h_{184}^{}
 - 6 h_{186}^{}
 + 6 h_{189}^{}
 - 6 h_{191}^{}
 + 24 h_{192}^{} -  h_{200}^{}
 -  h_{201}^{}
 + h_{202}^{}
 + h_{203}^{}
 -  h_{205}^{}
 - 4 h_{206}^{}),\\
\bar{h}_{167}^{}={}&\frac{1}{9} (18 h_{125}^{}
 + 3 h_{130}^{}
 - 6 h_{179}^{}
 - 6 h_{186}^{}
 + 6 h_{189}^{}
 - 6 h_{191}^{}
 + 6 h_{192}^{}
 + h_{201}^{}+ h_{205}^{}
 + h_{206}^{}),\\
\bar{h}_{168}^{}={}&\frac{1}{9} (18 h_{124}^{}
 + 3 h_{130}^{}
 - 6 h_{178}^{}
 - 6 h_{185}^{}
 + 6 h_{190}^{}
 + 6 h_{191}^{}
 - 6 h_{192}^{}
 + h_{201}^{} + h_{205}^{}
 + h_{206}^{}),\\
\bar{h}_{169}^{}={}&\frac{1}{9} (-18 h_{124}^{}
 - 18 h_{125}^{}
 + 3 h_{130}^{}
 + 6 h_{178}^{}
 + 6 h_{179}^{}
 + 6 h_{185}^{}
 + 6 h_{186}^{}
 - 6 h_{189}^{} - 6 h_{190}^{}\nonumber\\
& + h_{201}^{}
 + h_{205}^{}
 + h_{206}^{}),\\
\bar{h}_{170}^{}={}&\frac{1}{108} (27 h_{108}^{}
 - 12 h_{114}^{}
 + 72 h_{121}^{}
 - 72 h_{125}^{}
 + 6 h_{129}^{}
 - 6 h_{130}^{}
 + 9 h_{160}^{}
 - 9 h_{163}^{}+ 9 h_{164}^{}
 - 4 h_{169}^{}
 + 4 h_{172}^{}
\nonumber\\
&  - 4 h_{173}^{} + 48 h_{179}^{}
 - 24 h_{180}^{}
 + 24 h_{182}^{}
 - 24 h_{184}^{}+ 24 h_{186}^{}
 - 24 h_{189}^{}
 + 24 h_{191}^{}
 - 96 h_{192}^{}- 2 h_{200}^{}
 - 2 h_{201}^{}
\nonumber\\
& + 2 h_{202}^{} + 2 h_{203}^{}- 2 h_{205}^{}
 - 8 h_{206}^{}),\\
\bar{h}_{171}^{}={}&\frac{1}{108} (27 h_{109}^{}
 - 12 h_{115}^{}
 + 72 h_{122}^{}
 - 72 h_{124}^{}
 - 6 h_{130}^{}
 + 9 h_{159}^{}
 - 9 h_{164}^{}
 - 4 h_{168}^{}+ 4 h_{173}^{}
 + 24 h_{177}^{}
 + 24 h_{178}^{}\nonumber\\
&  + 24 h_{185}^{}
 - 48 h_{187}^{}
 - 24 h_{188}^{}
 - 24 h_{190}^{}- 96 h_{191}^{}
 + 24 h_{192}^{}
 - 2 h_{201}^{}
 - 8 h_{205}^{}
 - 2 h_{206}^{}),\\
\bar{h}_{172}^{}={}&\frac{1}{108} (-54 h_{135}^{}
 + 24 h_{141}^{}
 - 144 h_{144}^{}
 + 144 h_{145}^{}
 + 72 h_{147}^{}
 - 72 h_{149}^{}
 + 72 h_{150}^{}+ 72 h_{151}^{}
 - 72 h_{152}^{}
 + 12 h_{157}^{}
\nonumber\\
&   + 18 h_{163}^{}- 9 h_{164}^{}
 - 8 h_{172}^{}
 + 4 h_{173}^{}
 + 48 h_{180}^{}+ 48 h_{184}^{}
 + 24 h_{187}^{}
 + 24 h_{188}^{}
 + 48 h_{189}^{}
 + 24 h_{190}^{} - 24 h_{191}^{}\nonumber\\
& 
 + 24 h_{192}^{}+ 4 h_{200}^{}
 + 2 h_{204}^{}
 - 6 h_{206}^{}),\\
\bar{h}_{173}^{}={}&\frac{1}{108} (54 h_{135}^{}
 - 24 h_{141}^{}
 + 144 h_{144}^{}
 - 144 h_{145}^{}
 - 72 h_{147}^{}
 + 72 h_{149}^{}
 - 72 h_{150}^{} - 72 h_{151}^{}
 + 72 h_{152}^{}
 - 12 h_{157}^{}\nonumber\\
& + 9 h_{163}^{}
 + 9 h_{164}^{}
 - 4 h_{172}^{}
 - 4 h_{173}^{}
 + 24 h_{180}^{} + 24 h_{184}^{}
 + 48 h_{187}^{}
 + 48 h_{188}^{}
 + 24 h_{189}^{}
 + 48 h_{190}^{} + 24 h_{191}^{}\nonumber\\
& - 24 h_{192}^{} + 2 h_{200}^{}
 + 4 h_{204}^{}
 - 6 h_{205}^{}),\\
\bar{h}_{174}^{}={}&\frac{1}{108} (-54 h_{135}^{}
 + 24 h_{141}^{}
 - 144 h_{144}^{}
 + 144 h_{145}^{}
 + 72 h_{147}^{}
 - 72 h_{149}^{}
 + 72 h_{150}^{} + 72 h_{151}^{}
 - 72 h_{152}^{}
 + 12 h_{157}^{}
\nonumber\\
&  - 9 h_{163}^{}+ 18 h_{164}^{}
 + 4 h_{172}^{}
 - 8 h_{173}^{}
 - 24 h_{180}^{}- 24 h_{184}^{}
 + 24 h_{187}^{}
 + 24 h_{188}^{}
 - 24 h_{189}^{}
 + 24 h_{190}^{}
\nonumber\\
& + 48 h_{191}^{}
 - 48 h_{192}^{} - 2 h_{200}^{}
 + 2 h_{204}^{}
 - 6 h_{205}^{}
 + 6 h_{206}^{}),\\
\bar{h}_{175}^{}={}&\frac{1}{27} (-24 h_{142}^{}
 - 36 h_{146}^{}
 - 18 h_{147}^{}
 + 36 h_{148}^{}
 + 18 h_{149}^{}
 - 18 h_{150}^{}
 - 18 h_{153}^{}+ 18 h_{154}^{}
 - 6 h_{158}^{}
 + 8 h_{174}^{}\nonumber\\
&  - 4 h_{175}^{}
 + 12 h_{178}^{}
 + 12 h_{179}^{}
 + 6 h_{185}^{}
 + 6 h_{186}^{}- 12 h_{193}^{}
 - 6 h_{194}^{}
 + 6 h_{195}^{}
 - 6 h_{196}^{}
 -  h_{201}^{}
 + h_{203}^{}
 - 3 h_{208}^{}),\\
\bar{h}_{176}^{}={}&- \frac{2}{3} (3 h_{148}^{}
 - 3 h_{150}^{}
 + 3 h_{154}^{}
 - 2 h_{193}^{}
 + h_{195}^{}),\\
\bar{h}_{177}^{}={}&\frac{1}{27} (24 h_{142}^{}
 + 36 h_{146}^{}
 + 18 h_{147}^{}
 - 36 h_{148}^{}
 - 18 h_{149}^{}
 + 18 h_{150}^{}
 + 18 h_{153}^{} - 18 h_{154}^{}
 + 6 h_{158}^{}
 + 4 h_{174}^{} + 4 h_{175}^{}\nonumber\\
& + 6 h_{178}^{}
 + 6 h_{179}^{}
 + 12 h_{185}^{}
 + 12 h_{186}^{}- 6 h_{193}^{}
 - 12 h_{194}^{}
 - 6 h_{195}^{}
 + 6 h_{196}^{}
 - 2 h_{201}^{}
 -  h_{203}^{}
 - 3 h_{207}^{}),\\
\bar{h}_{178}^{}={}&- \frac{2}{3} (6 h_{146}^{}
 + 3 h_{147}^{}
 + 3 h_{153}^{}
 - 2 h_{194}^{}
 + h_{196}^{}),\\
\bar{h}_{179}^{}={}&\frac{1}{3} (6 h_{195}^{}
 + h_{207}^{}),\\
\bar{h}_{180}^{}={}&\frac{1}{3} (6 h_{148}^{}
 - 6 h_{150}^{}
 + 6 h_{154}^{}
 + 2 h_{193}^{}
 - 4 h_{195}^{}
 + h_{207}^{}),\\
\bar{h}_{181}^{}={}&\frac{1}{3} (6 h_{196}^{}
 + h_{208}^{}),\\
\bar{h}_{182}^{}={}&\frac{1}{3} (12 h_{146}^{}
 + 6 h_{147}^{}
 + 6 h_{153}^{}
 + 2 h_{194}^{}
 - 4 h_{196}^{}
 + h_{208}^{}),\\
\bar{h}_{183}^{}={}&\frac{1}{27} (-24 h_{142}^{}
 - 36 h_{146}^{}
 - 18 h_{147}^{}
 + 36 h_{148}^{}
 + 18 h_{149}^{}
 - 18 h_{150}^{}
 - 18 h_{153}^{} + 18 h_{154}^{}
 - 6 h_{158}^{}
 - 4 h_{174}^{}
 + 8 h_{175}^{}\nonumber\\
& - 6 h_{178}^{}
 - 6 h_{179}^{}
 + 6 h_{185}^{}
 + 6 h_{186}^{}+ 6 h_{193}^{}
 - 6 h_{194}^{}
 - 12 h_{195}^{}
 + 12 h_{196}^{}
 -  h_{201}^{}
 - 2 h_{203}^{}
 - 3 h_{207}^{}
 + 3 h_{208}^{}),\\
\bar{h}_{184}^{}={}&\frac{1}{3} (-6 h_{148}^{}
 + 6 h_{150}^{}
 - 6 h_{154}^{}
 - 2 h_{193}^{}
 - 2 h_{195}^{}
 + h_{207}^{}),\\
\bar{h}_{185}^{}={}&\frac{1}{3} (-12 h_{146}^{}
 - 6 h_{147}^{}
 - 6 h_{153}^{}
 - 2 h_{194}^{}
 - 2 h_{196}^{}
 + h_{208}^{}),\\
\bar{h}_{186}^{}={}&\frac{1}{27} (12 h_{117}^{}
 + 18 h_{120}^{}
 - 18 h_{127}^{}
 - 3 h_{131}^{}
 - 4 h_{170}^{}
 - 4 h_{175}^{}
 + 6 h_{180}^{}
 + 6 h_{181}^{}- 12 h_{185}^{}- 6 h_{186}^{}
 + 6 h_{187}^{}\nonumber\\
& + 6 h_{194}^{}
 + 24 h_{195}^{}
 - 6 h_{196}^{}
 + h_{200}^{}
 + h_{204}^{}- 4 h_{207}^{}
 -  h_{208}^{}),\\
\bar{h}_{187}^{}={}&\frac{1}{27} (12 h_{116}^{}
 + 18 h_{123}^{}
 + 18 h_{126}^{}
 + 18 h_{127}^{}
 - 3 h_{129}^{}
 - 3 h_{131}^{}
 - 4 h_{171}^{}
 - 4 h_{174}^{} + 4 h_{175}^{}
 - 6 h_{178}^{}
 - 6 h_{179}^{}
 + 6 h_{183}^{}\nonumber\\
& + 12 h_{184}^{}
 + 6 h_{188}^{}
 + 6 h_{193}^{}
 - 6 h_{195}^{} + 24 h_{196}^{}
 + 2 h_{200}^{}
 -  h_{202}^{}
 -  h_{203}^{}
 + h_{204}^{}
 -  h_{207}^{}
 - 4 h_{208}^{}),\\
\bar{h}_{188}^{}={}&\frac{1}{9} (18 h_{126}^{}
 + 18 h_{127}^{}
 + 3 h_{131}^{}
 + 6 h_{184}^{}
 + 6 h_{188}^{}
 + 6 h_{193}^{}
 - 6 h_{195}^{}
 + 6 h_{196}^{} -  h_{200}^{}
 -  h_{204}^{}
 + h_{207}^{}
 + h_{208}^{}),\\
\bar{h}_{189}^{}={}&\frac{1}{9} (-18 h_{127}^{}
 + 3 h_{131}^{}
 + 6 h_{180}^{}
 + 6 h_{187}^{}
 + 6 h_{194}^{}
 + 6 h_{195}^{}
 - 6 h_{196}^{}
 -  h_{200}^{}-  h_{204}^{}
 + h_{207}^{}
 + h_{208}^{}),\\
\bar{h}_{190}^{}={}&\frac{1}{9} (-18 h_{126}^{}
 + 3 h_{131}^{}
 - 6 h_{180}^{}
 - 6 h_{184}^{}
 - 6 h_{187}^{}
 - 6 h_{188}^{}
 - 6 h_{193}^{}
 - 6 h_{194}^{} -  h_{200}^{} -  h_{204}^{}
 + h_{207}^{}
 + h_{208}^{}),\\
\bar{h}_{191}^{}={}&\frac{1}{108} (27 h_{110}^{}
 - 12 h_{116}^{}
 - 72 h_{123}^{}
 - 72 h_{126}^{}
 - 72 h_{127}^{}
 - 6 h_{129}^{}
 - 6 h_{131}^{}- 9 h_{162}^{}
 - 9 h_{165}^{}
 + 9 h_{166}^{}
 + 4 h_{171}^{}\nonumber\\
& 
 + 4 h_{174}^{}
 - 4 h_{175}^{}
 + 24 h_{178}^{}
 + 24 h_{179}^{} - 24 h_{183}^{}
 - 48 h_{184}^{}
 - 24 h_{188}^{}
 - 24 h_{193}^{}
 + 24 h_{195}^{}
 - 96 h_{196}^{} + 4 h_{200}^{}\nonumber\\
& - 2 h_{202}^{}
 - 2 h_{203}^{}
 + 2 h_{204}^{}
 - 2 h_{207}^{}
 - 8 h_{208}^{}),\\
\bar{h}_{192}^{}={}&\frac{1}{108} (27 h_{111}^{}
 - 12 h_{117}^{}
 - 72 h_{120}^{}
 + 72 h_{127}^{}
 - 6 h_{131}^{}
 - 9 h_{161}^{}
 - 9 h_{166}^{}
 + 4 h_{170}^{} + 4 h_{175}^{}
 - 24 h_{180}^{}
 - 24 h_{181}^{}
\nonumber\\
&  + 48 h_{185}^{}+ 24 h_{186}^{}
 - 24 h_{187}^{}
 - 24 h_{194}^{} - 96 h_{195}^{}
 + 24 h_{196}^{}
 + 2 h_{200}^{}
 + 2 h_{204}^{}
 - 8 h_{207}^{}
 - 2 h_{208}^{}),\\
\bar{h}_{193}^{}={}&\frac{1}{108} (-54 h_{136}^{}
 + 24 h_{142}^{}
 + 144 h_{146}^{}
 + 72 h_{147}^{}
 - 144 h_{148}^{}
 - 72 h_{149}^{}
 + 72 h_{150}^{}+ 72 h_{153}^{}
 - 72 h_{154}^{}
 - 12 h_{158}^{}
\nonumber\\
&  + 18 h_{165}^{} - 9 h_{166}^{}
 - 8 h_{174}^{}
 + 4 h_{175}^{}
 - 48 h_{178}^{}- 48 h_{179}^{}
 - 24 h_{185}^{}
 - 24 h_{186}^{}
 + 48 h_{193}^{}
 + 24 h_{194}^{}
 - 24 h_{195}^{}
 \nonumber\\
 & + 24 h_{196}^{}- 2 h_{201}^{}
 + 2 h_{203}^{}
 - 6 h_{208}^{}),\\
\bar{h}_{194}^{}={}&\frac{1}{108} (54 h_{136}^{}
 - 24 h_{142}^{}
 - 144 h_{146}^{}
 - 72 h_{147}^{}
 + 144 h_{148}^{}
 + 72 h_{149}^{}
 - 72 h_{150}^{} - 72 h_{153}^{}
 + 72 h_{154}^{}
 + 12 h_{158}^{}
 \nonumber\\
& + 9 h_{165}^{}+ 9 h_{166}^{}
 - 4 h_{174}^{}
 - 4 h_{175}^{}
 - 24 h_{178}^{}- 24 h_{179}^{}
 - 48 h_{185}^{}
 - 48 h_{186}^{}
 + 24 h_{193}^{}
 + 48 h_{194}^{} + 24 h_{195}^{}\nonumber\\
&  - 24 h_{196}^{} - 4 h_{201}^{}
 - 2 h_{203}^{}
 - 6 h_{207}^{}),\\
\bar{h}_{195}^{}={}&\frac{1}{108} (-54 h_{136}^{}
 + 24 h_{142}^{}
 + 144 h_{146}^{}
 + 72 h_{147}^{}
 - 144 h_{148}^{}
 - 72 h_{149}^{}
 + 72 h_{150}^{} + 72 h_{153}^{}
 - 72 h_{154}^{}
 - 12 h_{158}^{}\nonumber\\
& - 9 h_{165}^{}
 + 18 h_{166}^{}
 + 4 h_{174}^{}
 - 8 h_{175}^{}
 + 24 h_{178}^{} + 24 h_{179}^{}
 - 24 h_{185}^{}
 - 24 h_{186}^{}- 24 h_{193}^{} + 24 h_{194}^{} + 48 h_{195}^{}\nonumber\\
&  - 48 h_{196}^{} - 2 h_{201}^{}
 - 4 h_{203}^{}
 - 6 h_{207}^{}
 + 6 h_{208}^{}),\\
\bar{h}_{196}^{}={}&\frac{1}{3} (6 h_{145}^{}
 + 6 h_{148}^{}
 + 6 h_{155}^{}
 - 4 h_{197}^{}
 - 2 h_{198}^{}
 + h_{209}^{}),\\
\bar{h}_{197}^{}={}&\frac{1}{3} (6 h_{197}^{}
 + h_{209}^{}),\\
\bar{h}_{198}^{}={}&\frac{1}{27} (-24 h_{143}^{}
 - 18 h_{144}^{}
 + 18 h_{145}^{}
 - 18 h_{146}^{}
 + 18 h_{148}^{}
 + 18 h_{155}^{}
 + 6 h_{156}^{} + 4 h_{176}^{}
 + 6 h_{182}^{} + 6 h_{183}^{}
\nonumber\\
& - 12 h_{197}^{} - 6 h_{198}^{}
 + h_{202}^{}
 - 3 h_{209}^{}),\\
\bar{h}_{199}^{}={}&\frac{1}{3} (-6 h_{145}^{}
 - 6 h_{148}^{}
 - 6 h_{155}^{}
 - 2 h_{197}^{}
 + 2 h_{198}^{}
 + h_{209}^{}),\\
\bar{h}_{200}^{}={}&\frac{1}{27} (48 h_{143}^{}
 + 36 h_{144}^{}
 - 36 h_{145}^{}
 + 36 h_{146}^{}
 - 36 h_{148}^{}
 - 36 h_{155}^{}
 - 12 h_{156}^{}+ 4 h_{176}^{}
 + 6 h_{182}^{}\nonumber\\
&  + 6 h_{183}^{}
 - 12 h_{197}^{}
 - 6 h_{198}^{}
 + h_{202}^{}
 - 3 h_{209}^{}),\\
\bar{h}_{201}^{}={}&\frac{2}{3} (3 h_{145}^{}
 + 3 h_{148}^{}
 + 3 h_{155}^{}
 + h_{197}^{}
 + 2 h_{198}^{}),\\
\bar{h}_{202}^{}={}&\frac{2}{3} (3 h_{128}^{}
 -  h_{177}^{}
 -  h_{181}^{}
 - 2 h_{197}^{}
 -  h_{198}^{}),\\
\bar{h}_{203}^{}={}&\frac{1}{27} (-12 h_{118}^{}
 + 18 h_{121}^{}
 + 18 h_{123}^{}
 + 18 h_{128}^{}
 + 3 h_{129}^{}
 + 4 h_{168}^{}
 + 4 h_{170}^{}+ 4 h_{176}^{}
 - 6 h_{177}^{}
 - 6 h_{178}^{}
 - 6 h_{180}^{}\nonumber\\
&  - 6 h_{181}^{}
 + 6 h_{182}^{}
 + 6 h_{183}^{}
 - 30 h_{197}^{}- 6 h_{198}^{}
 -  h_{200}^{}
 + h_{202}^{}
 + h_{203}^{}
 + 3 h_{209}^{}),\\
\bar{h}_{204}^{}={}&\frac{1}{108} (-54 h_{137}^{}
 + 24 h_{143}^{}
 + 72 h_{144}^{}
 - 72 h_{145}^{}
 + 72 h_{146}^{}
 - 72 h_{148}^{}
 - 72 h_{155}^{}+ 12 h_{156}^{}
 + 9 h_{167}^{}
 - 4 h_{176}^{}
 \nonumber\\
& - 24 h_{182}^{} - 24 h_{183}^{}
 + 48 h_{197}^{}
 + 24 h_{198}^{}
 + 2 h_{202}^{}- 6 h_{209}^{}),\\
\bar{h}_{205}^{}={}&\frac{1}{108} (108 h_{137}^{}
 - 48 h_{143}^{}
 - 144 h_{144}^{}
 + 144 h_{145}^{}
 - 144 h_{146}^{}
 + 144 h_{148}^{} + 144 h_{155}^{}
 - 24 h_{156}^{}
 + 9 h_{167}^{}
\nonumber\\
&  - 4 h_{176}^{}- 24 h_{182}^{}
 - 24 h_{183}^{}
 + 48 h_{197}^{}+ 24 h_{198}^{}
 + 2 h_{202}^{}
 - 6 h_{209}^{}),\\
\bar{h}_{206}^{}={}&\frac{1}{108} (-27 h_{112}^{}
 + 12 h_{118}^{}
 - 72 h_{121}^{}
 - 72 h_{123}^{}
 - 72 h_{128}^{}
 + 6 h_{129}^{}
 + 9 h_{159}^{}+ 9 h_{161}^{}
 + 9 h_{167}^{}
 - 4 h_{168}^{}\nonumber\\
&  - 4 h_{170}^{}
 - 4 h_{176}^{}
 + 24 h_{177}^{}
 + 24 h_{178}^{}
 + 24 h_{180}^{} + 24 h_{181}^{}
 - 24 h_{182}^{}
 - 24 h_{183}^{}
 + 120 h_{197}^{} + 24 h_{198}^{}\nonumber\\
& - 2 h_{200}^{}
 + 2 h_{202}^{} + 2 h_{203}^{}
 + 6 h_{209}^{}),\\
\bar{h}_{207}^{}={}&2 h_{199}^{},\\
\bar{h}_{208}^{}={}&\frac{1}{27} (12 h_{119}^{}
 - 18 h_{124}^{}
 - 18 h_{125}^{}
 - 18 h_{126}^{}
 - 3 h_{130}^{}
 - 3 h_{131}^{}
 + 4 h_{172}^{}
 + 4 h_{174}^{} + 6 h_{178}^{}
 + 6 h_{179}^{}
 - 6 h_{180}^{}
\nonumber\\
&  - 6 h_{184}^{}+ 6 h_{185}^{}
 + 6 h_{186}^{}
 - 6 h_{187}^{}
 - 6 h_{188}^{} - 6 h_{189}^{}
 - 6 h_{190}^{}
 - 6 h_{193}^{}
 - 6 h_{194}^{}
 + 18 h_{199}^{}
 + h_{200}^{}
 -  h_{201}^{} \nonumber\\
&+ h_{204}^{}-  h_{205}^{} -  h_{206}^{}
 -  h_{207}^{}
 -  h_{208}^{}),\\
\bar{h}_{209}^{}={}&\frac{1}{108} (27 h_{113}^{}
 - 12 h_{119}^{}
 + 72 h_{124}^{}
 + 72 h_{125}^{}
 + 72 h_{126}^{}
 - 6 h_{130}^{}
 - 6 h_{131}^{}+ 9 h_{163}^{}
 + 9 h_{165}^{}
 - 4 h_{172}^{}
 - 4 h_{174}^{}\nonumber\\
&  - 24 h_{178}^{}
 - 24 h_{179}^{}
 + 24 h_{180}^{}
 + 24 h_{184}^{}- 24 h_{185}^{}
 - 24 h_{186}^{}
 + 24 h_{187}^{}
 + 24 h_{188}^{}
 + 24 h_{189}^{}
 + 24 h_{190}^{}
 + 24 h_{193}^{} \nonumber\\
&
 + 24 h_{194}^{}
 - 72 h_{199}^{}
 + 2 h_{200}^{}- 2 h_{201}^{}
 + 2 h_{204}^{}
 - 2 h_{205}^{}
 - 2 h_{206}^{}
 - 2 h_{207}^{}- 2 h_{208}^{}).
\end{align}

\section{Coefficients of the vector and axial sectors in the cubic MAG Lagrangian}\label{appendix:l}

Following the Lagrangian density~\eqref{EqVectorDes}, which includes the vector and axial sectors of the gravitational action of MAG that is reduced to GR in the absence of torsion and nonmetricity, we define the following Lagrangian coefficients:
\begin{align}
l_{1}{}={}&\frac{1}{9} (4 c_{1}{}
 + c_{2}{}
 + 2 d_{1}^{})\,,\\
l_{2}{}={}&\frac{1}{2} h_{1}^{}
 + h_{2}^{}\,,\\
l_{3}{}={}&h_{1}^{}\,,\\
l_{4}{}={}&- \frac{2}{3} h_{2}^{}\,,\\
l_{5}{}={}&-2 h_{2}^{}\,,\\
l_{6}{}={}&-\frac{1}{72} (4 c_{1}{}
 +  c_{2}{}
 +  d_{1}^{})\,,\\
l_{7}{}={}&\frac{1}{72} \bigl[4 c_{1}{}
 + c_{2}{}
 + 36 (h_{3}^{}
 + 2 h_{4}^{})\bigr]\,,\\
l_{8}^{}={}&\frac{1}{9} c_{1}{}
 + \frac{1}{36} c_{2}{}
 + h_{3}^{}\,,\\
l_{9}^{}={}&\frac{1}{24} h_{4}^{}\,,\\
l_{10}^{}={}&\frac{1}{24} (c_{2}{}-2 c_{1}{})\,,\\
l_{11}^{}={}&d_{1}^{}- a_{10}^{}
 -  a_{12}^{}
 + 4 a_{14}^{}
 + 2 a_{15}^{}
 + 2 a_{2}^{}
 + a_{5}^{}
 -  a_{9}^{}\,,\\
l_{12}^{}={}&\frac{1}{2} (h_{47}^{}
 + h_{48}^{}
 + 2 h_{49}^{})\,,\\
l_{13}^{}={}&h_{47}^{}
 + h_{48}^{}\,,\\
l_{14}^{}={}&- \frac{3}{2} h_{49}^{}\,,\\
l_{15}^{}={}&3 h_{49}^{}\,,\\
l_{16}^{}={}&\frac{1}{16} (21 a_{10}^{}
 + 5 a_{11}^{}
 - 2 a_{12}^{}
 + 14 a_{2}^{}
 + 9 a_{5}^{}
 - 2 a_{6}^{}
 + 2 a_{9}^{}
 + 2 c_{1}{}
 -  c_{2}{}
 + 2 d_{1}^{})\,,\\
l_{17}^{}={}&\frac{1}{8} (5 a_{12}^{}-6 a_{10}^{}
 - 11 a_{11}^{}
 - 8 a_{2}^{}
 - 4 a_{6}^{}
 - 5 a_{9}^{}
 + 4 c_{1}{}
 - 2 c_{2}{}
 - 5 d_{1}^{}
 + 4 h_{50}^{}+ 4 h_{51}^{}
 + 8 h_{52}^{})\,,\\
l_{18}^{}={}&c_{1}{}
 -  \frac{1}{2} c_{2}{}
 - 2 d_{1}^{}-2 a_{11}^{}
 + 2 a_{12}^{}
 - 2 a_{2}^{}
 -  a_{6}^{}
 - 2 a_{9}^{}
 + h_{50}^{}
 + h_{51}^{}\,,\\
l_{19}^{}={}&\frac{3}{256} (78 a_{10}^{}
 + 111 a_{11}^{}
 - 33 a_{12}^{}
 + 72 a_{2}^{}
 + 36 a_{6}^{}
 + 33 a_{9}^{}
 - 36 c_{1}{}
 + 18 c_{2}{}
 + 33 d_{1}^{}- 16 h_{50}^{}
 + 80 h_{51}^{}
 + 24 h_{52}^{})\,,\\
l_{20}^{}={}&\frac{1}{2} (d_{1}^{}-2 a_{10}^{}
 -  a_{11}^{}
 -  a_{12}^{}
 + a_{9}^{})\,,\\
l_{21}^{}={}&-2 (2 a_{10}^{}
 + a_{11}^{}
 + a_{12}^{}
 -  a_{9}^{}
 -  d_{1}^{})\,,\\
l_{22}^{}={}&\frac{3}{64} (62 a_{10}^{}
 + 79 a_{11}^{}
 - 17 a_{12}^{}
 + 48 a_{2}^{}
 + 24 a_{6}^{}
 + 17 a_{9}^{}
 - 24 c_{1}{}
 + 12 c_{2}{}
 + 17 d_{1}^{}- 16 h_{50}^{}
 + 16 h_{51}^{}
 - 48 h_{52}^{})\,,\\
l_{23}^{}={}&\frac{3}{8} (6 a_{10}^{}
 + 7 a_{11}^{}
 -  a_{12}^{}
 + 4 a_{2}^{}
 + 2 a_{6}^{}
 + a_{9}^{}
 - 2 c_{1}{}
 + c_{2}{}
 + d_{1}^{})\,,\\
l_{24}^{}={}&\frac{1}{648} \bigl[16 c_{2}{}-8 c_{1}{}
 + 9 (48 h_{13}^{}
 + 8 h_{14}^{}
 + 16 h_{3}^{}- h_{1}^{})\bigr]\,,\\
l_{25}^{}={}&-\frac{1}{648} \bigl[16 c_{1}{}
 + 4 c_{2}{}
 - 9 (h_{1}^{}
 - 8 h_{14}^{}
 + 3 h_{2}^{}
 - 16 h_{3}^{}
 - 48 h_{4}^{})\bigr]\,,\\
l_{26}^{}={}&\frac{1}{18} \bigl[2 c_{1}{}
 -  c_{2}{}
 - 3 (6 h_{13}^{}
 + h_{14}^{}
 -  h_{15}^{})\bigr]\,,\\
l_{27}^{}={}&-\frac{1}{108} \bigl[4 c_{1}{}
 +  c_{2}{}
 + 9 (2 h_{14}^{}
 -  h_{15}^{}
 + 4 h_{3}^{}
 + 24 h_{4}^{})\bigr]\,,\\
l_{28}^{}={}&\frac{1}{54} \bigl[2 c_{1}{}
 - 4 c_{2}{}
 - 9 (6 h_{13}^{}
 + h_{14}^{}
 + 4 h_{3}^{})\bigr]\,,\\
l_{29}^{}={}& \frac{2}{3} (h_{14}^{}
 -  h_{15}^{})\,,\\
l_{30}^{}={}&\frac{1}{3} (2 a_{10}^{}
 -  a_{11}^{}
 + a_{12}^{}
 - 2 a_{15}^{}
 - 2 a_{16}^{}
 - 2 a_{5}^{}
 + a_{6}^{}
 + a_{9}^{}
 - 4 c_{1}{}
 - 3 d_{1}^{})\,,\\
l_{31}^{}={}&\frac{1}{2} (h_{108}^{}
 + h_{109}^{}
 + h_{110}^{}
 + h_{111}^{}
 + 2 h_{113}^{})\,,\\
l_{32}^{}={}&h_{108}^{}
 + h_{109}^{}
 + h_{110}^{}
 + h_{111}^{}\,,\\
l_{33}^{}={}&2 h_{2}^{}- \frac{2}{3} h_{113}^{}\,,\\
l_{34}^{}={}&\frac{1}{18} (9 h_{1}^{}
 + 6 h_{108}^{}
 + 6 h_{109}^{}
 + 6 h_{110}^{}
 + 6 h_{111}^{}
 + 36 h_{113}^{}
 + 4 h_{47}^{}
 + 4 h_{48}^{})\,,\\
l_{35}^{}={}&-\frac{1}{18} (9 h_{1}^{}
 + 6 h_{108}^{}
 + 6 h_{109}^{}
 + 6 h_{110}^{}
 + 6 h_{111}^{}
 + 27 h_{2}^{}
 + 4 h_{47}^{}
 + 4 h_{48}^{}
 + 12 h_{49}^{})\,,\\
l_{36}^{}={}&2 h_{49}^{}- \frac{3}{2} h_{113}^{}\,,\\
l_{37}^{}={}&\frac{1}{3} (h_{109}^{}+h_{111}^{}+6 h_{113}^{}- h_{108}^{}
 -  h_{110}^{})\,,\\
l_{38}^{}={}&-\frac{1}{6} (3 h_{108}^{}
 + 3 h_{109}^{}
 + 3 h_{110}^{}
 + 3 h_{111}^{}
 + 18 h_{113}^{}
 + 4 h_{47}^{}
 + 4 h_{48}^{})\,,\\
l_{39}^{}={}&\frac{1}{12} (3 h_{109}^{}+ 12 h_{112}^{}-9 h_{108}^{}
 - 3 h_{110}^{}
 - 3 h_{111}^{}
 - 4 h_{47}^{}
 - 4 h_{48}^{}
 - 24 h_{49}^{})\,,\\
l_{40}^{}={}&h_{109}^{}
 + 2 h_{112}^{}- h_{108}^{}
 - 3 h_{113}^{}\,,\\
l_{41}^{}={}&\frac{1}{6} \bigl[3 h_{1}^{}
 + 2 (h_{109}^{}
 + h_{111}^{}
 + 9 h_{2}^{})\bigr]\,,\\
l_{42}^{}={}&\frac{1}{3} (3 h_{1}^{}
 + h_{108}^{}
 + h_{109}^{}
 + h_{110}^{}
 + h_{111}^{}
 + 6 h_{113}^{})\,,\\
l_{43}^{}={}&\frac{1}{12} (6 a_{5}^{}+ 7 a_{9}^{}
 + 12 c_{1}{}
 + 7 d_{1}^{}- 3 a_{6}^{}-2 a_{10}^{}
 - 7 a_{11}^{}
 -  a_{12}^{})\,,\\
l_{44}^{}={}&\frac{1}{6} (4 a_{10}^{}
 + 2 a_{11}^{}
 + 2 a_{12}^{}
 - 2 a_{9}^{}
 - 2 d_{1}^{}
 + 3 h_{114}^{}
 + 3 h_{115}^{}
 + 3 h_{116}^{}
 + 3 h_{117}^{}+ 6 h_{119}^{})\,,\\
l_{45}^{}={}&\frac{8}{3} a_{10}^{}
 + \frac{4}{3} a_{11}^{}
 + \frac{4}{3} a_{12}^{}
 -  \frac{4}{3} a_{9}^{}
 -  \frac{4}{3} d_{1}^{}
 + h_{114}^{}
 + h_{115}^{}
 + h_{116}^{}
 + h_{117}^{}\,,\\
l_{46}^{}={}&\frac{1}{18} (8 a_{10}^{}
 + 4 a_{11}^{}
 + 4 a_{12}^{}
 - 4 a_{9}^{}
 - 4 d_{1}^{}
 - 9 h_{1}^{}
 - 12 h_{119}^{}
 - 27 h_{2}^{}
 + 18 h_{27}^{})\,,\\
l_{47}^{}={}&\frac{1}{288} (288 a_{10}^{}
 + 208 a_{11}^{}
 + 80 a_{12}^{}
 + 64 a_{2}^{}
 + 32 a_{6}^{}
 - 80 a_{9}^{}
 - 32 c_{1}{}
 + 16 c_{2}{}
 - 80 d_{1}^{}- 99 h_{1}^{}\nonumber\\
& - 120 h_{114}^{}
 - 120 h_{115}^{}
 - 24 h_{116}^{}
 - 24 h_{117}^{}
 - 432 h_{119}^{}
 + 432 h_{27}^{}+ 64 h_{50}^{}
 + 64 h_{51}^{})\,,\\
l_{48}^{}={}&\frac{1}{288} (144 a_{10}^{}
 + 200 a_{11}^{}
 - 56 a_{12}^{}
 + 128 a_{2}^{}
 + 64 a_{6}^{}
 + 56 a_{9}^{}
 - 64 c_{1}{}
 + 32 c_{2}{}+ 56 d_{1}^{}
 + 45 h_{1}^{}\nonumber\\
& - 24 h_{114}^{}
 - 24 h_{115}^{}
 + 168 h_{116}^{}
 + 168 h_{117}^{}
 + 81 h_{2}^{}
 - 108 h_{27}^{}- 64 h_{50}^{}
 - 64 h_{51}^{}
 - 192 h_{52}^{})\,,\\
l_{49}^{}={}&\frac{1}{32} (62 a_{10}^{}
 + 79 a_{11}^{}
 - 17 a_{12}^{}
 + 48 a_{2}^{}
 + 24 a_{6}^{}
 + 17 a_{9}^{}
 - 24 c_{1}{}
 + 12 c_{2}{}
 + 17 d_{1}^{}\nonumber\\
& - 6 h_{114}^{}
 - 6 h_{115}^{}
 + 30 h_{116}^{}
 + 30 h_{117}^{}
 + 9 h_{119}^{}
 - 16 h_{50}^{}
 + 16 h_{51}^{}
 - 48 h_{52}^{})\,,\\
l_{50}^{}={}&\frac{1}{3} (2 d_{1}^{}-4 a_{10}^{}
 - 2 a_{11}^{}
 - 2 a_{12}^{}
 + 2 a_{9}^{}
 -  h_{114}^{}
 + h_{115}^{}
 -  h_{116}^{}
 + h_{117}^{}
 + 6 h_{119}^{})\,,\\
l_{51}^{}={}&\frac{1}{24} (8 c_{1}{}
 - 4 c_{2}{}
 + 20 d_{1}^{}-72 a_{10}^{}
 - 52 a_{11}^{}
 - 20 a_{12}^{}
 - 16 a_{2}^{}
 - 8 a_{6}^{}
 + 20 a_{9}^{}\nonumber\\
& + 15 h_{114}^{}
 + 15 h_{115}^{}
 + 3 h_{116}^{}
 + 3 h_{117}^{}
 + 54 h_{119}^{}
 - 16 h_{50}^{}
 - 16 h_{51}^{})\,,\\
l_{52}^{}={}&\frac{1}{48} (72 a_{10}^{}
 + 68 a_{11}^{}
 + 4 a_{12}^{}
 + 32 a_{2}^{}
 + 16 a_{6}^{}
 - 4 a_{9}^{}
 - 16 c_{1}{}
 + 8 c_{2}{}
 - 4 d_{1}^{}
 - 3 h_{114}^{}\nonumber\\
& - 3 h_{115}^{}
 + 39 h_{116}^{}
 + 3 h_{117}^{}
 - 16 h_{50}^{}
 - 16 h_{51}^{}
 - 96 h_{52}^{})\,,\\
l_{53}^{}={}&\frac{1}{4} (4 c_{1}{}
 - 2 c_{2}{}
 - 2 d_{1}^{}-12 a_{10}^{}
 - 14 a_{11}^{}
 + 2 a_{12}^{}
 - 8 a_{2}^{}
 - 4 a_{6}^{}
 - 2 a_{9}^{}
 + 3 h_{114}^{}+ 3 h_{115}^{}
 - 6 h_{117}^{}
 + 9 h_{119}^{})\,,\\
l_{54}^{}={}&\frac{1}{24} (16 a_{10}^{}
 + 8 a_{11}^{}
 + 8 a_{12}^{}
 - 8 a_{9}^{}
 - 8 d_{1}^{}
 - 15 h_{1}^{}
 + 8 h_{115}^{}
 + 8 h_{117}^{}
 - 54 h_{2}^{}+ 12 h_{27}^{})\,,\\
l_{55}^{}={}&\frac{1}{3} (d_{1}^{}-2 a_{10}^{}
 -  a_{11}^{}
 -  a_{12}^{}
 + a_{9}^{})\,,\\
l_{56}^{}={}&\frac{1}{12} (8 d_{1}^{}-16 a_{10}^{}
 - 8 a_{11}^{}
 - 8 a_{12}^{}
 + 8 a_{9}^{}
 + 3 h_{1}^{}
 + 4 h_{114}^{}
 + 4 h_{115}^{}
 + 4 h_{116}^{}+ 4 h_{117}^{}
 + 24 h_{119}^{}
 - 24 h_{27}^{})\,,\\
l_{57}^{}={}&\frac{4}{3} (2 a_{10}^{}
 + a_{11}^{}
 + a_{12}^{}
 -  a_{9}^{}
 -  d_{1}^{})\,,\\
l_{58}^{}={}&\frac{1}{72} (4 c_{2}{}-2 c_{1}{}
 - 24 h_{132}^{}
 + 24 h_{133}^{}
 - 12 h_{134}^{}
 + 36 h_{3}^{}
 -  h_{47}^{}
 -  h_{48}^{})\,,\\
l_{59}^{}={}&\frac{1}{72} (h_{47}^{}
 + h_{48}^{}+ 3 h_{49}^{}-4 c_{1}{}
 -  c_{2}{}
 - 12 h_{132}^{}
 + 12 h_{133}^{}
 + 12 h_{134}^{}
 - 36 h_{3}^{}
 - 108 h_{4}^{})\,,\\
l_{60}^{}={}&\frac{1}{12} \bigl[c_{2}{}-2 c_{1}{}
 - 2 (2 h_{132}^{}
 - 2 h_{133}^{}
 + h_{134}^{}
 + h_{135}^{}
 + h_{136}^{})\bigr]\,,\\
l_{61}^{}={}&\frac{1}{72} \bigl[4 c_{1}{}
 + c_{2}{}
 + 6 (2 h_{132}^{}
 - 2 h_{133}^{}
 - 2 h_{134}^{}
 -  h_{135}^{}
 -  h_{136}^{}
 + 6 h_{3}^{}
 + 36 h_{4}^{})\bigr]\,,\\
l_{62}^{}={}&\frac{1}{18} (2 c_{2}{}- c_{1}{}
 - 6 h_{132}^{}
 + 6 h_{133}^{}
 - 3 h_{134}^{}
 + 18 h_{3}^{})\,,\\
l_{63}^{}={}&-(2 h_{133}^{}
 + h_{134}^{}
 + 2 h_{135}^{}
 + 4 h_{137}^{})\,,\\
l_{64}^{}={}&\frac{1}{576} (8 c_{2}{}+12 d_{1}^{}- 29 c_{1}{}
 + 48 h_{138}^{}-24 a_{10}^{}
 - 48 a_{11}^{}
 - 48 a_{12}^{}
 + 40 a_{2}^{}
 + 12 a_{5}^{}
 - 14 a_{6}^{}\nonumber\\
& - 24 a_{9}^{}- 240 h_{139}^{}
 + 72 h_{140}^{}
 + 144 h_{141}^{}
 - 144 h_{142}^{}
 + 864 h_{28}^{}- 198 h_{3}^{}
 - 8 h_{50}^{}
 - 8 h_{51}^{})\,,\\
l_{65}^{}={}&\frac{1}{1152} (12 a_{10}^{}
 + 78 a_{11}^{}
 + 78 a_{12}^{}
 - 128 a_{2}^{}
 + 48 a_{5}^{}
 + 16 a_{6}^{}
 + 66 a_{9}^{}
 + 28 c_{1}{}
 - 13 c_{2}{}- 6 d_{1}^{}
 + 624 h_{138}^{}\nonumber\\
& + 336 h_{139}^{}- 144 h_{140}^{}
 - 288 h_{141}^{}
 + 288 h_{142}^{}
 - 432 h_{28}^{}+ 180 h_{3}^{}
 + 324 h_{4}^{}
 + 16 h_{50}^{}
 + 16 h_{51}^{}
 + 48 h_{52}^{})\,,\\
l_{66}^{}={}&\frac{1}{48} (4 a_{10}^{}
 - 2 a_{11}^{}
 + 2 a_{12}^{}
 + 2 a_{9}^{}
 + 6 c_{1}{}
 - 3 c_{2}{}
 - 2 d_{1}^{}
 - 16 h_{138}^{}
 + 16 h_{139}^{}- 8 h_{140}^{}
 - 8 h_{141}^{}
 - 8 h_{142}^{})\,,\\
l_{67}^{}={}&\frac{1}{288} (8 a_{10}^{}
 - 8 a_{11}^{}
 + 4 a_{12}^{}
 + 20 a_{5}^{}
 - 10 a_{6}^{}
 + 8 a_{9}^{}
 - 12 c_{1}{}
 - 13 c_{2}{}
 - 4 d_{1}^{}+ 48 h_{138}^{}\nonumber\\
&- 48 h_{139}^{}
 - 48 h_{140}^{}
 - 24 h_{141}^{}
 - 24 h_{142}^{}
 + 144 h_{28}^{}
 - 180 h_{3}^{}- 648 h_{4}^{})\,,\\
l_{68}^{}={}&\frac{1}{72} (4 a_{10}^{}
 + 2 a_{11}^{}
 + 2 a_{12}^{}
 - 2 a_{5}^{}
 + a_{6}^{}
 - 2 a_{9}^{}
 + 3 c_{1}{}
 - 5 c_{2}{}
 - 2 d_{1}^{}\nonumber\\
& + 24 h_{139}^{}- 24 h_{138}^{}
 - 12 h_{140}^{}
 - 144 h_{28}^{}
 + 18 h_{3}^{})\,,\\
l_{69}^{}={}&\frac{1}{16} (9 a_{11}^{}
 + 3 a_{12}^{}
 + 5 a_{6}^{}
 + 3 a_{9}^{}
 + 4 c_{1}{}
 + 4 c_{2}{}
 + 3 d_{1}^{}+ 24 h_{139}^{}
 + 12 h_{140}^{}
 + 24 h_{142}^{}- 8 a_{2}^{}
 - 6 a_{5}^{}-6 a_{10}^{})\,,\\
l_{70}^{}={}&\frac{1}{4} (4 a_{10}^{}
 -  a_{11}^{}
 -  a_{12}^{}
 - 6 a_{16}^{}
 - 6 a_{5}^{}
 + 6 a_{6}^{}
 - 5 a_{9}^{}
 - 6 c_{1}{}
 + 3 c_{2}{}
 - 5 d_{1}^{})\,,\\
l_{71}^{}={}&\frac{1}{2} (d_{1}^{}-2 a_{10}^{}
 -  a_{11}^{}
 -  a_{12}^{}
 + a_{9}^{}
 + h_{68}^{}
 + h_{69}^{}
 + h_{70}^{}
 + h_{71}^{}
 + 2 h_{73}^{})\,,\\
l_{72}^{}={}&\frac{3}{4}(d_{1}^{}+a_{9}^{}
 +  h_{47}^{}
 + 3 h_{49}^{}
 - h_{48}^{}
 - 2 h_{73}^{}
 -2 a_{10}^{}
 - a_{11}^{}
 - a_{12}^{})\,,\\
l_{73}^{}={}&\frac{1}{32} (72 a_{10}^{}
 + 52 a_{11}^{}
 + 20 a_{12}^{}
 + 16 a_{2}^{}
 + 8 a_{6}^{}
 - 20 a_{9}^{}
 - 8 c_{1}{}
 + 4 c_{2}{}
 - 20 d_{1}^{}- 11 h_{47}^{}\nonumber\\
& + 37 h_{48}^{}
 + 16 h_{50}^{}
 + 16 h_{51}^{}
 + 20 h_{68}^{}
 + 20 h_{69}^{}
 + 4 h_{70}^{}
 + 4 h_{71}^{}+ 72 h_{73}^{})\,,\\
l_{74}^{}={}&2 d_{1}^{}-4 a_{10}^{}
 - 2 a_{11}^{}
 - 2 a_{12}^{}
 + 2 a_{9}^{}
 + h_{68}^{}
 + h_{69}^{}
 + h_{70}^{}
 + h_{71}^{}\,,\\
l_{75}^{}={}&\frac{1}{32} (36 a_{10}^{}
 + 50 a_{11}^{}
 - 14 a_{12}^{}
 + 32 a_{2}^{}
 + 16 a_{6}^{}
 + 14 a_{9}^{}
 - 16 c_{1}{}
 + 8 c_{2}{}
 + 14 d_{1}^{}+ 5 h_{47}^{}\nonumber\\
&- 7 h_{48}^{}
 + 9 h_{49}^{}
 - 16 h_{50}^{}
 - 16 h_{51}^{}
 - 48 h_{52}^{}
 + 4 h_{68}^{}
 + 4 h_{69}^{}
 - 28 h_{70}^{}- 28 h_{71}^{})\,,\\
l_{76}^{}={}&\frac{3}{64} (24 c_{1}{}- 12 c_{2}{}
 - 17 d_{1}^{}
+ 17 a_{12}^{}
+ 16 h_{50}^{}
+ 48 h_{52}^{}
+ 20 h_{70}^{}
 + 20 h_{71}^{}
 + 6 h_{73}^{}\nonumber\\
&-62 a_{10}^{}- 79 a_{11}^{}
- 48 a_{2}^{}
 - 24 a_{6}^{}
 - 17 a_{9}^{}
 - 16 h_{51}^{}
 - 4 h_{68}^{}
 - 4 h_{69}^{})\,,\\
l_{77}^{}={}&\frac{1}{2} (3 d_{1}^{}-6 a_{10}^{}
 - 3 a_{11}^{}
 - 3 a_{12}^{}
 + 3 a_{9}^{}
 + 2 h_{68}^{}
 - 2 h_{69}^{}
 - 4 h_{72}^{}
 - 6 h_{73}^{})\,,\\
l_{78}^{}={}&\frac{1}{8} (36 a_{10}^{}
 + 26 a_{11}^{}
 + 10 a_{12}^{}
 + 8 a_{2}^{}
 + 4 a_{6}^{}
 - 10 a_{9}^{}
 - 4 c_{1}{}
 + 2 c_{2}{}- 10 d_{1}^{}
 + 8 h_{50}^{}\nonumber\\
& + 8 h_{51}^{}
 + 5 h_{68}^{}
 + 5 h_{69}^{}
 + h_{70}^{}
 + h_{71}^{}
 + 18 h_{73}^{})\,,\\
l_{79}^{}={}&\frac{1}{16} (8 c_{1}{}
 - 4 c_{2}{}
 + 2 d_{1}^{}-36 a_{10}^{}
 - 34 a_{11}^{}
 - 2 a_{12}^{}
 - 16 a_{2}^{}
 - 8 a_{6}^{}+ 2 a_{9}^{}
 + 8 h_{50}^{}+ 8 h_{51}^{}\nonumber\\
&  + 48 h_{52}^{}
 -  h_{68}^{}
 -  h_{69}^{}
 + h_{70}^{}
 + 13 h_{71}^{})\,,\\
l_{80}^{}={}&\frac{1}{4} (6 d_{1}^{}-12 a_{10}^{}
 - 6 a_{11}^{}
 - 6 a_{12}^{}
 + 6 a_{9}^{}
 + h_{47}^{}
 - 7 h_{48}^{}
 - 2 h_{68}^{}
 - 2 h_{69}^{}
 - 2 h_{70}^{}- 2 h_{71}^{}
 - 12 h_{73}^{})\,,\\
l_{81}^{}={}&\frac{3}{4} (6 a_{10}^{}
 + 7 a_{11}^{}
 -  a_{12}^{}
 + 4 a_{2}^{}
 + 2 a_{6}^{}
 + a_{9}^{}
 - 2 c_{1}{}
 + c_{2}{}
 + d_{1}^{}
 + h_{68}^{}
 + h_{69}^{}
 - 2 h_{70}^{}+ 3 h_{73}^{})\,,\\
l_{82}^{}={}&\frac{1}{8} (12 a_{10}^{}
 + 6 a_{11}^{}
 + 6 a_{12}^{}
 - 6 a_{9}^{}
 - 6 d_{1}^{}
 - 5 h_{47}^{}
 -  h_{48}^{}
 - 18 h_{49}^{}
 + 2 h_{68}^{}
 - 6 h_{69}^{}- 2 h_{70}^{}
 - 2 h_{71}^{}
 - 8 h_{72}^{})\,,\\
l_{83}^{}={}&\frac{1}{2} (2 a_{10}^{}
 + a_{11}^{}
 + a_{12}^{}
 -  a_{9}^{}
 -  d_{1}^{})\,,\\
l_{84}^{}={}&-2 (2 a_{10}^{}
 + a_{11}^{}
 + a_{12}^{}
 -  a_{9}^{}
 -  d_{1}^{})\,,\\
l_{85}^{}={}&\frac{1}{216} \bigl[8 c_{1}{}
 - 16 c_{2}{}
 - 3 (h_{108}^{}
 + h_{109}^{}
 + h_{110}^{}
 + h_{111}^{}
 + 72 h_{13}^{}
 - 16 h_{132}^{}
 + 16 h_{133}^{}- 8 h_{134}^{}
 + 12 h_{14}^{}
 + 48 h_{3}^{})\bigr]\,,\\
l_{86}^{}={}&\frac{1}{216} \bigl[16 c_{1}{}
 + 4 c_{2}{}+3 (h_{108}^{}
 + h_{109}^{}
 + h_{110}^{}
 + h_{111}^{}
 + 3 h_{113}^{}+ 8 h_{132}^{}- 8 h_{133}^{}- 8 h_{134}^{}
 \nonumber\\
 &  + 12 h_{14}^{} + 48 h_{3}^{}
 + 144 h_{4}^{})\bigr]\,,\\
l_{87}^{}={}&\frac{1}{12}(12 h_{14}^{}
 + 12 h_{33}^{}
 + 48 h_{35}^{}
 + h_{109}^{}
 + h_{111}^{}
 - 24 h_{15}^{}
 - 24 h_{34}^{}
 - h_{108}^{}
 - h_{110}^{})\,,\\
l_{88}^{}={}&\frac{1}{3} (6 h_{15}^{}
 + 6 h_{34}^{}
 + 2 h_{133}^{}
 + 2 h_{134}^{}
 + 2 h_{135}^{}
 + 2 h_{136}^{}
 - 3 h_{14}^{}
 - 3 h_{33}^{}
 - 12 h_{35}^{}
 - 2 h_{132}^{})\,,\\
l_{89}^{}={}&\frac{1}{216} (10 c_{2}{}
 + 4 d_{1}^{}- 6 c_{1}{}-8 a_{10}^{}
 - 4 a_{11}^{}
 - 4 a_{12}^{}
 + 4 a_{5}^{}
 - 2 a_{6}^{}
 + 4 a_{9}^{}
 - 3 h_{114}^{}- 3 h_{115}^{}
 - 3 h_{116}^{}
 - 3 h_{117}^{}\nonumber\\
& + 162 h_{13}^{}
+ 48 h_{138}^{}
 - 48 h_{139}^{}
 + 27 h_{14}^{}
 + 24 h_{140}^{}- 54 h_{15}^{}
 + 288 h_{28}^{}
 - 36 h_{3}^{}
 - 36 h_{33}^{}
 - 108 h_{34}^{})\,,\\
l_{90}^{}={}&\frac{1}{216} (2 a_{10}^{}
 + a_{11}^{}
 + a_{12}^{}
 + 8 a_{5}^{}
 - 4 a_{6}^{}
 -  a_{9}^{}
 - 12 c_{1}{}
 - 7 c_{2}{}
 -  d_{1}^{}
 + 3 h_{114}^{}
 + 3 h_{115}^{}+ 3 h_{116}^{}
 + 3 h_{117}^{}
 + 9 h_{119}^{}\nonumber\\
& + 24 h_{138}^{}
 - 24 h_{139}^{}
 - 27 h_{14}^{}
 - 24 h_{140}^{}
 + 54 h_{15}^{}- 72 h_{28}^{}
 - 72 h_{3}^{}
 + 90 h_{33}^{}
 + 108 h_{34}^{}
 - 324 h_{4}^{})\,,\\
l_{91}^{}={}&\frac{1}{12} (  h_{117}^{}
 + h_{115}^{}
 - h_{114}^{}
 - h_{116}^{}
 - 9 h_{14}^{}
 - 9 h_{33}^{}
 - 18 h_{34}^{})\,,\\
l_{92}^{}={}&\frac{1}{12} (2 c_{2}{}
 + 2 d_{1}^{}
 + 2 a_{6}^{}
 + 2 a_{11}^{}
 + 9 h_{14}^{}
 + 9 h_{33}^{}
 + 18 h_{34}^{}
 + 8 h_{139}^{}+ 8 h_{140}^{}
 + 8 h_{141}^{}
 + 8 h_{142}^{}
 \nonumber\\
&  - 4 a_{5}^{}
 - 2 a_{9}^{}
 - 4 a_{10}^{}
 - 2 a_{12}^{}
 - 8 h_{138}^{})\,,\\
l_{93}^{}={}&\frac{1}{24} (24 a_{10}^{}
 + 12 a_{11}^{}
 + 12 a_{12}^{}
 - 12 a_{9}^{}
 - 12 d_{1}^{}
 + 3 h_{108}^{}
 + 3 h_{109}^{}
 - 21 h_{110}^{}- 21 h_{111}^{}
 - 12 h_{114}^{}
 - 12 h_{115}^{}\nonumber\\
& - 12 h_{116}^{}
 - 12 h_{117}^{}
 - 36 h_{119}^{}
 - 8 h_{47}^{}
 - 16 h_{48}^{}- 36 h_{49}^{}
 - 8 h_{68}^{}
 - 8 h_{69}^{}
 - 8 h_{70}^{}
 - 8 h_{71}^{})\,,\\
l_{94}^{}={}&\frac{1}{36} (12 d_{1}^{}+12 a_{9}^{}
-24 a_{10}^{}
 - 12 a_{11}^{}
 - 12 a_{12}^{}
 + 18 h_{1}^{}
 - 3 h_{108}^{}
 - 3 h_{109}^{}+ 21 h_{110}^{}
 + 21 h_{111}^{}
 - 12 h_{114}^{}\nonumber\\
& - 12 h_{115}^{}
- 12 h_{116}^{}
 - 12 h_{117}^{}
 + 81 h_{2}^{}
 + 18 h_{27}^{}- 8 h_{68}^{}
 - 8 h_{69}^{}
 - 8 h_{70}^{}
 - 8 h_{71}^{}
 - 24 h_{73}^{})\,,\\
l_{95}^{}={}&\frac{1}{24} (48 a_{10}^{}
 + 24 a_{11}^{}
 + 24 a_{12}^{}
 - 24 a_{9}^{}
 - 24 d_{1}^{}
 + 15 h_{108}^{}
 + 15 h_{109}^{}
 + 3 h_{110}^{}
 + 3 h_{111}^{}
 + 54 h_{113}^{}
 + 12 h_{114}^{}\nonumber\\
& + 12 h_{115}^{}
 + 12 h_{116}^{}
 + 12 h_{117}^{}
 - 4 h_{47}^{}
 + 28 h_{48}^{}
 + 8 h_{68}^{}
 + 8 h_{69}^{}
 + 8 h_{70}^{}
 + 8 h_{71}^{}
 + 48 h_{73}^{})\,,\\
l_{96}^{}={}&\frac{1}{96} (32 c_{1}{}
 - 16 c_{2}{}+ 80 d_{1}^{}-288 a_{10}^{}
 - 208 a_{11}^{}
 - 80 a_{12}^{}
 - 64 a_{2}^{}
 - 32 a_{6}^{}
 + 80 a_{9}^{}- 33 h_{108}^{}
 - 33 h_{109}^{}\nonumber\\
 &+ 111 h_{110}^{}+ 111 h_{111}^{}
 + 60 h_{114}^{}
 + 60 h_{115}^{} + 12 h_{116}^{}
 + 12 h_{117}^{}
 + 216 h_{119}^{}
 - 64 h_{50}^{}
 - 64 h_{51}^{}
 - 40 h_{68}^{}\nonumber\\
& - 40 h_{69}^{}
 - 8 h_{70}^{}- 8 h_{71}^{}
 - 144 h_{73}^{})\,,\\
l_{97}^{}={}&\frac{1}{36} (24 d_{1}^{}
-48 a_{10}^{}
 - 24 a_{11}^{}
 - 24 a_{12}^{}
 + 24 a_{9}^{}
 + 9 h_{1}^{}
 - 15 h_{108}^{}
 - 15 h_{109}^{}- 3 h_{110}^{}
 - 3 h_{111}^{}
 - 54 h_{113}^{}\nonumber\\
&+ 12 h_{114}^{}
 + 12 h_{115}^{}
 + 12 h_{116}^{}
 + 12 h_{117}^{}
 + 72 h_{119}^{} - 72 h_{27}^{}
 + 8 h_{68}^{}
 + 8 h_{69}^{}
 + 8 h_{70}^{}
 + 8 h_{71}^{})\,,\\
l_{98}^{}={}&\frac{1}{96} (64 c_{1}{}
 - 32 c_{2}{}- 56 d_{1}^{}
 -144 a_{10}^{}
 - 200 a_{11}^{}
 + 56 a_{12}^{}
 - 128 a_{2}^{}
 - 64 a_{6}^{}
 - 56 a_{9}^{}
 + 15 h_{108}^{}
 + 15 h_{109}^{}\nonumber\\
& - 21 h_{110}^{}
 - 21 h_{111}^{}
 + 27 h_{113}^{}
 + 12 h_{114}^{}
 + 12 h_{115}^{}
 - 84 h_{116}^{}
 - 84 h_{117}^{}
 + 64 h_{50}^{}
 + 64 h_{51}^{}+ 192 h_{52}^{}\nonumber\\
&- 8 h_{68}^{}
 - 8 h_{69}^{}
 + 56 h_{70}^{}+ 56 h_{71}^{})\,,\\
l_{99}^{}={}&\frac{1}{24} (3 h_{110}^{}
 + 39 h_{111}^{}- 16 h_{68}^{}
 - 16 h_{71}^{}-3 h_{108}^{}
 - 3 h_{109}^{})\,,\\
l_{100}^{}={}&\frac{1}{8} (13 h_{110}^{}
 + h_{111}^{}
 + 12 h_{114}^{}
 - h_{108}^{}
 -  h_{109}^{}
 - 4 h_{115}^{}
 + 4 h_{116}^{}
 + 4 h_{117}^{} - 16 h_{118}^{})\,,\\
l_{101}^{}={}&2 a_{10}^{}
 + a_{11}^{}
 + a_{12}^{}
 -  a_{9}^{}
 -  d_{1}^{}
 + \frac{5}{8} h_{108}^{}
 + \frac{5}{8} h_{109}^{}
 + \frac{1}{8} h_{110}^{}
 + \frac{1}{8} h_{111}^{}
 + \frac{9}{4} h_{113}^{} -  \frac{2}{3} h_{69}^{}
 -  \frac{2}{3} h_{70}^{}\,,\\
l_{102}^{}={}&\frac{1}{24} (24 d_{1}^{}-48 a_{10}^{}
 - 24 a_{11}^{}
 - 24 a_{12}^{}
 + 24 a_{9}^{}
 - 3 h_{108}^{}
 - 3 h_{109}^{}
 + 39 h_{110}^{}+ 3 h_{111}^{}
 - 8 h_{68}^{}\nonumber\\
&  - 8 h_{69}^{}
 - 8 h_{70}^{}
 - 8 h_{71}^{}
 - 48 h_{73}^{})\,,\\
l_{103}^{}={}&\frac{1}{8} (16 a_{10}^{}
 + 8 a_{11}^{}
 + 8 a_{12}^{}
 - 8 a_{9}^{}
 - 8 d_{1}^{}
 + 5 h_{108}^{}
 + 5 h_{109}^{}
 + h_{110}^{}
 + h_{111}^{}+ 18 h_{113}^{}
 - 4 h_{114}^{}\nonumber\\
&  + 12 h_{115}^{}
 + 4 h_{116}^{}
 + 4 h_{117}^{}
 + 16 h_{118}^{})\,,\\
l_{104}^{}={}&\frac{1}{8} (8 d_{1}^{}-16 a_{10}^{}
 - 8 a_{11}^{}
 - 8 a_{12}^{}
 + 8 a_{9}^{}
 -  h_{108}^{}
 -  h_{109}^{}
 + h_{110}^{}
 + 13 h_{111}^{} + 4 h_{114}^{}
 + 4 h_{115}^{}\nonumber\\
& + 4 h_{116}^{}
 + 4 h_{117}^{}
 + 24 h_{119}^{})\,,\\
l_{105}^{}={}&\frac{1}{144} (12 c_{1}{}
 + 7 c_{2}{}
 + d_{1}^{}
 -2 a_{10}^{}
 -  a_{11}^{}
 -  a_{12}^{}
 - 8 a_{5}^{}
 + 4 a_{6}^{}
 + a_{9}^{}
 + 78 h_{132}^{}+ 42 h_{133}^{}
 - 18 h_{134}^{}
 - 36 h_{135}^{}
 + 36 h_{136}^{}\nonumber\\
& - 24 h_{138}^{}
 + 24 h_{139}^{}
 + 24 h_{140}^{} + 72 h_{28}^{}
 + 72 h_{3}^{}
 + 324 h_{4}^{}
 + 2 h_{68}^{}
 + 2 h_{69}^{}
 + 2 h_{70}^{}
 + 2 h_{71}^{}
 + 6 h_{73}^{})\,,\\
l_{106}^{}={}&\frac{1}{72} (4 a_{10}^{}
 + 2 a_{11}^{}
 + 2 a_{12}^{}
 - 2 a_{5}^{}
 + a_{6}^{}
 - 2 a_{9}^{}
 + 3 c_{1}{}
 - 5 c_{2}{}
 - 2 d_{1}^{}
 + 6 h_{132}^{}- 30 h_{133}^{}
 + 9 h_{134}^{}
 + 18 h_{135}^{}\nonumber\\
& - 18 h_{136}^{}- 24 h_{138}^{}
 + 24 h_{139}^{}
 - 12 h_{140}^{} - 144 h_{28}^{}
 + 18 h_{3}^{}
 -  h_{68}^{}
 -  h_{69}^{}
 -  h_{70}^{}
 -  h_{71}^{})\,,\\
l_{107}^{}={}&\frac{1}{12} (h_{69}^{}
 +  h_{70}^{}
 - h_{68}^{}
 - h_{71}^{}
 -18 h_{133}^{}
 - 9 h_{134}^{}
 - 18 h_{136}^{})\,,\\
l_{108}^{}={}&\frac{1}{4} ( 2 a_{5}^{}+3 a_{9}^{}
+ 2 a_{10}^{}
+  a_{11}^{}
 + 3 a_{12}^{}
 + 2 a_{16}^{}
 - 3 a_{6}^{}
 - 2 a_{15}^{}
 - 2 c_{2}{}
 - d_{1}^{} + 6 h_{133}^{}
 + 3 h_{134}^{}\nonumber\\
& + 6 h_{136}^{}
 - 8 h_{139}^{}
 - 4 h_{140}^{}
 - 8 h_{141}^{}
 - 16 h_{143}^{})\,,\\
l_{109}^{}={}&\frac{1}{8} (h_{109}^{}
 + h_{110}^{}- h_{108}^{}
 -  h_{111}^{}).
\end{align}

\section{Coefficients of the cubic Lagrangian providing Reissner-Nordström-like black hole solutions}\label{appendix:RN}

From the Lagrangian~\eqref{Theory} of cubic MAG, it is possible to find the general Reissner-Nordström-like black hole solution shown in Sec.~\ref{sec:RN} by setting a total number of $123$ Lagrangian coefficients $h_i$. First of all, the stability conditions~\eqref{MAG1}-\eqref{MAG6} set $36$ of these coefficients. Then, the presence of the shear charge $\kappa_{\rm sh}$ in the solution requires additional $45$ coefficients to satisfy the following expressions:
\begin{align}
h_{8}{}={}&2 \bigl(2 h_{5}{}
 + h_{6}{}
 -  h_{7}{}
 + 2 (h_{9}{}
 + h_{10}{})
 + h_{11}{}\bigr),\\
h_{12}{}={}&\frac{1}{12} (- h_{9}{}
 + h_{10}{}
 -  h_{11}{}),\\
h_{17}{}={}&\frac{1}{2} h_{16}{},\\
h_{19}{}={}&0,\\
h_{29}{}={}&-2 h_{5}{}
 -  h_{6}{}
 - 2 (h_{9}{}
 + h_{10}{})
 -  h_{11}{},\\
h_{31}{}={}&\frac{1}{2} (- h_{10}{}
 + h_{11}{}),\\
h_{32}{}={}&\frac{1}{2} h_{11}{},\\
h_{37}{}={}&- h_{36}{},\\
h_{38}{}={}&\frac{1}{4} \bigl(- h_{16}{}
 + 2 (h_{18}{}
 + h_{36}{})\bigr),\\
h_{39}{}={}&0,\\
h_{50}{}={}&\frac{1}{48} \Bigl[606 a_{2}{}
 + 303 a_{6}{}
 - 108 h_{53}{}
 + 108 h_{55}{}
 - 108 h_{56}{}
 - 108 h_{57}{}- 324 h_{59}{}
 + 324 h_{61}{}
 - 3537 h_{65}{}
 + 1815 h_{66}{}\nonumber\\
& 
 + 288 h_{90}{}
 + 288 h_{91}{}+ 78 h_{95}{}
 + 447 h_{96}{}
 - 111 h_{97}{}
 + 135 h_{99}{}
 + 135 h_{101}{} + 135 h_{102}{} - 468 h_{103}{}
 - 738 h_{104}{}\nonumber\\
&  - 5 \bigl(54 h_{105}{}
 - 4 h_{174}{}
 - 6 h_{189}{}
 - 6 h_{190}{} + 6 (h_{193}{}
 + h_{194}{})
 - 7 h_{201}{}\bigr)
 - 333 (2 a_{2}{}
 + a_{6}{}) N_{3}{}\Bigr],\\
h_{62}{}={}&\frac{1}{216} \biggl[1110 a_{2}{}
 + 555 a_{6}{}
 - 2 \Bigl(54 h_{53}{}
 - 54 h_{55}{}
 + 54 h_{56}{}
 + 54 h_{57}{}
 + 54 h_{59}{}- 108 h_{60}{}
 - 54 h_{61}{}
 + 447 h_{65}{}
 + 477 h_{66}{}\nonumber\\
&  - 2 \bigl(72 h_{90}{}
 + 72 h_{91}{}
 - 18 h_{95}{}+ 39 h_{96}{}
 + 45 h_{97}{}
 + 27 h_{99}{}
 + 27 h_{101}{}
 + 27 h_{102}{}
 - 144 h_{103}{}
 - 198 h_{104}{}- 54 h_{105}{}
\nonumber\\
&  + 26 h_{174}{} + 39 (h_{189}{}
 + h_{190}{}
 -  h_{193}{}
 -  h_{194}{})\bigr)
 - 91 h_{201}{}\Bigr) - 711 (2 a_{2}{}
 + a_{6}{}) N_{3}{}\biggr],\\
h_{63}{}={}&\frac{1}{1296} \biggl[906 a_{2}{}
 + 453 a_{6}{}
 - 2 \Bigl(54 h_{53}{}
 + 54 h_{54}{}
 - 162 h_{56}{}
 - 108 h_{57}{} + 54 h_{58}{}
 + 216 h_{59}{}
 + 216 h_{60}{}
 + 324 h_{61}{}
 + 1107 h_{65}{}\nonumber\\
&
 - 267 h_{66}{} + 2 \bigl(72 h_{90}{}
 + 72 h_{91}{}
 - 18 h_{95}{}
 + 39 h_{96}{}
 + 87 h_{97}{}
 - 99 h_{99}{}
 + 27 h_{101}{}+ 27 h_{102}{}
 + 108 h_{103}{}
 + 54 h_{104}{}\nonumber\\
&  - 54 h_{105}{}
 - 58 h_{174}{}
 - 87 (h_{189}{}
 + h_{190}{} -  h_{193}{}
 -  h_{194}{})\bigr)
 - 203 h_{201}{}\Bigr)
 - 801 (2 a_{2}{}
 + a_{6}{}) N_{3}{}\biggr],\\
h_{64}{}={}&\frac{1}{7} \Bigl[23 h_{65}{}
 - 9 h_{66}{}
 - 2 \bigl(h_{96}{}
 + 3 h_{99}{}
 - 6 h_{103}{}
 - 6 h_{104}{}
 + 2 h_{174}{}
 + 3 (h_{189}{} + h_{190}{}
 -  h_{193}{}
 -  h_{194}{})\bigr)
 - 7 h_{201}{}\nonumber\\
 &
 + 12 a_{2}{} (-2
 + 3 N_{3}{})
 + 6 a_{6}{} (-2+ 3 N_{3}{})\Bigr],\\
h_{67}{}={}&0,\\
h_{83}{}={}&\frac{1}{7} (h_{75}{}
 + h_{76}{}
 - 15 h_{78}{}
 - 17 h_{79}{}
 -  h_{80}{}
 + h_{81}{})
 -  \frac{8}{3} h_{84}{},\\
h_{85}{}={}&- \frac{6}{7} (h_{78}{}
 + h_{79}{}
 + h_{80}{}
 + h_{81}{})
 -  h_{84}{},\\
h_{87}{}={}&\frac{1}{144} \Bigl[-7440 a_{2}{}
 - 3720 a_{6}{}
 - 648 h_{53}{}
 + 216 h_{54}{}
 + 864 h_{55}{}
 - 540 h_{56}{} - 324 h_{57}{}
 + 216 h_{58}{}
 - 2484 h_{59}{}
 \nonumber\\
&- 864 h_{60}{} + 2268 h_{61}{}
 - 16929 h_{65}{} + 11175 h_{66}{}
 + 144 h_{86}{}
 - 576 h_{90}{}
 - 864 h_{91}{}
 - 288 h_{92}{}
 - 102 h_{95}{}\nonumber\\
& + 123 h_{96}{}
 - 963 h_{97}{}
 - 1233 h_{99}{}
 + 783 h_{101}{}
 + 783 h_{102}{}
 + 3132 h_{103}{}+ 1566 h_{104}{}
 - 1566 h_{105}{}+ 6066 (2 a_{2}{}
 + a_{6}{}) N_{3}{}
\nonumber\\
&  - 391 \bigl(4 h_{174}{}
 + 6 h_{189}{}
 + 6 h_{190}{}
 - 6 (h_{193}{}+ h_{194}{})
 + 7 h_{201}{}\bigr)
 \Bigr],\\
h_{88}{}={}&6 h_{53}{}
 - 6 h_{55}{}
 + 6 h_{56}{}
 + 6 h_{57}{}
 + 18 h_{59}{}
 - 18 h_{61}{}
 + \frac{393}{2} h_{65}{}
 -  \frac{605}{6} h_{66}{}-  h_{86}{}
 + 2 h_{91}{}
 + 2 h_{92}{}
 + \frac{8}{3} h_{95}{}\nonumber\\
&  - 5 h_{96}{}
 + \frac{22}{3} h_{97}{}
 -  \frac{15}{2} h_{99}{}
 -  \frac{15}{2} h_{101}{} -  \frac{15}{2} h_{102}{}
 + 26 h_{103}{}
 + 41 h_{104}{} + a_{2}{} (- \frac{101}{3}
 + 37 N_{3}{})\nonumber\\
&
 + \frac{5}{18} \bigl(54 h_{105}{}
 - 4 h_{174}{}
 - 6 h_{189}{} - 6 h_{190}{}
 + 6 (h_{193}{}
 + h_{194}{})
 - 7 h_{201}{}\bigr)
+ \frac{1}{6} a_{6}{} (-101
 + 111 N_{3}{}),\\
h_{93}{}={}&- h_{90}{}
 -  h_{91}{}
 -  h_{92}{}
 -  \frac{7}{6} (h_{96}{}
 + h_{97}{}),\\
h_{98}{}={}&\frac{1}{9} \Bigl[90 h_{65}{}
 - 6 h_{66}{}
 - 3 h_{96}{}
 + 3 h_{97}{}
 - 9 h_{99}{}
 + 2 \bigl(18 h_{103}{}
 + 18 h_{104}{} - 4 h_{174}{}
 - 6 h_{189}{}
 - 6 h_{190}{}
 + 6 (h_{193}{}
 + h_{194}{})
 - 7 h_{201}{}\bigr)
\nonumber\\
& + 24 a_{2}{} (-2 + 3 N_{3}{})
 + 12 a_{6}{} (-2
 + 3 N_{3}{})\Bigr],\\
h_{100}{}={}&\frac{1}{18} \Bigl[-63 h_{65}{}
 - 21 h_{66}{}
 + 3 h_{95}{}
 + 3 h_{96}{}
 - 3 h_{97}{}
 - 2 \bigl(9 h_{99}{}
 + 9 h_{101}{} + 9 h_{102}{}
 + 18 h_{103}{}
 + 18 h_{104}{}
 - 4 h_{174}{}
 - 6 h_{189}{}
\nonumber\\
&  - 6 h_{190}{}+ 6 (h_{193}{} + h_{194}{})
 - 7 h_{201}{}\bigr)
 + a_{2}{} (48
 - 72 N_{3}{})
 + a_{6}{} (24
 - 36 N_{3}{})\Bigr],\\
h_{106}{}={}&\frac{1}{9} \bigl[-21 h_{65}{}
 - 21 h_{66}{}
 - 9 h_{103}{}
 - 9 h_{104}{}
 - 9 h_{105}{}
 + 4 h_{174}{}
 + 6 h_{189}{} + 6 h_{190}{}
 - 6 (h_{193}{}
 + h_{194}{})
 + 7 h_{201}{}
\nonumber\\
& + a_{2}{} (24
 - 36 N_{3}{})
 + 6 a_{6}{} (2 - 3 N_{3}{})\bigr],\\
h_{120}{}={}&\frac{1}{3} \bigl(-3 h_{123}{}
 - 2 a_{2}{} (1
 + 3 N_{3}{})
 -  a_{6}{} (1
 + 3 N_{3}{})\bigr),\\
h_{122}{}={}&\frac{1}{3} \bigl(-3 h_{121}{}
 - 2 a_{2}{} (1
 + 3 N_{3}{})
 -  a_{6}{} (1
 + 3 N_{3}{})\bigr),\\
h_{125}{}={}&- h_{124}{}
 -  \frac{1}{6} a_{2}{} (1
 + 3 N_{3}{})
 -  \frac{1}{12} a_{6}{} (1
 + 3 N_{3}{}),\\
h_{126}{}={}&\frac{1}{12} (2 a_{2}{}
 + a_{6}{}) (1
 + 3 N_{3}{}),\\
h_{127}{}={}&\frac{1}{12} \bigl[6 (h_{121}{}
 + h_{123}{}
 - 2 h_{124}{})
 - 10 a_{2}{} (1
 + 3 N_{3}{})
 - 5 a_{6}{} (1
 + 3 N_{3}{})\bigr],\\
h_{129}{}={}&0,\\
h_{130}{}={}&0,\\
h_{165}{}={}&- h_{163}{},\\
h_{166}{}={}&- \frac{1}{2} h_{160}{}
 -  \frac{1}{2} h_{162}{}
 + 7 h_{163}{}
 + h_{164}{},\\
h_{168}{}={}&\frac{1}{8} \bigl[-8 h_{169}{}
 + 12 h_{174}{}
 + 45 h_{178}{}
 + 42 h_{179}{}
 - 3 (17 h_{180}{}
 - 8 h_{181}{} + 8 h_{182}{}
 + 10 h_{184}{}
 - 14 h_{185}{} - 20 h_{186}{}\nonumber\\
& + 10 h_{187}{}
 + 12 h_{188}{}
 - 30 h_{189}{}- 34 h_{190}{}
 + 29 h_{191}{}
 + 31 h_{192}{}
 - 2 h_{193}{}
 + 2 h_{194}{}
 + 19 h_{195}{}
 + 17 h_{196}{})\nonumber\\
& + 77 h_{201}{}
 + 132 a_{2}{} (2
 - 3 N_{3}{})
 + 66 a_{6}{} (2
 - 3 N_{3}{})\bigr],\\
h_{170}{}={}&\frac{1}{8} \bigl[-8 h_{171}{}
 + 4 h_{174}{}
 - 3 (15 h_{178}{}
 + 14 h_{179}{}
 - 17 h_{180}{}
 + 8 h_{181}{}
 - 8 h_{182}{}- 10 h_{184}{}
 + 14 h_{185}{}
 + 20 h_{186}{}
 - 10 h_{187}{}
\nonumber\\
&  - 12 h_{188}{}
 + 14 h_{189}{} + 18 h_{190}{}
 - 5 h_{191}{}
 - 7 h_{192}{}
 + 18 h_{193}{}
 + 14 h_{194}{}
 - 43 h_{195}{}
 - 41 h_{196}{} - 21 h_{201}{})\nonumber\\
&
 + 108 a_{2}{} (2
 - 3 N_{3}{})
 + 54 a_{6}{} (2
 - 3 N_{3}{})\bigr],\\
h_{172}{}={}&- h_{174}{},\\
h_{175}{}={}&- \frac{1}{2} h_{169}{}
 -  \frac{1}{2} h_{171}{}
 + h_{173}{}
 - 7 h_{174}{},\\
h_{177}{}={}&\frac{1}{24} \bigl[4 h_{174}{}
 + 3 (h_{178}{}
 - 6 h_{179}{}
 + h_{180}{}
 - 8 h_{181}{}
 - 6 h_{184}{}
 + 2 h_{185}{} - 4 h_{186}{}
 + 2 h_{187}{}
 - 4 h_{188}{}
 + 2 h_{189}{} - 2 h_{190}{}\nonumber\\
&
 - 19 h_{191}{}
 - 17 h_{192}{}- 2 h_{193}{}
 + 2 h_{194}{}
 + 19 h_{195}{}
 + 17 h_{196}{})
 + 7 h_{201}{}
 + a_{2}{} (24
 - 36 N_{3}{})+ 6 a_{6}{} (2
 - 3 N_{3}{})\bigr],\\
h_{183}{}={}&\frac{1}{8} \bigl[4 h_{174}{}
 -  h_{178}{}
 - 2 h_{179}{}
 -  h_{180}{}
 - 8 h_{182}{}
 - 2 h_{184}{}
 + 6 h_{185}{}
 + 4 h_{186}{} + 6 h_{187}{}
 + 4 h_{188}{}
 + 14 h_{189}{}
 + 18 h_{190}{}
 \nonumber\\
&- 5 h_{191}{}
 - 7 h_{192}{}
 - 14 h_{193}{}- 18 h_{194}{}
 + 5 h_{195}{}
 + 7 (h_{196}{}
 + h_{201}{})
 + a_{2}{} (24
 - 36 N_{3}{})
 + 6 a_{6}{} (2- 3 N_{3}{})\bigr],\\
h_{199}{}={}&\frac{1}{6} (h_{178}{}
 + h_{179}{}
 -  h_{180}{}
 -  h_{184}{}
 + h_{185}{}
 + h_{186}{}
 -  h_{187}{}
 -  h_{188}{}
 -  h_{189}{} -  h_{190}{}
 -  h_{193}{}
 -  h_{194}{}),\\
h_{200}{}={}&h_{201}{}
 -  h_{204}{},\\
h_{202}{}={}&2 (h_{205}{}
 -  h_{207}{}),\\
h_{203}{}={}&-2 h_{201}{}
 -  h_{204}{}
 - 2 h_{205}{}
 + 2 h_{207}{},\\
h_{206}{}={}&\frac{1}{2} (h_{201}{}
 - 2 h_{205}{}),\\
h_{208}{}={}&- \frac{1}{2} h_{201}{}
 -  h_{207}{},\\
h_{89}{}={}&\frac{1}{48} \biggl[-36 h_{53}{}
 - 72 h_{54}{}
 - 36 h_{55}{}
 - 72 h_{56}{}
 - 144 h_{57}{}
 - 72 h_{58}{}
 - 36 h_{59}{}+ 288 h_{60}{}
 + 108 h_{61}{}
 - 987 h_{65}{}
 + 1197 h_{66}{}
 - 48 h_{86}{}\nonumber\\
& 
 + 192 h_{90}{} + 192 h_{91}{}
 - 2 h_{95}{}
 + 3 \Bigl(59 h_{96}{}
 - 3 h_{97}{}
 + 239 h_{99}{}
 + 15 h_{101}{}
 + 15 h_{102}{} - 388 h_{103}{}
 - 418 h_{104}{}
 - 30 h_{105}{}\nonumber\\
& + 25 \bigl(4 h_{174}{}
 + 6 h_{189}{}
 + 6 h_{190}{}- 6 (h_{193}{}
 + h_{194}{})
 + 7 h_{201}{}\bigr)\Bigr)
 + 450 a_{6}{} (2
 - 3 N_{3}{})
 - 900 a_{2}{} (-2 + 3 N_{3}{})\biggr]\,.
\end{align}
On the other hand, for the dilation charge $\kappa_{\rm d}$, the following $18$ coefficients must be set as follows:
\begin{align}
h_{40}{}={}&- \frac{1}{2} h_{16}{}
 + h_{36}{},\\
h_{76}{}={}&\frac{1}{2} \bigl(-2 h_{74}{}
 + a_{6}{} (1
 + 3 N_{3}{})
 + a_{2}{} (2
 + 6 N_{3}{})\bigr),\\
h_{77}{}={}&\frac{1}{2} \bigl(-2 h_{75}{}
 + a_{6}{} (1
 + 3 N_{3}{})
 + a_{2}{} (2
 + 6 N_{3}{})\bigr),\\
h_{81}{}={}&\frac{1}{4} \bigl(2 (h_{74}{}
 -  h_{75}{}
 - 2 h_{78}{})
 + 6 a_{2}{} (1
 + 3 N_{3}{})
 + a_{6}{} (3
 + 9 N_{3}{})\bigr),\\
h_{82}{}={}&\frac{1}{4} \bigl(a_{6}{}
 + 2 h_{74}{}
 - 2 h_{75}{}
 + 3 a_{6}{} N_{3}{}
 + a_{2}{} (2
 + 6 N_{3}{})\bigr),\\
h_{84}{}={}&0,\\
h_{107}{}={}&\frac{1}{60} \biggl[2 \Bigl(18 h_{53}{}
 - 18 h_{55}{}
 + 18 h_{56}{}
 + 18 h_{57}{}
 + 54 h_{59}{}
 - 54 h_{61}{}
 + 532 h_{65}{} - 280 h_{66}{}
 + 12 h_{91}{}
 + 12 h_{92}{}
 + 12 h_{94}{} \nonumber\\
& + 8 h_{95}{}- 9 h_{96}{}
 + 23 h_{97}{}
 - 3 \bigl(5 h_{99}{}+ 5 h_{101}{}
 + 5 h_{102}{}
 - 4 (4 h_{103}{}
 + 9 h_{104}{}
 + 5 h_{105}{})\bigr)\Bigr)
\nonumber\\
&+ 162 a_{2}{} (-1
 + N_{3}{})+ 81 a_{6}{} (-1
 + N_{3}{})\biggr],\\
h_{128}{}={}&\frac{1}{2} (h_{121}{}
 + h_{123}{}),\\
h_{159}{}={}&- h_{160}{},\\
h_{161}{}={}&- h_{162}{},\\
h_{163}{}={}&0,\\
h_{167}{}={}&\frac{1}{2} (- h_{160}{}
 -  h_{162}{}),\\
h_{176}{}={}&\frac{1}{2} (- h_{169}{}
 -  h_{171}{}
 - 8 h_{174}{}),\\
h_{198}{}={}&\frac{1}{8} (- h_{178}{}
 - 2 h_{179}{}
 -  h_{180}{}
 - 2 h_{184}{}
 - 2 h_{185}{}
 - 4 h_{186}{}
 - 2 h_{187}{}
 - 4 h_{188}{}+ 2 h_{189}{}
 - 2 h_{190}{}\nonumber\\
&  -  h_{191}{}
 + h_{192}{}
 - 2 h_{193}{}
 + 2 h_{194}{}
 + h_{195}{}
 -  h_{196}{}
 - 4 h_{197}{}),\\
h_{209}{}={}&\frac{1}{70} \bigl[606 a_{2}{}
 + 303 a_{6}{}
 - 108 h_{53}{}
 + 108 h_{55}{}
 - 108 h_{56}{}
 - 108 h_{57}{}
 - 324 h_{59}{} + 324 h_{61}{}
 - 3537 h_{65}{}
 + 1815 h_{66}{}\nonumber\\
& - 72 h_{91}{}
 - 72 h_{92}{}
 - 72 h_{94}{}
 - 48 h_{95}{} + 69 h_{96}{}
 - 153 h_{97}{}
 + 135 h_{99}{}
 + 135 h_{101}{}
 + 135 h_{102}{}
 - 468 h_{103}{}
 - 738 h_{104}{}\nonumber\\
& - 10 (27 h_{105}{}
 - 2 h_{174}{}
 - 3 h_{189}{}
 - 3 h_{190}{}
 + 3 h_{193}{}
 + 3 h_{194}{}
 + 7 h_{205}{}
 - 7 h_{207}{}) - 333 (2 a_{2}{}
 + a_{6}{}) N_{3}{}\bigr],\\
h_{79}{}={}&\frac{1}{8} \bigl(-8 h_{78}{}
 + a_{6}{} (1
 + 3 N_{3}{})
 + a_{2}{} (2
 + 6 N_{3}{})\bigr),\\
h_{80}{}={}&\frac{1}{8} \bigl(-4 h_{74}{}
 + 4 h_{75}{}
 + 8 h_{78}{}
 - 14 a_{2}{} (1
 + 3 N_{3}{})
 - 7 a_{6}{} (1
 + 3 N_{3}{})\bigr),\\
h_{201}{}={}&\frac{1}{35} \bigl[108 h_{53}{}
 - 108 h_{55}{}
 + 108 h_{56}{}
 + 108 h_{57}{}
 + 324 h_{59}{}
 - 324 h_{61}{}
 + 3537 h_{65}{} - 1815 h_{66}{}
 + 72 h_{91}{}
 \nonumber\\
& + 72 h_{92}{}
 + 72 h_{94}{}
 + 48 h_{95}{}- 69 h_{96}{}
 + 153 h_{97}{} - 135 h_{99}{}
 - 135 h_{101}{}
 - 135 h_{102}{}
 + 468 h_{103}{}
 + 738 h_{104}{}
\nonumber\\
& + 10 (27 h_{105}{} - 2 h_{174}{}
 - 3 h_{189}{}
 - 3 h_{190}{}
 + 3 h_{193}{}
 + 3 h_{194}{})
 + a_{6}{} (-303
 + 333 N_{3}{})+ a_{2}{} (-606
 + 666 N_{3}{})\bigr].
\end{align}
Finally, in order to incorporate the spin charge $\kappa_{\rm s}$, a last set of $24$ coefficients must acquire the following form:
\begin{align}
h_{7}{}={}&\frac{1}{8} \bigl[-15 d_{1}{}
 + 4 (4 h_{5}{}
 + 8 h_{6}{}
 + h_{9}{})
 - 30 h_{25}{}\bigr],\\
h_{10}{}={}&- h_{9}{},\\
h_{11}{}={}&- \frac{3}{2} d_{1}{}
 + 3 h_{6}{}
 + h_{9}{}
 - 3 h_{25}{},\\
h_{13}{}={}&0,\\
h_{16}{}={}&0,\\
h_{18}{}={}&0,\\
h_{20}{}={}&\frac{1}{8} (d_{1}{}
 - 2 h_{23}{}
 + 2 h_{25}{}),\\
h_{22}{}={}&\frac{1}{6480} \biggl[-18375 d_{1}{}
 + 4 \Bigl\{8000 h_{5}{}
 + 16000 h_{6}{}
 + 2000 h_{9}{}
 - 2625 h_{23}{} + 6 \bigl(-2200 h_{25}{}
 - 4500 h_{28}{}
 + 875 h_{41}{}
 + 1145 h_{42}{}\nonumber \\
 &
 + 990 h_{43}{}
 + 1095 h_{44}{} + 1365 h_{45}{}
 + 1116 h_{53}{}
 - 1116 h_{55}{}
 + 1116 h_{56}{}
 + 1116 h_{57}{}
 + 3348 h_{59}{}- 3348 h_{61}{}
 + 36549 h_{65}{}\nonumber \\
 &
 - 18755 h_{66}{}
 + 744 h_{91}{}
 + 744 h_{92}{}
 + 744 h_{94}{} + 496 h_{95}{}
 - 713 h_{96}{}
 + 1581 h_{97}{}
 - 1395 h_{99}{}
 - 1395 h_{101}{}
 - 1395 h_{102}{}
\nonumber \\
 & + 4836 h_{103}{}+ 7626 h_{104}{}
 + 5 (558 h_{105}{}
 - 216 h_{144}{}
 - 108 h_{145}{}
 + 216 h_{146}{} + 108 h_{148}{}
 + 270 h_{151}{}
 - 18 h_{152}{}
 + 270 h_{153}{}
\nonumber \\
 &  - 18 h_{154}{}- 10 h_{178}{}
 - 40 h_{179}{} - 10 h_{180}{}
 - 4 h_{181}{}
 - 164 h_{182}{}
 - 4 h_{184}{}
 - 20 h_{185}{}
 + 26 h_{186}{}
 + 144 h_{187}{} + 54 h_{188}{}
 \nonumber \\
 &+ 65 h_{189}{}+ 115 h_{190}{}
 - 112 h_{191}{}
 - 152 h_{192}{}
 - 99 h_{193}{} - 129 h_{194}{})\bigr)\Bigr\}
 + 60264 a_{6}{} (-1
 + N_{3}{})\biggr]
 + \frac{93}{5} a_{2}{} (-1
 + N_{3}{}),\\
h_{24}{}={}&\frac{1}{2} (h_{23}{}
 + h_{25}{}),\\
h_{36}{}={}&-2 d_{1}{}
 + \frac{8}{3} h_{5}{}
 + \frac{16}{3} h_{6}{}
 + \frac{2}{3} h_{9}{}
 + h_{23}{}
 - 5 h_{25}{}
 + 6 h_{33}{}
 - 2 (h_{41}{}
 + h_{42}{}
 - 3 h_{43}{}),\\
h_{46}{}={}&\frac{1}{3240} \biggl[-8985 d_{1}{}
 + 16000 h_{5}{}
 + 32000 h_{6}{}
 + 4000 h_{9}{}
 - 6 \Bigl\{1080 h_{21}{}
 + 1415 h_{23}{}+ 4265 h_{25}{}
 + 2 \bigl(4500 h_{28}{}
 - 875 h_{41}{}
 \nonumber\\
&  - 1415 h_{42}{}
 - 855 h_{43}{}
 - 1230 h_{44}{} - 1230 h_{45}{}- 1116 h_{53}{}
 + 1116 h_{55}{}
 - 1116 h_{56}{}
 - 1116 h_{57}{}
 - 3348 h_{59}{}+ 3348 h_{61}{}
\nonumber\\
& - 36549 h_{65}{}
 + 18755 h_{66}{}
 - 744 h_{91}{}
 - 744 h_{92}{}
 - 744 h_{94}{} - 496 h_{95}{}
 + 713 h_{96}{}
 - 1581 h_{97}{}
 + 1395 h_{99}{}
 + 1395 h_{101}{}
\nonumber\\
& + 1395 h_{102}{} - 4836 h_{103}{}
 - 7626 h_{104}{}
 - 5 (558 h_{105}{}
 - 216 h_{144}{}
 - 108 h_{145}{}
 + 216 h_{146}{}+ 108 h_{148}{}
 + 270 h_{151}{}
 - 18 h_{152}{}
\nonumber\\
& + 270 h_{153}{}
 - 18 h_{154}{}
 - 10 h_{178}{}
 - 40 h_{179}{}- 10 h_{180}{}
 - 4 h_{181}{}
 - 164 h_{182}{}
 - 4 h_{184}{}
 - 20 h_{185}{}
 + 26 h_{186}{}
 + 144 h_{187}{}\nonumber\\
&+ 54 h_{188}{}
 + 65 h_{189}{}
 + 115 h_{190}{}
 - 112 h_{191}{}
 - 152 h_{192}{}
 - 99 h_{193}{}- 129 h_{194}{})\bigr)\Bigr\}
 + 30132 a_{6}{} (N_{3}{}-1
 )\biggr]\nonumber\\
 &
 + \frac{93}{5} a_{2}{} ( N_{3}{}-1
 ),\\
h_{155}{}={}&\frac{1}{48} \Bigl[2 \bigl(6 d_{1}{}
 - 8 h_{5}{}
 - 16 h_{6}{}
 - 2 h_{9}{}
 - 3 h_{23}{}
 + 15 h_{25}{}
 + 6 (h_{41}{}
 + h_{42}{}
 - 3 h_{43}{}-  h_{74}{}
 + h_{75}{}
 + 4 h_{78}{}
 + 12 h_{132}{} - 2 h_{144}{}\nonumber\\
&  -  h_{145}{}
 - 2 h_{146}{}
 -  h_{148}{}
 + h_{151}{}-  h_{152}{}
 -  h_{153}{}
 + h_{154}{}
 + h_{162}{})\bigr)
 - 42 a_{2}{} (1
 + 3 N_{3}{})
 - 21 a_{6}{} (1
 + 3 N_{3}{})\Bigr],\\
h_{164}{}={}&\frac{1}{4} (3 h_{160}{}
 -  h_{162}{}),\\
h_{195}{}={}&\frac{1}{2700} \biggl[7800 d_{1}{}
 - 10400 h_{5}{}
 - 20800 h_{6}{}
 - 2600 h_{9}{}
 + 3 \Bigl\{260 h_{23}{}
 + 6500 h_{25}{} - 520 h_{41}{}
 - 520 h_{42}{}
 - 4680 h_{43}{}
 - 1908 h_{53}{}\nonumber\\
 &
 + 1908 h_{55}{}
 - 1908 h_{56}{}- 1908 h_{57}{}
 - 5724 h_{59}{}
 + 5724 h_{61}{}
 - 62487 h_{65}{}
 + 32065 h_{66}{}
 - 1272 h_{91}{}- 1272 h_{92}{}
 - 1272 h_{94}{}\nonumber\\
 &
 - 848 h_{95}{}
 + 1219 h_{96}{}
 - 2703 h_{97}{}
 + 2385 h_{99}{}+ 2385 h_{101}{}
 + 2385 h_{102}{}
 - 2 \Big(4134 h_{103}{}
 + 6519 h_{104}{}
\nonumber\\
 & + 5 (477 h_{105}{}
 + 45 h_{151}{} + 45 h_{152}{}
 + 45 h_{153}{}
 + 45 h_{154}{}
 + 71 h_{178}{}
 - 40 h_{179}{}
 - 85 h_{180}{}
 - 22 h_{181}{}- 134 h_{182}{}
 - 10 h_{184}{}
\nonumber\\
 & + 64 h_{185}{}
 + 80 h_{186}{}
 + 42 h_{187}{}
 - 24 h_{188}{}
 + 41 h_{189}{} + 121 h_{190}{}
 - 46 h_{191}{}
 - 164 h_{192}{}
 - 93 h_{193}{}
 - 69 h_{194}{})\Big)\Bigr\}
\nonumber\\
 & - 25758 a_{2}{} ( N_{3}{}-1 )
 - 12879 a_{6}{} ( N_{3}{}-1
 )\biggr],\\
h_{196}{}={}&\frac{1}{36} \biggl[
 + 3 \Bigl\{-4 h_{23}{}
 - 100 h_{25}{}
 + 8 h_{41}{}+ 8 h_{42}{}
 + 72 h_{43}{} + 36 h_{53}{}
 - 36 h_{55}{}
 + 36 h_{56}{}+ 36 h_{57}{}
 + 108 h_{59}{}
 - 108 h_{61}{}\nonumber\\
& 
 + 1179 h_{65}{} - 605 h_{66}{}
 + 24 h_{91}{}
 + 24 h_{92}{}
 + 24 h_{94}{}
 + 16 h_{95}{}
 - 23 h_{96}{}
 + 51 h_{97}{} - 45 h_{99}{}
 - 45 h_{101}{}
 - 45 h_{102}{}\nonumber\\
& + 2 \bigl(78 h_{103}{}
 + 123 h_{104}{}
 + 45 h_{105}{}
 + 9 h_{151}{}+ 9 h_{152}{}
 + 9 h_{153}{}
 + 9 h_{154}{}
 + 5 h_{178}{}- 4 h_{179}{}
 - 7 h_{180}{}
 + 2 h_{181}{}\nonumber\\
& - 14 h_{182}{} + 2 h_{184}{}
 + 4 h_{185}{}
 + 8 h_{186}{}
 + 6 h_{187}{}
 + 5 h_{189}{}
 + 13 h_{190}{}
 - 4 h_{191}{}
 - 14 h_{192}{}- 9 (h_{193}{}
 + h_{194}{})\bigr)\Bigr\}\nonumber\\
& -120 d_{1}{}
 + 160 h_{5}{}
 + 320 h_{6}{}
 + 40 h_{9}{} + 486 a_{2}{} (N_{3}{}-1
 )
 + 243 a_{6}{} (N_{3}{}-1
 )\biggr],\\
h_{204}{}={}&\frac{1}{3150} \Bigl[
  3 \Big\{310 h_{23}{}
 - 3650 h_{25}{} - 25200 h_{28}{}
 - 620 h_{41}{}
 - 620 h_{42}{}
 + 3420 h_{43}{}
 + 1200 h_{44}{}
 + 1200 h_{45}{}+ 1692 h_{53}{} \nonumber\\
& - 1692 h_{55}{}
 + 1692 h_{56}{} + 1692 h_{57}{}
 + 5076 h_{59}{}
 - 5076 h_{61}{}+ 55413 h_{65}{}
 - 28435 h_{66}{}
 + 1128 h_{91}{}
 + 1128 h_{92}{}
\nonumber\\
& + 1128 h_{94}{}
 + 752 h_{95}{} - 1081 h_{96}{}
 + 2397 h_{97}{}
 - 2115 h_{99}{}
 - 2115 h_{101}{}
 - 2115 h_{102}{}
 + 7332 h_{103}{}+ 11562 h_{104}{}\nonumber\\
& + 4230 h_{105}{}
 - 20 (90 h_{152}{}
 + 90 h_{154}{}
 - 315 h_{158}{}
 + 15 h_{174}{}- 37 h_{178}{}
 - 40 h_{179}{}+ 35 h_{180}{}
 + 14 h_{181}{}
 - 2 h_{182}{}\nonumber\\
&  + 50 h_{184}{}
 - 38 h_{185}{} - 55 h_{186}{}
 + 36 h_{187}{}
 + 63 h_{188}{}
 - 7 h_{189}{}
 - 2 h_{190}{}
 + 32 h_{191}{}
 + 28 h_{192}{} - 54 h_{193}{}
 - 57 h_{194}{})\Big\}
\nonumber\\
&-1500 d_{1}{}
 + 5600 h_{5}{}
 + 11200 h_{6}{}
 + 1400 h_{9}{} + 54 a_{2}{} (-523
 + 573 N_{3}{})
 + 27 a_{6}{} (-523
 + 573 N_{3}{})\Bigr],\\
h_{207}{}={}&\frac{1}{12600} \biggl[ -110052 a_{2}{}
 - 55026 a_{6}{}
 + 1650 d_{1}{}
 + 5600 h_{5}{}
 + 11200 h_{6}{}
 + 1400 h_{9}{}
 + 60426 (2 a_{2}{} + a_{6}{}) N_{3}{}\nonumber\\
 &- 3 \Big\{2315 h_{23}{}
 + 7325 h_{25}{}
 + 2 \Big(31500 h_{28}{}
 - 2315 h_{41}{}
 - 2315 h_{42}{}
 + 915 h_{43}{} + 2550 h_{44}{}
 + 2550 h_{45}{}
 - 3276 h_{53}{}
 \nonumber\\
&+ 3276 h_{55}{}
 - 3276 h_{56}{}
 - 3276 h_{57}{} - 9828 h_{59}{}
 + 9828 h_{61}{}
 - 107289 h_{65}{}
 + 55055 h_{66}{}
 - 2184 h_{91}{}
 - 2184 h_{92}{}\nonumber\\
& - 2184 h_{94}{}
 - 1456 h_{95}{}
 + 2093 h_{96}{}
 - 4641 h_{97}{}
 + 4095 h_{99}{}
 + 4095 h_{101}{} + 4095 h_{102}{}
 - 14196 h_{103}{}
 - 22386 h_{104}{}\nonumber\\
& - 5 (1638 h_{105}{}
 - 180 h_{152}{} - 180 h_{154}{}
 + 1260 h_{158}{}
 - 120 h_{174}{}
 + 74 h_{178}{}
 + 80 h_{179}{}
 - 70 h_{180}{}
 - 28 h_{181}{} + 4 h_{182}{}\nonumber\\
& - 100 h_{184}{}
 + 76 h_{185}{}
 + 110 h_{186}{}
 - 72 h_{187}{}
 - 126 h_{188}{}
 - 121 h_{189}{} - 131 h_{190}{}
 - 64 h_{191}{}- 56 h_{192}{}
\nonumber\\
&+ 243 h_{193}{}
 + 249 h_{194}{})\Big)\Big\}\biggr],\\
h_{26}{}={}&\frac{1}{6480} \biggl[
  4 \Big\{-8000 h_{5}{}
 - 16000 h_{6}{}
 - 2000 h_{9}{}
 + 1620 h_{21}{}
 + 3435 h_{23}{} + 6 \Big(2065 h_{25}{}
 + 4500 h_{28}{}
 - 875 h_{41}{}
 - 1145 h_{42}{}
\nonumber\\
& - 990 h_{43}{}
 - 1095 h_{44}{} - 1365 h_{45}{}
 - 1116 h_{53}{}
 + 1116 h_{55}{}
 - 1116 h_{56}{}
 - 1116 h_{57}{}
 - 3348 h_{59}{}+ 3348 h_{61}{}
 - 36549 h_{65}{}\nonumber\\
&  + 18755 h_{66}{}
 - 744 h_{91}{}
 - 744 h_{92}{}
 - 744 h_{94}{} - 496 h_{95}{}
 + 713 h_{96}{}
 - 1581 h_{97}{}
 + 1395 h_{99}{}
 + 1395 h_{101}{}
 + 1395 h_{102}{}\nonumber\\
& - 4836 h_{103}{}
 - 7626 h_{104}{}
 - 5 (558 h_{105}{}
 - 216 h_{144}{}
 - 108 h_{145}{}
 + 216 h_{146}{} + 108 h_{148}{}
 + 270 h_{151}{}
 - 18 h_{152}{}
 + 270 h_{153}{}
 \nonumber\\
& - 18 h_{154}{}- 10 h_{178}{}
 - 40 h_{179}{} - 10 h_{180}{}
 - 4 h_{181}{}
 - 164 h_{182}{}
 - 4 h_{184}{}
 - 20 h_{185}{}
 + 26 h_{186}{}
 + 144 h_{187}{} + 54 h_{188}{}\nonumber\\
&+ 65 h_{189}{}
 + 115 h_{190}{}
 - 112 h_{191}{}
 - 152 h_{192}{}
 - 99 h_{193}{} - 129 h_{194}{})\Big)\Big\}
+18375 d_{1}{} - 60264 a_{6}{} ( N_{3}{}-1
 )\biggr]
\nonumber\\
&-  \frac{93}{5} a_{2}{} (N_{3}{}-1
 ),\\
h_{124}{}={}&\frac{1}{72} \Big[-24 d_{1}{}
 + 32 h_{5}{}
 + 64 h_{6}{}
 + 8 h_{9}{}
 + 12 h_{23}{}
 - 60 h_{25}{}
 - 24 h_{41}{}
 - 24 h_{42}{} + 18 (4 h_{43}{}
 + h_{121}{}
 + h_{123}{})
\nonumber\\
&- 30 a_{2}{} (1
 + 3 N_{3}{})
 - 15 a_{6}{} (1
 + 3 N_{3}{})\Bigr],\\
h_{138}{}={}&\frac{1}{21600} \Biggl[  54 a_{6}{} (-5503
 + 5253 N_{3}{}) + 108 a_{2}{} (-5053
 + 5253 N_{3}{})+17025 d_{1}{}
 - 1400 h_{5}{}
 - 2800 h_{6}{}
 - 350 h_{9}{}\nonumber\\
&  + 17385 h_{23}{}
 - 34725 h_{25}{}- 6 \Big\{70200 h_{28}{}
 + 5795 h_{41}{}
 + 5795 h_{42}{}
 - 5445 h_{43}{}
 + 11700 h_{44}{}
 + 11700 h_{45}{}- 20412 h_{53}{}\nonumber\\
&  + 2250 h_{54}{}
 + 21762 h_{55}{}
 - 20412 h_{56}{}
 - 19062 h_{57}{}
 + 1800 h_{58}{} - 63486 h_{59}{}
 - 3600 h_{60}{}
 + 62586 h_{61}{}
 - 685968 h_{65}{}
\nonumber\\
& + 337060 h_{66}{}
 - 1800 h_{90}{}- 16608 h_{91}{}
 - 13608 h_{92}{}
 - 14208 h_{94}{}
 - 8872 h_{95}{}
 + 11966 h_{96}{}
 - 29942 h_{97}{} + 23640 h_{99}{}\nonumber\\
&+ 29940 h_{101}{}
 + 21540 h_{102}{}
 - 81252 h_{103}{}
 - 126432 h_{104}{} - 5 \Big(11556 h_{105}{}
 + 810 h_{144}{}
 + 405 h_{145}{}
 - 1350 h_{146}{}\nonumber\\
 &
 - 810 h_{147}{}
 - 675 h_{148}{}
 - 810 h_{149}{}
 + 1305 h_{151}{}
 + 1575 h_{152}{}
 - 1125 h_{153}{}
 - 315 h_{154}{}
 + 4 \bigl(945 h_{158}{}- 15 h_{171}{}
 \nonumber\\
&- 45 h_{174}{}- 83 h_{178}{}
 - 125 h_{179}{}
 + 55 h_{180}{}
 + 91 h_{181}{}
 + 2 h_{182}{} + 70 h_{184}{}
 - 7 h_{185}{}
 + 25 h_{186}{}
 + 39 h_{187}{}
 - 3 h_{188}{}
\nonumber\\
&  - 98 h_{189}{}- 28 h_{190}{} + 43 h_{191}{}
 - 13 h_{192}{}
 - 6 (h_{193}{}
 + 8 h_{194}{})\bigr)\Big)\Big\}\Biggr],\\
h_{150}{}={}&\frac{1}{1350} \biggl[-1200 d_{1}{}
 + 1900 h_{5}{}
 + 3800 h_{6}{}
 + 475 h_{9}{}
 + 3 \Big\{65 h_{23}{}
 - 1375 h_{25}{} - 900 h_{28}{}
 - 130 h_{41}{}
 - 130 h_{42}{}
 + 1080 h_{43}{}\nonumber\\
&
 - 150 h_{44}{}
 - 150 h_{45}{}
 + 468 h_{53}{} - 468 h_{55}{}
 + 468 h_{56}{}
 + 468 h_{57}{}
 + 1404 h_{59}{}
 - 1404 h_{61}{}
 + 15327 h_{65}{}- 7865 h_{66}{}\nonumber\\
& 
 + 312 h_{91}{}
 + 312 h_{92}{}
 + 312 h_{94}{}
 + 208 h_{95}{}
 - 299 h_{96}{}
 + 663 h_{97}{}- 585 h_{99}{}
 - 585 h_{101}{}
 - 585 h_{102}{}+ 2028 h_{103}{}\nonumber\\
&  
 + 3198 h_{104}{}
 + 5 \Big(234 h_{105}{} - 90 h_{144}{}
 - 45 h_{145}{}
 + 90 h_{146}{}
 + 45 h_{147}{}
 + 45 h_{148}{}
 - 45 h_{149}{}
 + 2 (45 h_{151}{}+ 45 h_{153}{}\nonumber\\
& 
 + h_{178}{}
 - 5 h_{179}{}
 - 5 h_{180}{}
 - 2 h_{181}{}
 - 34 h_{182}{}
 - 5 h_{184}{}
 -  h_{185}{} + 10 h_{186}{}
 + 27 h_{187}{}
 + 6 h_{188}{}
 + 16 h_{189}{}
 + 26 h_{190}{}
\nonumber\\
& - 26 h_{191}{}
 - 34 h_{192}{} - 18 h_{193}{}
 - 24 h_{194}{})\Big)\Big\}
 + 6318 a_{2}{} (N_{3}{}-1
 )
 + 3159 a_{6}{} (N_{3}{}-1
 )\biggr],\\
h_{157}{}={}&\frac{1}{18900} \Big[  972 a_{2}{} (N_{3}{}-1
 )
 + 486 a_{6}{} ( N_{3}{}-1 )+75 d_{1}{}
 + 5600 h_{5}{}
 + 11200 h_{6}{}
 + 1400 h_{9}{}
 + 5655 h_{23}{}
 - 18825 h_{25}{}\nonumber\\
& + 6 \Big\{-31500 h_{28}{}
 - 1885 h_{41}{}
 - 1885 h_{42}{}
 + 3285 h_{43}{}
 - 2550 h_{44}{}
 - 2550 h_{45}{}+ 36 h_{53}{}
 - 36 h_{55}{}
 + 36 h_{56}{} + 36 h_{57}{}\nonumber\\
&  + 108 h_{59}{}
 - 108 h_{61}{}
 + 1179 h_{65}{} - 605 h_{66}{}
 + 24 h_{91}{}
 + 24 h_{92}{}
 + 24 h_{94}{}
 + 16 h_{95}{}
 - 23 h_{96}{}
 + 51 h_{97}{}
 - 45 h_{99}{}\nonumber\\
& - 45 h_{101}{}
 - 45 h_{102}{}
 + 156 h_{103}{}
 + 246 h_{104}{}
 + 5 (18 h_{105}{}
 - 180 h_{152}{} - 180 h_{154}{}
 + 630 h_{158}{}
 + 74 h_{178}{}
 + 80 h_{179}{}\nonumber\\
& - 70 h_{180}{}
 - 28 h_{181}{}
 + 4 h_{182}{} - 100 h_{184}{}
 + 76 h_{185}{}
 + 110 h_{186}{}
 - 72 h_{187}{}
 - 126 h_{188}{}
 + 59 h_{189}{}
 + 49 h_{190}{}\nonumber\\
& - 64 h_{191}{}
 - 56 h_{192}{}
 + 63 h_{193}{}
 + 69 h_{194}{})\Big\}\Big],\\
h_{173}{}={}&  \frac{594}{25} a_{2}{} (N_{3}{}-1
 )
 + \frac{297}{25} a_{6}{} ( N_{3}{}-1)- \frac{55}{24} d_{1}{}
 + \frac{32}{9} h_{5}{}
 + \frac{64}{9} h_{6}{}
 + \frac{8}{9} h_{9}{}
 + \frac{1}{300} \Big[-80 h_{23}{}
 - 2000 h_{25}{}
 - 5400 h_{28}{} + 160 h_{41}{}\nonumber\\
&
 + 160 h_{42}{}
 + 1440 h_{43}{}
 + 1584 h_{53}{}
 - 1584 h_{55}{}
 + 1584 h_{56}{}+ 1584 h_{57}{}
 + 4752 h_{59}{}
 - 4752 h_{61}{}
 + 51876 h_{65}{}
 - 26620 h_{66}{}
\nonumber\\
& + 1056 h_{91}{} + 1056 h_{92}{}
 + 1056 h_{94}{}
 + 704 h_{95}{}
 - 1012 h_{96}{}
 + 2244 h_{97}{}
 - 1980 h_{99}{} - 1980 h_{101}{}
 - 1980 h_{102}{}
 + 6864 h_{103}{}\nonumber\\
&
 + 10824 h_{104}{}
 + 5 \Big\{792 h_{105}{}
 + 630 h_{151}{}+ 630 h_{152}{}
 + 630 h_{153}{}
 + 630 h_{154}{}
 + 45 h_{169}{}
 - 15 h_{171}{}
 + 210 h_{174}{}\nonumber\\
& - 8 (43 h_{178}{}
 + 55 h_{179}{}
 - 35 h_{180}{}
 - 26 h_{181}{}
 + 38 h_{182}{}
 - 65 h_{184}{}
 + 47 h_{185}{}+ 25 h_{186}{}
 - 69 h_{187}{}
 - 57 h_{188}{}
 + 28 h_{189}{}
 \nonumber\\
& + 8 h_{190}{}
 - 38 h_{191}{}
 - 22 h_{192}{}+ 66 h_{193}{}
 + 78 h_{194}{})\Big\}\Big],\\
h_{205}{}={}&\frac{1}{12600} \Big[- 60426 \bigl(2 a_{2}{}+ a_{6}{}\bigr) N_{3}{}+110052 a_{2}{}
 + 55026 a_{6}{}
 - 1650 d_{1}{}
 - 5600 h_{5}{}
 - 11200 h_{6}{}
 - 1400 h_{9}{}\nonumber\\
& - 18255 h_{23}{}
 + 21975 h_{25}{}
 + 6 \Big\{31500 h_{28}{}
 + 6085 h_{41}{}
 + 6085 h_{42}{}
 - 7485 h_{43}{}+ 2550 h_{44}{}
 + 2550 h_{45}{}
 - 3276 h_{53}{}\nonumber\\
&  + 3276 h_{55}{}
 - 3276 h_{56}{}
 - 3276 h_{57}{}- 9828 h_{59}{}
 + 9828 h_{61}{}
 - 107289 h_{65}{}
 + 55055 h_{66}{}
 - 2184 h_{91}{}
 - 2184 h_{92}{}\nonumber\\
& - 2184 h_{94}{}
 - 1456 h_{95}{}
 + 2093 h_{96}{}
 - 4641 h_{97}{}
 + 4095 h_{99}{}
 + 4095 h_{101}{} + 4095 h_{102}{}
 - 14196 h_{103}{}
 - 22386 h_{104}{}\nonumber\\
&
 - 5\Big(1638 h_{105}{}
 - 180 h_{152}{} - 180 h_{154}{}
 + 1260 h_{158}{}
 - 120 h_{174}{}
 + 74 h_{178}{}
 + 80 h_{179}{}
 - 70 h_{180}{}
 - 28 h_{181}{}\nonumber\\
& + 4 h_{182}{}
 - 100 h_{184}{}
 + 76 h_{185}{}
 + 110 h_{186}{}
 - 72 h_{187}{}
 - 126 h_{188}{}
 - 121 h_{189}{} - 131 h_{190}{}
 - 64 h_{191}{}\nonumber\\
&
 - 56 h_{192}{}
 + 243 h_{193}{}
 + 249 h_{194}{}\Big)\Big\}\Big].
\end{align}

\bibliographystyle{utphys}
\bibliography{references}

\end{document}